\documentclass[final,english]{phdulb}

\usepackage{myphdstyle}   
\usepackage{kantlipsum} 
\usepackage{natbib}
\usepackage{epigraph} 
\usepackage{pdflscape}
\setsecnumdepth{subsubsection}
\makeevenhead{headings}{\rightmark}{}{}
\makeoddhead{headings}{}{}{\leftmark}

\makeevenfoot{headings}{\thepage}{}{}
\makeoddfoot{headings}{}{}{\thepage}

\definecolor{bookColor}{cmyk}{0,0,0,1}
\color{bookColor}

\begin{document}
	
	\includepdf[pages=-]{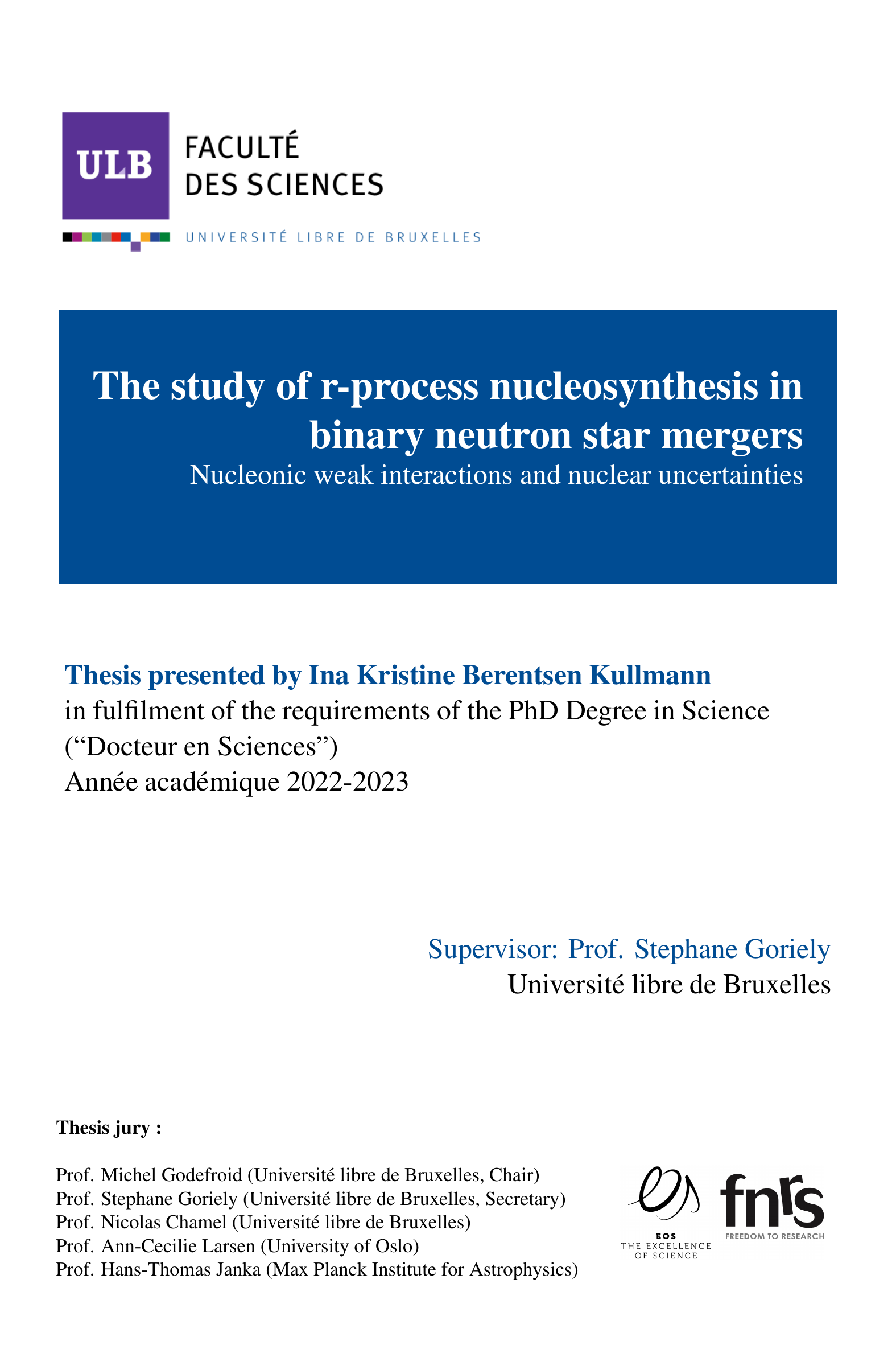}
	
 	\frontmatter        

    \thispagestyle{empty}
\vspace*{\stretch{1}}
\begin{flushright}
    \emph{Til min kj{\ae}re Trond}
\end{flushright}
\vspace*{\stretch{3}}

\chapter*{Abstract}

A long-standing scientific puzzle has been to explain the origin of the heaviest elements in the Universe and, more particularly, the production of the elements heavier than iron up to uranium. The rapid neutron capture process (or r-process) is known to synthesize about 50\% of these heavy elements and the long-lived actinides observed in our solar system and so-called metal-poor r-process-enhanced stars. This thesis aims to study the r-process nucleosynthesis and some of the uncertainties that still govern our predictions, namely, nuclear yields and heating rates. 

Our focus will be on the r-process in neutron star (NS) mergers, which is in the spotlight after recent ``multi-messenger'' observations, including the combined detection of gravitational waves from a NS-NS merger event and its subsequent electromagnetic counterpart. In this work, we base our nucleosynthesis calculations on hydrodynamical simulations of NS merger systems, which estimate the amount of ejected mass and its properties during ejection. Through our r-process calculations, we estimate the composition of this gravitationally unbound material, which can contribute to the r-process enrichment of the Galaxy.

This work is divided into two studies, each addressing a remaining open question regarding the r-process nucleosynthesis.
First, a coherent study of the impact of neutrino interactions on the r-process element nucleosynthesis and the heating rate produced by the radioactive decay of nuclei synthesized in the dynamical ejecta of NS-NS mergers is presented. 
We have studied the material ejected from four NS-NS merger systems based on hydrodynamical simulations, which handle neutrino effects in an elaborate way by including neutrino equilibration with matter in optically thick regions and re-absorption in optically thin regions. 
Second, we present an in-depth study of the nuclear physics uncertainties that still affect the r-process results by systematically and coherently varying theoretical nuclear input models that describe the experimentally unknown neutron-rich nuclei. 
This includes two frameworks for calculating the radiative neutron capture rates and 14 different models for nuclear masses, $\beta$-decay rates and fission properties.
Our r-process calculations are based on detailed hydrodynamical simulations of dynamically ejected material from NS-NS or NS-black hole (BH) binary mergers plus the secular ejecta from BH-torus systems.  

In the first study, we find that the neutron richness of the dynamical ejecta is significantly affected by the neutrinos emitted at the time of the coalescence, in particular when compared to a case neglecting all neutrino interactions. Our nucleosynthesis results show that a solar-like distribution of r-process elements with mass numbers $A \ga 90$ is produced, including a significant enrichment in Sr and a reduced production of actinides compared to simulations without including the nucleonic weak processes.
The composition of the dynamically ejected matter and the corresponding rate of radioactive decay heating are found to be rather independent of the system mass asymmetry and the adopted equation of state. This approximate degeneracy in abundance pattern and heating rates can be favourable for extracting the ejecta properties from kilonova observations, at least if the dynamical component dominates the overall ejecta. 

The impact of nuclear uncertainties on the r-process abundance distribution and the early radioactive heating rate is found to be modest (within a factor of $\sim 20$ for individual $A>90$ abundances and a factor of 2 for the heating rate). However, the impact on the late-time heating rate is more significant and depends strongly on the contribution from fission. 
We witness significantly larger sensitivity to the nuclear physics input if only a single trajectory is used compared to considering ensembles of a few hundred trajectories.
Our models for the total ejecta reproduce the solar system distribution well for $A>90$ nuclei and yield a significant amount of Th and U, irrespective of the adopted nuclear physics model.
We use the predicted Th/U ratio to estimate the cosmochronometric age of six metal-poor stars to set a lower limit of the age of the Galaxy and find the impact of the nuclear uncertainties to be up to 2 Gyr.

\chapter*{Résumé}

\noindent L’une des questions scientifiques de longue date concerne l’origine des éléments les plus lourds dans l’Univers, et plus particulièrement la production des éléments plus lourds que le fer jusqu’à l’uranium. Le processus de capture rapide de neutrons (ou processus r) est connu pour synthétiser près de 50\% de ces éléments lourds ansi que les actinides à longue durée de vie observés dans notre système solaire et dans les étoiles de faible métallicité enrichies en éléments r. Cette thèse a pour but l’étude de la nucléosynthèse par le processus r et certaines des incertitudes qui affectent encore nos prédictions des abondances nucléaires et les taux d’échauffement. Nous nous focaliserons sur le processus r lors de la coalescence d’étoiles à neutrons qui est sous les projecteurs après de récentes observations multi-messagers, incluant la detection combinée d’ondes gravitationnelles de la coalescence de deux étoiles à neutrons et de leur contrepartie électromagnétique. Dans ce travail, nous basons nos calculs nucléosynthétiques sur des simulations hydrodynamiques de systèmes d’étoiles à neutrons en coalescence qui permettent d’estimer la quantité de matière éjectée ainsi que leurs propriétés. Par nos calculs de processus r, nous estimons la composition de la matière non-liée gravitationnellement qui peut ainsi contribuer à l’enrichissement de la Galaxie par le processus r.

Ce travail est divisé en deux parties principales, chacune traitant une question encore ouverte concernant la nucléosynthèse par le processus r. Premièrement, une étude cohérente est présentée sur l’impact des interactions neutriniques sur la nucléosynthèse des éléments r et du taux d’échauffement par décroissance radioactive des noyaux synthétisés dans l’éjecta dynamique de deux étoiles à neutrons en coalescence. Nous avons étudié la matière ejectée à partir de quatre systèmes de fusion d’étoiles à neutrons sur base de simulations hydrodynamiques qui traitent les effets neutriniques de façon élaborée en y incluant la mise à l’équilibre avec la matière dans les regions optiquement épaisses et la ré-absorption dans les regions optiquement fines. Deuxièmement, nous présentons une étude approfondie des incertitudes nucléaires qui affectent encore les calculs de processus r, ceci en variant de façon systématique et cohérente les modèles théoriques nucléaires qui décrivent les noyaux riches en neutrons inconnus expérimentalement. Ceci inclut deux approches pour le calcul des taux de capture radiative de neutrons et 14 modèles différents pour les masses nucléaires, les taux de désintégration beta et les propriétés de fission. Nos calculs de processus r sont basés sur des simulations hydrodynamiques détaillées de la matière éjectée dynamiquement lors de la coalescence de deux étoiles à neutrons ou d’une étoile à neutron et d’un trou noir, ainsi que celle éjectée après formation du système trou noir et tore.

Dans la première étude, nous trouvons que l’enrichissement en neutrons de l’éjecta dynamique est affecté de manière significative par les neutrinos émis lors de la coalescence, en particulier en comparaison avec un cas où les interactions neutriniques sont négligées. Nos résultats nucléosynthétiques montrent qu’une distribution proche de celle du soleil est obtenue pour les éléments r avec un nombre de masse A>90, incluant un enrichissement important en Sr et une production réduite d’actinides, ceci en comparaison aux simulations négligeant les interactions faibles entre les nucléons. La composition de la matière éjectée dynamiquement et le taux correspondant d’échauffement radioactif restent relativement independants de l’asymétrie en masse du système considéré et de l’équation d’état adoptée. Une telle dégénérescence approximative sur les distributions d’abondance et les taux d'échauffement peuvent être favorables à l’extraction des propriétés de l’éjecta à partir d’observations de kilonovae, du moins si la composante dynamique domine l’éjecta dans son ensemble.

L'impact des incertitudes nucléaires sur la distribution des abondances du processus r et le taux d’échauffement radioactif dans les premiers instants s’avère être modeste (d'un facteur 20 environ pour certaines abondances individuelles pour des noyaux à A>90 et d'un facteur 2 pour les taux d’échauffement). En revanche, l’impact sur les taux d’échauffement tardifs est plus significatif et dépend fortement de la contribution de la fission.  Nous trouvons une sensibilité significativement plus importante aux données nucléaires si une seule trajectoire est considérée au lieu d’un ensemble de plusieurs centaines de trajectoires. Nos modèles combinés de l’éjecta reproduisent bien la distribution solaire pour les noyaux à A>90 et produisent une quantité significative de Th et U, indépendamment du modèle nucléaire adopté. Nous utilisons les prédictions du rapport Th/U pour estimer par cosmochronométrie l’âge de six étoiles pauvres en métaux afin de poser une limite inférieure sur l’âge de la Galaxie et trouvons un impact des incertitudes nucléaires limité à 2 milliards d'années.






\chapter*{Acknowledgements}

Firstly, I would like to express my sincere gratitude to my supervisor Prof. Stephane Goriely for his invaluable advice, knowledge and support during my PhD. Without your project proposal during the conference in Longyearbyen, Svalbard, I would not have ended up here finalizing a PhD in my dream topic. Besides my advisor, I would like to thank the remainder of my thesis committee: Prof. Michel Godefroid and Prof. Sophie Van Eck, for their insightful comments and encouragement and for always asking me how I am doing. I would also like to extend my gratitude to the jury members not mentioned above, including Prof. Nicolas Chamel, Prof. Ann-Cecilie Larsen and Prof. Hans-Thomas Janka, for accepting to read and evaluate my dissertation. I also thank Dr Oliver Just and Dr Andreas Bauswein for our fruitful collaboration and Prof. Lionel Siess for the helpful chats about scientific programming, his suggestions and Linux assistance. My gratitude extends to the Fonds de la Recherche Scientifique (FNRS, Belgium) for the funding to undertake my studies at Université Libre de Bruxelles. 

As for the rest of my colleagues at Institut d'Astronomie et d'Astrophysique, Nancy, I am thankful for your help, particularly during my settlement in Brussels, doing paperwork, going to the bank and lending me your bike. Guillaume and Takako, I am grateful for our (french) lunches and day trips exploring new places; I hope we can see you and little Thomas again soon! A special thanks goes to my first office mate, Shreeya; I missed your smiley face and our break-time chats when you weren't working in the office and during the last two years. Thanks to the ``post-pandemic group'' for our lunchtime discussions, afternoon tea or coffee breaks, the after-work beers and countless move-in parties. I am grateful to Wouter and Nikolai for reading my draft. To my current office mates: I appreciate the good atmosphere we have created, and Pawel, thanks for bringing the (quite large) sofa; it has helped me get through some tough days. Finally, Lysandra, my little french friend, I will miss our small breaks, chats, and our shared love for exploring and sitting outside. 

When I moved here, I didn't dare to imagine my Brussels life would become so rich and filled with such great people and activities. To the UBHC handball team and Sofie: I dearly wish I had found the club sooner; your inclusive atmosphere made me instantly feel welcome. To all BigM'ers, I have appreciated all our MTB rides in Foret de Soignes and elsewhere in Belgium. Chris, David and Judit, thanks for always offering your help (or bikes!) to get back into riding; you are a super group of people and always up for some excellent post-ride Belgian beers. To my new friend Ina, I hope we can continue to be squared. Silje \& Zoran, you are my extended family; I appreciate the meals you cook, our VR nights and the trips we took together (and Silje, the lock-down walks). Ana, my Brussels-best friend, knuffelcontact, hiking partner, and coffee friend, I am thankful for our moments in the once ``too-quiet'' times and our in-between pauses in the current busy life. To my Norwegian friends Karoline, Kine, Astrid \& Alex, Morten \& Silje, it meant a lot to me that you came to visit. Likewise, I am grateful to Ellen \& Ponzie for the fun times (I'll find you soon!) and Kristine \& S. Now for unforgettable mountain adventures.

I would like to express my gratitude to Trond's family for allowing me to stay with them during the pandemic and holidays. You always made me feel welcome and cooked wonderful meals so I could focus on the thesis. My dear family, including uncles and cousins, thanks for supporting and encouraging me during my studies. Mum, dad (\& Turid!) and Carl-Erik, thanks for always being there for me whenever I needed you, in particular during the difficulties of the pandemic. My niece and nephew Celine and Adrian, you have grown up so fast; I hope to see you more with a shorter travel time. Finally, my dear twin Annelise, our almost daily talks make me happy and keep me sane; I look forward to more face-to-face time. 

Last but not least, Trond, to which this thesis is dedicated, thanks for your continuous love, support and everlasting patience. Without your wish, together with mine, to go abroad and sometimes invisible help, this PhD would not have happened. With you by my side, it has been an amazing experience.

\vskip\onelineskip
\begin{flushleft}
    \textbf{Ina K. B. Kullmann}
    \\
    Brussels, October 2022
\end{flushleft}





    \cleartorecto
    \tableofcontents    
    \cleartorecto
    \listoffigures      
    \cleartorecto
    \listoftables       

    \mainmatter         


\chapter{Introduction}
\label{ch:intro}

\epigraph{\itshape ``Sit back and allow the words to wash around you, like music.''}{--- \textup{Roald Dahl}, Matilda}

\todo{[all text from paper 2 (gray command ptwo) updated according to changes before 19.july (however fig captions might not be updated??) --> but now also text are updated here but not in paper.....]}

Scientific research is driven forward by human curiosity concerning our place in the Universe and the desire to better explain and understand the world around us. 
Although profound advances have been made in the previous centuries and years, our knowledge is far from complete, and many questions remain unresolved.
This thesis is motivated by one of the remaining open questions \citep{arnould2007,arnould2020,cowan2021}, which can be phrased as,
what is the origin of the chemical elements that our Universe, including the earth, plants, animals and humans, consists of? 
Both astrophysics and nuclear physics knowledge is required to answer this question, giving rise to the multidisciplinary field of nuclear astrophysics. In this field, scientists work on astronomical observations, nuclear physics experimental data, and theoretical modelling 
to explain how the elements were synthesized, also referred to as nucleosynthesis. 
This thesis will contribute to the latter by performing nucleosynthesis simulations.

\begin{figure}[tbp]
\begin{center}
\includegraphics[width=\textwidth]{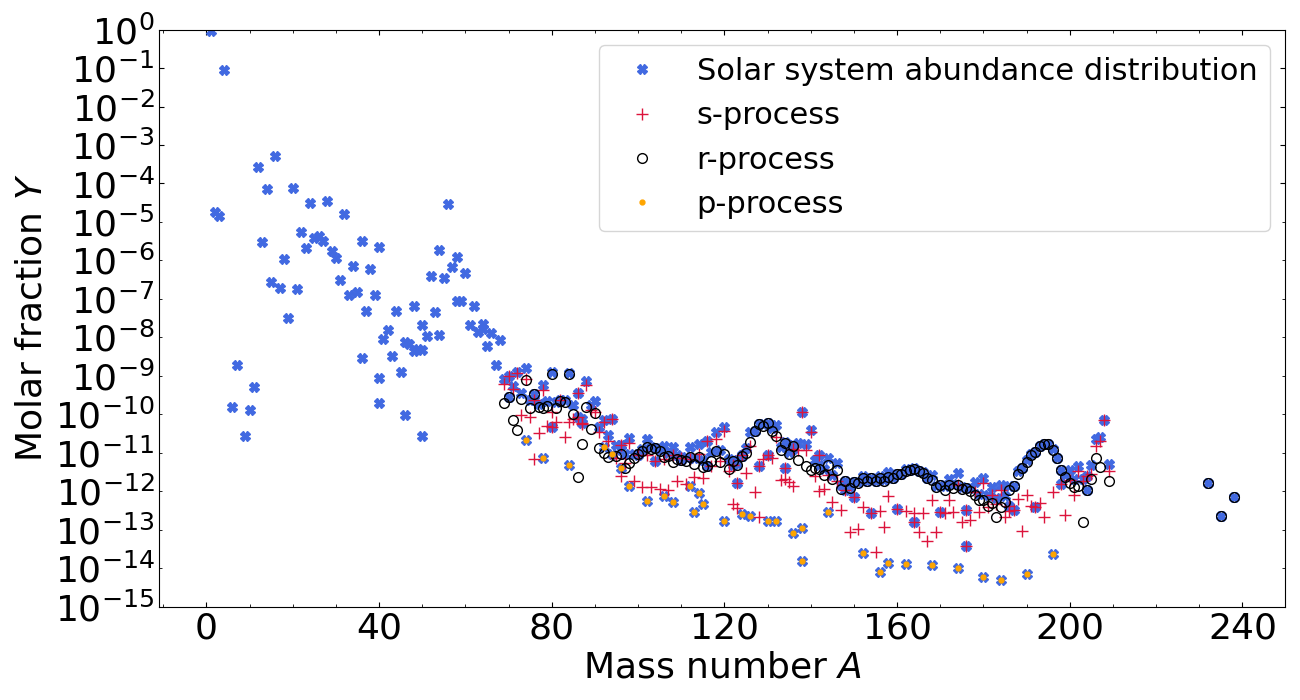}
\caption[Solar system isotopic abundance distribution.]{Solar system isotopic abundance distribution from \citet{lodders2003} decomposed by \citet[][see also \citet{goriely1999}]{arnould2007} into the contribution from the s-, r- and p-processes. 
\todo{make larger dots, tight layout? fontsize legend?}
}
\label{fig_solar_srp}
\end{center}
\end{figure}

\todo{Consider adding this ref for the age of the earth: https://www.lyellcollection.org/doi/abs/10.1144/gsl.sp.2001.190.01.14}

Chemical elements (also called elements) are the basic building blocks of all matter. 
They are defined by the number of protons in the nucleus (often referred to as atomic number) and cannot be decomposed into simpler substances by ordinary chemical reactions.
What we regard as heavy elements in this work will usually be elements with a proton number larger than iron (which has 26 protons in its nucleus).
When discussing the abundance of a nucleus $i$ with $Z$ protons and $N$ neutrons, we often refer to either the molar fraction $Y_i$ or the mass fraction $X_i$, which is related to the molar fraction through
\begin{equation}
X_i=A_iY_i,
\end{equation}
where $A=Z+N$ is the mass number of the nucleus and $\sum X_i\equiv 1$. The molar fraction can be defined in terms of the number density:
\begin{equation}
n_i = \frac{\rho Y_i}{m_u},
\label{eq:nrdens}
\end{equation}
where $\rho$ is the density of the medium and $m_u$ is the atomic mass unit. 
About 94 naturally\footnote{In addition, about 24 elements have been synthesized artificially in labs.} occurring elements are observed in different amounts on earth and in the solar system (see \cref{fig_solar_srp}).
The present solar system abundance distribution is thought to be representative of the composition at the time the solar system formed some 4.6~Gyr ago, assuming that chemical and geological processes have only modified the distribution in a minor way. 
The elemental abundances are obtained from solar spectra, and the isotopic abundances are from primitive meteorites and terrestrial values \citep{lodders2009}. 
The solar system abundance distribution is also used as a reference when comparing abundances derived from stars within or outside our Galaxy. 
The distribution of our solar system is assumed to be representative of the Galactic abundances, though many variations in the stellar surface abundances are observed \citep{arnould2020}.

\begin{figure}[tbp]
\begin{center}
\includegraphics[width=\textwidth]{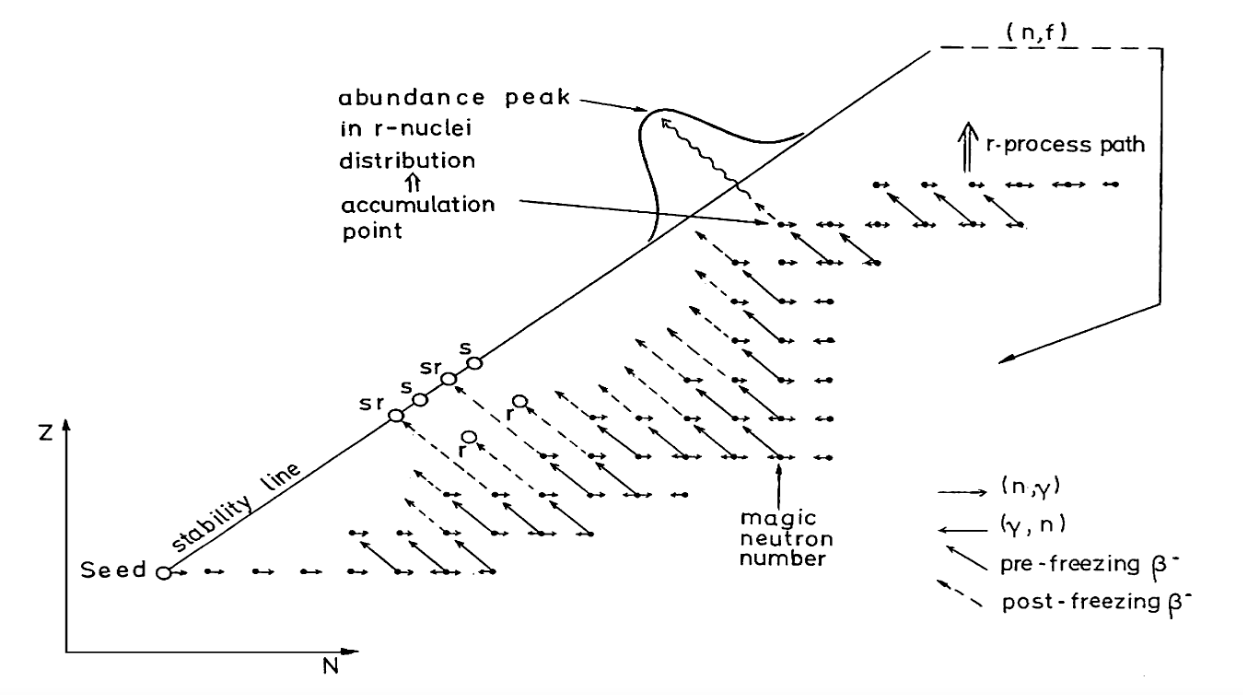}
\caption[Illustration of the r-process flow.]{
An illustration of the r-process flow in the $(N,Z)$-plane adapted from \citet{arnould2007}. See the text for more details. 
}
\label{fig_sch_rpro}
\end{center}
\end{figure}

As can be seen in \cref{fig_solar_srp}, the lightest elements have the largest abundances, in particular hydrogen and helium, which were synthesized in the first few seconds after the Big Bang (often called Big Bang nucleosynthesis (BBN)\footnote{In addition to helium and hydrogen, trace amounts of lithium were also produced in the BBN.}).
The remaining elements (except Li, Be and B) up to and including the iron peak around $A\sim 60$ were produced in different stellar burning phases. 
As initially proposed by \citet{burbidge1957,cameron1957}, the production of the nuclides heavier than iron can mainly be attributed to two neutron-capture processes: the slow and rapid neutron-capture processes referred to as s- and r-process, respectively \citep{arnould2007,arnould2020,cowan2021,Siegel2022}.
The focus of this work will be on r-process simulations and understanding how the heaviest elements were synthesized in nature. When discussing the r-process yields from a given simulation, results are often compared to the so-called solar r-process abundance distribution (see \cref{fig_solar_srp}). This distribution is derived from the solar system abundance distribution by subtracting the (model dependent) yields from the s-process \citep[i.e., see][]{goriely1999}. 
In \cref{fig_solar_srp}, we can see that the s- and r-process yields together sum up to the total solar abundance distribution and each account for the production of about 50\% of the heaviest elements. 
In addition to the s- and r-processes, there are a few  p-nuclei (or p-process nuclei) which lie on the proton-rich side of the valley of stability (i.e., to the left of the line illustrated in \cref{fig_sch_rpro}). The p-nuclei account for less than $~\sim1$~\% of the s- and r-abundances in \cref{fig_solar_srp}, and are thought to be produced through the destruction of pre-existing s- or r-nuclei through $(\gamma,n)$, $(\gamma,p)$ or $(\gamma,\alpha)$ reactions \citep[e.g., see][for more details]{arnould2003,arnould2020}.
In recent years the intermediate process, or i-process, has gained much attention since it can explain 
abundances observed in some metal-poor stars that could, otherwise, only be explained by a combination of the s- and r-processes \citep[e.g., see][and references therein]{arnould2020,Choplin2021}. 

For decades there existed no direct evidence of where in the Universe the r-process could occur \citep{arnould2007,cowan2021}. The earliest r-process studies used the so-called canonical model \citep[e.g.,][and references within]{goriely1996} where $^{56}$Fe was usually taken as the starting point (often referred to as a ``seed'' nucleus) for the r-process (see \cref{sec:nuc_net} for details about the approach used in this work).
A simplified r-process flow in the $(N,Z)$ plane is illustrated in \cref{fig_sch_rpro}. 
Assuming a high neutron density in the environment, a sequence of $(n,\gamma)$ reactions on the seed nucleus will move the abundance towards the neutron-rich side of the nuclear chart since the radiative neutron capture rates (right arrows) are faster than the $\beta$-decay rates (solid, diagonal arrows).
With increasing neutron excess or equivalently lower neutron separation energies, the rates for the inverse photo-disintegrations $(\gamma,n)$ increase (left arrows), limiting the flow towards even more neutron-rich nuclei. 
Eventually, the nuclear flow proceeds to higher $Z$ elements through $\beta$-decays.

The proton ($p$) or neutron ($n$) separation energy for a nucleus with $Z$ protons and $N$ neutrons with mass $m(A,Z)$ is defined as:
\begin{equation}
\begin{aligned}
S_p &= m(A-1,Z-1) + m_p - m(A,Z),\\
S_n &= m(A-1,Z) + m_n - m(A,Z).
\end{aligned}
\end{equation}
Here, $m_x$ is the mass of either the proton or neutron $x\in [p,n]$ and $m(A-1,Z-1)$ or $m(A-1,Z)$ is the mass of the nucleus with one less nucleon.
When scanning the nuclear chart along constant proton or neutron numbers, discontinuities in the proton or neutron separation energies are found at proton or neutron numbers equal to 2, 8, 20, 28, 50, 82, and 126. 
Nuclei at these so-called magic numbers are thus more stable against neutron capture or photoneutron emission then the neighbouring nuclei. 
The discontinuities in the proton or neutron separation energies can be attributed to the closing of shells in the nucleus, analogous to atomic orbitals in atomic physics.
What is referred to as magic nuclei are those nuclei that either have a neutron or proton number corresponding to a magic number, and nuclei that are magic in both neutron and proton numbers are called doubly magic. 
These nuclear magic numbers are known \citep{burbidge1957,cameron1957} to give their imprint on the solar r-process abundance distribution and lead to the r-process peaks visible in \cref{fig_solar_srp}. 
The low $S_n$ values beyond the neutron magic numbers hinder the flow towards more neutron-rich species so that $\beta$-decays move the abundance to larger $Z$ at practically constant $N$ (see \cref{fig_sch_rpro}).  As the material gets closer to the valley of stability, the $\beta$-decays generally become slower, and consequently, the abundance accumulates at nuclei with a magic neutron number.
Closer to the stability line, the $S_n$ value for the magic nuclei also increases. At one point, the $(n,\gamma)$ reactions become faster than 
the $(\gamma,n)$ photo-disintegrations so that the abundance once again can flow towards the more neutron-rich nuclei.  
Later, when the neutron density falls off, often referred to as freeze-out,  the material decays back to the valley of stability (dotted diagonal lines in \cref{fig_sch_rpro}). 
The previous accumulation of material at neutron closed shells $N=50$, 82 and 126 leads to the formation of the peaks in the r-process abundance distribution at $A\sim 80$, 130 and 195, respectively. 
\cref{fig_sch_rpro} also shows how the abundance of some nuclei can originate only from the s-process when a stable r-only nucleus shields the s-only isobar from decays of r-process material.  

In contrast to the r-process, the s-process nuclear flow remains within the valley of $\beta$-stability and terminates at $^{209}$Bi. Therefore, only the r-process can produce the long-lived actinides such as Th and U and reach the $\alpha$- and fission-unstable nuclei above the Pb-region.
\todo{can i-process reach there?? if so add footnote?} 
If the r-process enters the trans-Pb region, which includes fission-unstable nuclei, neutron-induced, spontaneous and $\beta$-delayed fissions recycle material back to the $A\sim 140$ region as illustrated by the top right arrow in \cref{fig_sch_rpro}.
When the conditions allow for the fission fragments to start capturing neutrons again, i.e., when a large number of free neutrons are available before freeze-out, the abundance can once again flow into the trans-Pb region, which starts another ``cycle'' of fissioning. This process is therefore referred to as ``fission recycling'' \citep{goriely2011,goriely2013, goriely2015,goriely2015b,Mendoza-Temis2015,Vassh2019}. 
In most cases, when $\alpha$-unstable species are produced, the nuclear flow follows a sequence of $\alpha$-decays until nuclei in the Pb region are reached (unless an $\alpha$-decay daughter is unstable with respect to fission or $\beta$-decay).

The illustration in \cref{fig_sch_rpro} only gives an overview of the nuclear mechanisms at stake during the r-process. In order to study the details of the r-process, the time-dependent ordinary differential equations (ODE) of the abundance flow have to be solved numerically, as described in \cref{sec:nuc_net}.  
The set of these coupled equations constitutes a ``reaction network''. 
The network includes production and destruction terms on all relevant nuclei for all reactions considered, i.e., charged-particle fusion reactions, radiative neutron captures, photodisintegrations, $\beta$- and $\alpha$-decay, fission and so on.
Several thousand neutron-rich nuclei are involved in the r-process, which makes solving the ODEs a computationally demanding task (see \cref{sec:HPC}). 
The abundances themselves typically range over several orders of magnitude.
In addition, nuclear reaction time scales vary by orders of magnitude, making the systems of equations ``stiff'', which poses further numerical challenges \citep{Hix1999,Timmes1999}. 

\begin{figure}[tbp]
\begin{center}
\includegraphics[width=0.99\textwidth]{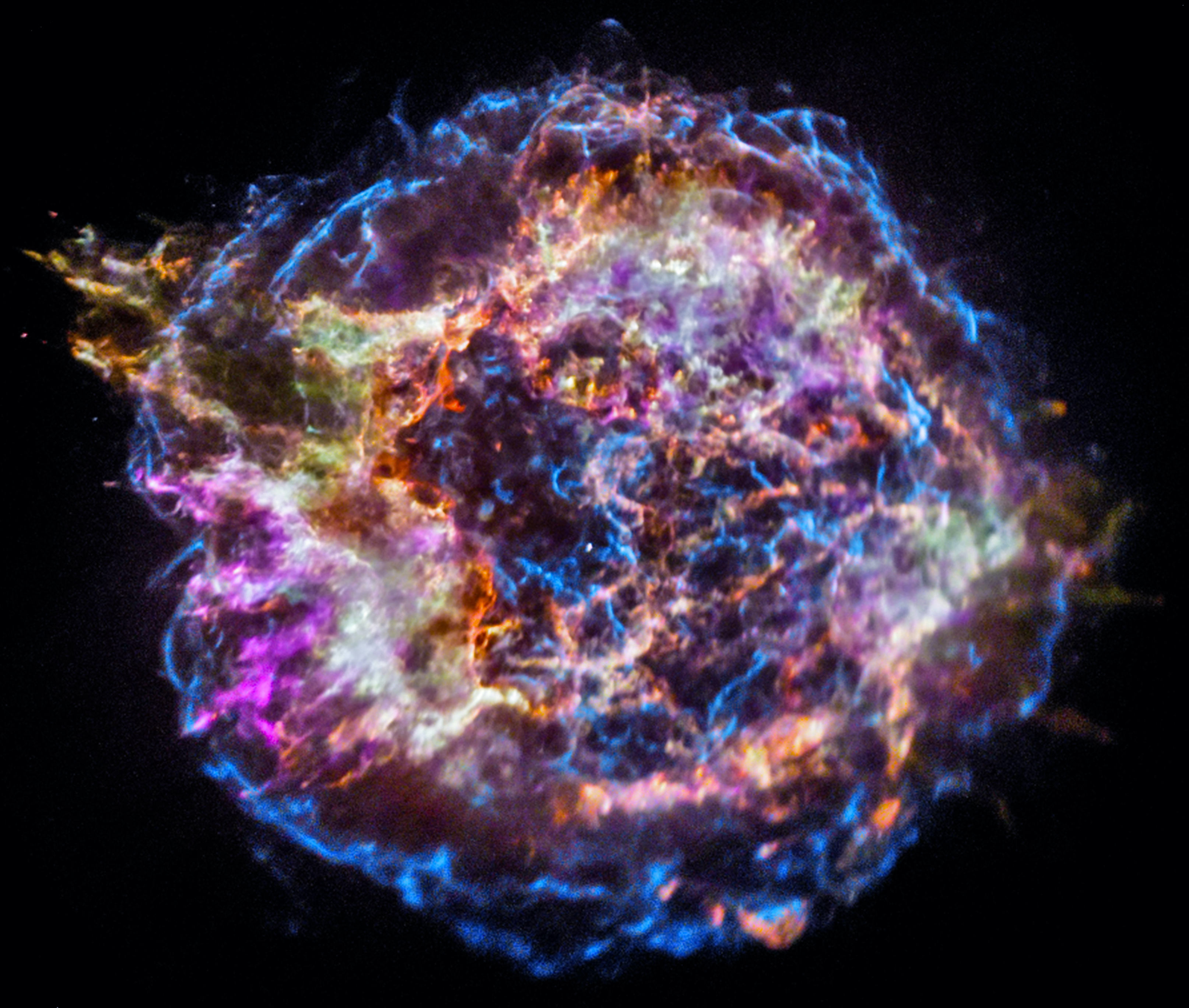}
\includegraphics[width=0.99\textwidth]{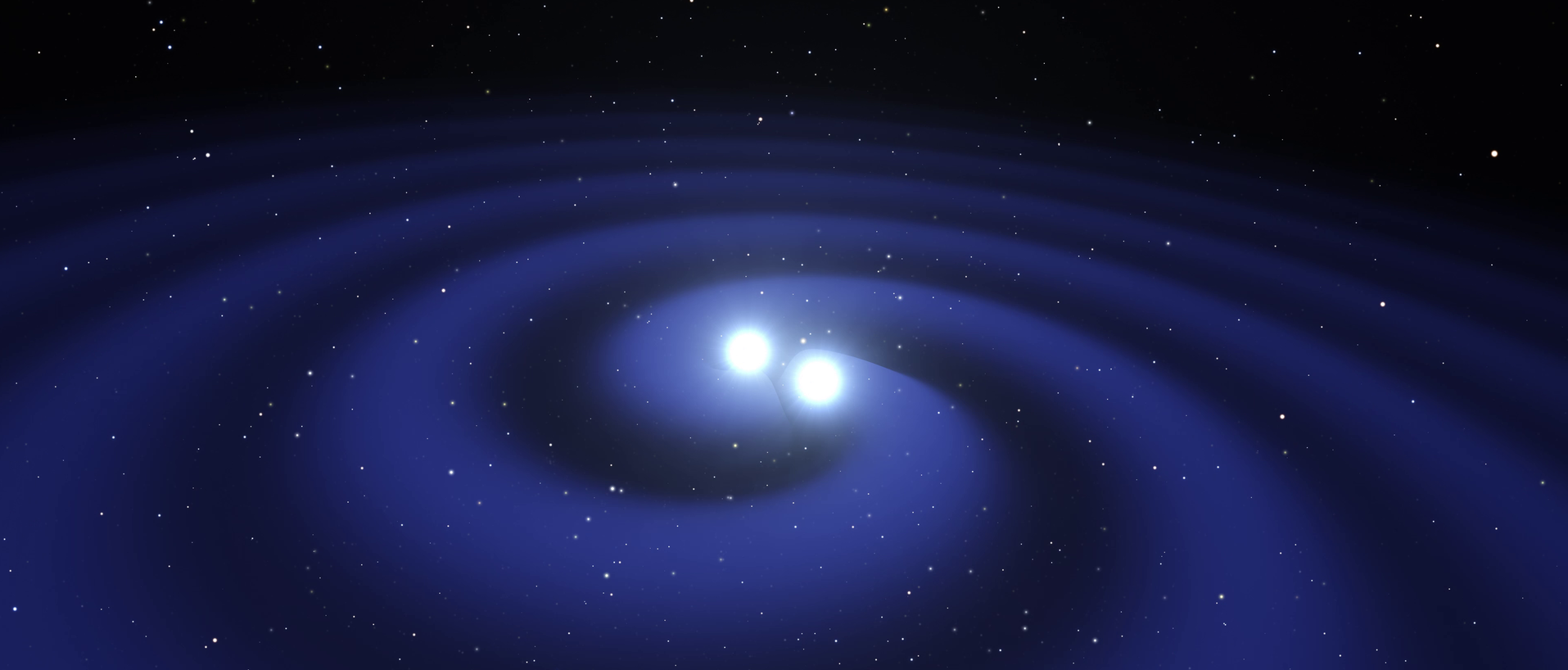}
\caption[The Cassiopeia SN remnant and NS-NS coalescence.]{
Top: The SN remnant of a massive star that exploded about 300 years ago called Cassiopeia A observed by \citet{SN_remnant}. The small white dot in the centre comes from a point-like X-ray source which is presumed to be a NS surrounded by an expanding shell of hot gas produced by the explosion. 
Bottom: An artist's impression of two NSs spiralling towards each other, adapted from \citet{NSmerger}.
}
\label{fig_SNremnant}
\end{center}
\end{figure}

Recently, our understanding of the astrophysical site for the r-process and the nucleosynthesis has improved drastically both in terms of observational data but also due to computational advancements of site-specific simulations \citep{arnould2020,cowan2021}.
Such simulations of the astrophysical environment are often referred to as hydrodynamical simulations, where the name originates from the fact that the astrophysical plasma is approximated as a fluid.
The astrophysical site(s) of the r-process have been debated since the initial proposal by \citet{burbidge1957} and \citet{cameron1957}. 
Until a decade ago, the so-called neutrino-driven wind launched during a core-collapse supernova (CCSN) \citep[e.g.,][]{takahashi1994,qian1996,hoffman1997,Arcones2007} was a very popular candidate for the r-process site.
However, until now, a successful r-process cannot be obtained \textit{ab initio} without tuning the relevant parameters (neutron excess, entropy, expansion timescale) in a way that is not supported by increasingly more sophisticated existing models \citep{witti1994,takahashi1994,roberts2010,wanajo2011, janka2012,Mirizzi2016,wanajo2018c}.
Although these scenarios still remain promising, especially in view of their potential to contribute to the Galactic chemical enrichment significantly, they remain affected by large uncertainties associated mainly with the still complex mechanism responsible for the supernova explosion and the persistent difficulties in obtaining suitable r-process conditions in self-consistent dynamical explosion and neutron-star (NS) cooling models \citep{janka2012,hudepohl2010,fischer2010,Janka2016,Muller2017, Bollig2017,Bolling2021,Pinedo2012}. 
Moreover, a subclass of core-collapse supernovae, the so-called jet-driven or jet-associated collapsars\footnote{These systems might also be the origin of observed long $\gamma$-ray bursts \citep{Woosley1993,macfadyen1999}.} corresponding to the death of rapidly rotating, massive stars, could represent another promising r-process site. In these systems, r-process viable conditions may be found in magnetically driven jets \citep{cameron2003,winteler2012,Nishimura2015,Mosta2018, Grimmett2020,Reichert2021,Reichert2022} or in the ejecta launched from the central black-hole (BH) accretion disk \citep{Cameron2001,Pruet2004, Surman2004,Siegel2019,Siegel2022, miller2019a,just2021,Just2022}.

The focus shifted towards binary NS mergers after hydrodynamical models managed to demonstrate that a significant amount of material ($10^{-3}-10^{-2}$~\Msun, where \Msun$\approx2\cdot 10^{30}$~kg refer to the solar mass.) can become unbound in the first mass ejection phase during the merging of two NSs \citep[e.g.,][]{Ruffert1997, Ruffert1999,Freiburghaus1999,rosswog1999, Ruffert2001, Janka2002, oechslin2007,goriely2011, hotokezaka2013, Bauswein2013,Just2015,radice2018b,foucart2016}. 
The original idea proposed by \citet{lattimer1974,lattimer1977} and \citet{meyer1989} is that decompressed neutron-rich material (which gives suitable conditions for the r-process) would be ejected when two binary NSs (or a NS and a BH) merge, polluting the interstellar medium with freshly synthesized r-process-rich material (see \cref{fig_SNremnant}). 
Later studies found that the neutron-rich conditions in the ejected material enable a robust r-process, producing a solar-like abundance distribution up to the third r-process peak and the actinides \citep[e.g.,][]{Freiburghaus1999, goriely2011, korobkin2012, wanajo2014, Just2015, radice2018b}.

The observational confirmation that r-process material is synthesized in NS-NS mergers came in 2017 with the first gravitational wave detection from a NS-NS merger, GW170817 \citep[e.g.,][]{abbott2017c} accompanied by an electromagnetic signal, AT2017gfo \citep{abbott2017d,kasen2017}. 
Such a signal in the optical electromagnetic spectrum in the aftermath of a NS merger event is often referred to as a ``kilonova'' \citep[e.g.,][]{li1998, roberts2011, metzger2010, goriely2011, barnes2013,Kulkarni2005,Tanaka2013} and is powered by the decay of freshly produced radioactive r-process elements.
Fits to the observed kilonova light curve provided a strong indication that r-process elements have been synthesized in this event \citep[e.g.,][]{kasen2017,tanaka2017a}. 
Recently, \citet{watson2019} also reported the spectroscopic identification of strontium, providing the first direct evidence that NS mergers eject material enriched in heavy elements. 
However, it is not observationally settled yet whether binary NS mergers can also produce the heaviest of the solar r-process elements such as uranium and whether they are the only r-process site, or if other sites contribute to the Galactic enrichment (and therefore also the enrichment of the solar system). 

Modelling kilonova light curves requires knowledge of the energy released by the newly synthesized radioactive species in the ejecta, i.e., the rate of heat released through $\beta$-decays, fission, and $\alpha$-decays. 
The heat generated by the various decay components thermalizes in the material with different efficiencies. The modelling of thermalization efficiencies often depends on the ejecta's total mass and average velocity in addition to the photon opacity.
Before they escape, kilonova photons will be absorbed and re-emitted by atomic transitions of the elements in the opaque inner region of the ejecta.
A major source of opacity is believed to be due to the presence of lanthanides ($Z = 57-71$) 
and actinides ($Z=90-100$) in the ejecta \citep[][and references within]{Mendoza-Temis2015}, which has thousands of possible transition lines between atomic energy levels, of which many are experimentally unknown. 
The most detailed opacity models are directly based on atomic-physics considerations \citep[e.g.,][]{Kasen2013,Wollaeger2018,tanaka2020}, while others use a more parametric approach \citep[e.g.,][]{just2021} or (piecewise) constant opacities \citep[e.g.,][]{perego2017,Villar2017,Rosswog2018,Hotokezaka2020}.
If lanthanides are present in the ejecta, the light curve luminosity peak might be delayed, 
and the peak frequency also shifted to the red \citep[][]{kasen2017}. 
For this reason, one of the most important input quantities for kilonova models is the lanthanide and actinide fraction of the ejecta, which r-process calculations can provide. 
Another detectable feature of the kilonova light curve is the decay of free neutrons left after the r-process finishes (a feature present in the fast ejecta), which releases additional heat that peaks hours after the merger in the light curve \citep{metzger2015,Mendoza-Temis2015,kullmann2021}.

In addition to explaining the solar r-process abundance, the r-process is expected to explain the observed abundance of so-called metal-poor r-process-enhanced stars \citep[e.g.,][]{sneden2003c,frebel2007,Mello2013,Hill2017,Placco2017,holmbeck2018}, which are found to be over-abundant in r-process elements\footnote{The elemental abundance distribution of several of these metal-poor r-process-enhanced stars is also found to resemble the solar r-process abundance distribution for elements in the $56 \leq Z \leq 76$ range.} (also referred to as neutron capture elements) compared to other metal-poor stars.
These stars are some of the oldest objects in the Galaxy and must have been enriched by r-process elements early in Galactic history. 
This implies that at least one of the, or the r-process site, must be able to operate soon after the formation of the first stars. 
For NS-NS mergers, the question of the earliest possible onset of r-process nucleosynthesis 
 comes with significant uncertainties. 
One problem may be that the time required to form the first NS binary systems plus their in-spiral phase before merging is too long to explain the presence of Eu, which is an r-process element, in metal-poor stars in the Galactic halo, and, therefore, another r-process site might be needed \citep{Hotokezaka2018,Haynes2018,Siegel2019,cote2019,Zevin2019}. This conclusion, however, has been weakened by several Galactic chemical evolution models \citep[e.g.,][]{Shen2015,RamirezRuiz2015,Bartos2019, dvorkin2020,Voort2020,Voort2022}. More work in this direction is required to settle this debate. 

Several observational methods exist to estimate the age of stars.
Cosmochronometry \citep{Butcher1987} is an independent method from other observational age-dating methods. If applied to (old) metal-poor stars, it can set a lower limit on the age of our Galaxy. 
Since the Universe must be as old as the objects within, a lower limit on the age of the Galaxy also sets a lower limit on the age of the Universe. 
Therefore, cosmochronometry can provide valuable comparisons to other age estimates, for example, the current estimate of 13.8~Gyr for the age of the Universe by \citet{Planck2020}. 
Cosmochronometry can be applied to stars that allow for the derivation of long-lived radioactive elements (such as Th or U) in their stellar spectra. 
The method can be used to estimate the time from when a metal-poor r-process-enhanced star (or the material that formed the star) was polluted by an r-process event and until now (when r-process elements are detected in the stellar spectra).
In short, the method combines the effort of stellar observations with theoretical estimates of the initial r-process yields of the star (which originate from an r-process event like, for example, NS-NS mergers) and the knowledge of the radioactive decay time of the detected elements. 
However, the r-process simulations still contain significant uncertainties which propagate into the cosmochronometric age estimates. As discussed above, some of these uncertainties are related to the debate regarding the r-process site. The focus of this work is those uncertainties that concern the r-process in NS mergers, as discussed below.  

One of the important open questions regarding the r-process concerns the initial neutron richness of the material that undergoes r-process nucleosynthesis, in particular in the so-called dynamical (i.e., early-phase) ejecta of NS-NS mergers. Traditionally the neutron richness has been quantified by the electron fraction $Y_e$ defined as the total number of protons ($p$) to the total number of nucleons ($n+p$): \todo{[SG: is this notation inconsistent with the rest of the thesis or ok?]}
\begin{equation}
Y_e = \frac{p}{n+p}.
\label{eq:Y_e}
\end{equation}
Early studies \citep[e.g.,][]{goriely2011,korobkin2012,Mendoza-Temis2015} found very neutron-rich ($Y_e\la 0.1$) dynamical outflows and a significant solar-like production of the $A>140$ elements. 
However, the impact of weak processes on the ejecta was neglected in those simulations\footnote{In some scenarios neglecting weak interactions could be justified due to the dominant mass of the so-called tidal ejecta, which included cold material that escaped the remnant at high velocities, not allowing for the reprocessing of the material by weak nucleonic interactions \citep{korobkin2012}.} and, under such an approximation, the r-process nucleosynthesis yields were shown to be rather insensitive to the initial astrophysical conditions, such as the masses of the binary NS or the nuclear equation of state (EoS), due to fission recycling \citep{goriely2011}.  

This picture changed when weak nucleonic interactions started to be included in both the hydrodynamical simulations as well as the nucleosynthesis calculations \citep{wanajo2014, goriely2015a, martin2018, radice2016,foucart2016,kullmann2021}. 
The weak reactions that are most important are the $\beta$-interactions of electron neutrinos ($\nu_e$) with free neutrons ($n$) and electron antineutrinos ($\bar{\nu}_e$) with free protons ($p$) as well as their inverse reactions,
\begin{equation}
\begin{aligned}
 & \nu_e+ n \rightleftharpoons p + e^- , \label{eq:betareac}  \\
 & \bar{\nu}_e + p \rightleftharpoons n + e^+  .
\end{aligned}
\end{equation}
In studies incorporating the above neutrino emission/absorption reactions, the neutron richness of the dynamical ejecta was shown to be substantially altered by the neutrinos emitted by the post-merger remnant (i.e., the newly formed central NS), in particular in the polar regions. 
However, the exact impact of these weak nucleonic processes in NS mergers remains unclear for several reasons. Firstly, there is a large spread in the applied approximations for the neutrino treatment in the literature or simply many studies do not take neutrino absorption into account \citep[e.g.,][]{wanajo2014, Palenzuela2015,lehner2016,sekiguchi2015,foucart2016,bovard2017,radice2016,martin2018}. Secondly, the quantitative effects of weak reactions may depend on the details of the merger dynamics and the strength of the shock heating at the collision interface between the NSs during merging, and therefore on the adopted EoS \citep{sekiguchi2015} and possibly, to some extent, on numerics. In addition, flavour conversions of the neutrino species can also affect the neutron richness, hence the nucleosynthesis in the merger ejecta \citep[e.g.,][and also \citet{Just2022b} for a BH-torus system]{wu2017,george2020}. 
Since the r-process nucleosynthesis is very sensitive to the initial composition, in particular to the neutron richness of the material, it is critical to base r-process calculations on hydrodynamical simulations, which include such neutrino reactions. 

In addition to the astrophysical modelling uncertainties, r-process nucleosynthesis calculations rely on nuclear models and experimental data for thousands of nuclei\footnote{Depending on the nuclear mass model applied, there are about $\sim5000$ nuclei between the valley of stability and the neutron drip line.}, of which only a tiny fraction of all possible experimental data are available.
Thus, theoretical models are crucial to predict fundamental nuclear properties such as nuclear masses, $\alpha$- and $\beta$-decay rates, radiative neutron capture rates and fission probabilities, all of which enter into the r-process reaction network as input. 
Despite much progress, these nuclear models are still affected by a variety of uncertainties, in particular for the complex description of exotic neutron-rich nuclei.
In addition to the radiative neutron captures and the reverse photodisintegrations, all charged-particle fusion reactions on light and medium mass elements and $\beta$-delayed neutron emission probabilities become important during the nucleosynthesis. If the r-process reaches the fissile region, fission processes such as neutron-induced, spontaneous and $\beta$-delayed fission, together with the corresponding fission fragment distribution, have to be taken into account for all fissioning nuclei.  
Previous works have shown that the nuclear uncertainties can significantly affect the results for the r-process abundances  \cite[e.g.,][]{Caballero2014,Mendoza-Temis2015,Eichler2015,goriely2015,Martin2016,Liddick2016,Mumpower2016,Nishimura2016,Bliss2017, Denissenkov2018, Vassh2019,Nikas2020,sprouse2020a,McKay2020,Giuliani2020,kullmann2022} and, therefore, for the radioactive heating rate that gives rise to the observable kilonova emission \cite[e.g.,][]{Rosswog2017,zhu2018,wu2019,Even2020,Zhu2021,Barnes2021}.
However, the conclusions of these studies may change with, for example, the inclusion of weak nucleonic interactions in the dynamical ejecta or the ability for the r-process to reach the fissile region or the nuclear physics model considered, as discussed in \citet{Lemaitre2021,kullmann2021,kullmann2022}. 

Currently, the most sophisticated and realistic r-process calculations rely on advanced hydrodynamical simulations, which dictate the conditions of the astrophysical environment where the nucleosynthesis occurs, and a full nuclear reaction network with regularly improved nuclear physics input is applied in a post-processing step (due to computational constraints). 
Therefore, studying the r-process in this way requires collaborations between researchers that develop nuclear physics models, scientists that work on the hydrodynamical simulations, and those that run the r-process calculations, as in a nuclear astrophysics symbiosis.

This work will focus on some of the remaining open questions and uncertainties related to the r-process in NS mergers. 
Our first main goal is to contribute to the question of whether a more accurate description of neutrino interactions can significantly affect the r-process abundance yields and the subsequent heating rate produced in the dynamical ejecta of NS mergers, as also presented in \citet{kullmann2021,just2021}. 
The second main goal is to study uncertainties related to theoretical nuclear models on the r-process nucleosynthesis yields and heating rates by varying the nuclear physics input systematically and coherently between global models, as discussed in \citet{kullmann2022}.
To this end, r-process nucleosynthesis calculations based on state-of-the-art hydrodynamical simulations will be performed.
The astrophysical models include the dynamical ejecta of four NS-NS and one NS-BH merger systems, separately, and two BH-torus simulations of the post-merger phase, where the models of the NS-NS merger ejecta include weak interactions (\cref{eq:betareac}). 
The impact on the r-process nucleosynthesis and the subsequent heating rate produced in the NS-NS dynamical ejecta will be studied by comparing our results including weak interactions to a case that ignores such interactions. 
For completeness, kilonova light curves, which are not calculated herein, but based on the r-process yields and heating rates of this work, will be shown, as also presented in \citet{just2021}.
The nucleosynthesis composition of the total ejecta will be estimated by combining the appropriate hydrodynamical models for the dynamical and post-merger ejecta. 
Based on these models (i.e., the total ejecta), the uncertainties related to theoretical nuclear models on the r-process nucleosynthesis will be studied by varying the input systematically between global models that have been adjusted on the full set of available experimental data.
By using the Th/U (molar fraction) ratios obtained in our r-process calculations, we will apply the Th/U cosmochronometer (described in \cref{sec_cosmoc}) to estimate the age of six metal-poor r-process-enriched stars that allowed for accurate derivation of both U and Th lines.
These ages will include an estimate of the uncertainties stemming from the nuclear physics input of the r-process calculations. 


\section*{Outline}



The thesis is organized as follows:
\begin{description}
    \item[\Cref{ch:methods}] introduces the topics relevant to NS binary mergers and r-process nucleosynthesis and describes the methods and models applied in this work. In particular, the different mechanisms responsible for the mass ejection from NS-NS or NS-BH mergers and remnant systems will be introduced as well as the hydrodynamic models that the r-process calculations are based upon. A description of the conditions of the ejected material that can undergo r-processing, the r-process network, the kilonova model, the method for estimating the age of stars, and the nuclear physics models used as input and varied in this work will be given. 
    
    \item[\Cref{ch:sensitivity}]
    discusses and tests the sensitivity to the assumptions and approximations made in \cref{ch:methods}. In particular, tests regarding the size of the network, the resolution of the abundance evolution, the expansion and temperature evolution, and the inclusion of additional reaction channels in the network will be performed. In addition, an analysis of the physical observables related to the r-process results and input will be conducted.
    
    \item[\Cref{ch_dynweak}] presents the results of the nucleosynthesis calculations for four NS-NS merger models of the dynamical ejecta, including nucleonic weak reactions, and studies the impact of including such reactions. The composition, the radioactive decay heat and the kilonova light curve of the ejected material will be discussed with special attention paid to the angular and velocity dependence of the results.
    
    \item[\Cref{ch_nucuncert}] studies the impact of consistently propagating nuclear uncertainties into the r-process calculations, including abundance distributions, heating rates and cosmochronometric age estimates. The nucleosynthesis calculations in this study are based on a model for the total ejecta of NS-NS or NS-BH mergers, which combines hydrodynamical models of the dynamical ejecta (also studied in \cref{ch_dynweak}) with post-merger BH-torus models. 
    
    \item[\Cref{ch:concl}]
    gives concluding remarks and a future outlook. 
    
    
\end{description}




\chapter{Modelling and physical input}
\label{ch:methods}

\epigraph{\itshape \$ pip install PhD }{--- \textup{Unknown}}



The r-process remains the most complex nucleosynthetic process to model from the astrophysics as well as nuclear-physics points of view. 
In order to perform the most realistic r-process simulations possible, a vast amount of input is required. Firstly the amount of material ejected from the NS-NS or NS-BH binary system must be determined including the thermodynamic conditions during and after the merger. In addition, a variety of nuclear physics data and models are required to evolve the ejecta's abundances and determine the final r-process yields and heating rate.
\cref{sec_intNS} introduces the topic of NS mergers.
In \cref{sec:astro_mods}, the models of the astrophysical environment and their relevance to the r-process simulations will be discussed, followed by an overview of the initial conditions and abundances in \cref{sec_init_cond}. 
The details of the r-process nuclear network simulations will be outlined in \cref{sec:nuc_net}. Then, the kilonova light curve model and the cosmochronometry method will briefly be discussed in \cref{sec_kilo,sec_cosmoc}, respectively.
An overview of the nuclear physics input requirements, nuclear theories and models will be given in \cref{sec_nucin}.
Finally, a short description of the computer facilities and some high-performance computing techniques are given in \cref{sec:HPC}.

\todo{[all text from paper 2 (ptwo) updated according to changes before 19.july (however fig captions might not be updated??)]}

\section{Introduction to NS mergers}
\label{sec_intNS}

A large portion of the stars in our Galaxy, particularly the heaviest ones, are found to exist in binary systems \citep{Raghavan2010,Sana2012}.
Consequently, those binary systems that contain massive stars ($\gtrsim 8$~\Msun) can form NS binary systems if both stars undergo CCSN (see \cref{fig_SNremnant}). 
The explosion which forms the first NS may destroy the companion star, but if the binary survives, the second star can continue its evolution and produce the second NS (or a BH).
Due to energy and angular momentum lost through gravitational waves (GW), the orbit of the NSs slowly shrinks, until they finally merge. 
The evolution of binary massive stars can be described with population synthesis models \citep[e.g.][]{Andrews2015,Mink2015,Belczynski2018,Chruslinska2018, Artale2019,Chruslinska2018b,Giacobbo2019}.
The probability distribution for the merging to occur at a given time following the formation of the binary system are called the delayed time distribution, of which many different models exist \citep{Chruslinska2018,cote2019,dvorkin2020}.
These models vary in their predictions, but generally the in-spiral period for NS-NS systems are believed to be on the order of hundreds of Myr and up to several Gyr \citep{Chruslinska2017,Belczynski2018,Giacobbo2019}.
The explosion(s) giving birth to the NSs are thought to give the stars large velocities, often referred to as a ``natal kick'', and such kicks have been found to significantly modify the coalescence timescales of NS-NS and BH-NS systems, and may also lead the NS to escape from the binary system \citep{Banerjee2020b,Voort2022}. 
Due to the small sample of known binary NSs, there are large uncertainties related to NS natal kicks, delayed time distributions and merger rates, which affect Galactic chemical evolution simulations that model the r-process enrichment through Galactic history. 




\todo{want to add something about pulsars constrain radius and mass of NSs? Improved by GW detections? }

\begin{landscape}
\begin{figure}[p]
\begin{center}
\includegraphics[width=1.6\textwidth]{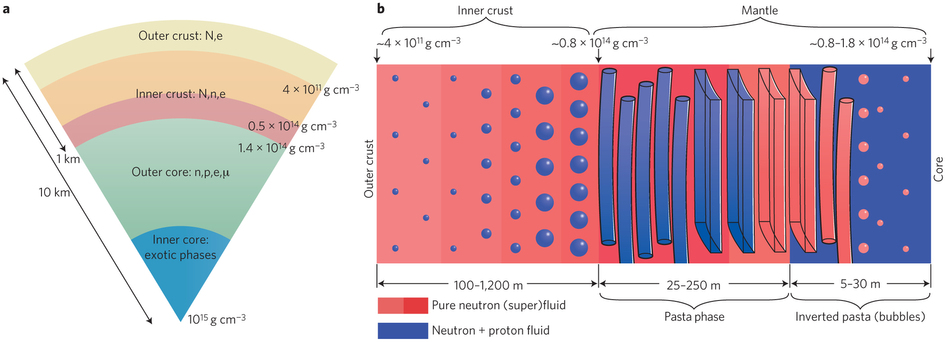}
\caption[Illustration of the NS interior.]{
Illustration of the NS interior by \citet{NS_interior}. 
a) The main regions of a NS, roughly given by the density and radius indicated on the right and left side, respectively. The symbols $N$, $n$, $p$, $e$ and $\mu$ correspond to nuclei, neutrons, protons, electrons and muons, respectively. The composition of the inner core is uncertain and several hypotheses include hyperons and deconfined quark matter. 
b)
Detailed composition of the inner crust and mantle, i.e., the dark orange and red region in a). 
For densities $\rho\lesssim 0.8\cdot 10^{14}$~g~cm$^{-3}$, a lattice of superheavy, neutron-rich nuclei is immersed in a fluid of neutrons and a relativistic electron gas. 
At higher densities ``nuclear pasta'' phases may exist in the mantle, which may include droplets (gnocchi), rods (spaghetti), cross-rods (waffles), slabs (lasagna), tubes (bucatini) and bubbles (swiss cheese).
}
\label{fig_NS_interior}
\end{center}
\end{figure}
\end{landscape}

NSs are the most compact stars known in the Universe \citep{Haensel2007,Burgio2021}.
NSs typically have masses around $1.4$~\Msun\ and radii $R\sim 13$~km, giving a mean mass density several times the density of nuclear matter in heavy atomic nuclei. 
The nuclear EoS gives a theoretical description of high-density matter, i.e., the dependence of pressure on density, and also affects the composition of the NS interior.
A NS can be divided into the inner/outer core and crust depending on the density within the star, as illustrated by \cref{fig_NS_interior}. 
Generally, the ejected material of NS-NS (or NS-BH) mergers originates from the outer and inner crust of the in-spiralling NSs \citep[][see also \cref{fig_snapshot}]{Bauswein2013}.
The outer crust consist of nuclei in a charge compensating electron gas.
Towards higher densities the limit between the outer and the inner crust is reached at the so-called ``neutron drip density'' $\rho_\mathrm{drip}=4.2\times 10^{11}$~g~cm$^{-3}$, where unbound neutrons appear, forming a neutron liquid filled with heavy, neutron-rich nuclear clusters and electrons. 
A variety of EoSs exist in the literature \citep[e.g.,][for recent reviews]{Oertel2017,Burgio2021,Raduta2021}, which make different assumptions about the description of the nuclear interactions and the composition of matter at extreme densities.  
Several EoSs \citep[e.g.,][see also \citet{Chamel2008,Watanabe2007} and references within]{Ravenhall1983,Hashimoto1984,Williams1985,Oyamatsu1993,Gogelein2007,Caplan2017,Newton2022} find that so-called ``nuclear pasta'' exists prior to the transition to the core (this part of the inner crust is sometimes referred to as the mantle, see \cref{fig_NS_interior}b). 
The crust-core interface is reached at even higher densities ($\rho \sim 1.4\cdot 10^{14}$~g~cm$^{-3}$), and beyond this limit the nuclei starts to disappear, and muons might exist in a mix of protons and electrons in a sea of neutrons.

There is a one-to-one correspondence between the EoS and the predicted NS mass-radius relation. 
Generally, ``stiff'' EoSs predict larger NS radius for the same mass, while lower mass objects with smaller radius are expected within ``soft'' EoSs.
Only EoSs which are compatible with observational NS mass-radius constraints should be applied in astrophysical simulations, i.e., have a predicted maximum mass above\footnote{For example, some of the ``softest'' EoSs does not reach a maximum mass of 2~\Msun.} $\sim2.1$~\Msun \citep[][]{Cromartie2020}, and also preferably the observational constraints from the GW170817 event \citep{abbott2017} and the NICER mission \citep{Riley2019,Miller2019b}. The EoSs applied in this work fulfil the aforementioned criteria. 
The compactness of the NSs has a direct impact on the collision dynamics, and therefore also on the amount of gravitationally unbound material which can contribute to the Galactic enrichment of r-process elements. 


\subsection{Mass ejection mechanisms and phases}
\label{subsec_eject}

\begin{figure}[tbp]
\begin{center}
\includegraphics[width=0.95\textwidth]{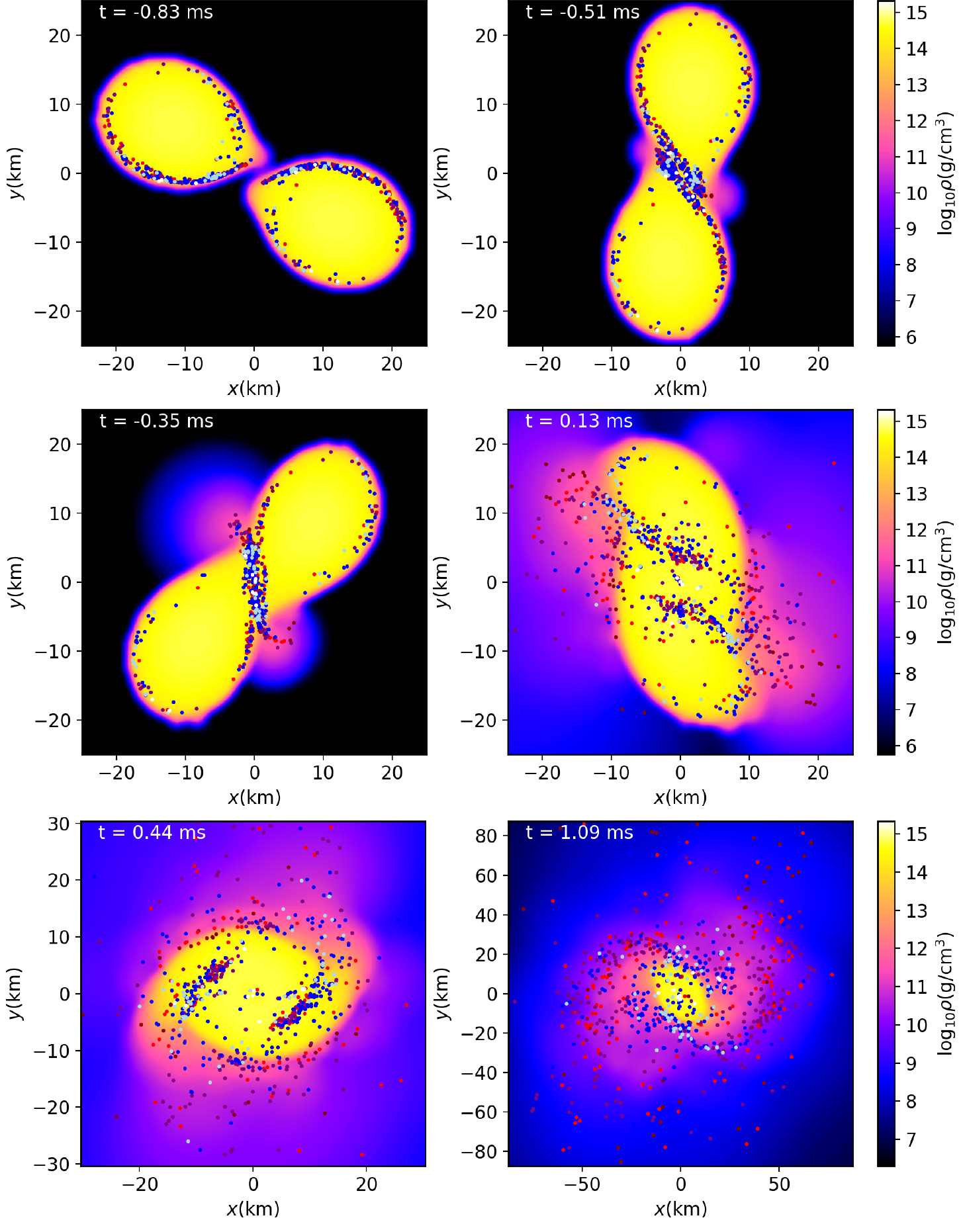}
\caption[2D density snapshots of a NS-NS merger.]{
2D snapshots of the density distributions in the equatorial plane at six different times from a NS-NS merger simulation using equal NS masses of 1.35~\Msun\ and the DD2 EoS (see \cref{subsec_dynsim} and \cref{tab:astromods} for more details).
The locations of the 783 finally ejected mass elements are indicated by coloured dots, projected onto the equatorial plane, where the final $Y_e$-values (\cref{eq:Y_e}) are colour coded with dark red, red, purple, blue and light blue for $Y_e$-bins of 0.1 and the last bin of $Y_e \ge 0.5$ is white, respectively. 
Note that 
the last snapshot at $t=1.09$~ms shows the system at a larger radius than the other snapshots. Image credit: A. Bauswein \citep{kullmann2021}.
}
\label{fig_snapshot}
\end{center}
\end{figure}
\begin{figure}[tbp]
\begin{center}
\includegraphics[width=\columnwidth]{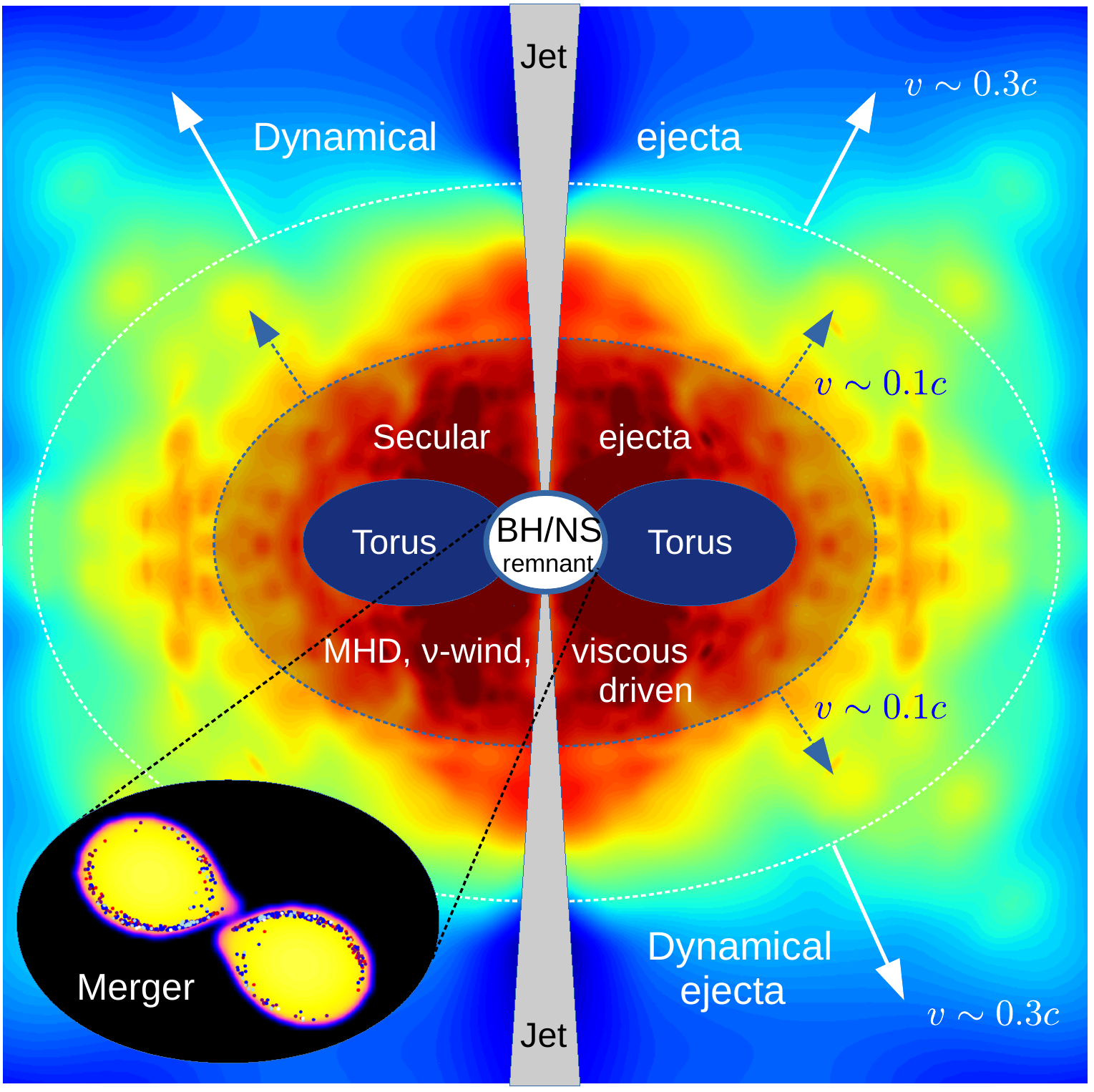}
\caption[Illustration of NS-NS or NS-BH merger evolution and ejecta mechanisms.]{
A compact object (NS-NS or NS-BH) merger leads to either a NS or BH remnant producing ejecta via different mechanisms. During the first few ms after merging, the dynamical ejecta (outermost sphere) become unbound due to tidal torques and shock heating from the contact interface (bottom left insert) with typical ejecta velocities of $\sim0.3$~c (where c is the speed of light). The material ejected on the dynamical time scale surrounds the remnant plus torus system that forms at later times and its corresponding secular ejecta with velocities of $\sim0.1$~c. A jet is also expected to form, drilling through the ejecta and producing a short $\gamma$-ray burst. Image credit: A. Bauswein \citep{kullmann2022}.
}
\label{fig_ejphases}
\end{center}
\end{figure}

During and shortly after the merging, the NS-NS system undergoes several mass ejection phases (see \cref{fig_snapshot} and \cref{fig_ejphases}), which depend on the binary system properties (such as for example the NSs masses) and the properties of the nuclear EoS. 
\cref{fig_snapshot} displays the evolution of a NS-NS merger system where the NSs have equal mass, i.e., a so-called symmetrical system, where the coloured dots show the mass elements that will become gravitationally unbound later in the simulation.
In the first ejection phase, some of the material becomes gravitationally unbound due to tidal torques created by the dynamical movement of the NSs after they interact. 
For the merger in \cref{fig_snapshot}, the tidal tail is small, however it is possible to notice the elongation and disruption of the system in the snapshots at $t=0.13$~ms and $t=0.44$~ms.
A portion of the ejected matter also originates from the contact interface between the NSs (see the central mass elements in the snapshots at $t=-0.51$~ms and $t=-0.35$~ms in \cref{fig_snapshot}). This material is shock-heated to large temperatures\footnote{Hydrodynamical simulations of NS binary mergers should therefore apply an EoS which also includes finite temperatures.} (typically several tens of GK), while the tidal ejecta remains cold. 
The material which becomes gravitationally unbound due to these two ejection mechanisms are called the dynamical ejecta, and this phase lasts on ms time scales. 
The ratio between the cold tidal and hot shocked ejecta mainly depends on the mass ratio between both stars $q=M_1/M_2$ (with $M_2\ge M_1$) and the nuclear EoS, where the relative amount of cold tidal ejecta typically grows with smaller $q$ and stiffer EoSs \citep{Bauswein2013}.

Similar to NS-NS mergers, a significant amount of neutron-rich material can also be ejected from NS-BH mergers \citep[e.g.,][]{Janka1999,Rosswog2005,Etienne2008,Etienne2009,korobkin2012,Kyutoku2013, Foucart2014,Bauswein2014,Just2015,Kiuchi2015, Fernandez2017,Fernandez2020,Kruger2020} so that they are also viable candidates for the r-process site. 
The evolution from two orbiting, in-spiralling compact objects to a BH or a long-lived NS remnant depends on the initial conditions of the system. 
In NS-BH mergers with sufficiently small BH masses and NS-NS mergers with sufficiently high total masses and/or a sufficiently soft EoS, a BH-torus system is formed right after the merger. Alternatively, if the NS remnant can be stabilized against a prompt collapse in a NS-NS merger, it will evolve as a (hyper- or super-) massive NS (HMNS or SMNS) and either undergo a delayed collapse after some time or remain indefinitely stable.
The evolution of NS-BH mergers also depend on the initial BH properties like spin and mass \citep{Bauswein2014}.

\cref{fig_ejphases} shows a schematic overview of the NS-NS or NS-BH merger ejecta components viewed perpendicular to the rotational axis, i.e., along the $z$-axis of figure \cref{fig_snapshot}. 
The insert in the bottom left corner of \cref{fig_ejphases} shows the two NSs right before they touch, i.e., the first snapshot in \cref{fig_snapshot}. 
The NSs form a merger remnant which is a massive NS or a BH surrounded by an accretion torus. 
The r-nuclide enrichment is predicted to originate from both the dynamical (prompt) material expelled during the merger phase and from the outflows generated during the evolution of the remnant (called secular ejecta).
The dynamical ejecta is expelled first, and it typically has larger escape velocities ($v\sim0.3$~c) than the material ejected at later stages ($v\sim0.1$~c). Therefore, the dynamical ejecta will be enclosing the secular ejecta, visualized by the two dotted spheres. 
During the secular phase, material can be expelled due to viscous energy dissipation and turbulent angular momentum transport, magnetic pressure, nucleon-recombination heating and neutrino-energy deposition
\citep[e.g.,][]{Metzger2014,Siegel2018,HosseinNouri2018,Christie2019,miller2019, Fernandez2019,Fujibayashi2018,Fujibayashi2020b,Fujibayashi2020, perego2014,Perego2021,Just2015,Just2022,Fahlman2022}. 
As can be seen in \cref{fig_ejphases}, the material is not ejected in an isotropic way, thus the conditions of the ejecta vary with the angle relative to the rotational plane, but also the ejection velocity. 
The composition of the dynamical ejecta have been found to be in particular angle dependent if neutrino reactions (\cref{eq:betareac}) are included in the hydrodynamical simulations
\citep[][see also \cref{fig_ye_ileas}]{wanajo2014, goriely2015a, martin2018, radice2016,foucart2016,kullmann2021}, however the uncertainties related to the treatment of neutrino interactions remain large \citep[e.g.,][]{kullmann2021}.
These uncertainties are related to the composition of the dynamical ejecta and come in addition to still unknown details with respect to the secular ejecta composition and ejection mechanisms. 
In addition, it is not yet settled which component of the dynamical or secular ejecta that contributes the most to the total ejecta mass from NSs \citep{Just2015}. Several simulations \citep[e.g.,][]{Metzger2014,perego2014,Just2015,Fujibayashi2018b,foucart2020} yield a dominant contribution from the secular ejecta (if the remnant system remains stable against collapse into a BH). 
Higher $Y_e$-values have also been found in these simulations, which can lead r-process yields from the secular ejecta to differ from those of the dynamical ejecta significantly. 
The details regarding the composition and ejection mechanisms have to be clarified before conclusions regarding the role of NS mergers in the Galactic chemical evolution can be made. 

\subsection{Properties of the ejecta}
\label{subsec_NScond}

\begin{figure}[tbp]
\begin{center}
\includegraphics[width=0.8\textwidth]{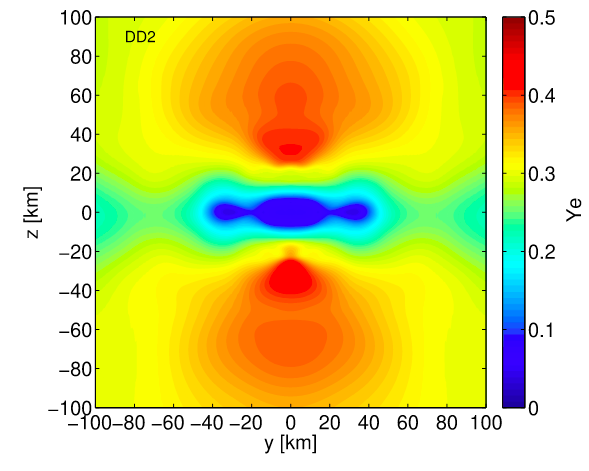}
\caption[Vertical slice of a NS-NS merger remnant showing the electron fraction.]{
Vertical slice of the remnant of a NS-NS merger showing the angular averages of the electron fraction adapted from \citet{ardevol-pulpillo2019} 5~ms after merger. 
The NS merger system corresponds to the same model as in \cref{fig_snapshot} and \cref{fig_trajs_evo}, which applied the DD2 EoS and initially merged two 1.35~\Msun\ NSs (see \cref{subsec_dynsim} and \cref{tab:astromods} for more details).
}
\label{fig_ye_ileas}
\end{center}
\end{figure}

\begin{figure}[tbp]
\begin{center}
\includegraphics[width=0.99\textwidth]{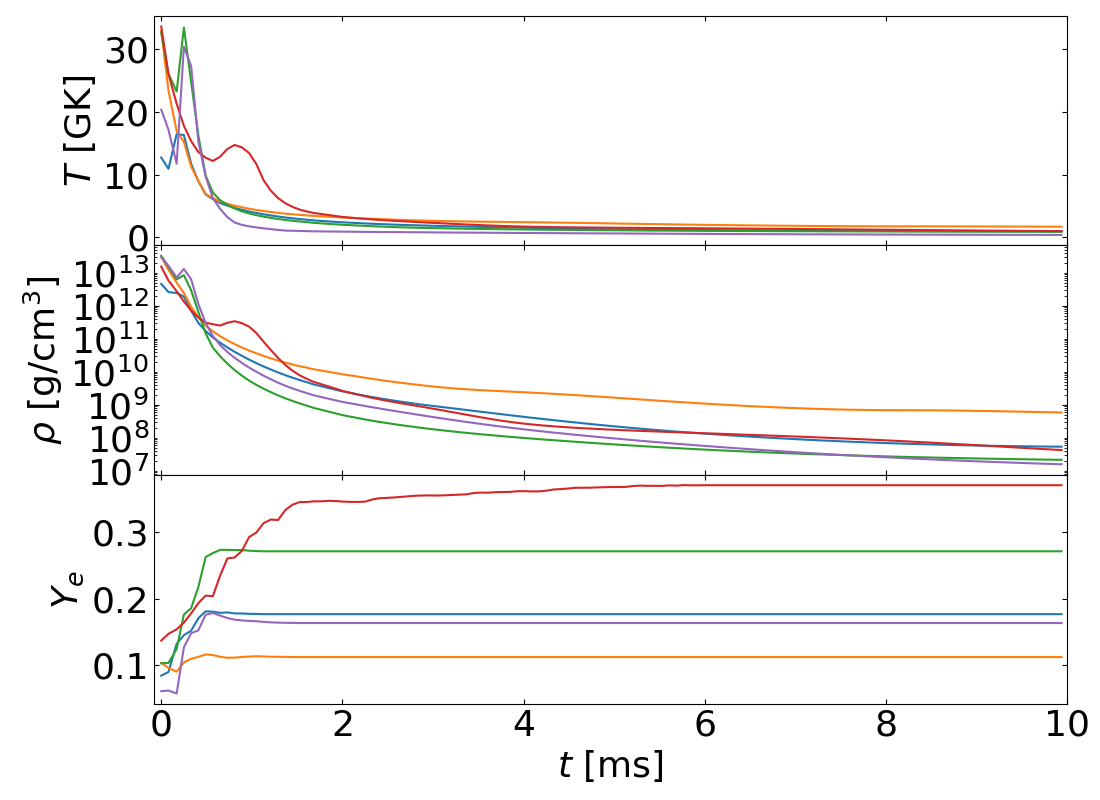}
\caption[The temperature, density and electron fraction evolution of five trajectories.]{
Illustration of the time evolution of the temperature (top), density (middle) and electron fraction (bottom) of five ejected mass elements, i.e., trajectories. 
The trajectories are from the same NS-NS merger simulation as in \cref{fig_snapshot}, which uses a 1.35~\Msun-1.35~\Msun\ configuration and the DD2 EoS (see \cref{subsec_dynsim} and \cref{tab:astromods} for more details).
}
\label{fig_trajs_evo}
\end{center}
\end{figure}

After formation and during the in-spiral time of the binary system the NSs gradually cool down to relatively low temperatures \citep[e.g.,][]{Yakovlev2004}.
The NS material is further assumed to remain in $\beta$-equilibrium. Therefore, the NSs (or the one NS in the case of a NS-BH merger) are usually approximated to be made of so called ``cold catalyzed matter'', i.e. matter in its ground state at zero temperature ($T=0$).
Under these assumptions the electron fraction $Y_e$ will approach an equilibrium value \citep{Arcones2010,goriely2015a,Just2022b} when sufficient time has passed at constant temperature and density. 
Thus, the $Y_e$-values within the NSs before merging are determined by $\beta$-equilibrium.
However, temperature effects become important during merging and at later stages, and the material will not be in $\beta$-equilibrium any longer. 
Since full reaction networks are computationally expensive (e.g., see \cref{sec:HPC}), the hydrodynamical simulations usually only follow the abundances of a few nuclei, i.e., those that are involved in the main energy generation reactions \citep{Bodenheimer2006}.
Typically only the evolution of free nucleons, electrons and often a single or a few heavy nuclei are traced, and the electron fraction $Y_e$ is advected with the fluid. 
If the weak reactions in \cref{eq:betareac} are included in the hydrodynamical simulations the overall $Y_e$ value in the ejecta can change significantly, and have to be followed by the hydrodynamical simulations (see also \cref{fig_trajs_evo}).

As an example of the conditions in NS-NS mergers, \cref{fig_ye_ileas} shows the electron fraction 5~ms after merging in a hydrodynamical simulation including the weak nucleonic reactions of  \cref{eq:betareac}. The remnant is visible in the centre where $Y_e<0.1$, and the surrounding dynamical ejecta typically have $Y_e>0.2$.
The effect of including the weak interactions in the simulations are in particular large in the polar regions (along the $z$-axis) due to the neutrinos emitted by the newly formed remnant NS, driving the electron fraction up to $\sim0.3-0.4$ and even $\sim0.5$. 
This is not the case when such reactions are omitted, where a nearly uniform angular distribution is found for the $Y_e$ of the dynamical ejecta (see \cref{sec_nuc} for a detailed discussion).

Most of the ejecta (except for the tidal tail) is shock-heated to large temperatures, generally to more than 10~GK, during merging. 
At such high temperatures and densities ($T\gtrsim10$~GK and $\rho \gtrsim 10^{6}$~g~cm$^{-3}$), a balance is established between photodissociations of nuclei into individual nuclei and rapid nuclear reactions which builds up nuclei. This equilibrium between the strong and electromagnetic reactions is called nuclear statistical equilibrium (NSE).
If NSE is assumed, the estimation of the ejecta composition depends only on the density, temperature and electron fraction of the plasma (see \cref{sec_init_cond} for more details). 
After being re-heated to high temperatures during the collision, the ejecta will start to expand freely, and the temperatures and densities will decrease accordingly. When the material falls out of NSE (roughly below $T\lesssim7$~GK and $\rho \lesssim 10^{6}$~g~cm$^{-3}$), the abundance evolution have to be followed by a full nuclear reaction network (starting from the initial abundances calculated by assuming NSE). 

When the time evolution of thermodynamic quantities such as the temperature, density and entropy is provided for an unbound mass element in an hydrodynamical simulation (i.e., the coloured dots in \cref{fig_snapshot}) we often refer to it as a ``trajectory''.
The collection of these trajectories provides the basis of the r-process nucleosynthesis calculations, which are performed in a post-processing step (see \cref{sec:nuc_net} for more details).
As an example, the typical time evolution of five trajectories from a simulation of the dynamical ejecta are shown in \cref{fig_trajs_evo}. 
Since weak interactions are included in this simulation, the $Y_e$-value changes substantially over time, as can be seen in the bottom panel of \cref{fig_trajs_evo}. 
As done in many studies \citep[e.g.,][]{lippuner2015,Giuliani2020, Zhu2021,Barnes2021}, it is also possible to produce trajectories that mimic the conditions predicted by hydrodynamical simulations in a parametric way. All r-process calculations in this thesis have been based on trajectories from hydrodynamical models provided by collaborators (see \cref{sec:astro_mods} for more details about the models).

\todo{nikas2020? for refs above if need more?}

\section{Astrophysical models of NS mergers}
\label{sec:astro_mods}

\epigraph{\itshape ``There is no point in using the word `impossible' to describe something that has clearly happened.''}{--- \textup{Douglas Adams}, Dirk Gently's Holistic Detective Agency }

Great efforts have been put into hydrodynamical simulations of different NS-NS or NS-BH merger configurations in the last decade.
An important goal for the community is to develop astrophysically consistent models which can estimate the total ejecta mass, either by combining models for the ejecta components consistently or, ideally, by covering the entire evolution of the merging system. 
The first hydrodynamical models of NS-NS mergers focused on the dynamical ejecta were performed within Newtonian approximations
\citep[e.g.,][]{Davies1994,Ruffert1996,rosswog1999,Janka1999,Ruffert2001,roberts2011,korobkin2012,rosswog2013, Piran2013}.
Later, models including general relativistic (GR) approximations \citep[e.g.,][]{oechslin2007,goriely2011,hotokezaka2013,Bauswein2013} and also what is referred to as ``full GR'' models became available \citep[e.g.,][]{Sekiguchi2011,sekiguchi2015,sekiguchi2016,lehner2016,foucart2016a, bovard2017,radice2016,radice2018a,Palenzuela2015,Palenzuela2022}. \todo{consider adding: Oechslin R, Rosswog S, Thielemann FK. Phys. Rev. D 65:103005 (2002)
39. Oechslin R, Uryu ̄ K, Poghosyan G, Thielemann FK. Mon. Not. R. Astron. Soc. 349:1469
(2004)}
These studies showed that the inclusion of GR effects fundamentally changed the merger dynamics and the relative contribution of ejected mass from the tidal versus shock-heated ejecta \citep{Bauswein2013}.
Simulations of the long-term evolution of merger remnants \citep[][]{Setiawan2004,Shibata2007,Metzger2008,Metzger2009,Metzger2018,Dessart2009,Lee2009, Fernandez2013,Metzger2014,perego2014, Siegel2014,Just2015,martin2015,Wu2016,Fujibayashi2017,Fujibayashi2018,lippuner2017, Siegel2017,Siegel2018, HosseinNouri2018,Janiuk2019,Fernandez2019,miller2019a,Fujibayashi2020b,Just2022b} have shown that magnetically, neutrino-, and viscously driven secular outflows complement the dynamical ejecta, and can in most cases even dominate the total outflow mass. 
Recently, improved models including fully relativistic approaches (i.e., full GR) have managed to cover the dynamical phase and the late-time evolution of the remnant and secular ejecta \citep{Kawaguchi2021,Hayashi2021}\footnote{Note that the NS-NS simulations of \citet{Kawaguchi2021} involve a mapping between different codes; see also \citet{Just2015} for an early attempt to combine different ejecta components in a ``system-preserving'' (thus attempting consistency) manner.}.
However, due to computational constraints, most models only cover one phase of the evolution, i.e., either the dynamical or the secular phase, as is the case for the models considered in this work (see \cref{sec:astro_mods}).

The numerical methods applied to evolve the fluid in hydrodynamical simulations can be divided into so-called grid-based techniques \citep[e.g., see][for a review]{Teyssier2015} or smoothed-particle-hydrodynamics (SPH) introduced by \citep{Gingold1977,Lucy1977}. 
As implied from the name, grid-based codes uses a grid with a chosen geometry to discretize the computational volume into fixed volume elements.
In order to follow the time evolution of the conserved quantities such as the mass, the net fluxes of the plasma through the computational cell faces have to be calculated at each time step in grid-based codes. 
The mesh points can be regularly or irregularly spaced, and even vary depending on, for example, the density through Adaptive Mesh Refinement (AMR) methods \citep[][and references within]{Berger1984}. 
The SPH method is a mesh-free approach which discretizes the fluid quantities into ``particles'' of fixed mass that move with the fluid velocity. In this method, the density is computed as a weighted sum over neighbouring particles, where the weight decreases with distance from the sample point according to a scale factor. 
A big advantage for the SPH method is that the mass is an exactly conserved quantity since particles cannot lose, gain or diffuse mass, and the resolution naturally follows the particles. This is in contrast to grid based codes which might not resolve low-density regions with the same resolution as high-density regions (unless AMR or other grid-refinement techniques are applied). 
SPH is also well suited for problems with free boundaries and geometries without significant symmetries, however grid-based codes are considered preferrable to particle methods such as SPH for the treatment of shocks \citep{Bodenheimer2006}.


The amount of ejected mass, i.e., the material which becomes gravitationally unbound from the system and therefore may enrich the surrounding environment with freshly synthesized r-process elements, has to be determined by hydrodynamical simulations. 
Determining the amount of ejected mass is not a trivial task \citep[e.g., see][]{Foucart2021}, since different criteria for gravitationally unbound conditions can be considered, and furthermore, the adopted criterion can be applied at different times or radii. 
Several computational challenges related to the ejection mechanisms have to be overcome in order to model the conditions of the unbound material realistically, but also the treatment of neutrino interactions, in particular for the dynamical ejecta, can have a large impact on the ejecta composition, as discussed earlier. 


Our nucleosynthesis calculations are based on advanced computational models of the NS-NS or NS-BH merger systems, in addition to the BH-torus remnant simulations. 
The total ejecta is modelled as the combination of the separate ejecta components, since fully consistent models, which include weak nucleonic reactions and cover all evolution phases over a wide range of binary parameters, are not available at the present time.
Therefore, the BH-torus models presented in \cref{subsec_BH-t_sim} are those that approximately match the remnant configurations obtained after merger in \cref{subsec_dynsim}, so that they can be combined (using the same EoS) into the total NS-NS (NS-BH) merger ejecta, as described in \cref{subsec_combsim}.
In this thesis only BH-torus systems as possible merger remnants have been considered. 
This is assumed to be a reasonable approximation \citep{Just2015} since the NS-NS or NS-BH models described in \cref{subsec_dynsim} are not expected to support a long-lived HMNS remnant, and thus the remnant will collapse into a BH soon after merging. 

Note that all of the hydrodynamical simulations presented here have been performed by collaborators of the Max-Planck-Institut f\"ur Astrophysik (Garching, Germany) and GSI Helmholtzzentrum f\"ur Schwerionenforschung (Darmstadt, Germany) \citep{Just2015,ardevol-pulpillo2019} which provide simulation datasets containing the trajectories that we use as input for the r-process calculations. However, the data analysis of the trajectories (i.e., \cref{subsec_hydro_evo,sec_init_cond,sec_analysis}) is not provided by collaborators but performed herein. 

\subsection{Simulations of NS-NS and NS-BH dynamical ejecta}
\label{subsec_dynsim}

\todo{Note that some of the hydro stuff is too detailed and outside of my expertise and should take care to prepare before defence? spess the footnote}

The NS-NS merger simulations that produce the dynamical ejecta are based on a relativistic SPH code coupled to the so-called improved leakage-equilibration-absorption scheme (ILEAS) for neutrino transport \citep{ardevol-pulpillo2019}.
The simulations include two systems with a total mass of $M_1+M_2=2.7$~\Msun; one symmetric (1.35~\Msun - 1.35~\Msun) and one asymmetric (1.25~\Msun - 1.45~\Msun) merger. 
The SPH code assumes a conformally flat spatial metric to reduce the complexity of the Einstein equations and the NSs are modelled with about 150 000 SPH particles per star. 
As shown for instance in \citet{Ciolfi20}, the impact of magnetic fields on the properties of the dynamical ejecta can be neglected at early times, i.e., during the dynamical phase.

ILEAS is designed to follow the neutrino effects in the hot and dense astrophysical plasma \citep{ardevol-pulpillo2019}. This approximation improves the lepton number and energy losses of traditional leakage descriptions by a novel prescription of the diffusion time-scale based on a detailed energy integral of the flux-limited diffusion equation. The leakage module is also supplemented by a neutrino-equilibration treatment that ensures the proper evolution of the total lepton number and medium plus neutrino energies as well as neutrino-pressure effects in the neutrino-trapping domain. In addition, a simple and straightforwardly applicable ray-tracing algorithm is used for including re-absorption of escaping neutrinos especially in the decoupling layer and during the transition to semitransparent conditions. All details can be found in \citet{ardevol-pulpillo2019} where it was shown in particular that ILEAS can satisfactorily reproduce local losses and re-absorption of neutrinos as found in more sophisticated transport calculations on the level of 10\%\footnote{The tests in \citet{ardevol-pulpillo2019} were performed for a spherically symmetric proto-NS configuration and a BH-torus. The accuracy for the merger case is likely similar though not exactly known due to the lack of detailed reference solutions.}.
\todo{flavour conversions are ignored}

At the start of the hydrodynamical simulation, a few orbits before the merging of the two NSs, equilibrium is assumed to hold between electrons, positrons and neutrinos. 
The Cartesian grid covering the NS in which the neutrino interactions are calculated expands 100~km in all six directions from the centre of mass of the system, with a resolution of 0.738~km\footnote{
Although detailed resolution tests have yet to be conducted, the rather modest resolution of 0.738~km can be justified by three main arguments. First, the density gradients at the neutron star surfaces are steep only prior to the collision of the two stars, i.e., at times when the neutrino emission is still low. Steep density gradients develop again at late post-merger times after considerable neutrino cooling of a stable merger remnant. Such a late evolution phase, however, is not reached in the present set of merger models. Second, the neutrino treatment is not based on numerically determined gradients but employs space integrals that are usually less sensitive to fine spatial structures. Third, the possibility of resolving steep density gradients or local density variations on small scales is essentially set by the numerical limits from SPH hydrodynamics rather than from the neutrino treatment.}.
After the merger, as the remnant torus expands, the grid size is progressively increased keeping the whole remnant covered at all times and the same resolution (cell size) throughout the simulation \citep{ardevol-pulpillo2019}.
These relativistic hydrodynamic simulations also provide the temperature evolution to which no post-processing is applied \cite[in contrast to][]{goriely2011} to retain consistency between the neutrino properties and the other thermodynamical properties.
In these simulations, either the temperature-dependent DD2 \citep{hempel2010,typel2010} or SFHo EoS \citep{steiner2013} is adopted.
For clarity, from now on the DD2 1.35-1.35~\Msun\ and 1.25-1.45~\Msun\ merger models will be referred to by DD2-135-135 and DD2-125-145, respectively,  and similarly the SFHo 1.35-1.35~\Msun\ and 1.25-1.45~\Msun\ models as SFHo-135-135 and SFHo-125-145, respectively. 

The NS-BH binary system consists of a $1.1$~\Msun\ NS and a $2.3$~\Msun\ BH, also modelled with SPH and the SFHo EoS \citep[referred to as sfho\_1123 in][]{Just2015}, from now on named SFHo-11-23 (this model does not include neutrino effects which may be a decent approximation for the mostly tidally ejected material; ignoring neutrino emission and absorption may be a valid approximation because of the rapid expansion of mostly tidally ejected material in NS-BH mergers). 
\todo{ is the below discussion about determination of ejected particles also valid for the BHNS model? - Yes see fig 2 in Just2015}

\begin{figure}[tbp]
\begin{center}
\includegraphics[width=\textwidth]{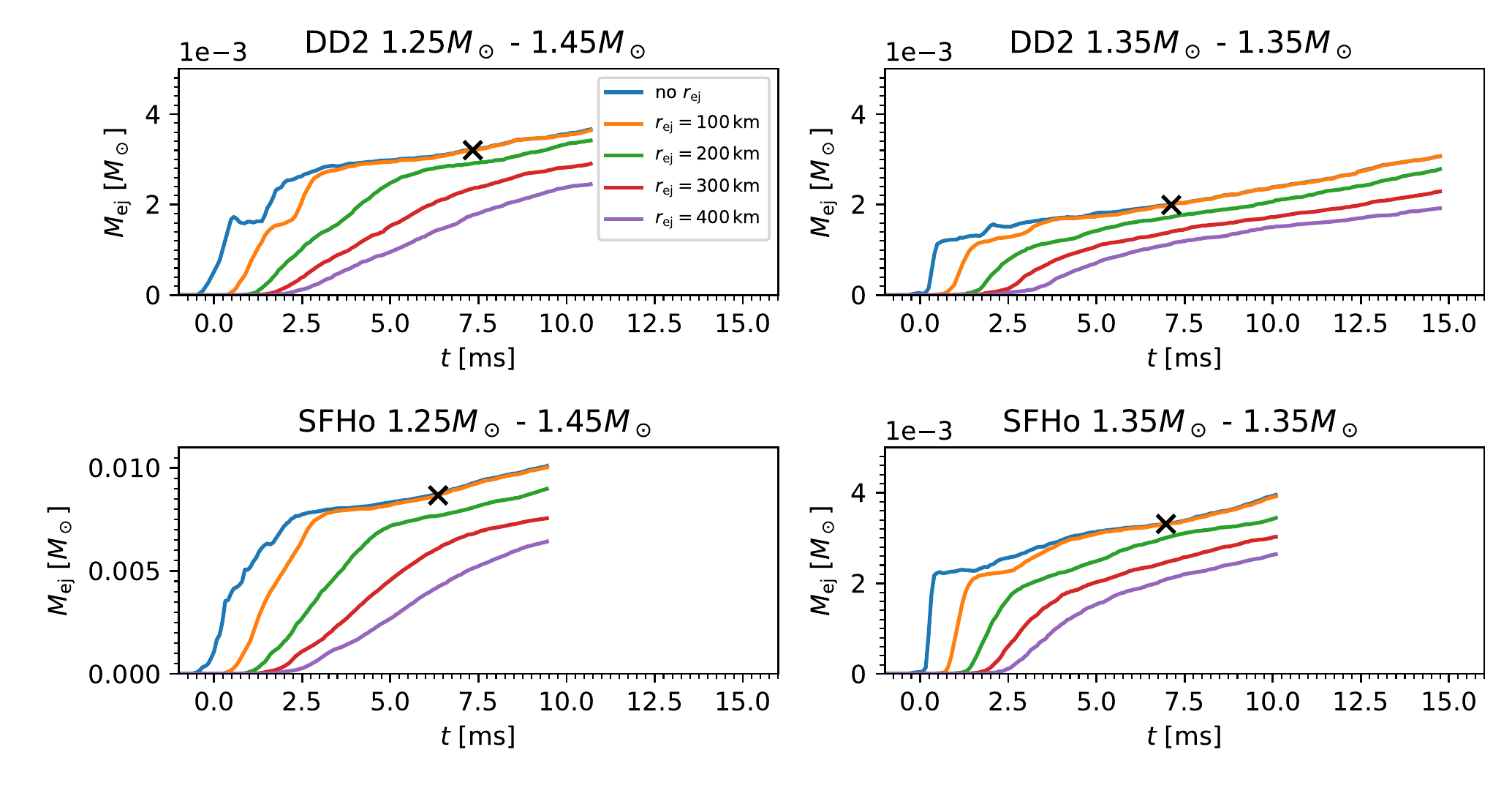}
\caption[Dynamically ejected masses versus time for the four NS-NS merger models.]{ 
Time evolution of the dynamically ejected masses for the four NS-NS merger models as measured by the criterion $\varepsilon_{\rm stationary}>0$ applied beyond radii of $r_{\rm ej}=0$, 100~km, 200~km, 300~km, 400~km. The $t=0$ reference time corresponds to the phase of the merger when the remnant is maximally compressed about 1~ms after the stars first touched. The fact that lines for different values of $r_{\rm ej}$ are similar to each other indicates that material flagged
unbound at low radii keeps expanding and truly ends up as ejected
material. The crosses at time $t=t_{\rm ej}\sim 7$~ms for lines with $r_{\rm ej}=100$~km correspond to the ejecta material that is adopted for nucleosynthesis calculations in this study (see also \cref{tab:astromods}).
Fig.~2 in \citet{Just2015} shows a similar figure for our NS-BH merger model. Image credit: \citet{kullmann2021}.
}
\label{fig_mejec}
\end{center}
\end{figure}

For the five SPH models, the number of unbound mass elements are 783, 1263,  1290, 4398 and 13175 with a total ejected mass of $2\times 10^{-3}$~\Msun, $3.2\times10^{-3}$~\Msun, $3.3\times10^{-3}$~\Msun\, $8.6\times10^{-3}$~\Msun\ and $40.4\times10^{-3}$~\Msun\ for DD2-135-135, DD2-125-145, SFHo-135-135, SFHo-125-145 and SFHo-11-23, respectively, see \cref{tab:astromods}.
The positive stationary energy criterion ($\varepsilon_{\rm stationary}>0$) is employed to determine the amount of mass dynamically ejected from the systems. It states that a mass element becomes unbound if at a chosen time $t_{\rm ej}$ and beyond a certain radius $r_{\rm ej}$, the kinetic plus thermal energy exceeds the gravitational one \citep[modulo relativistic corrections, see, e.g.,][for a detailed derivation of this criterion]{oechslin2007}.
In \cref{fig_mejec} we compare for each NS-NS model the growth of $M_{\rm ej}$ with time, where $M_{\rm ej}(t)$ is the mass of all particles that fulfill $\varepsilon_{\rm stationary}>0$ and $r >
r_{\rm ej}$, and $r_{\rm ej}$ is varied between 0, 100~km, 200~km, 300~km, and 400~km. Independent of the chosen value of $r_{\rm ej}$, the total ejected mass is found to grow with time and not to saturate at the end of the simulations.  In order to select ejecta particles for our nucleosynthesis calculations, an ejection
radius of 100~km and time of 7~ms are adopted, as shown by the crosses in \cref{fig_mejec}.
A finite value of $r_{\rm ej}$ is chosen to ensure the correct
identification of expanding material, however, as visible in \cref{fig_mejec}, the choice of $r_{\rm ej}=100$~km is equivalent to applying no radius criterion at all. The similar, nearly parallel, line shape for different values of $r_{\rm ej}$ indicates the robustness of using a low value of $r_{\rm ej}$ in the sense that a larger value of $r_{\rm ej}$ would count the same ejecta material but just at a correspondingly later time $t_{\rm ej}$ (the time needed for material to reach that particular  $r_{\rm ej}$). We indeed verified that, for model DD2-135-135, nearly the same mass elements are counted when using $t_{\rm ej}=10$~ms and $r_{\rm ej}=220$~km instead of 7~ms and 100~km, respectively. 
Note that the time $t_{\rm ej}$ is chosen in the middle of the shallow, nearly linear growth since at later times the ejecta are more likely to be affected by magneto-hydrodynamical and secular effects in the merger remnant. The exact physical reasons behind such a continuous increase in the mass loss has not been determined, but it could be associated with various mechanisms, such as a growing 
$m=1$ mode \citep{nedora2019}, neutrino-driven winds, or viscous effects and angular momentum redistribution in the merger remnant.
The ejected masses for the different models are summarized in \cref{tab:astromods}. 
See \cref{sect_comp} for a comparison to other works.

For the NS-NS and NS-BH merger systems, a delayed collapse of the HMNS remnant to a BH is expected to take place soon after the dynamical merger phase, i.e., some $\sim10$~ms after the stars touch.
Although the gravitational collapse did not yet occur at the end of the simulations at about $\sim10$~ms after merging, we do not consider ejecta launched from the HMNS on time-scales longer than $\sim10$~ms in this study. Even though the amount of matter ejected during the HMNS evolution can, in principle, be comparable to the other ejecta components, the ejecta masses and other properties still carry large uncertainties, mainly because the magnetohydrodynamic effects responsible for angular momentum transport are yet poorly understood \citep[see, e.g.,][for discussions of HMNS ejecta]{perego2014,Fujibayashi2018,Palenzuela2022,Shibata2021b}.
In our cases, the HMNS phase will be relatively short. Thus, neglecting the HMNS ejecta is an acceptable approximation and is assumed to introduce uncertainties at the same level as the uncertainties presently existing in hydrodynamical simulations.  

\subsection{Simulations of BH-torus remnant systems}
\label{subsec_BH-t_sim}


The simulations of the BH-torus systems are performed with a finite-volume neutrino-hydrodynamics code described in \citet{Obergaulinger2008} and \citet{Just2015b}, which employ Newtonian hydrodynamics with a modified gravitational potential \citep{Artemova1996} to approximately model some relativistic effects like an innermost stable circular orbit.
The discs are modelled assuming axisymmetry and use $416\times 160$ grid cells to cover the domain $(r,\theta) \in [10\ \mathrm{km},2.5
\cdot 10^4\ \mathrm{km}] \times [0,\theta]$. 
To initialize the fluid configuration, the initial electron fraction is set to $Y_e=0.1$ (roughly guided by post-merger results) everywhere in the disc, and the volume surrounding the torus is filled with an ambient medium that has 1.5 times the floor density, see \citet{Just2015} for more details. 
The baryonic species are assumed to be in NSE and the neutrino transport is described by a truncated two-moment scheme \citep{Just2015b} so that $\beta$-processes with free nucleons and scattering of neutrinos off free nucleons as well as annihilation of neutrino-antineutrino pairs are included.
The material which cross the radius of $10^4$~km from the BH within the simulation time is defined as ejected from the system (see Fig.~7 in \citet{Just2015} for more details).

For the post-merger phase, we used a BH mass of 3~\Msun\ and torus masses of 0.1 and 0.3~\Msun, named models M3a8m1a5 and M3a8m3a5-v2 in \citet{Just2015}.
To include the effects of turbulent angular momentum transport, two $\alpha$-viscosity approaches called ``type 1'' and ``type 2'' are applied in model M3a8m1a5 and M3a8m3a5-v2, respectively. See \citet{Just2015} for a detailed discussion of the simulations and the adopted parameters. 

\begin{landscape}
\begin{table}[p]
\centering
\caption[Summary of the ejecta properties for the seven hydrodynamical models considered.]{Summary of the ejecta properties for the seven models considered in the present study: the total ejected mass $M_{\rm ej}$, total number of trajectories $N_\mathrm{tot}$, number of trajectories in the subset $N_\mathrm{sub}$, fraction of outflow mass in the subset $M_{\mathrm{sub}}/M_{\mathrm{ej}}$, the mass of the high-velocity ejecta $M_{\rm ej}^{v \ge 0.6c}$, the weighted mean velocity $\langle v/c \rangle$, the weighted mean $Y_e$ and the weighted mean mass fraction of free neutrons $X_n^0$. 
The three latter quantities have been extracted when the reaction network is initiated, i.e., at the density $\rho=\rho_{\rm net}$ defined in \cref{sec_init_cond}. When calculating the weighted mean values, the mass of each trajectory is used as weight (see \cref{eq:mass_avr} for more details).
A trajectory subset have only been extracted for the models which are included in the nuclear uncertainties study (\cref{ch_nucuncert}), i.e., all models excluding those using the DD2 EoS. 
}
\begin{tabular}{lcccccccc}
\hline \hline
Model & $M_{\rm ej}$ & $N_\mathrm{tot}$ & $N_\mathrm{sub}$ & $\frac{M_{\mathrm{sub}}}{M_{\mathrm{ej}}}$ & $M_{\rm ej}^{v \ge 0.6c}$ & $\langle v/c \rangle$ & $\langle Y_e\rangle$ & $\langle X_n^0 \rangle$ \\
& [$10^{-3}$~\Msun] & & & & [$10^{-4}$~\Msun]  &   &   &   \\
\hline
  DD2-125-145 & 3.20 & 1263 & - & - & 1.77   & 0.25   & 0.22 & 0.58 \\
\hline
  DD2-135-135 & 1.99 & 783 & - & - & 0.87   & 0.25   & 0.27 & 0.53 \\
\hline
 SFHo-125-145 & 8.67 & 4398 & 266 & 0.15 & 2.56   & 0.24   & 0.24 & 0.56 \\
\hline
 SFHo-135-135 & 3.31 & 1290 & 256 & 0.24 & 1.53  & 0.29   & 0.26 & 0.54 \\
\hline
SFHo-11-23   & 40.37 & 13175 & 150 & 0.01 & 2.7 & 0.17 & 0.04 & 0.85  \\ 
\hline
M3A8m1a5     & 23.26 & 4150 & 296 & 0.50 & 0 & 0.05 & 0.23 & 0.58  \\ 
\hline
M3A8m3a5-v2  & 70.13 & 2116 & 177 & 0.20 & 0.38 & 0.05 & 0.24 & 0.56  \\  
\hline \hline
\end{tabular}
\label{tab:astromods}
\end{table}
\end{landscape}

\subsection{Time evolution of the thermodynamic quantities}
\label{subsec_hydro_evo}

\begin{figure}[tbp]
\begin{center}
\includegraphics[width=0.5\textwidth]{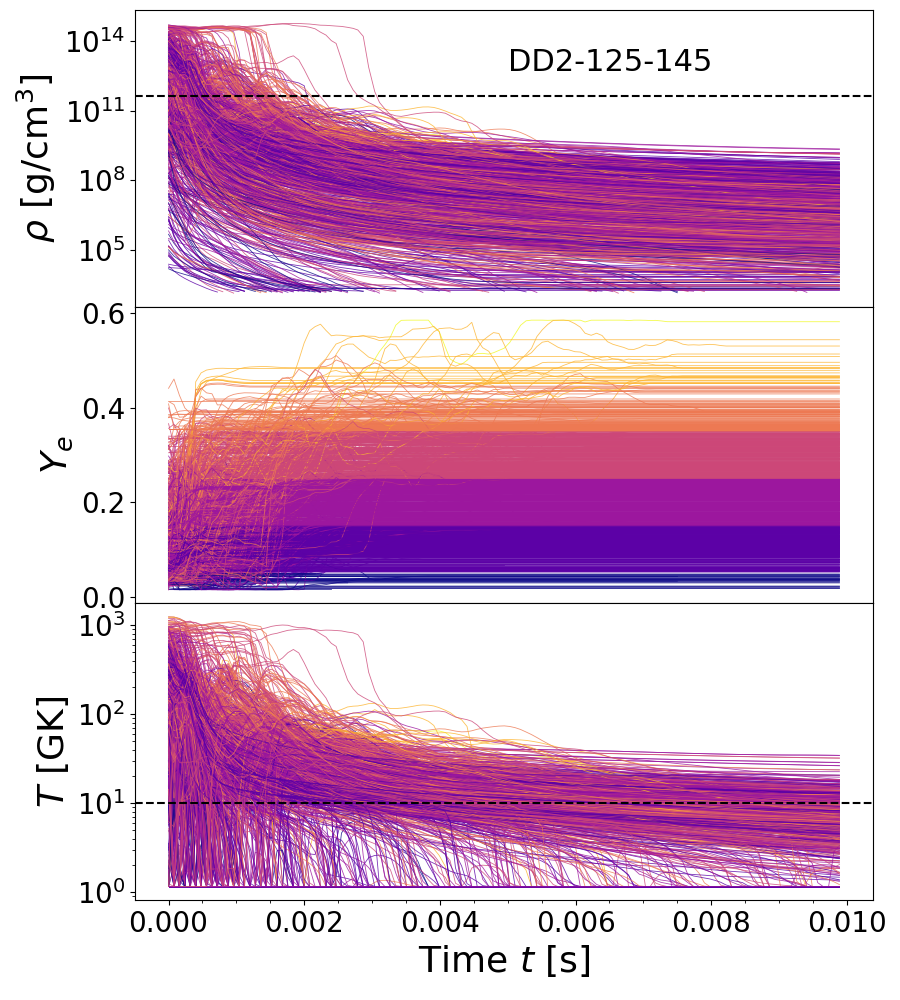}%
\includegraphics[width=0.5\textwidth]{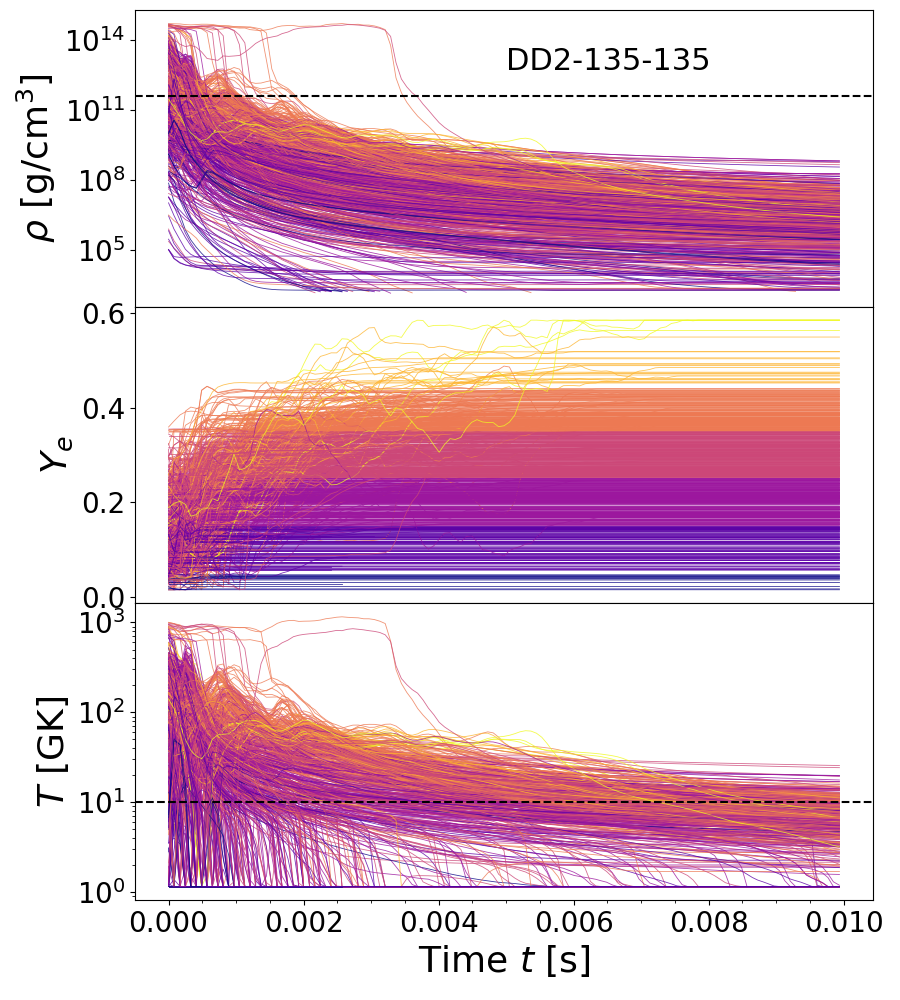}
\caption[Density, electron fraction and temperature versus time for all trajectories of the NS-NS models using the DD2 EoS.]{
Density $\rho$ (top), electron fraction $Y_e$ (middle) and temperature $T$ (bottom) versus time for all trajectories of the NS-NS models using the DD2 EoS: DD2-125-145 (left) and DD2-135-135 (right). 
The line colours correspond to the $Y_e$-value at the last time step taken by the hydrodynamical simulation at $t\sim10$~ms (after this time an extrapolation is performed, see \cref{sec_init_cond}).
The black dotted lines indicate $\rho=\rho_\mathrm{drip}$ and $T=10$~GK relevant for the definition of the network initiation time defined in \cref{sec_init_cond}, i.e., the r-process network will be initiated when the trajectory falls below both lines. \todo{merge the y-axes horizontally, add inward ticks etc!?}
}
\label{fig_hyd_trajdd2}
\end{center}
\end{figure}

\begin{figure}[tbp]
\begin{center}
\includegraphics[width=0.5\textwidth]{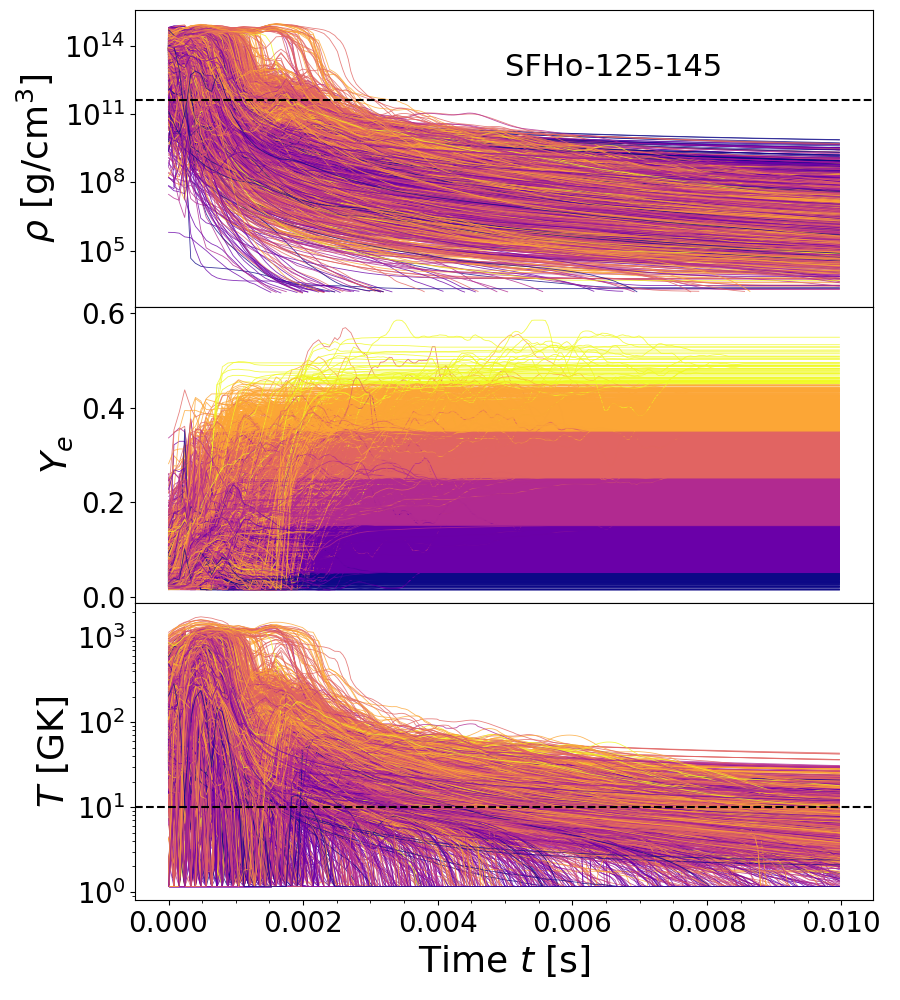}%
\includegraphics[width=0.5\textwidth]{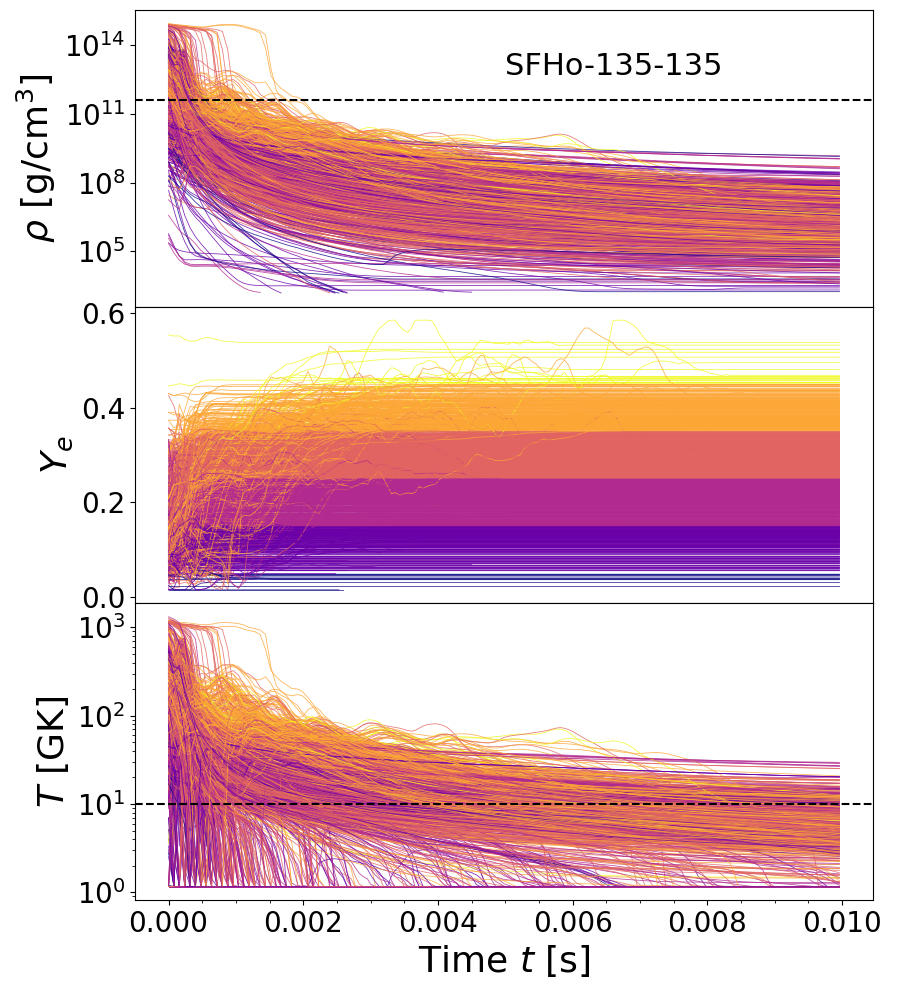}
\caption[Density, electron fraction and temperature versus time for all trajectories of the NS-NS models using the SFHo EoS.]{
Same as \cref{fig_hyd_trajdd2} for the NS-NS models using the SFHo EoS: SFHo-125-145 (left) and SFHo-135-135 (right). 
}
\label{fig_hyd_trajsfho}
\end{center}
\end{figure}

\begin{figure}[tbp]
\begin{center}
\includegraphics[width=0.5\textwidth]{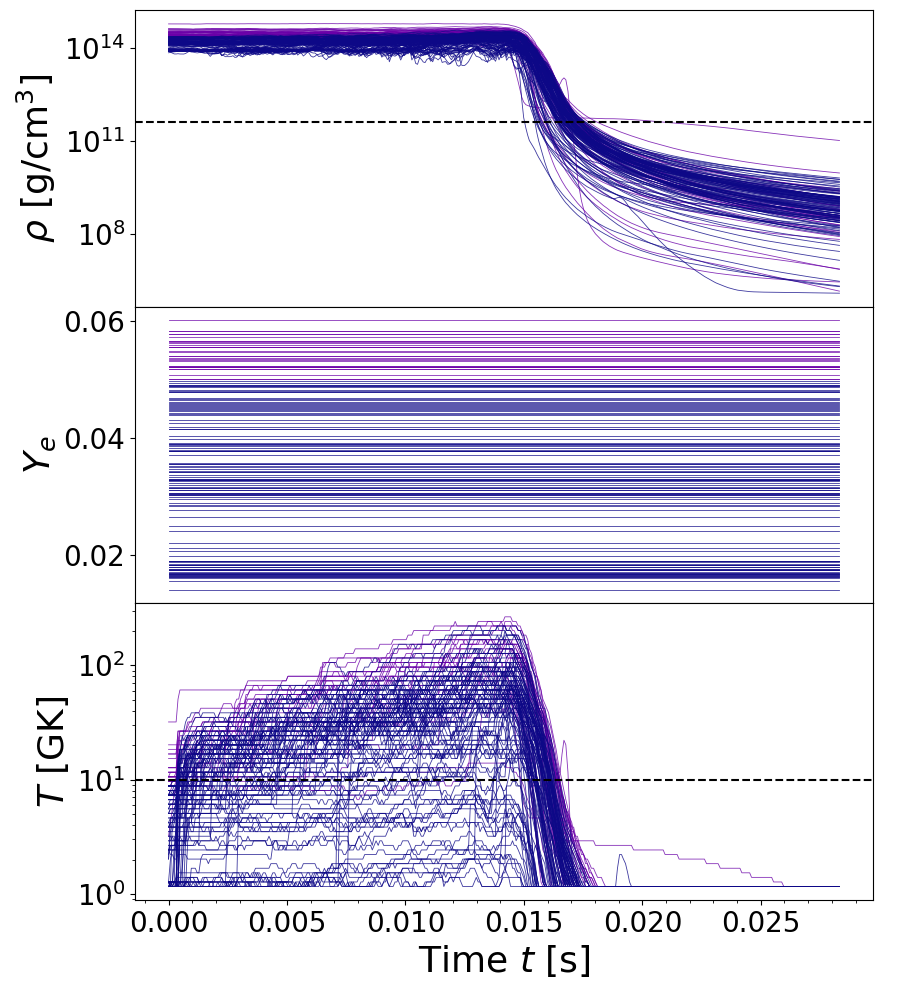}
\caption[Density, electron fraction and temperature versus time for the trajectories of the NS-BH dynamical ejecta model.]{
Same as \cref{fig_hyd_trajdd2} for 150 NS-BH dynamical ejecta trajectories from model SFHo-11-23. 
}
\label{fig_hyd_trajsbhns}
\end{center}
\end{figure}

The time evolution of the density, electron fraction and temperature for each trajectory during the simulation time of the hydrodynamical models for the NS-NS mergers are plotted in \cref{fig_hyd_trajdd2,fig_hyd_trajsfho}.
The colour of each line highlights the electron fraction at the last time step (i.e., the values shown in \cref{fig_xn0_ye_distr}). 
It is difficult to see the individual history of each trajectory since one to several thousand trajectories are plotted at once, however, some general trends become apparent (see also the discussion in \cref{sec_init_cond} about the conditions at the time the r-process network is initiated). 
As expected, both the temperatures and densities start at initially high values and generally fall off. However, many trajectories experience shocks which lead to an increase in the temperature (typically within the first few ms). The hydrodynamical simulations for the NS-NS mergers have a lower minimum temperature set at $\sim1$~GK, and a few (typically less than 10) trajectories start the simulation with very low densities (see  \cref{sec_init_cond} for more details). 
For all models in \cref{fig_hyd_trajdd2,fig_hyd_trajsfho}, the electron fraction is affected by neutrino reactions (\cref{eq:betareac}) which lead to rapid changes in the $Y_e$-value within the first few ms before it flattens out to a constant value when the trajectory has moved sufficiently far away from the neutrino source (i.e., the NS remnant).
As noted in \cref{subsec_dynsim}, no neutrino reactions are included in the NS-BH model SFHo-11-23, and thus the electron fraction remains constant throughout the hydrodynamical simulation, as shown in \cref{fig_hyd_trajsbhns}.
In contrast to the NS-NS merger models in \cref{fig_hyd_trajdd2,fig_hyd_trajsfho}, the density evolution of the NS-BH trajectories show little diversity in terms of evolutionary tracks where all trajectories start with initial high densities $\rho\sim10^{14}$~g/cm$^3$, and fall off after $t\sim0.15$~s. For the temperature a larger spread is found, however the temperature drops down towards $T\sim1$~GK after $t\sim0.015$~s for all trajectories. 

\begin{figure}[tbp]
\begin{center}
\includegraphics[width=0.5\textwidth]{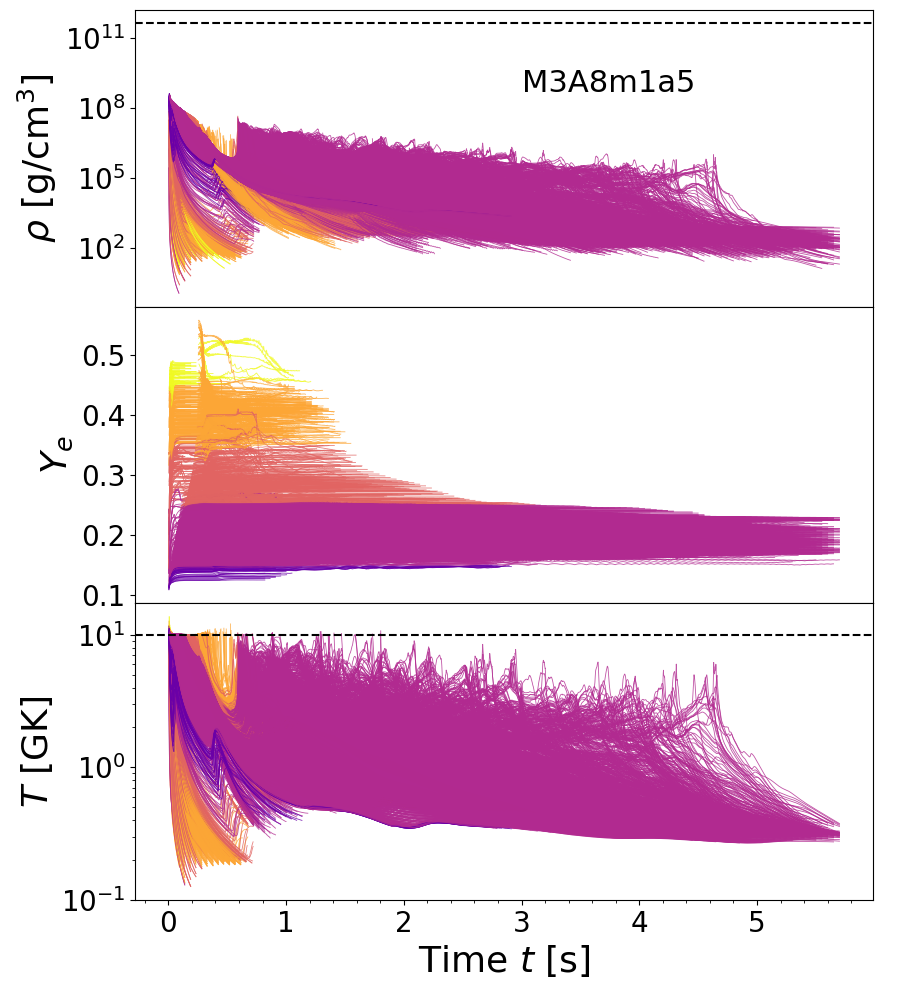}%
\includegraphics[width=0.5\textwidth]{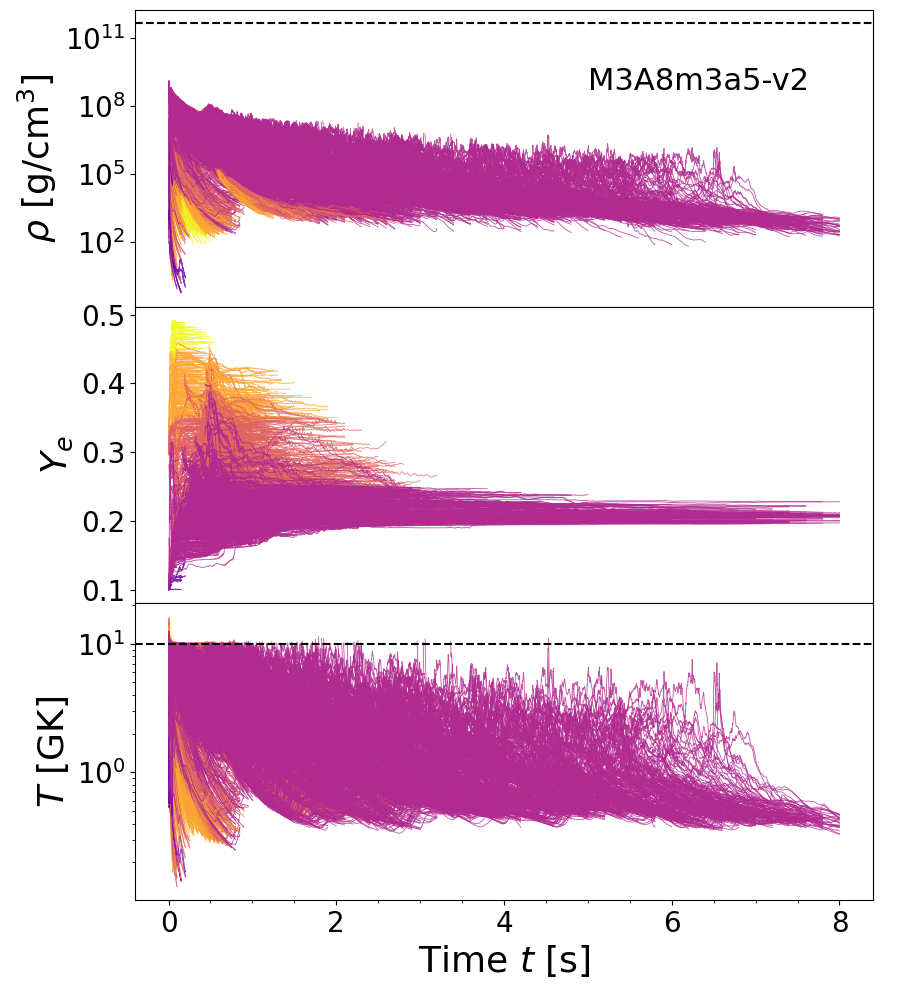}
\caption[Density, electron fraction and temperature versus time for all trajectories of the BH-torus models.]{
Same as \cref{fig_hyd_trajdd2,fig_hyd_trajsfho} for the BH-torus models: M3A8m1a5 (left) and M3A8m3a5-v2 (right). 
\todo{add inward ticks etc!}
}
\label{fig_hyd_wm}
\end{center}
\end{figure}

The time evolution of the density, electron fraction and temperature is shown in \cref{fig_hyd_wm} for the BH-torus models.
We can see that the BH-torus system is simulated over longer time scales compared to the NS-NS merger models in \cref{fig_hyd_trajdd2,fig_hyd_trajsfho}, and the initial density of the torus has similar order or magnitude as the dynamical ejecta after a few ms. 
The BH-Torus trajectories start at temperatures $T\sim10$~GK. Therefore, the first time step in \cref{fig_hyd_wm} will be at the network initiation time defined in \cref{sec_init_cond}.


Each trajectory in \cref{fig_hyd_trajdd2,fig_hyd_trajsfho,fig_hyd_trajsbhns,fig_hyd_wm} represents a given chunk of ejected material with a unique mass and therefore contributes to the total ejecta mass. 
To study how r-process elements synthesized by a given ejecta model contribute to the enrichment of the Galaxy, we must relate all ejecta properties (and therefore also r-process yields) from each trajectory to its corresponding contribution to the total ejecta. 
In this context, whenever discussing the mean of a given quantity $K$ we will apply the mass-weighted mean defined as:
\begin{equation}
\langle K \rangle = \frac{\sum_j K_j m_j}{\sum_j m_j},
\label{eq:mass_avr}
\end{equation}
where $m_j$ is the mass of trajectory $j$ and the sum runs over all trajectories considered. 
This ensures that when displaying the distribution of an ejecta quantity, the trajectories which contribute the most to the Galactic enrichment are highlighted. 
Some properties of the ejecta are summarized in \cref{tab:astromods}, including the total ejected mass, the number of trajectories, the mass of the high-velocity component (i.e., with $v\ge 0.6c$), and the mass-averaged velocities, electron fractions and initial mass fractions of neutrons. 

\todo{[: why are the simulation times varying so much between the trajs? same criteria for rho\_net? start at t=0 ]}

\subsection{Combining dynamical and secular ejecta}
\label{subsec_combsim}

\begin{table*}
\centering
\caption[Parameters for the combined ejecta models.]{Parameters of the models which will be combined into the total ejecta: model name, mass ratio $q=M_1/M_2$ for the NS masses $M_1$ and $M_2$, BH mass, BH spin and torus mass.
}
 \begin{tabular}{lcccc}
 \hline  \hline 
 Model name & $q$ &  $M_{\mathrm{BH}}$ & $A_{\mathrm{BH}}$ & $M_{\mathrm{torus}}$  \\ 
                       &  & [\Msun] & & [\Msun] \\ 
 \hline 
 SFHo-125-145 & 0.86 & 2.40 & 0.80 & 0.17 \\ 
 SFHo-135-135 & 1.00 & 2.45 & 0.83 & 0.09 \\ 
 \hline
 SFHo-11-23  & 0.48  & 3.09 & 0.82 & 0.26 \\ 
 \hline
 M3A8m1a5    & -     & 3.00 & 0.80 & 0.10 \\ 
 M3A8m3a5-v2 & -     & 3.00 & 0.80 & 0.30 \\ 
 \hline  \hline 
 \end{tabular} 
\label{tab_astro_cmods}
\end{table*}

Given the challenge to obtain fully consistent models covering the entire merger evolution and the large uncertainties related to the viscosity of the disk models, we have chosen two BH-torus systems that approximately match the configurations obtained after the merger.
To consistently use the same EoS for the combined ejecta models, only two NS-NS models from \cref{subsec_dynsim} are included, i.e., those that use the SFHo EoS. 
We combine model SFHo-125-145 with M3A8m3a5-v2, SFHo-135-135 with M3A8m1a5 and SFHo-11-23 with M3A8m3a5-v2.
\cref{tab_astro_cmods} lists the relevant models and their properties, including the BH and torus mass. 
For the models covering the dynamical ejecta, the values refer to those extracted at the end of the simulation time ($t\sim10$~ms after merging).
In these simulations, the merger remnant did not yet collapse to form a BH, and the torus mass is estimated by determining which fraction of matter is rotationally supported, assuming a collapse took place (see \citet{oechslin2007} for details). 
As can be seen in \cref{tab:astromods}, the total outflow mass of the dynamical ejecta in the NS-NS models are 7-20 times less massive than the BH-torus ejecta; hence the total r-process abundance distribution is dominated by the BH-torus component. In contrast, in the NS-BH system, the dynamical and secular ejecta masses are of the same order of magnitude.

\section{Initial conditions and abundances}
\label{sec_init_cond}

\epigraph{\itshape \enquote{Head Waiter: This is a vegetarian restaurant -- we serve no meat of any kind. We're not only proud of that, we're smug about it.}}{--- \textup{Monty Python's Flying Circus}}

\begin{figure}[tbp]
\begin{center}
\includegraphics[width=0.99\textwidth]{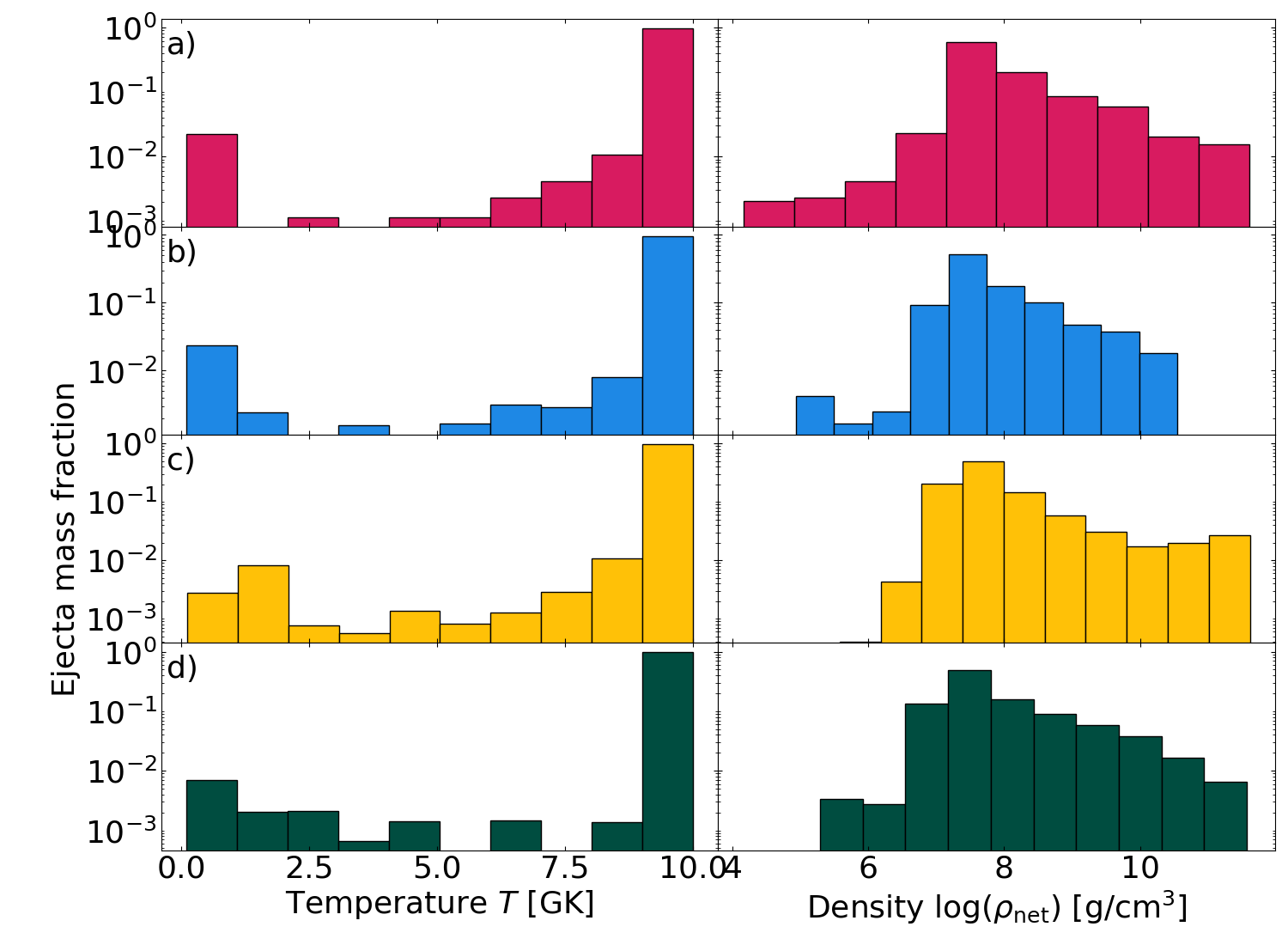}
\includegraphics[width=0.99\textwidth]{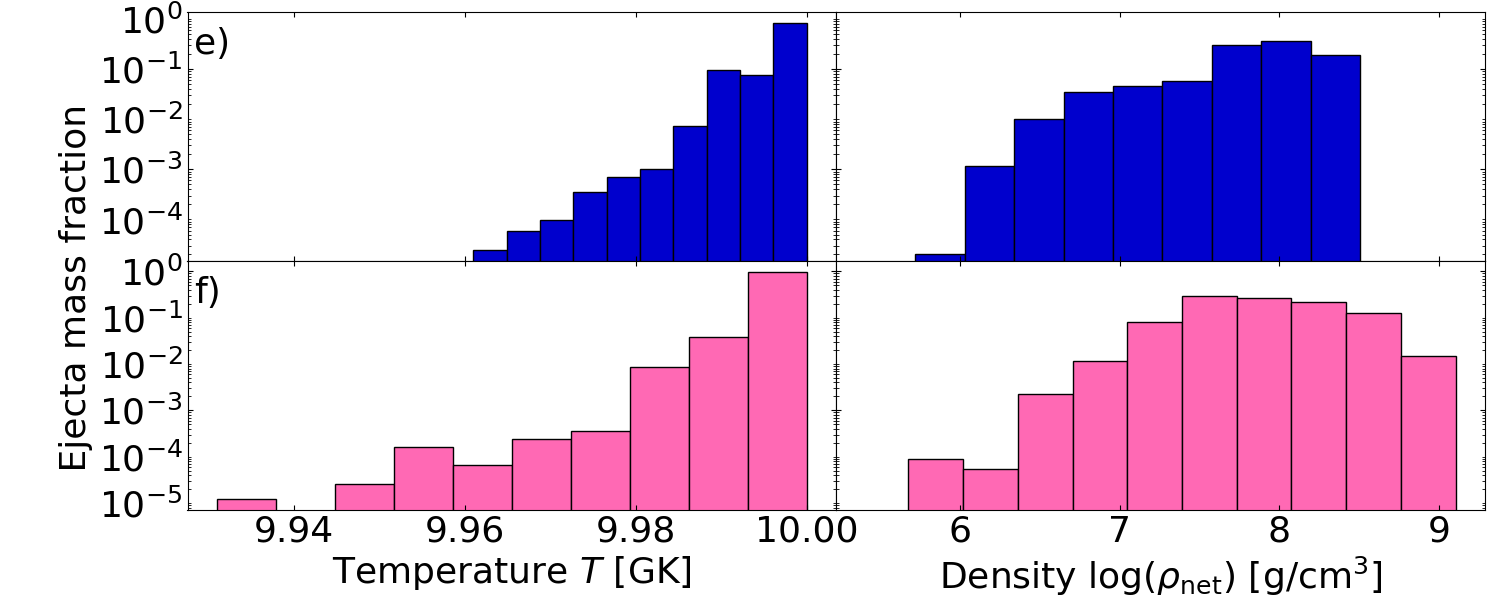}
\includegraphics[width=0.99\textwidth]{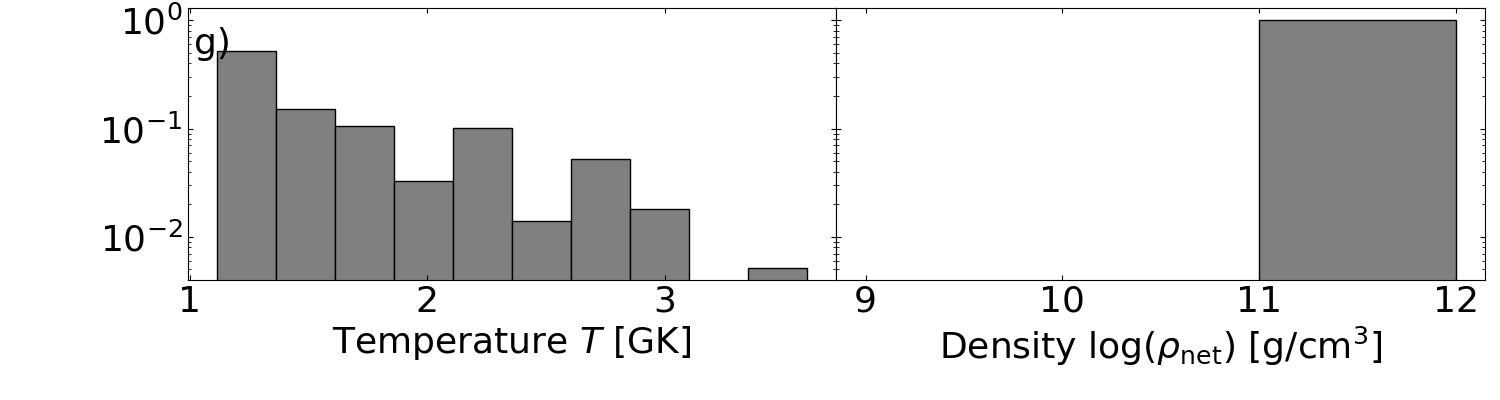}
\caption[Temperature and density mass distributions at the network initiation time.]{
Fractional mass distributions of the matter ejected as a function of the temperature $T$ (left column) and density $\rho_\mathrm{net}$ (right column) at the network initiation time for the models: a) DD2-125-145, b) DD2-135-135, c) SFHo-125-145, d) SFHo-135-135, e) M3A8m1a5, f) M3A8m3a5-v2 and g) SFHo-11-23.
Note that $\rho_\mathrm{net}=\rho_\mathrm{drip}$ for all trajectories of NS-BH model SFHo-11-23.
}
\label{fig_distr_t9_rho}
\end{center}
\end{figure}
\begin{figure}[tbp]
\begin{center}
\includegraphics[width=\textwidth]{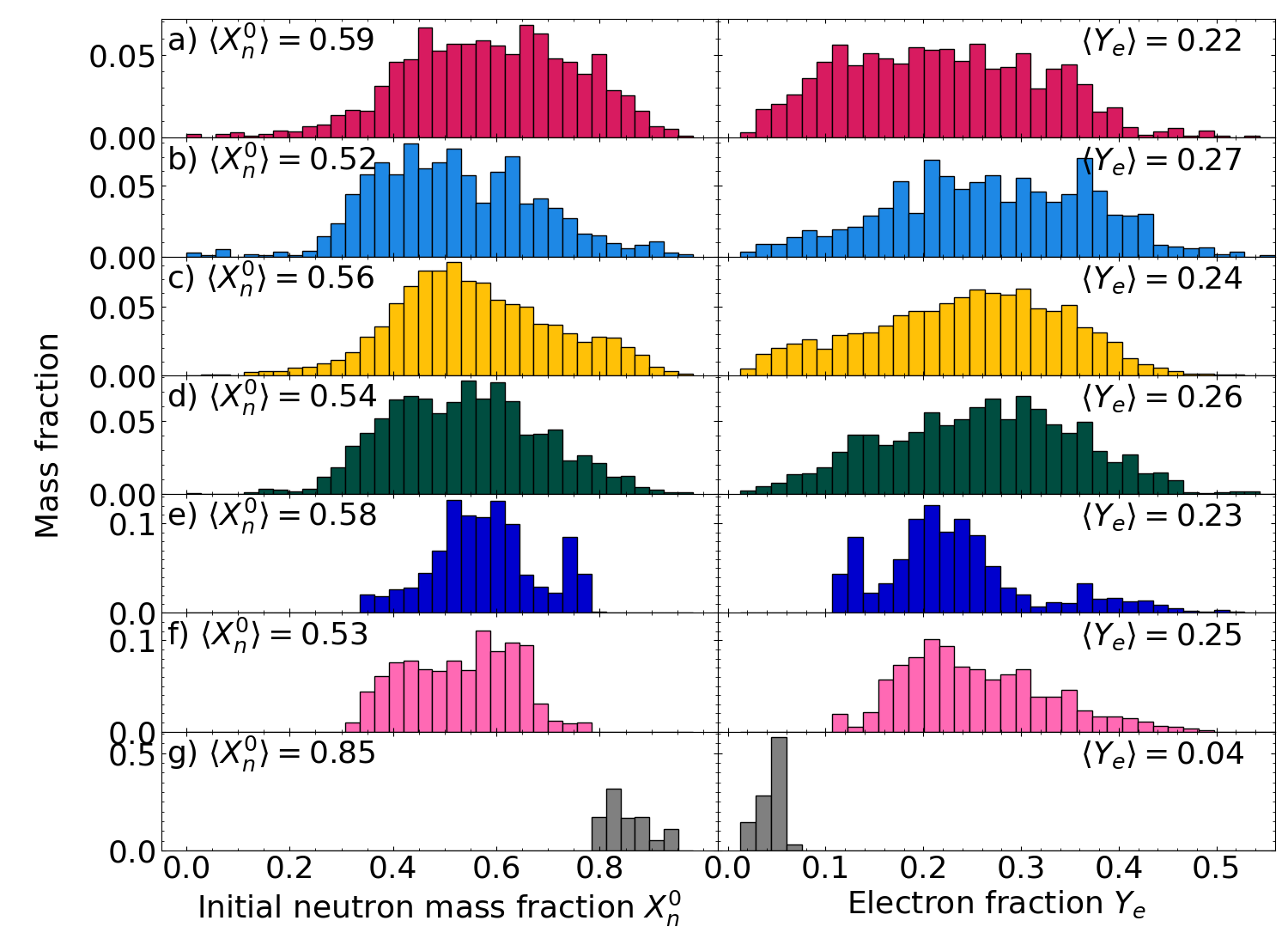}
\caption[$X_n^0$ and $Y_e$ mass distributions at the network initiation time.]{
Fractional mass distributions of the matter ejected as a function of the mass fraction of free neutrons $X_n^0$ (left column) and electron fraction $Y_e$ (right column) at the network initiation time. The mean electron fraction $\langle Y_e \rangle$ and the mean initial neutron fraction $\langle X_n^0 \rangle$ are given for each merger and BH-torus models: a) DD2-125-145, b) DD2-135-135, c) SFHo-125-145, d) SFHo-135-135, e) M3A8m1a5, f) M3A8m3a5-v2 and g) SFHo-11-23.
}
\label{fig_xn0_ye_distr}
\end{center}
\end{figure}

The astrophysical simulations provide the detailed evolution of, among other variables, the temperature, density and entropy of the ejecta up to a few tens of ms or seconds, depending on the model. After that, the ejecta is assumed to expand according to a homologous expansion, i.e., with a constant velocity, 
and the densities of the ejecta clumps decrease proportionally to $1/t^3$ (see \cref{sec_extrap_sens} for sensitivity tests of the r-process results to the assumed expansion model). 
Therefore, what we refer to as trajectories in the following contain the time evolution predicted by the hydrodynamical simulations for the first few ms or seconds plus the subsequent free expansion followed up to one year. 

We define the network initiation time to be the first time when both the temperature and density of a given trajectory drop below the threshold values of $T=10\times 10^{9}$~K and $\rho_\mathrm{drip}\simeq 4.2\times 10^{11}$~g~cm$^{-3}$, respectively. 
Depending on the specific history of the trajectory followed and the astrophysical scenario modelled (i.e., \cref{fig_hyd_trajdd2,fig_hyd_trajsfho,fig_hyd_wm}), the temperature may already be lower than 10~GK at the first time step, and in these cases, the network calculation starts as soon as the density drops below the drip density $\rho_\mathrm{drip}$.
The opposite may also be true (in particular for the BH-torus models) where the densities start initially below $\rho_\mathrm{drip}$, thus the network will be initiated when the temperature falls below 10~GK.
The density at which the full reaction network is initiated is referred to as $\rho_\mathrm{net}$.
In the following, whenever discussing properties related to the r-process, ``initial'' indicates the network initiation time (defined here at $\rho_\mathrm{net}$), not the conditions at the start of the hydrodynamical simulations.

Mass weighted histograms of the temperature and density at the network initiation time are shown in \cref{fig_distr_t9_rho} for all models (note that the $y$-axis is logarithmic to show all bins of the distributions). The density $\rho_\mathrm{net}$ is in particular high for the NS-BH dynamical ejecta where all trajectories have  $\rho_\mathrm{net}\sim \rho_\mathrm{drip}$. For the other models, i.e., the NS-NS dynamical ejecta or the BH-torus models the density at the network initiation time is generally high, where only a few trajectories representing less than five per cent of the ejected mass have $\rho_\mathrm{net}<10^{-7}$~g/cm$^3$. 
Similarly, less than three per cent of the NS-NS dynamical ejecta have temperatures below 5~GK. All trajectories of the BH-torus models initiate the network at temperatures close to 10~GK while the opposite is true for the NS-BH trajectories which all have $T<5$~GK. 

As described in \cref{sec:astro_mods}, the electron fraction $Y_e$ is followed by the network and its value at $\rho_\mathrm{net}$ corresponds to the value estimated by the hydrodynamical model.
The hydrodynamical simulations for the NS-NS dynamical ejecta (i.e., the ILEAS simulations including weak interactions in \cref{subsec_dynsim}) lasts for about 10~ms after merger. The evolution of the $Y_e$ rapidly reaches a plateau before $\rho_\mathrm{net}$ (which for most trajectories are before the end of the simulation, e.g., see \cref{fig_hyd_trajdd2,fig_hyd_trajsfho}) showing that the weak interactions of nucleons (\cref{eq:betareac}) no longer play a significant role. For this reason these neutrino reactions are neglected after $\rho_\mathrm{net}$ and during the extrapolation. 
Generally $\rho_\mathrm{net}$ is reached before the end of the ILEAS simulations, however for the few trajectories where this is not the case the $Y_e$ value at 10~ms after merger is adopted at $\rho_\mathrm{net}$. 

The ejecta distributions of the electron fraction $Y_e$ and mass fraction of free neutrons $X_n^0$ at the network initiation time (see \cref{subsec_NSE} for the calculation of $X_n^0$) for the seven merger and remnant models are shown in \cref{fig_xn0_ye_distr}, where the masses have been integrated over the simulated evolution time and normalized to the total ejected mass for each individual model. 
The initial neutron mass fraction $X_n^0$ provides a better rough indicator of the efficiency of the r-process than the $Y_e$ alone \citep[which is commonly used in the literature when discussing the conditions characterizing the r-process efficiency, e.g., see][]{Lemaitre2021}. 
By definition, a lower $Y_e$ gives a larger $X_n^0$; however, a higher initial entropy will also give a larger $X_n^0$ due to the release of free neutrons through photodissociation \citep[see, for example,][for a discussion of the interplay between the $Y_e$, entropy, expansion time-scale and a successful r-process]{meyer1989,hoffman1997,Otsuki2000}.
For merger ejecta conditions with similar expansion time scales, it is only possible to produce the heaviest elements with a high $X_n^0$, while it is not always safe to conclude that heavy elements are only produced for $Y_e\lesssim 0.25$ because it is possible to have a successful r-process for $Y_e\gtrsim 0.25$ if the entropy is sufficiently high. 
Therefore, for the remainder of this thesis either the $X_n^0$ or the $Y_e$-value at the network initiation time will be used when discussing the ejecta composition.

Each of the distributions in \cref{fig_xn0_ye_distr} are used as initial conditions to estimate the final composition of the ejecta through a full reaction network (see \cref{sec:nuc_net}). 
As can be seen in \cref{fig_xn0_ye_distr}(g), the entire $X_n^0$ distribution lies above 0.8 for the NS-BH ejecta, which contrasts to the wide $X_n^0$ distribution obtained in the NS-NS simulations (\cref{fig_xn0_ye_distr}(a-d)) that include neutrino interactions (see below for a comparison to a ``no-neutrino'' case). Therefore, the NS-BH scenario is well suited to study the impact of fission uncertainties on the final r-process abundance distribution since the r-process flow will reach the fissile region for most of the ejected material in this astrophysical scenario \citep[see][for a discussion of fission and the relation to the $X_n^0$ value]{Lemaitre2021}.
The BH-torus ejecta generally has larger $Y_e$-values than the dynamical ejecta, where the lowest value of the distributions is around $\sim 0.1$.
See also the summary of ejecta properties in \cref{tab:astromods}.

\subsubsection*{The no neutrino case}

\begin{figure}[tbp]
\centering
\includegraphics[width=0.8\columnwidth]{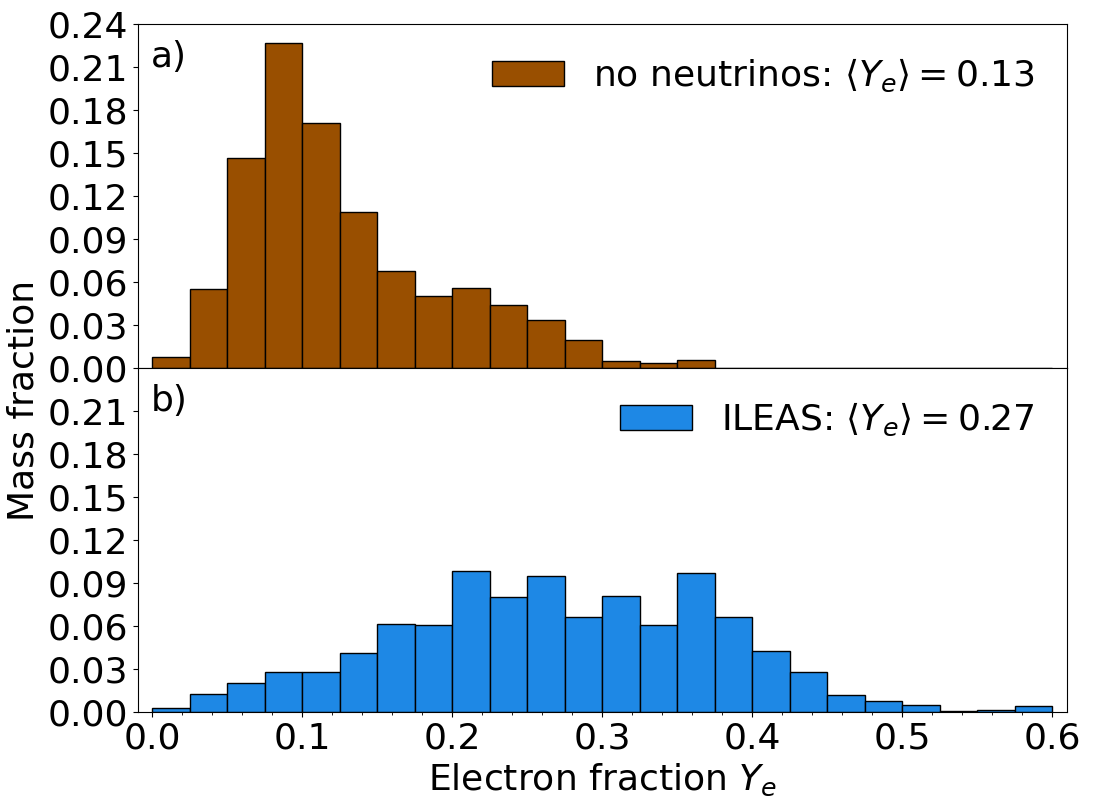}
\caption[Same as \cref{fig_xn0_ye_distr} for the ``no-neutrino'' case.]{Same as \cref{fig_xn0_ye_distr} for the two cases, without (a) or with (b) weak nucleonic interactions, of the DD2-135-135 model discussed in the text.  Note that the ILEAS case with neutrinos corresponds to \cref{fig_xn0_ye_distr}b.
  }
\label{fig_yedist_cases}
\end{figure} 

To study the impact of neglecting weak nucleonic reactions for the NS-NS dynamical ejecta, an approximation referred to as the `no neutrino' case was implemented.
This model follows the same density and temperature evolution as the original DD2-135-135 model, but considers the electron fraction at the time $t=0$, as estimated from the hydrodynamical simulation, and assumes no more weak interaction (\cref{eq:betareac}) with nucleons to take place afterwards. 
Thus, $Y_e$ is assumed to remain unaffected by nucleonic weak processes in all ejected mass elements after $t = 0$. 
In \cref{fig_yedist_cases}, the $Y_e$ distributions at $\rho=\rho_\mathrm{net}$ together with the mean values $\langle Y_e \rangle$ are displayed for the DD2-135-135 cases with and without neutrinos.  It should be noted that the $Y_e$ distribution without neutrinos does not correspond to the one found in cold NSs \citep[which was assumed in][]{goriely2011} since matter has been affected by neutrino absorption and emission between the start of the simulation and time $t=0$. In particular, a significant amount of ejected material has a $Y_e >0.1$, as seen in \cref{fig_yedist_cases}a.


\subsection{Nuclear statistical equilibrium }
\label{subsec_NSE}

\begin{figure}[tbp]
\begin{center}
\includegraphics[width=\columnwidth]{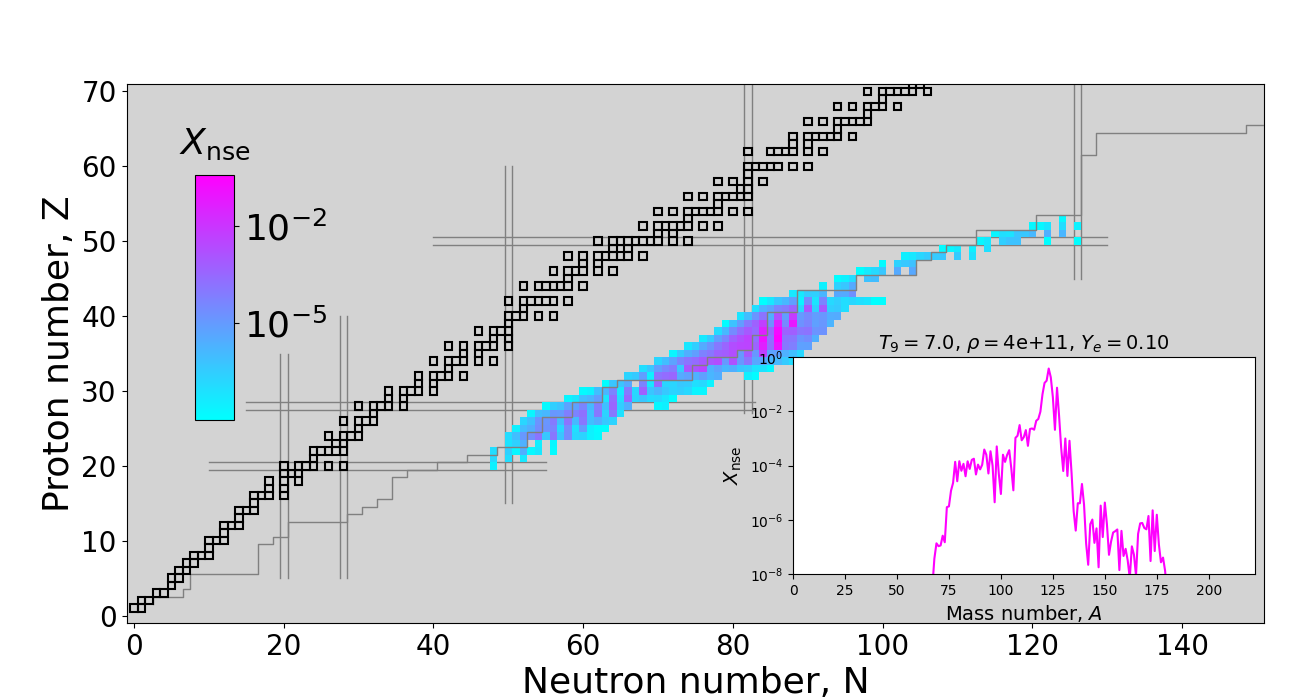}
\includegraphics[width=\columnwidth]{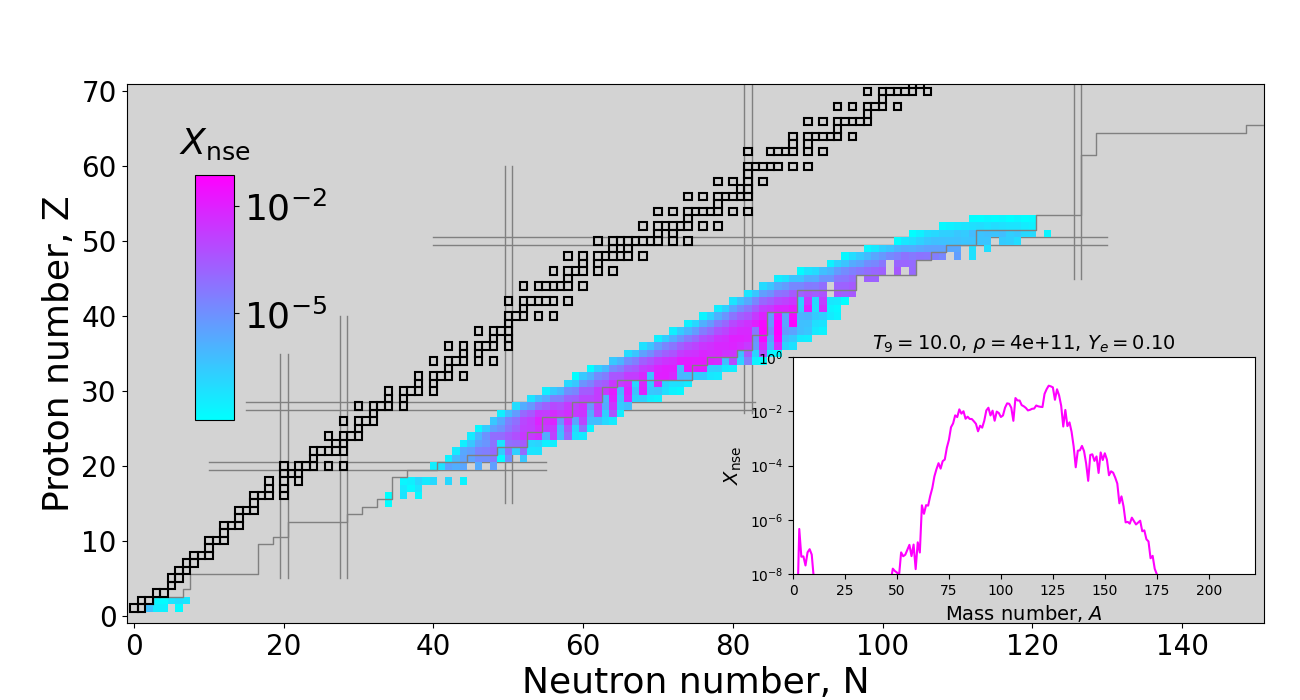}
\caption[NSE mass fractions in the $(N,Z)$ plane at two temperatures.]{
Representation in the $(N,Z)$ plane of the NSE mass fraction $X_\mathrm{nse}$ calculated with \cref{eq:Ynse} at $Y_e=0.1$ and $\rho=4\cdot 10^{11}$~g~cm$^{-3}$ for two temperatures: 7~GK (top) and 10~GK (bottom). 
The abundance $X_\mathrm{nse}$ versus mass number $A$ is also shown as an insert. 
Mass model HFB-21 (see \cref{sec_nucin} for details) is used for all NSE calculations and the corresponding neutron drip line is shown as a grey line to the right of the stable nuclei (black squares). 
The double solid lines depict the neutron and proton magic numbers. Note that abundances below $10^{-8}$ are not shown. 
}
\label{fig_xnse}
\end{center}
\end{figure}

Since the density and temperature at $\rho_\mathrm{net}$ are typically high, NSE can be used to estimate the initial network abundances.
Assuming that the nuclei follow Maxwell-Boltzmann statistics, the abundance of the nuclei in NSE can be calculated by \citep{goriely2011a,Bravo1999,Arcones2010}
\begin{equation}
Y_i = \frac{G_i(T)}{\rho/m_u }\Bigg( \frac{k_bTAm_u}{2\pi \hbar^2} \Bigg)^{3/2}  e^{(N\eta_\mathrm{n} + Z\eta_\mathrm{p})} e^{(B_i/k_bT)},
\label{eq:Ynse}
\end{equation}
where $G_i$ is the temperature dependent partition functions of the nuclei normalized to their ground state \citep[see][]{Aikawa2005,Arnould2006,goriely2008} defined as
\begin{equation}
G_i(T) = \frac{1}{2J_i^0 + 1} \sum_\mu (2J_i^\mu + 1) \exp{ \Big( -\frac{\epsilon_i^\mu}{k_bT} \Big)}.
\label{eq:partition}
\end{equation}
Here, the summation extends over the states $\mu$ of nuclei $i$ with excitation energy $\epsilon_i^\mu$ and spin-parity $J_i^\mu$ ($\mu=0$ for the ground state), and $k_b$ is the Boltzmann constant.
The nuclear binding energy of nucleus $i$ with mass $m_i$ is given by:
\begin{equation}
B_i = B(Z,A) = \Big( Zm_\mathrm{p} + Nm_\mathrm{n} - m_i \Big)c^2 + Z\mu_\mathrm{C,p} - \mu_\mathrm{C}(Z,A),
\label{eq:B_i}
\end{equation}
which includes the coulomb corrections to the chemical potentials $\mu_\mathrm{C,p}$ and $\mu_\mathrm{C}(Z,A)$ \citep[e.g., see][]{goriely2011a,Bravo1999,Haensel2007}\todo{read Haensel2007 (p.75 comment from nsedeg->lattice ilat=3)}
due to the interaction of nucleus $i$ with the electron background.
The inclusion of coulomb effects usually shifts the abundance distribution to favour heavier nuclei. \todo{see Bravo1999 reffers to Mochkovitch \& Nomoto 1986} 
Since neutrons become degenerate for $\rho \gtrsim 10^{11}$~g~cm$^{-3}$ \citep{Arcones2010}, the degeneracy parameters for neutrons and protons, $\eta_n$ and $\eta_p$, respectively, have been included in \cref{eq:Ynse}.  
Due to the high densities, we use Fermi-Dirac statistics to describe the number densities of both neutrons and protons,
\begin{equation}
n_i = \frac{8\pi \sqrt{2}}{h^3}m_i^3c^3\beta_i \big(\mathcal{F}_{1/2}(\eta_i,\beta_i) + \beta_i\mathcal{F}_{3/2}(\eta_i,\beta_i) \big),
\end{equation}
where $\beta=kT/mc^2$ is the relativistic parameter and $\mathcal{F}_k$ are the Fermi functions:
\begin{equation}
\mathcal{F}_k(\eta,\beta) = \int_0^\infty \frac{x^k(1 + \frac{1}{2}\beta x )}{e^{-\eta + x} + 1} \mathrm{d}x.
\end{equation}
Then, the degeneracy parameters $\eta_n$ and $\eta_p$ can be determined through charge neutrality and mass conservation:
\begin{equation}
\begin{aligned}
\sum Z_i Y_i &= Y_e, \label{eq:ch_mass_cons}\\
\sum A_i Y_i &= 1,
\end{aligned}
\end{equation}
for a given electron fraction $Y_e$. Thus, the NSE abundances for each trajectory are found by solving \cref{eq:Ynse,eq:ch_mass_cons}, which only depend on the density ($\rho_\mathrm{net}$), temperature ($T_{\rho_\mathrm{net}}$) and electron fraction ($Y_{e,\rho_\mathrm{net}}$) for the given trajectory. 
\cref{fig_xnse} shows an example of the abundances in the $(N,Z)$ plane under NSE conditions for two temperatures: 7~GK (top) and 10~GK (bottom), at constant density $\rho=4\cdot 10^{11}$~g~cm$^{-3}$ and electron fraction $Y_e=0.1$.  For both temperature cases, the most abundant nuclei are found around neutron magic number $N=82$, however, the distribution also reaches down to some $N=50$ nuclei for $T=10$~GK. 
Generally, for increasing temperatures (and constant $\rho$ and $Y_e$),  the $X_\mathrm{nse}(A)$ distribution widens even more, and an increasing number of light nuclei ($A\lesssim6$) are produced until the distribution is dominated by the lightest species. 
Similarly, higher densities tend to favour nuclei with larger mass numbers (for constant $T$ and $Y_e$), and generally higher (lower) $Y_e$ values shifts the distribution to less (more) neutron-rich nuclei (for constant $T$ and $\rho$). 
The NSE calculation also depend on the nuclear masses of the involved nuclei (see \cref{eq:Ynse,eq:B_i}). When experimental masses are not available, mass model HFB-21 (defined in \cref{sec_nucin}) is used for all NSE calculations. 

\begin{figure}[tbp]
\begin{center}
\includegraphics[width=\columnwidth]{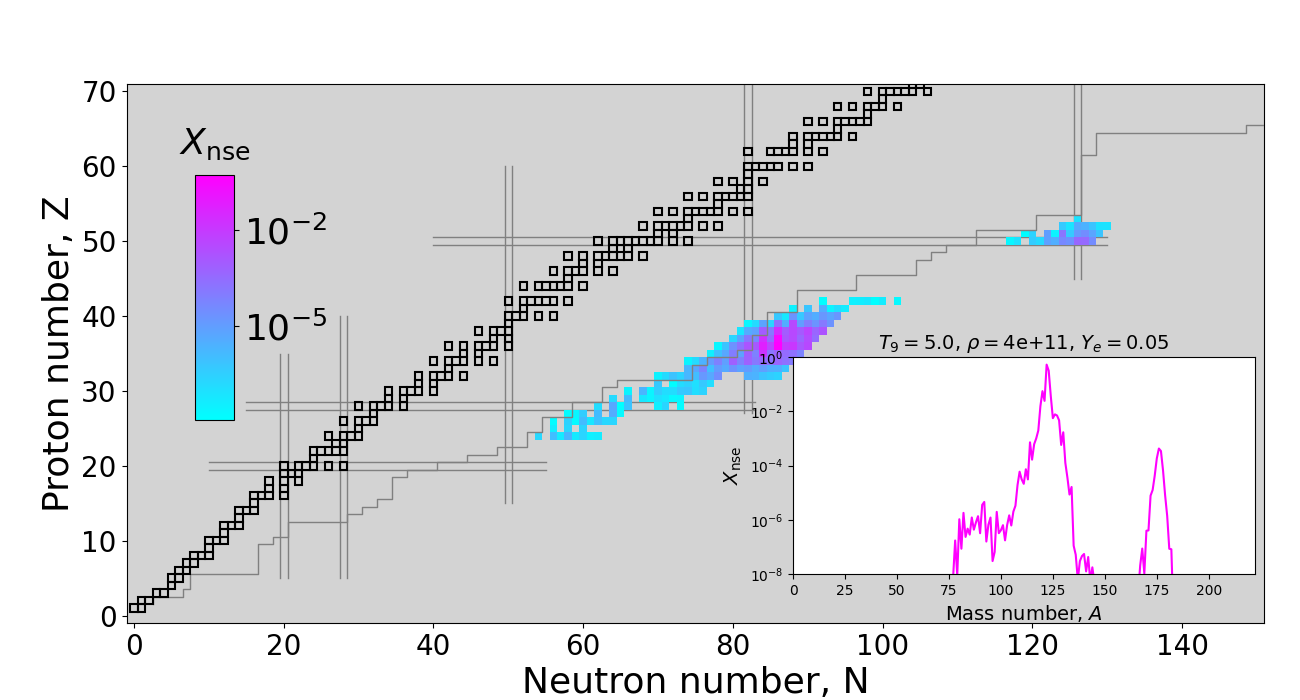}
\caption[NSE mass fractions in the $(N,Z)$ plane at low temperature and high density.]{
Same as \cref{fig_xnse} for initial conditions similar to those for the NS-BH merger model with $Y_e=0.05$ and $\rho=4\cdot 10^{11}$~g~cm$^{-3}$ where the minimum NSE temperature $T=5$~GK is assumed. 
}
\label{fig_xnse_lowT}
\end{center}
\end{figure}

To ensure high enough temperatures in the NSE calculation for all trajectories a minimum temperature of 5~GK is imposed. 
The impact of this approximation is small since the trajectories which have $T<5$~GK at $\rho_\mathrm{net}$ correspond to less than three per cent of the ejecta mass for the NS-NS merger models, and all BH-torus trajectories have high temperatures at the network initiation time. 
A different scenario is found for the NS-BH model where all trajectories have initial low temperatures ($T<5$~GK at $\rho_\mathrm{net}$) and high densities, see \cref{fig_hyd_trajsbhns,fig_distr_t9_rho}. Thus, this material originates from the inner NS crust and the NSE calculations will estimate an initial composition similar to what is found at the edge between the inner and outer NS crust (i.e., at $\rho_\mathrm{drip}$) as shown in \cref{fig_xnse_lowT} \citep[see also][for a discussion about NS crust NSE composition estimated at different temperatures.]{goriely2011a}.
The material outside of the neutron drip line in \cref{fig_xnse_lowT} is assumed to instantaneously emit neutrons before the network is started. 

\newpage
\section{Nuclear network simulations}
\label{sec:nuc_net}

\epigraph{\itshape ``Talk is cheap. Show me the code.''}{--- \textup{Linus Torvalds}}

For each of the hydrodynamical simulations introduced in \cref{sec:astro_mods}, a post-processing reaction network is used to calculate the composition and radioactive decay heat of the ejected material.
In \cref{sec:ab_evo} the equations which govern the time dependent abundance evolution will be discussed. The methods applied to calculate the nuclear heat generated by the radioactive r-process species will be discussed in \cref{sec:heat}. 
The nuclear physics input plays a major role in the r-process calculations and will be discussed separately in \cref{sec_nucin}.

\subsection{Abundance evolution and network equations}
\label{sec:ab_evo}

The nucleosynthesis is followed by a reaction network consisting of $\sim5000$ species, ranging from protons up to $Z\simeq110$ and including all isotopes from the valley of beta stability to the neutron drip line.
The location of the neutron drip line is defined by the applied mass model (see \cref{subsec_massmods}), and therefore the number of nuclei in the network vary with the nuclear physics input applied in the calculation.
For the mass models applied in this work the number of nuclei in the network ranges between 4808 and 4263. 
The relevant nuclear physics input for all involved reactions and nuclei in the network will be specified in \cref{sec_network_input}.
Some studies \citep[e.g.,][]{lippuner2017} also include nuclei on the proton rich side which increases the size of the network to $\sim6000$-8000 nuclei.
Elements with higher proton numbers than 110 are assumed to fission spontaneously with very short lifetimes. Therefore the production of elements with higher proton numbers is unlikely, and $Z=110$ is assumed to be a good termination point of the r-process network (see also the discussion about the so-called ``fission roof'' for $Z=110$ and $A\geq 130$ in \citet{Lemaitre2021}).
To reduce the computational demand of varying the nuclear physics input for all astrophysical models (i.e., \cref{ch_nucuncert}), some r-process calculations were run with $Z=100$ as the maximum limit for the network. 
\todo{repeat that this is mostly done for subset of NS-BH models..}
The impact of this limitation is tested in \cref{sec_netlimtol}, and found to be negligible while significantly reducing the CPU time. 
The network conserves the mass fraction and number of nucleons at all times, which includes the consistent recapturing of neutrons released through for example $(\gamma,n)$ reactions and fission.
If a neutron-rich isotopes is outside of the network limit (i.e., outside the neutron drip line) it is assumed to instantly emit neutrons to get back inside the drip line.
The abundances are followed by the full network up to $\sim 1$ year. After this time all unstable nuclei are assumed to instantaneously decay to their stable descendants, except for the long-lived Th and U isotopes. 
Thus, what we refer to as final abundances in the remainder of the thesis are those obtained after the decay of all radioactive species (except for the long-lived isotopes of Th and U).

Starting from the abundances at $\rho_\mathrm{net}$, the changes of individual nuclear abundances can be calculated by summing over all reactions that involve the production or destruction of that species.
In its most general form the molar fraction $Y_i$ of a nucleus $i$ can be evolved according to:
\begin{equation}
\begin{aligned}
\frac{\mathrm{d}Y_i}{\mathrm{d}t} =& -\sum_{j=i\ \mathrm{or}\ k=i} n_i \frac{Y_j^{n_j} Y_k^{n_k}}{n_j! n_k!} \Lambda_{jk,lm} (\rho,T) \\
&+ \sum_{l=i\ \mathrm{or}\ m=i} n_i \frac{Y_j^{n_j} Y_k^{n_k}}{n_j! n_k!} \Lambda_{jk,lm} (\rho,T),
\end{aligned}
\label{eq:ab_evo}
\end{equation}
where the first (second) term corresponds to the destruction (production) of the nucleus through the reactions $n_jj + n_kk = n_ll + n_mm$, where $j,k,l,m$ corresponds to the reacting particles and $n$ counts the number of particles for each reacting species, i.e., the stoechiometric number. 
The two sums runs over all species in the network that may be linked to species $i$, and therefore $i$ should replace either $j$ or $k$ ($l$ or $m$) to describe a destruction (production) reaction.
Each reaction $n_jj + n_kk = n_ll + n_mm$ has its corresponding density and temperature dependent reaction rate $\Lambda_{jk,lm}$
which defines the number of reactions taking place per time unit and volume unit.
For reacting particles $j+k\rightarrow l+m$, the Maxwellian-averaged (astrophysical) reaction rate is traditionally written as an energy (or velocity $v$) integral over the reaction cross-section $\sigma(E)$:
\begin{equation}
\begin{aligned}
\Lambda_{jk,lm} =& N_A \langle \sigma v \rangle_{jk,lm} = \bigg( \frac{8}{\pi M} \bigg)^{1/2} \frac{N_A}{(k_bT)^{3/2} G_j(T)} \label{eq:astrorate} \\ 
\times & \int_0^\infty \sum_\mu \frac{(2J_j^\mu +1)}{(2J_j^0 +1)} \sigma_{jk,lm}^\mu(E) E \exp{\bigg( -\frac{E + \epsilon_j^\mu}{k_bT} \bigg) \mathrm{d}E},
\end{aligned}
\end{equation}
where $M=\frac{m_jm_k}{m_j + m_k}$ is the reduced mass of the reacting particles, $N_A$ the Avogadro number and $G(T)$ the partition functions defined in \cref{eq:partition}.
The above expression takes into account that the target nucleus can exist in its ground state as well as excited states in the astrophysical plasma with temperature $T$.
Here, $J_j^\mu$ and $\epsilon_j^\mu$ is the spin and energy of the excited states $\mu$ of nucleus $j$ assuming the populations of the various levels obey a Maxwell-Boltzmann distribution. 
The cross-section $\sigma_{jk,lm}$ is a measure of the probability that the reaction takes place, and is either provided by experimental data or by theoretical models (e.g., see \cref{sec_nucin,subsec_ncap}).
The forward and reverse reaction cross-sections are not equal, but related by the principle of time-reversal invariance. 
Thus, if the forward reaction cross-section $\langle\sigma v \rangle_{jk,lm}$ is known, the cross-section $\langle\sigma v \rangle_{lm, jk}$ of the reverse reaction $l + m \rightarrow j + k$ can be calculated through detailed balance \citep{Angulo1999}:
\begin{equation}
\begin{aligned}
\frac{\langle\sigma v \rangle_{lm, jk}}{\langle\sigma v \rangle_{jk, lm}} =& \frac{(2J_j+1)(2J_k+1)(1+\delta_{lm})}{(2J_l+1)(2J_m+1)(1+\delta_{jk})} \\
&\times \bigg(\frac{G_j G_k}{G_l G_m}\bigg) \bigg(\frac{A_j A_k}{A_l A_m}\bigg)^{3/2} \exp{\bigg( -\frac{q_{jk, lm}}{k_bT} \bigg)},
\label{eq:det_bal}
\end{aligned}
\end{equation}
where $J_i$ is the spin of the interacting particles and the Kronecker delta $\delta_{jk}$ or $\delta_{lm}$ takes into account that the interacting nuclei can be identical. $A_i$ is the mass number, and the energy released (or absorbed) by a reaction\footnote{This is often referred to as the reaction $Q$-value, but here the use of the capital letter $Q$ is avoided since it is reserved for the radioactive decay heating rate defined in \cref{sec:heat}.} is defined as $q_{jk, lm}=(m_j+m_k-m_l-m_m)c^2$. 
Note that \cref{eq:det_bal} is valid for particles with rest mass as well as photons, i.e., either one of $j$, $k$, $l$, or $m$ can be replaced with $\gamma$. 
For radioactive decays the reaction rate is defined in terms of the half-life $T_{1/2}$ of the nucleus:
\begin{equation}
\lambda = \frac{\ln 2}{T_{1/2}},
\end{equation}
which is also either given by experimental data or theoretical models (see \cref{sec_nucin,subsec_betarates}). 

The reaction network described by \cref{eq:ab_evo} contains all charged particle reactions including the capture (or release) of nucleons and $\alpha$ particles, but also photodissociation reactions (and their reverse reactions), radioactive decays and fission processes. 
For temperatures typically below 2-3~GK, the probability for most charged particle reaction to occur are close to zero so that in these conditions the rates for $\beta$-decay, neutron capture and their inverse reactions in addition to $\alpha$-decay and fission rates dominate the right-hand side (RHS) of \cref{eq:ab_evo}.
Therefore, we can highlight the reactions which are the most important for the r-process by only considering the following reactions when evolving the molar fraction of nucleus $(Z,A)$ \citep{Hix1999,arnould2007}: 
\begin{equation}
\begin{aligned}
\frac{\mathrm{d}Y(Z,A)}{\mathrm{d}t} =& N_n \langle \sigma v \rangle_{Z,A-1} Y(Z,A-1) + \lambda_{\gamma,n}^{Z,A+1}Y(Z,A+1) \\
&+ \lambda_{\beta0}^{Z-1,A}Y(Z-1,A) + \sum_k \lambda_{\beta kn}^{Z-1,A+k}Y(Z-1,A+k) \\
&+ \lambda_\alpha^{Z+2,A+4}Y(Z+2,A+4) \\
&- \bigg[ N_n \langle \sigma v \rangle_{Z,A} + \lambda_{\gamma,n}^{Z,A} + \lambda_{\beta}^{Z,A} + \lambda_\alpha^{Z,A} + \lambda_f^{Z,A} \bigg] Y(Z,A) \\
& + \sum_f p_{Z_f,A_f}(Z,A) \lambda_{sf}^{Z_f,A_f} Y(Z_f,A_f) \\
& + \sum_f p_{Z_f,A_f}^\beta (Z,A) \lambda_{\beta f}^{Z_f-1,A_f} Y(Z_f-1,A_f) \\
& + \sum_f p_{Z_f,A_f}^n (Z,A) \lambda_{n f}^{Z_f,A_f-1} Y(Z_f,A_f-1).
\end{aligned}
\label{eq:ab_evobig}
\end{equation}
Here, $N_n$ is the neutron density, $\langle \sigma v \rangle$ is the reaction rate for radiative neutron capture, and the various $\lambda_i$ the rates for the following reactions: $\lambda_{\gamma,n}$ for inverse photoneutron emission, $\lambda_{\beta0}$ for $\beta$-decay (followed by no delayed neutron),  $\lambda_{\beta kn}$ for $\beta$-decay followed by the delayed emission of $k$ neutrons, $\lambda_\alpha$ for $\alpha$-decay, 
 $\lambda_\beta=\lambda_{\beta 0} + \sum_k \lambda_{\beta kn}$ for the total $\beta$-decay rate, $\lambda_f = \lambda_{sf} + \lambda_{\beta f} + \lambda_{n,f}$ for fission where $\lambda_{sf}$, $\lambda_{\beta f}$ and $\lambda_{n,f}$ are the rates for spontaneous, $\beta$-delayed and neutron induced fission, respectively.
The last three terms are relevant when a nucleus $(Z_f,A_f)$ in the network enters the trans-Pb region and nucleus $(Z,A)$ may be produced through the spontaneous, $\beta$-delayed or neutron-induced fission of nucleus $(Z_f,A_f)$,
where $p_{Z_f,A_f}(Z,A)$ is the probability for the production of nucleus $(Z,A)$ as a fragment when nucleus $(Z_f,A_f)$ fissions. 
The processes producing and destroying nuclei are correlated, i.e., the production term of nucleus $(Z,A)$ through neutron capture on nucleus $(Z,A-1)$ is identical to the neutron capture destruction term of nucleus $(Z,A-1)$, and therefore \cref{eq:ab_evo} or \cref{eq:ab_evobig} represent a set of coupled differential equations (coupled ODEs), which have to be solved simultaneously.  
Note that the full set of equations in \cref{eq:ab_evo} are always used to evolve the nuclear abundances, never the reduced set in \cref{eq:ab_evobig} which is included here for illustration purposes. 

\newpage
\subsubsection*{Numerical methods}

By using vectors (arrays) that contain the molar fraction \textit{\textbf{Y}} of all nuclei in the network, a compact version of \cref{eq:ab_evo} can be written as:
\begin{equation}
\frac{\mathrm{d}\textit{\textbf{Y}}}{\mathrm{d}t} = \textit{\textbf{f}}(\textit{\textbf{Y}},t),
\end{equation}
where the vector function \textit{\textbf{f}} summarizes the RHS of \cref{eq:ab_evo}.
As briefly discussed in \cref{ch:intro}, the time-scales of the reactions involved in the r-process (i.e., the entries of $\textit{\textbf{f}}(\textit{\textbf{Y}},t)$) span many orders of magnitude, and such systems are termed stiff. 
In addition, the abundances of the involved nuclei can vary by several orders of magnitude. 
Implicit time integration methods have to be applied when solving stiff ODE systems, which are more computationally involved than explicit integration methods \citep{Timmes1999,Hix1999}.
For example, the Jacobian matrix $\frac{\mathrm{d}\textit{\textbf{f}} }{\mathrm{d}\textit{\textbf{Y}}}$ have to be computed, which require the calculation of partial derivatives $\frac{\partial f_i}{\partial Y_j}$ for all ($i,j$) entries in the matrix.  

\todo{here it would be cool to show a visualisation of the "matrix"! as in Timmes1999 - possible to extract from jac() ??}

The double precision Variable-coefficient Ordinary Differential Equation solver (\texttt{DVODE}) \citep{Byrne1975,Hindmarsh1977, Hindmarsh1983,Jackson1980,Brown1989} have been applied to solve the system of first order ODEs.
As input, \texttt{DVODE} takes the abundances \textit{\textbf{Y}} and the RHS $\textit{\textbf{f}}(\textit{\textbf{Y}},t)$ at the time $t_\mathrm{in}$, and calculates the new abundances at the next time step $t_\mathrm{out}$.
After a successful calculation, the system can be evolved further by calling \texttt{DVODE} with the recently calculated abundances as the new abundances so that $t_\mathrm{in,new}=t_\mathrm{out,prev}$ and setting $t_\mathrm{out,new}=t_\mathrm{in,new} + \mathrm{d}t$.
Note that the time step size d$t$ increases logarithmically, with small time steps early during the neutron capture dominant phase, while larger time steps are taken after freeze-out (in particular at late time when the decay of longer-lived nuclei dominate).
As can be seen in \cref{eq:astrorate,eq:det_bal}, the forward and reverse reaction rates, i.e., the content of \textit{\textbf{f}}, depend exponentially on temperature, and therefore the rates have to be updated between each iteration as the temperature evolves.
The heat released by the nuclei in the network due to $\beta$-decays, fission processes, and $\alpha$-decays are also calculated at each time step, as described in \cref{sec:heat}.

The time interval $[t_\mathrm{in},t_\mathrm{out}]$ is subdivided into smaller time steps with size $h$:
\begin{equation}
t_n = t_\mathrm{in} + \sum_{i=1}^n h_i
\end{equation}
where the maximum number of steps $n$ needed to reach $t_\mathrm{out}$ can be chosen by the user and the time steps $h$ can be of unequal length. 
A benefit of using \texttt{DVODE} is that an algorithm within the program selects the step size $h$ used throughout the integration based on estimates of the local discretization error \citep{Byrne1975}. 
In practice this means that the user provides \texttt{DVODE} with the desired relative and absolute error in \textbf{\textit{Y}},
and the algorithm chooses a small enough time step to achieve the sought accuracy while also maintaining numerical stability. 
For stiff systems the limitation of the time step size is usually due to numerical stability rather than accuracy \citep{Timmes1999,Hix1999}.
\texttt{DVODE} will also indicate if the computation was successful or not through an integer value of the output. 
Some error flags are less severe than others and in those cases the r-process calculations will continue to take new time steps (i.e., call \texttt{DVODE} again), or the calculation might be restarted at $t_\mathrm{in}$ after adjusting the tolerances. 
However, severe or repeated errors are flagged and the program stopped. 
In this work we will set the default relative and absolute errors\footnote{The estimated local error in $Y(i)$ will be controlled so as to be roughly less (in magnitude) than $\mathrm{ewt}(i) = \mathrm{rtol}*\mathrm{abs}(Y(i)) + \mathrm{atol}$ \citep{Brown1989}.} to $10^{-3}$ and $10^{-11}$, respectively. A set of calculations run with smaller tolerances, i.e., higher resolution, will be performed in \cref{sec_netlimtol}.
For some trajectories it can be challenging for the integration method to converge between $t_\mathrm{in}$ and $t_\mathrm{out}$. These problems are often related to the size of the tolerances, and will be discussed further in \cref{sec_netlimtol}.


The time needed to solve a set of ODEs increases rapidly with the dimension of the system (i.e., number of coupled equations), and therefore the size of the r-process network is dynamically increased as heavier nuclei are synthesized.
By default, the network is initialized with a maximum proton number of $Z_\mathrm{max}=10$, or the $Z_\mathrm{max}$ given by the NSE initial abundances.
When the sum of the molar fractions of nuclei with $Z_\mathrm{max}$ protons are larger than typically $10^{-12}$ (which also can be the case for the initial abundances), the network is scaled up by increasing the value of $Z_\mathrm{max}$ by 20 protons.  
As discussed above, not all nuclei in the network interact with each other which leads to many zero-valued elements in \textbf{\textit{f}}, making it a so-called ``sparse'' matrix. \texttt{DVODE} have implemented several methods to handle sparse matrices like for example the use of ``banded'' matrices, however the full Jacobian matrix is used in the calculations presented herein. 

The execution flow is summarized in the following pseudocode:
\begin{lstlisting}[xleftmargin=0.9cm, xrightmargin=0.9cm, language=Python]
# calculation started at a given temperature T and density rho_net at time:
t=0
Zmax = 10


for all nuclei in network (at T & rho_net):
  fill Y with initial abundances
  fill f by applying initial reaction rates 

specify dvode parameters (tolerances, ...)

while t<1 yr:
  specify dt depending on t
  tout = t + dt
  if Y(Zmax) > 1e-12:
    # increase network size:
    Zmax = Zmax + 20
    update dvode parameters 
    update reaction rates
    update Y & f
  
  while t<tout:
    call dvode(Y,t,f,jac,..)
    # Note that dvode internally updates the
    # time t=t+h where h is the step size for 
    # the finer time grid, in addition to 
    # updating Y=Y(t+h)
    update temperature & density 
    for i nuclei in network:
      # electron fraction:
      Ye = Ye + Y(i)*Z(i)
      # heat released:
      Q = Q + Qnuc(i)
      # and similarly for 
      # Qalpha, Qbeta, Qfission
    update reaction rates
\end{lstlisting}


First the network and \texttt{DVODE} parameters are initialized, and the rates are updated at the temperature $T$ and density $\rho_\mathrm{net}$ (initially setting $t=0$ for the r-process calculation). 
The main time loop updates the time according to the chosen time step d$t$ and continues until the time reaches a year. For each step in this outer loop the abundances at $Z_\mathrm{max}$, i.e., $Y(Z_\mathrm{max})=\sum_i Y(i)Z(i)$, are checked against the threshold for enlarging the network.
\texttt{DVODE} is called in the second loop, which provides approximations to the solution $Y$ at each sub-time step $t+h$ (or only at $t_\mathrm{out}$ if that is desirable). 
The frequent input-output of \texttt{DVODE} allows the reaction rates to be updated according to the temperatures at the given $t$. 
The total heat released \texttt{Q} is calculated according to the description in \cref{sec:heat} from the heat released by each nuclei \texttt{Qnuc(i)}. 
As will be discussed in more detail in \cref{sec:heat}, the heat $Q$ is released into the ejecta, and may therefore affect the temperature evolution, hence, the r-process abundances. 
This is taken into account in the NS-NS and NS-BH merger models (i.e., not the BH-torus models).
This implies that some additional steps are required in the execution flow for these models after the calculation of $Q$. 

A valuable tool to study the time evolution of the abundance and its connection to other variables is animations of the mass fractions in the $(N,Z)$ plane (see \cref{fig_Zmax_absum} for an example of an animation snapshot). 
These animations can show the abundance evolution for a single trajectory, the mass-averaged abundances for a selection of trajectories or the mass-averaged abundances of the complete trajectory set. 
In addition, it is possible to display the mass fractions ratio of two calculations to highlight their differences.
The r-process animations related to this work are available for download or interactive viewing on the \citet{GitHubAnims}.

The Fortran code used for the r-process calculations (\texttt{rpro}) has been developed over many years by S. Goriely \citep[e.g.,][]{goriely1996,goriely2011}.
Many advancements have been made over the years by adding additional features to the code so that it can be used to study a variety of complex details related to r-process nucleosynthesis, and hence the code itself has grown in complexity. 
\texttt{rpro} was partly rewritten and modernized with the main goal of improving its readability and thus enhancing the understanding of the execution flow.
The version of \texttt{rpro} at the end of 2018 (and similar versions) will in the following be referred to as the ``original'' code, while the modernised version will be referred to as \texttt{rpro}.
The newest official Fortran standard is Fortran 2018, however, most modern Fortran is usually referred to as Fortran 90 or 95 \citep[see the][and also the \citet{FortranWiki}]{FortranCommunity}. 
The original \texttt{rpro} code is in a mix of FORTRAN 77\footnote{The numbers refer to the year the standard was set, i.e., 1977 for FORTRAN 77. The standards released after this also changed the spelling from FORTRAN to Fortran.} and Fortran 90/95 style code and uses what is referred to as fixed-format. For historical reasons\footnote{FORTRAN was used early in the development of computers where the instructions to the computer were given via punched cards (plastic cards with holes in them), see for example the \citet{FortranWikiPunch} for more information.}, fixed-format code limits the source code to a maximum of 72 characters per line, and the first 6 character positions give specific instructions to the compiler. 
These limitations are not present in Fortran 90 or later standards, so a part of the modernization of \texttt{rpro} was to convert it into free format code. 
Except for a few deleted features, all Fortran standards have retained compatibility with older forms of the language so that code written in FORTRAN 77 will also work with compilers written for the newest Fortran standard. However, a few implementations in for example FORTRAN 77 have become ``obsolete'' meaning that they should not be used for new code, and the use of some then popular features (which at the time often lacked good alternatives) has since been discouraged \citep[e.g.,][]{FortranBest,DoctorFortran,FortranWikiModern}.
The latter includes the use of ``\texttt{goto}'' statements which allows the program to jump to (almost) whichever line in the source code specified by the programmer through a numbered ``label'', and can lead to an execution flow which is hard to understand for humans. 
Therefore, reducing the number of \texttt{goto} statements and ``labels'' improved the readability of the code along with other efforts to control which sub-programs (i.e., subroutines) have access to global output variables. 

In summary, the work of updating the code included:
\begin{itemize}
\item shortening the code by removing unused (i.e., ``historical'') options no longer in use
\item modernizing and reorganizing the code
\item verification of results obtained using the new version of the code
\end{itemize}
which lead to a reduction of the number of lines in the source code from $\sim11000$ to $\sim8000$ lines in the original code and \texttt{rpro}, respectively. This is without counting some external subroutines for which the same versions are used by the original code and \texttt{rpro}.

\subsection{Radioactive decay heat}
\label{sec:heat}

In addition to changing the nuclear composition, nuclear reactions also absorb or release energy into the environment. 
This possible re-heating of the ejecta is included in the temperature evolution using the laws of thermodynamics \citep[e.g., see][]{meyer1989}. 
For the r-process, the most important reactions include $\beta$-decays, fission, and $\alpha$-decays.
While $(n,\gamma) \rightleftharpoons (\gamma,n)$ equilibrium conditions hold, the contribution from $(n,\gamma)$ and $(\gamma,n)$ reactions cancel each other out, and after freeze-out they become negligible.
The heat generated by radioactive decays per time per unit mass can be calculated as:
\begin{equation}
Q = Q_\beta + Q_{\mathrm{fis}} + Q_\alpha.
\label{eq:Q_m1}
\end{equation}
The contribution from $\beta$-decays are given by:
\begin{equation}
Q_\beta = Q_n + Q_{\beta 0} + \sum_k Q_{\beta kn},
\end{equation}
which includes the decay of free neutrons ($Q_n=q_n \lambda_n Y_n$), $\beta$-decays without the emission of a neutron ($Q_{\beta 0}$) and $\beta$-decays with the delayed emission of $k$ neutrons ($Q_{\beta kn}$).   
The contribution from neutrino energy loss $Q_\nu$ in each of the $\beta$-decays are removed following the prescription of \citet{Fowler1975},  since the energy carried by the neutrinos is lost from the environment and does not contribute to the heating of the ejecta. 
The heat released through $\beta$-decay, i.e.,  $m\in [\beta 0, \beta kn ]$, is found by summing over all $i$ species in the network:
\begin{equation}
Q_m = \sum_i q_{i,m} \lambda_{i,m} p_{i,m} Y_i,
\end{equation}
where the $\lambda$'s are the decay rates, the $q$'s the radioactive heat released, and the $p$'s the probability for species $i$ to decay through mode $m$. 
This information is provided by experimental data when available and otherwise the applied nuclear models (see \cref{sec_network_input}).
Similarly, the energy released through $\alpha$-decay is given by:
\begin{equation}
Q_\alpha = \sum_i q_{i,\alpha} \lambda_i Y_i,
\end{equation}
and fission by:
\begin{equation}
Q_\mathrm{fis} = Q_{sf} + Q_{\beta f} + Q_{nf},
\end{equation}
which includes the contribution from spontaneous fission $Q_{sf}= \sum_i q_{i,sf} \lambda_{sf} Y_i$, $\beta$-delayed fission $Q_{\beta f} = \sum_i q_{i,\beta f} p_{i,\beta f} \lambda_i Y_i$ and neutron-induced fission $Q_{nf}= \sum_i q_{i,nf} N_n \langle \sigma v \rangle_{(n,f)}  Y_i$. Photo-induced fission is not included in this work since its contribution is found to be insignificant due to the low temperatures found when fissile nuclei are produced by the r-process (this assumption is verified in \cref{subsec_photofis} for the models applied in this work).
Except at late times (i.e., $t\gtrsim 1$~d), the r-process heating rate is dominated by the heat generated from $\beta$-decays, and the general trend of the heating rate can be approximated by the empirical formula \citep{metzger2010} of 
\begin{equation}
Q_0[{\rm erg/g/s}]=10^{10}~t[{\rm d}]^{-1.3}.
\label{eq:Q_0}
\end{equation}

The released (or absorbed) heat from nuclear reactions (including radioactive decay) results from the difference between the sum of the binding energies of the incoming particles and the outgoing particles. 
Therefore, the total energy released per unit volume within a time step $\Delta t$ through nuclear reactions can also be calculated by following the abundance changes $\Delta Y_i$ \citep[e.g., see][]{Arnett1996}:
\begin{equation}
Q = \frac{1}{\Delta t} \sum_i \Delta Y_i B_i, 
\label{eq:Q_m2}
\end{equation}
where $B_i$ is the binding energy defined in \cref{eq:B_i}. 
The advantage of \cref{eq:Q_m2} is that only the knowledge of the abundance changes within the time step for each species and their masses are required, which is less computationally demanding compared to the approach of \cref{eq:Q_m1}. 
However, it is not possible to identify which decay mode, i.e., $\beta$-, $\alpha$-decay or fission, contributes the most to the radioactive heating rate through \cref{eq:Q_m2}, in contrast to \cref{eq:Q_m1} which by definition embeds this information.
As default, we apply \cref{eq:Q_m1} to calculate the nuclear re-heating, and a study to compare this approach to \cref{eq:Q_m2} is done in \cref{sec_sensheat}.

\todo{is this Mueller1986 another ref I have not seen?}

\section{Kilonova model}
\label{sec_kilo}

\epigraph{\itshape \enquote{I solemnly swear I am up to no good.}}{--- \textup{J. K. Rowling}, Harry Potter and the Prisoner of Azkaban}  

The light curve of a kilonova delivers unique information about the outflow mass, velocity, and composition to the observer.
Models of the kilonova observed in the aftermath of GW170817 have indicated the presence of (at least) two kilonova components, the so-called ``blue'' and ``red''  components\footnote{Note that the ``blue'' and ``red'' kilonova components did not exhibit spectral peaks in the blue optical and red optical bands, but rather in the red optical and near infrared bands, respectively.},  which are presumably associated with ejecta material having {a small or large fraction of lanthanides,} 
respectively \citep[e.g.,][]{kasen2017}.  However, given the complexity of modelling the merger and its remnant the identification of these kilonova components with the basic ejecta components (dynamical vs. secular ejecta) is not yet unambiguous. In particular, the role of the dynamical ejecta, e.g., whether they contribute more to the blue or more to the red kilonova component, is not finally settled \citep[e.g.,][]{kasen2017,Kawaguchi2018}.

Kilonova light curves depend strongly on the heat generated by the r-process elements through the various decay channels, thermalization efficiencies (defined below), the lanthanide and actinide content of the ejecta and the average ejecta mass and velocity.
This information can, for example, be provided by the combination of hydrodynamical models and r-process calculations (see \cref{sec:astro_mods,sec:heat}).
The kilonova model applied in this work is simulated on a velocity grid which is discretized in the radial and angular directions.
For each trajectory, all quantities predicted by the hydrodynamical models and the r-process that are relevant for the light curve evolution are mapped onto this grid.
Altogether five quantities are mapped, namely the mass density $\rho$, the electron fraction $Y_e(\rho_\mathrm{net})$ at the onset of network calculations, the specific heating rate $Q_\mathrm{heat}(t)$ (defined below), and the lanthanide plus actinide mass fraction $X_\mathrm{LA}(t)$ (from now on only referred to as the lanthanide mass fraction), as well as the average mass number $\langle A(t)\rangle$.
Note that $Y_e(\rho_\mathrm{net})$ is extracted merely for diagnostic purposes but otherwise not needed by the applied kilonova scheme\footnote{The dependence of the kilonova on the electron fraction is already included in the quantities $Q_\mathrm{heat}(t)$, $X_\mathrm{LA}$, and $\langle A\rangle$ provided by the nucleosynthesis calculations.}.
While $\rho$\footnote{For the time scales relevant for the kilonova, homologous expansion is assumed so that the density at any given time can be obtained using $\rho\propto t^{-3}$.} and $Y_e(\rho_\mathrm{net})$ only need to be mapped once, the remaining quantities are time-dependent and therefore need to be mapped for a sufficiently large number of discrete times, see \citet{just2021} for more details.


The fraction $f_\mathrm{th}$ of the total radioactive energy release rate ($Q + Q_\nu$) that is converted to thermal energy on time scales shorter than the evolution time scale is referred to as the effective heating rate:
\begin{equation}
Q_\mathrm{heat} \equiv f_\mathrm{th} (Q + Q_\nu),
\label{eq:Qheat}
\end{equation}
where $Q$ denotes the release rate without neutrinos (e.g., \cref{eq:Q_m1} or \cref{eq:Q_m2}), and $Q_\nu$ stands for the contribution that is carried away by neutrinos (and not available for thermal heating in any case). 
The heat generated by the various decay components thermalizes in the material in different ways, which is specified through $f_\mathrm{th}$ (which is model dependent).

More details about the adopted kilonova model can be found in \citet{just2021}, however, for completeness, some details of the description will be repeated herein.
Following the approach in \citet[][who adopted the formalism by \citet{barnes2016}, see also \citet{just2021}]{Rosswog2017}, the thermalization efficiency can be calculated as:
\begin{equation}
f_\mathrm{th} = \frac{Q_\beta [\zeta_\gamma f_\gamma + \zeta_e f_e] + Q_\alpha f_\alpha + Q_\mathrm{fis} f_\mathrm{fis}}{Q + Q_\nu}
\end{equation}
where $Q_\beta$, $Q_\alpha$, and $Q_\mathrm{fis}$ are the partial release rates from $\beta$-decay, $\alpha$-decay, and fission, and $f_i$ ($i \in \{\gamma, e, \alpha, \mathrm{fis}\}$) the thermalization efficiencies for photons, electrons, $\alpha$ particles, and fission products.
In this model, the thermalization efficiencies are functions of the thermalization time-scales which again only depend on the total mass and average velocity of the ejecta; see \citet{Rosswog2017} for the explicit expressions.
Since in our notation, $Q_\beta$ does not contain the energy release going into neutrinos, the weighting factors $\zeta_\gamma = 0.69$ and $\zeta_e = 0.31$ differ from the ones ($\zeta_\gamma = 0.45$ and $\zeta_\gamma = 0.2$) employed in \citet{Rosswog2017}. 

The kilonova model also requires a description of the opacities which are dominated mainly by thousands of possible transition lines between energy levels of lanthanide and actinide elements newly created by the r-process.
In our approximate kilonova solver \citep{just2021}, the opacity is expressed in a parametric fashion as a function of the lanthanide mass fraction and gas temperature as
\begin{equation}
\kappa(X_\mathrm{LA},T) = \kappa_\mathrm{LA}\kappa_T
\end{equation}
where the first term depends on the lanthanide mass fraction
\begin{equation}
\begin{aligned}
\kappa_\mathrm{LA} \equiv \left\{
  \begin{array}{ll}
    30\ \text{cm}^2\text{ g}^{-1} (X_\mathrm{LA}/10^{-1})^{0.1}, & X_\mathrm{LA} > 10^{-1}, \\
    3\ \text{cm}^2\text{ g}^{-1} (X_\mathrm{LA}/10^{-3})^{0.5}, & 10^{-3} < X_\mathrm{LA} < 10^{-1}, \\
    3\ \text{cm}^2\text{ g}^{-1} (X_\mathrm{LA}/10^{-3})^{0.3}, & 10^{-7} < X_\mathrm{LA} < 10^{-3}, \\
    0.2\ \text{cm}^2\text{ g}^{-1}, & X_\mathrm{LA} < 10^{-7}, 
  \end{array}
  \right.
\end{aligned}
\end{equation}
and the temperature-dependent part is 
\begin{equation}
\begin{aligned}
\kappa_\mathrm{LA} \equiv \left\{
  \begin{array}{ll}
    1, & T > 2000\ \text{K}, \\
    \big(\frac{T}{2000\ \text{K}}\big)^5, & T < 2000\ \text{K}.
  \end{array}
  \right.
\end{aligned}
\end{equation}
The above prescription was motivated by fits to bolometric light
curves from the atomic-physics based models of \citet{kasen2017}, see \citet{just2021} for more details. 

Together with an approximate radiation transport model, the above setup can predict the kilonova light curve viewed by an observer. 
Note that the calculation of the kilonova model discussed above is not performed herein, but by O. Just based on the r-process yields and heating rates of this work.
The kilonova model will be applied to the r-process results of the NS-NS dynamical ejecta models (\cref{subsec_dynsim}), and are presented within this thesis in \cref{sec_kilonova} for completeness (see also the companion papers \citet{kullmann2021} and \citet{just2021}).


\newpage
\section[Estimating the age of the Galaxy]{Estimating the age of the Galaxy: Cosmochronometers}
\label{sec_cosmoc}

\epigraph{\itshape \enquote{Your lack of fear is based on your ignorance.}}{--- \textup{ Liu Cixin}, The Three-Body Problem}  

After high-resolution spectra became available for metal-poor stars, it has become possible to derive the abundances of long-lived radioactive elements such as thorium and uranium in stellar spectra of r-process-enhanced metal-poor stars. This has opened the door for an independent age-dating method of individual stars and, therefore, also sets lower limits on the age of our Galaxy. 
The method, called cosmochronometry \citep{Butcher1987}, relies on the known radioactive decay time (i.e., half-life) of long-lived radioactive nuclei and the theoretical prediction of the initial\footnote{Initial here refers to the Th/U ratio at the time when the star was first enriched in r-process elements by an r-process event, here assumed to be a NS-NS merger. } abundance (before decay) of the involved nuclei.
By combining the known radioactive decay time for $^{238}$U and $^{232}$Th of $\tau(^{238}\mathrm{U}) = 6.45$~Gyr and $\tau(^{232}\mathrm{Th}) = 20.27$~Gyr, respectively, an age estimate of the star ($t^*_{\mathrm{U},\mathrm{Th}} $) can be calculated from 
\begin{equation}
\log \Big( \frac{\mathrm{Th}}{\mathrm{U}} \Big)_{\mathrm{obs}} = \log \Big( \frac{\mathrm{Th}}{\mathrm{U}} \Big)_\mathrm{r} + \log \Big[ \exp \Big( \frac{1}{\tau(\mathrm{U})} - \frac{1}{\tau(\mathrm{Th})} \Big)  \Big] \times t^*_{\mathrm{U},\mathrm{Th}} 
\label{eq_cosmo}
\end{equation}
where $(\mathrm{Th}/\mathrm{U})_{\mathrm{obs}}$ and $(\mathrm{Th}/\mathrm{U})_{\mathrm{r}}$ are the observationally derived and model-dependent r-process abundance ratios, respectively, of $^{232}$Th and $^{238}$U (after decay of their shorter-lived isotopes).
Here, we use the notation commonly used in astronomy Th/U for the abundance ratios, which is identical to $Y_{{}^{232}\mathrm{Th}}/Y_{{}^{238}\mathrm{U}}$ in the notation introduced herein.
Other cosmochronometers have been applied in the literature as, for example, the Th/Eu ratio since Eu is relatively easy to identify in the spectra of metal-poor stars. 
However, Eu and Th are widely separated in atomic mass, which makes the theoretical estimates more uncertain compared to U and Th, which are neighbouring elements \citep{Goriely2001,Goriely2016d}.
The uncertainty in Th/Eu is also related to the long half-life of Th.
With this method, we can use the Th/U ratio obtained in our r-process calculations to estimate the age of six metal-poor r-process enriched stars with both Th and U lines in their spectra.

\section{Nuclear physics input}
\label{sec_nucin}

\epigraph{\itshape ``I may not have gone where I intended to go, but I think I have ended up where I needed to be.''}{--- \textup{J. Allan Toogood}}

For all the nuclei in the network many nuclear ingredients and processes are required to calculate the abundance evolution. 
Our r-process reaction network includes all charged particle fusion reactions on light and medium mass isotopes, photodisintegrations, beta-delayed neutron emission probabilities and the rates of radiative neutron capture, $\beta$-decay and $\alpha$-decay. 
If the network reaches trans-Pb species, we also include neutron induced, spontaneous and $\beta$-delayed fission and their corresponding fission fragment distributions for all fissile nuclei.  Each fissioning parent is linked to the daughter nucleus, and the emitted neutrons may be recaptured by the nuclei present in the environment.
Whenever experimental data is available, it is included in the network. Reaction rates on light species are taken from the NETGEN library \citep{xu2013}, which includes the latest compilations of experimentally determined rates. 

\todo{if some info doesn't exist for one model fall back on rates from Talys62(?) and also for fission SPY 14 and 16(?) and HFB-14 barriers vs MS99}



The r-process involve many neutron-rich nuclei for which at best very little,  or no experimental data is available.
To describe fundamental nuclear properties such as nuclear masses, $\alpha$- and $\beta$-decay rates, radiative neutron capture rates and fission probabilities, theoretical models have to be applied. 
Despite much progress in the last decades,  nuclear models are still affected by large uncertainties, and one of the main goals of this thesis is to investigate the impact of such model uncertainties on the r-process yields (see \cref{ch_nucuncert}). 

Nuclear predictions are affected by systematic as well as statistical uncertainties. The former ones, also referred to as model uncertainties, are known to dominate for unstable nuclei \citep[e.g.,][]{Goriely2014a} since there is usually no or very little experimental information available to constrain the model on those nuclei. Whenever experimental data are available, either for rates or reaction model ingredients, they are considered in the theoretical modelling. In this case,  they also constrain the possible range of variation of the model parameters and reduce the impact on the model as well as parameter uncertainties. However, for exotic nuclei, models of different natures, ranging between macroscopic to microscopic approaches \citep[e.g.,][]{arnould2007,Goriely15b,Hilaire16, Goriely17a}, may provide rather different predictions.
For this reason, systematic global uncertainties affecting reaction or decay rates will be considered in the present study. Those systematic global uncertainties are propagated to the calculation of reaction rates, and consistently applied to the nucleosynthesis simulations to estimate their impact on the r-process yields and decay heat.

\begin{table}[tbp]
\centering
\caption[The nuclear physics input sets.]{An overview of the nuclear physics input varied in the r-process nuclear network calculations in our nuclear uncertainty study in \cref{ch_nucuncert}: model number, mass model, direct capture component (yes if included), the model for the $\beta$-decay rates, fission barriers and fission fragment distributions. In \cref{ch_nucuncert} we use input set 1 as ``default'' in our calculations, while input set 6 is used for all calculations in \cref{ch_dynweak}. 
See the text for the model references. 
 }
\begin{tabular}{llccccc}
\hline \hline 
Set & Mass mod. & DC & $\beta$ model & F. barr.  &  F. frag. \\ 
\hline 
1 & BSkG2 & no & RHB+RQRPA & HFB-14 & SPY\\ 
2 & FRDM12 & no & RHB+RQRPA & HFB-14 & SPY\\ 
3 & WS4 & no & RHB+RQRPA & HFB-14 & SPY\\ 
4 & D1M & no & RHB+RQRPA & HFB-14 & SPY\\ 
5 & HFB-21 & no & RHB+RQRPA & HFB-14 & SPY\\ 
6 & HFB-31 & no & RHB+RQRPA & HFB-14 & SPY\\ 
\hline 
7 & HFB-31 & no & HFB21+GT2 & HFB-14 & SPY\\ 
8 & HFB-31 & no & TDA   & HFB-14 & SPY\\ 
9 & HFB-31 & no & FRDM+QRPA & HFB-14 & SPY\\ 
10 & FRDM12 & no & HFB21+GT2 & HFB-14 & SPY\\ 
11 & FRDM12 & no & TDA   & HFB-14 & SPY\\ 
12 & FRDM12 & no & FRDM+QRPA & HFB-14 & SPY\\
\hline
13 & HFB-21 & no & RHB+RQRPA & MS99  & GEF \\ 
14 & BSkG2 & no & RHB+RQRPA & MS99  & GEF \\ 
\hline 
15 & HFB-21    & yes & RHB+RQRPA & HFB-14 & SPY\\
\hline \hline 
\end{tabular} 
\label{tab_nuc_mods}
\end{table}

\cref{tab_nuc_mods} lists the nuclear models considered in this work and described in the following sections. 
Note that all of the global models listed in  \cref{tab_nuc_mods} have proven their ability to describe experimental data with a high degree of accuracy (e.g., mass predictions should have an rms deviation below 0.8 MeV with respect to the 2457 $Z, N \ge 8$ experimentally known masses \citep{Wang2021}\footnote{The rms values listed herein are consistently calculated with respect to the 2457 $Z, N \ge 8$ experimentally known masses in \citet{Wang2021}, which in total include $\sim2500$ masses.}). Models that do not fulfil such a necessary requirement 
are not considered to be state-of-the-art and therefore excluded from our comparison study (\cref{sec_comp2}).
However, accurately describing experimental data is a necessary but not a sufficient condition for a model to be applied to the r-process nucleosynthesis \citep[for more details see][]{arnould2007}. In summary, in order to best meet the nuclear-physics needs of the r-process, we require the applied models to be both accurate with respect to experimental observables but also as reliable as possible, i.e., to be based on a physically sound model that is as close as possible to a microscopic description of the nuclear systems. This second criterion is fundamental for applications involving extrapolation away from experimentally known regions, in particular towards exotic neutron-rich nuclei of relevance for the r-process.
Only nuclear physics models that fulfil the aforementioned criteria will be referred to in the following as ``advisable'' models.
Not advisable are, for example, the masses obtained with the Hartree-Fock-Bogolyubov model based on the SLy4 Skyrme interaction \citep{Stoitsov2003} since they poorly reproduce the complete set of experimental masses with an rms deviation larger than 5 MeV. This is significantly worse than other Skyrme-HFB mass models which describe all known masses with an rms deviation smaller than 0.8 MeV.
Another example concerns some recent mass models based on machine learning algorithms \citep{Shelley2021}, which essentially consider mathematical rather than physical approaches and are consequently believed not to be suited for a reliable extrapolation far away from the experimentally known region.

\subsection{Radiative neutron capture rates}
\label{subsec_ncap}

Many nuclear structure properties are required to calculate the radiative neutron capture rates, including mass model,  nuclear level density, nuclear potential, $\gamma$-strength function, and for the non-resonant region, even the energy, spin and parity of discrete states are needed. 
In the following discussion, we will focus on two different methods, namely the statistical Hauser-Feshbach method and the direct capture (DC) model for the estimation of radiative neutron capture rates.   

\subsubsection*{Compound nucleus and Direct capture}

The statistical Hauser-Feshbach reaction model is used as default in astrophysical applications to calculate the radiative neutron capture cross-sections for experimentally unknown nuclei. 
Within the Hauser-Feshbach method, the capture process is assumed to be a two-step process. First, a so-called compound nucleus (CN) is formed as an intermediate step. Second, the nucleus de-excites to the ground state of the residual nucleus by emitting a particle or a $\gamma$-ray.
In this model, the CN is assumed to be in thermodynamic equilibrium so that the energy of the incident particle is shared uniformly by all the nucleons before de-excitation. This assumption is expected to hold if we assume that the level density of the compound nucleus at the energy of the incident energy is large, and then the compound nucleus has an average statistical continuum superposition of available resonances at this energy. 
Although the Hauser-Feshbach method has proven to accurately reproduce cross sections for medium- and heavy-mass nuclei, the model suffers from uncertainties originating from the underlying theoretical nuclear models, and the validity of its CN assumption should be questioned for some light and neutron-rich nuclei for which only a few or no resonant states are available.

When the number of available states in the CN is relatively small, the neutron capture process may be dominated by the direct electromagnetic transition to a bound final state without forming a compound nucleus. 
The DC process has been shown to be non-negligible compared to the Hauser-Feshbach component and can even contribute up to 100 times more to the total cross-section for the most neutron-rich nuclei close to the neutron drip line \citep{sieja2021}. 
The DC model also suffers from large model uncertainties mainly due to the remarkable sensitivity of the cross-section to the few available final states, which for the very neutron-rich nuclei are unknown. Moreover, the energy, spin and parity of the discrete levels can modify the DC contribution by many orders of magnitude since the selection rules, which rely on the spin and parity differences, may switch on or off the DC component \citep{Xu2012}.

It is normally assumed that the statistical and direct processes may contribute to the radiative neutron capture rate in a non-exclusive way. It remains, however, fundamental to use the same set of nuclear-structure ingredients to estimate both contributions. In particular, the same optical potential needs to be adopted to ensure the same total reaction cross-section is found in both channels.
According to the study of \citet{Xu2014}, the experimental radiative neutron capture cross-sections are in good agreement with the DC model for the lightest nuclei, where the CN model overestimates the contribution.  For the experimentally unknown neutron-rich nuclei, the DC component contributes significantly to the total cross-section or even dominates over the CN contribution in some regions.

\begin{figure}[tbp]
\begin{center}
\includegraphics[width=\textwidth]{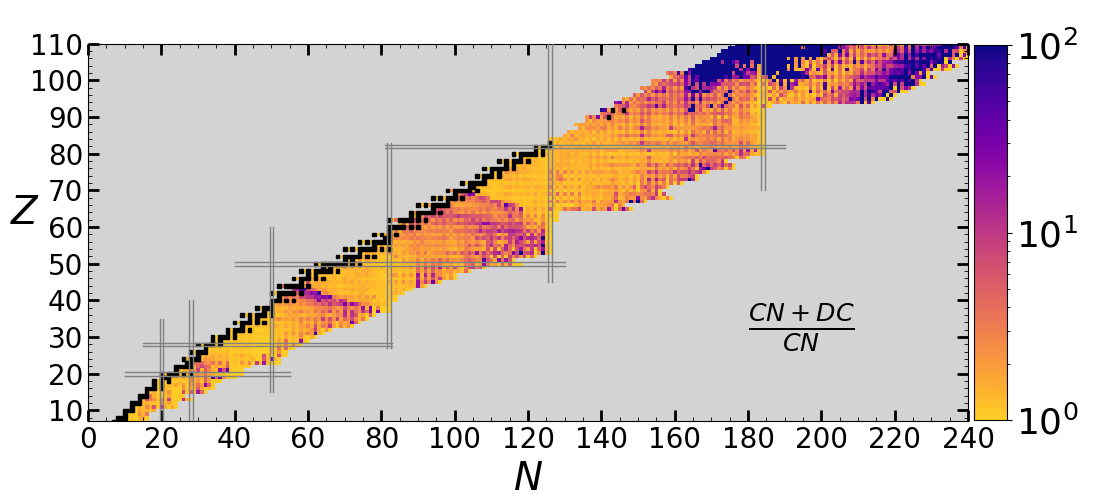}
\caption[Comparison of $(n,\gamma)$ rates calculated with the CN and DC methods in the ($N$,$Z$) plane.]{
Representation in the ($N$,$Z$) plane of the ratio between the Maxwellian-averaged $(n,\gamma)$ rates estimated within the CN statistical method only and when also including the DC component. The neutron capture rates are calculated with the TALYS code \citep{goriely2008,koning2012,Xu2014} using the HFB-21 mass model at $T = 10^9$ K for all nuclei from $Z=8$ up to $Z=110$ between the valley of $\beta$-stability and neutron drip line. The black squares correspond to the stable nuclei and the double solid lines depict the neutron and proton magic numbers.
Note that all values above $10^2$ are displayed in blue, which is relevant for the $Z>90$ region. 
}
\label{fig_ratesdiff}
\end{center}
\end{figure}

Both the DC and CN methods for calculating the theoretical radiative neutron capture rates are consistently included in the TALYS code \citep{goriely2008,koning2012,Xu2014}, which we use to estimate the rates. The reverse photoneutron emission rates, i.e., the $(\gamma,n)$ rates, are calculated using detailed balance (\cref{eq:det_bal}), which includes an exponential dependence on the $Q$-value of the nuclei involved. Therefore, a large part of the sensitivity to the radiative neutron capture rates originates directly from uncertainties in the prediction of the neutron separation energy, hence of nuclear masses.  Deviations in the mass predictions on the order of several MeV (see Sec.  \cref{subsec_massmods}) translate into deviations in the neutron capture rates up to 3-5 orders of magnitude in certain regions of the nuclear chart.
In \cref{fig_ratesdiff}, a comparison between the neutron capture rates obtained with the CN and CN plus DC component is displayed.  We can see that the largest contribution from the DC component is for the neutron-rich nuclei and in certain regions in between the nuclear magic numbers ($Z\sim 35-40$ and $Z\sim 50-70$).  There are also significant discrepancies in the fissile region ($Z>90$), where the CN+DC component is over $100$ times larger than the CN component. 

As default, all our r-process calculations are based on CN model rates, except for one variation of the nuclear inputs, which also includes the DC component (see \cref{tab_nuc_mods}, input set 15).
In this work, we have used the microscopic HFB plus combinatorial nuclear level densities  \citep{goriely2008} and the E1 and M1 D1M+QRPA  strength functions (with the inclusion of an empirical upbend) \citep{Goriely18a}.
The cross-section calculations are believed to be quite insensitive to varying the nuclear potential as long as they have the same volume integral per nucleon \citep{goriely1997b,Xu2012} and if the isovector imaginary potential remains large enough to ensure the neutron absorption \citep{Goriely07b}. We will therefore use the same nuclear potential, the Woods-Saxon-type optical potential \citep[KD;][]{koning2003} for all calculations in this work.




\subsection{Nuclear mass models}
\label{subsec_massmods}


The mass of the nucleus is one of the most crucial ingredients on which many nuclear properties depend.  Although great experimental progress has been made in the last decades to reach the exotic neutron-rich region and future large facilities such as the Facility for Rare Isotope Beams \citep{Surman18}, the Facility for Antiproton and Ion Research (FAIR) \citep{Walker2013} and the Radioactive Isotope Beam Factory (RIBF) at RIKEN \citep{Li2022} will certainly help to provide experimental masses deep inside the neutron-rich region,  it will not be feasible to measure the masses of the nuclei of most importance to the r-process in the years to come. Thus, our nucleosynthesis calculations will have to continue to rely on theoretical masses in the foreseeable future.

In order to compare mass models and measure their performance, it is common to calculate the root mean square (rms) deviation between a given mass model and all of the experimentally measured masses. 
Ideally, a mass model should be able to provide not only the mass but also the other nuclear properties like charge radii, quadrupole moments, fission barriers, shape isomers, as well as infinite nuclear matter properties. 
This way, it would be possible to evaluate the models performance not only based on its rms deviation from the known masses but also according to other nuclear properties and constraints.  The mass models considered here have rms deviations from about 0.3 to 0.8 MeV on all the 2457 $Z, N \ge 8$  known masses  \citep{Wang2021}.

Many different approaches exist in order to calculate nuclear masses ranging from the first macroscopic classical models (e.g., liquid drop model) to microscopic models only relying on first principles (e.g., shell model). In between these two extremes, there are many semi-empirical approaches where a theoretical description of the nucleus is combined with free parameters which are fitted to the known masses (and sometimes other nuclear properties). 
In the following, we will focus on global mass models that try to reproduce the masses of all nuclei lying between the proton and neutron drip lines of relevance for astrophysical applications. For a review, see, e.g., \citet{Pearson00,Lunney03}. Some of these models are described below.

\subsubsection*{Macroscopic-microscopic approach}

The classical liquid drop model describes the nucleus as consisting of nucleons that behave like the particles in a drop of liquid.  Based on this model, the semi-empirical mass formula (also known as the Bethe-Weizs\"acker formula) \citep{VonWeizsacker1935} was developed and shown to be quite successful in reproducing the general trends observed in nuclear data. However, it fails to describe quantum effects, and therefore several macroscopic-microscopic models have been proposed where microscopic corrections are added to the liquid drop part to account for the quantum shell and pairing correlation effects.
In this framework, the macroscopic and microscopic features are treated independently, both parts being connected exclusively by a parameter fit to experimental masses. The most sophisticated version of this macroscopic-microscopic mass formulas is the ``finite-range droplet model'' (FRDM) \citep{Moller2016}.  
The calculations are based on the finite-range droplet macroscopic model and the folded-Yukawa single-particle microscopic correction. The 31 independent mass-related parameters of the FRDM model
are determined directly from a least-squares adjustment to  the ground-state masses on all the masses available at that time. 
This fit (i.e., FRDM12) leads to a final rms deviation of 0.61~MeV for the 2457 $Z,N \geq 8$ nuclei with experimental masses.
Inspired by the Skyrme energy-density functional (see below), the so-called Weizs\"acker-Skyrme (WS) macroscopic-microscopic mass formula was proposed by \citet{Wang10,Wang2014,Liu11} with an rms deviation of about 0.3~MeV, from now on referred to as WS4. In such an approach, the mass formula is mathematically corrected by including a Fourier spectral analysis examining the deviations of nuclear mass predictions to the experimental data at the expense of a huge increase in the number of free parameters and potentially a decrease in the predictive power of the model.

Despite the great empirical success of the macroscopic-microscopic approach, it suffers from major shortcomings, such as the incoherent link between the macroscopic part and the microscopic correction, the instability of the mass prediction to different parameter sets, or the instability of the shell correction \citep{Pearson00,Lunney03}. 
The quality of the mass models available is traditionally estimated by the rms error obtained in the fit to experimental data and the associated number of free parameters. 
However, this overall accuracy does not imply a reliable extrapolation far away from the experimentally known region since models may achieve a small rms value by mathematically driven or unphysical corrections or can have possible shortcomings linked to the physics theory underlying the model.  
As discussed above, the reliability of the mass extrapolation should be considered as the second criterion of first importance when dealing with specific applications such as astrophysics, but also more generally for the predictions of experimentally unknown ground and excited state properties. For this reason, microscopic mass models have been developed, as discussed below.

\subsubsection*{Mean-field approach}

One of the most microscopic and successful models of the nucleus is the shell model, where the interaction between each individual nucleon (or each nucleon outside of a rigid  core) is taken into account. However, due to the impossible task of solving the many-body problem for a large number of nucleons, e.g., due to the extreme computational demand, such a model can only be applied to light systems. Instead of calculating the interaction between all nucleons exactly, effective potentials describe the interaction of a nucleon with the mean-field generated by all the other nucleons. 
By using microscopic nuclear many-body models as a basis, relativistic or non-relativistic mean-field approaches based on the density functional theory can calculate all the masses of the entire nuclear chart. However, fitting interaction parameters to essentially all mass data remains an extremely demanding task, and for this reason, most of these mean-field models have been adjusted on the properties of only a few nuclei. 
Such mean-field models may present very different accuracies on the global set of experimental masses. Consequently, extrapolations towards exotic neutron-rich nuclei become unreliable, and these models should therefore not be considered for r-process applications.

The underlying interactions of the microscopic models can also be used to calculate the EoS of infinite nuclear matter and neutron star material. This is an advantage for the mean-field  models since it gives another application of the theoretical framework and provides additional model constraints as, for example, maximum NS mass or mass-radius relations \citep[e.g., see][]{fantina2013,pearson2018}.
In the long run, this approach may establish a consistency between the model of the network calculations and the EoS used in the astrophysical simulations, which is currently not considered.

\begin{figure}[tbp]
\begin{center}
\includegraphics[width=\textwidth]{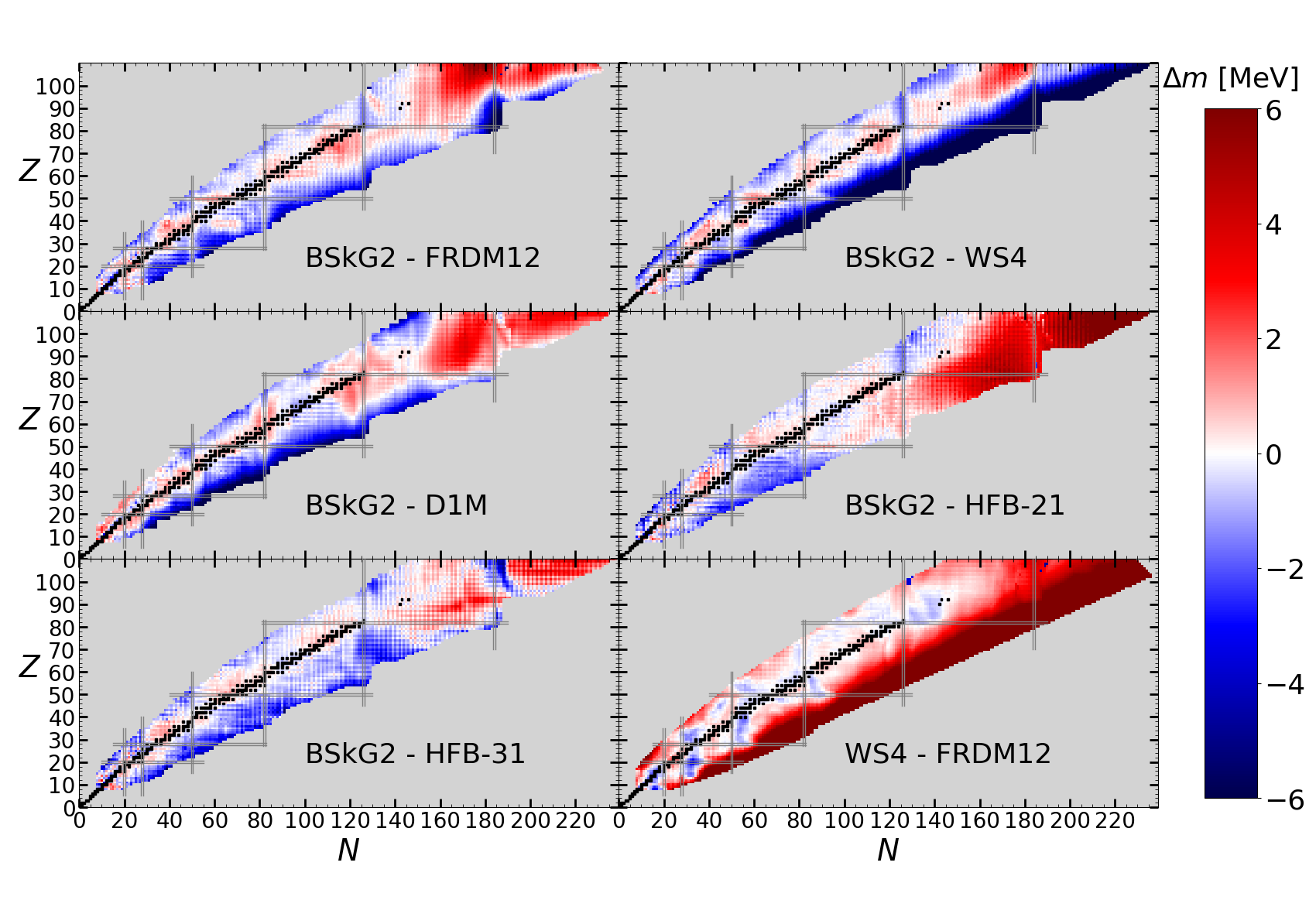}
\caption[Similar to \cref{fig_ratesdiff} for mass difference between six mass models. ]{
Similar to \cref{fig_ratesdiff} for mass difference $\Delta m$ between different mass models listed in  \cref{tab_nuc_mods}. Mass differences larger or smaller than the maximum or minimum values of the colour bar ($\pm 6$ MeV) are represented by red and blue, respectively.
}
\label{fig_massdiff}
\end{center}
\end{figure}

In this work, we consider mean-field mass models that have been fitted to essentially all known masses and that are based on two different types of effective interactions, namely the Skyrme \citep{Vautherin72} and finite-range Gogny \citep{Gogny73} interactions. More specifically, we adopt the Gogny-D1M mass model \citep{Goriely2009b}, which takes into account all the quadrupole correlations self-consistently and microscopically and reproduces the 2457 experimental masses  \citep{Wang2021} with an rms deviation of 0.81 MeV.
Many versions of the Brussels-Montr\'eal Skyrme-HFB mass models have been developed in a series of continuous improvements. 
We consider here the HFB-21 \citep{goriely2010} and HFB-31 \citep{Goriely2016c} mass models with an rms deviation of 0.59~MeV, respectively, since both have been used previously in nucleosynthesis calculations. In addition, we also consider the BSkG2 mass model \citep{Ryssens2022}, which is obtained through a three-dimensional coordinate-space representation of the single-particle wave functions allowing for both axial and triaxial deformations and treats nuclei with odd number of nucleons in the same way as the even-even nuclei by breaking time-reversal symmetry. Its rms deviation amount to 0.68 MeV with respect to the 2457 known masses. Unless otherwise specified, we use BSkG2 as our default mass model.

In \cref{fig_massdiff}, the mass differences between BSkG2 and the other five mass models, as well as between both droplet-type parametrizations,  are displayed.  The mass difference can be over 6 MeV in some regions, where the largest deviations generally lie close to the drip lines or in regions further from the neutron magic numbers for some models.

\subsection{$\beta$-decay rates}
\label{subsec_betarates}

$\beta$-decay rates remain crucial for the r-process nucleosynthesis since they define the time scale for the flow of abundance from one $Z$ to the next. At later times, after freeze-out, the $\beta$-decay and $\beta$-delayed neutron emission probabilities play an essential role in determining the flow back to the valley of $\beta$-stability and also the energy release relevant for the kilonova light curve. 
Although $\beta$-delayed neutron emission occurs throughout the duration of the r-process, it is in particular important during the late phases when the competition between neutron captures and $\beta$-decays shape the final r-process abundance distribution, in particular around the r-process peaks. 

As with the other nuclear input properties required by the r-process, the experimental $\beta$-decay rates are included when available \citep{kondev2021}, but for almost all nuclei produced during the neutron irradiation, we have to apply theoretical rates from one of the few available global models. 
Similar to the mass models, microscopic shell model calculations for $\beta$-decay exist. However, to the computational cost, they are restricted to nuclei near closed shells and cannot yet tackle the heavy, neutron-rich nuclei required for r-process applications.
One widely used model is Gross Theory, which allows for fast and reasonably accurate computation of the half-lives given the input of appropriate $Q_\beta$-values. The Gross Theory is based on $\beta$-decay one-particle strength functions to describe the general behaviour of $\beta$-decay. The one-particle strength functions and the pairing scheme has been improved in its GT2 version \citep{tachibana1990} based on the HFB-21 $Q_\beta$-values.

Several global semi-microscopic models are based on an effective interaction to describe the nuclear structure and adopt the random phase approximation (RPA) or its quasiparticle (QRPA) extension to include pairing interactions to describe allowed Gamow-Teller decay.
The FRDM+QRPA \citep{moller2003} model uses the FRDM $Q$-values and a separable residual interaction for the QRPA description of the Gamow-Teller transitions and also includes the Gross Theory for the first-forbidden transitions.
One of the most reliable models available is the spherical relativistic Hartree-Bogoliubov (RHB) model plus relativistic QRPA (RQRPA) \citep{marketin2016}  based on the DC3* residual interaction, which also includes first-forbidden transitions.  

Finally, the Tamm-Dancoff approximation (TDA) \citep{klapdor1984} is another global model available that provides a simple analytical solution to calculate the $\beta$-decay strength distribution by using a simplified Gamow-Teller residual interaction, neglecting the influence of first-forbidden transitions on the half-lives. 

\begin{figure}[tbp]
\begin{center}
\includegraphics[width=\textwidth]{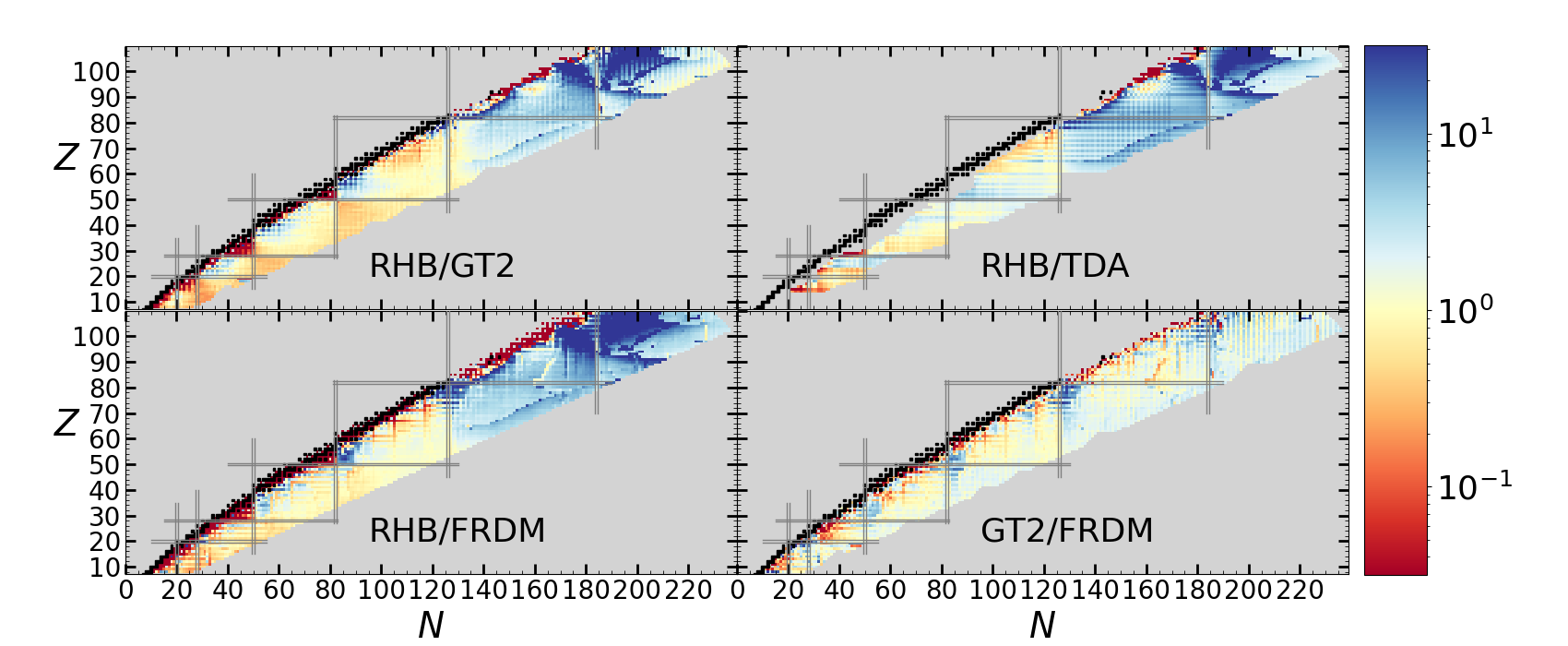}
\caption[Similar to \cref{fig_ratesdiff} for $\beta$-decay rates predicted by four models.]{
Similar to \cref{fig_ratesdiff} where the ratio between the $\beta$-decay rates predicted by the four models listed in  \cref{tab_nuc_mods} are plotted in the ($N$,$Z$) plane.  Differences larger or smaller than the maximum or minimum values of the colour bar ($10^{\pm 1.5}$ MeV) are represented by blue and red, respectively.  
}
\label{fig_betadiff}
\end{center}
\end{figure}

In \cref{fig_betadiff}, the ratio between the $\beta$-decay rates estimated by the relativistic mean-field (RMF), i.e., RHB+RQRPA, and the HFB21+GT2, TDA and FRDM+QRPA as well as the ratio between the HFB21+GT2 and FRDM+QRPA predictions are shown.  
Major differences can be seen for exotic neutron-rich nuclei, in particular for $N>126$, where, in particular, the RHB+RQRPA model is seen to predict larger rates.
We apply all four $\beta$-decay models in \cref{fig_betadiff} in our r-process calculations, where the RHB+RQRPA model is used as default (see  \cref{tab_nuc_mods}). 
Note that when we change the applied mass model for our r-process calculations, it directly impacts the estimated neutron capture rates but not the $\beta$-decay rates since they are estimated separately by the models discussed above.  Thus, due to the limited set of available global $\beta$-decay models, we often have an inconsistency between the mass model applied in the r-process calculations and the masses used to calculate the $\beta$-decay rates. Therefore, a future prospective is to develop fully consistent models for all nuclear physics input applied in r-process calculations.

\subsection{Fission models for r-process applications}
\label{subsec_fission}

In sufficiently neutron-rich astrophysical scenarios ($X_n^0$ typically larger than 0.7), the r-process can reach nuclei in the $Z \ga 90$ region, where fission may become the dominant decay mode. Therefore, if efficient, fission terminates the r-process because it stops the abundances from reaching even further up in the super-heavy region of the nuclear chart, and recycles material back to the $100\lesssim A\lesssim 180$ region. Given sufficient time before freeze-out, these daughter nuclei may again be involved in a series of neutron captures and $\beta$-decays before reaching the fissile region for the second time. 
In addition to producing daughter nuclei, fission also contributes to the r-process by releasing neutrons that may be recaptured by the various nuclei in the ejecta. This late neutron production can boost the neutron capture phase and impact the final abundance distribution after freeze-out. 
Another essential aspect of fission during the r-process is that it releases heat which can impact the kilonova light curve. 
In particular, the spontaneous fission of ${}^{254}$Cf with a half-life of 60 days has been shown to be important for energy generation and to impact the light curve at $t>100$ days \citep{wanajo2018,wu2019,zhu2018}.

\begin{landscape}
\begin{figure}[tbp]
\begin{center}
\includegraphics[width=\columnwidth]{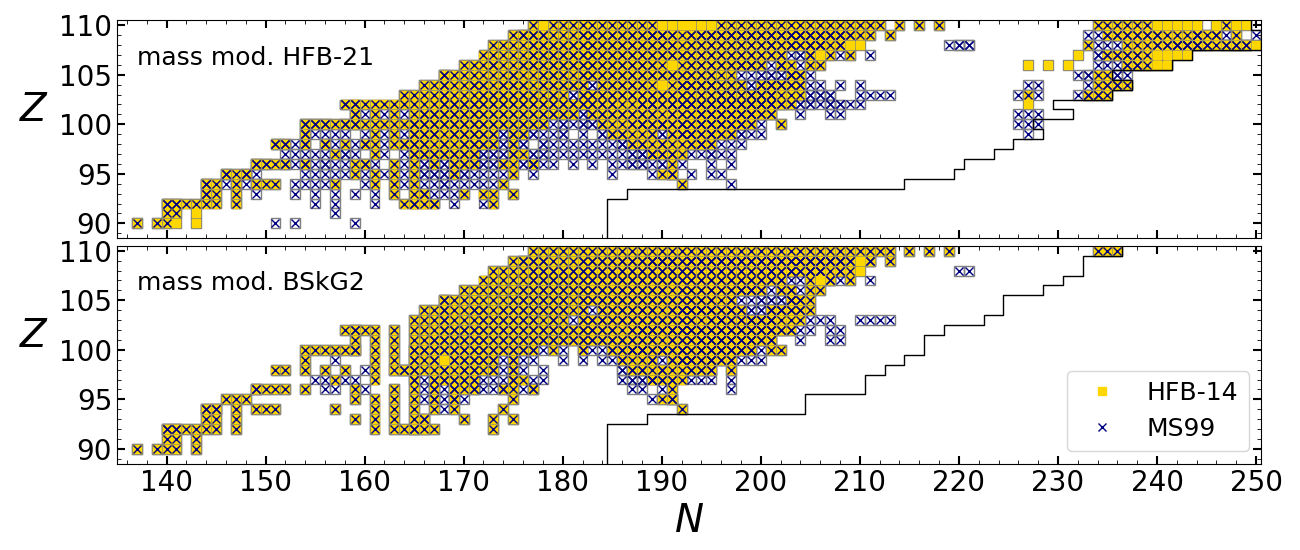}
\caption[The fission-dominated regions using the HFB-14 or MS99 fission barriers.]{
The fission-dominated regions using the HFB-14 (gold squares) or MS99 (dark blue crosses) fission barriers, as used for input sets 1, 6, 15 and 17 in  \cref{tab_nuc_mods}. For each nucleus, a marker is drawn whenever the probability (or rate) for fission is larger than any other decay or reaction mode (i.e., whenever the probability (or rate) for spontaneous fission or $\beta$-induced fission is larger than $\beta$-decay, or neutron-induced fission is larger than neutron capture). The top panel uses mass model HFB-21, and the bottom panel BSkG2 for calculating the neutron capture rates, where the corresponding neutron drip line is shown as a black line.
}
\label{fig_fiss}
\end{center}
\end{figure}
\end{landscape}

Fission is a very complex nuclear process, and a variety of nuclear structure inputs are required to achieve a successful model. For the r-process, we need models describing the probability for a nucleus to undergo neutron-induced, spontaneous and $\beta$-delayed fission, but also fission fragment distributions to determine the daughter nuclei and the number of free neutrons released. 
Photo-induced fission is assumed to be unimportant since the temperatures have typically fallen below 1 GK by the time the r-process network reaches the fissile region (see \cref{subsec_photofis} for a sensitivity analysis). 
In this work, we have used either the HFB-14 fission paths based on the BSk14 Skyrme interaction\footnote{Ideally, the fission paths should be calculated based on the same interaction as the mass models applied in the r-process calculations (i.e., HFB-21 and so on, see \cref{tab_nuc_mods}).  However, fission barrier calculations are very expensive; therefore, only a few global models suitable for astrophysical applications are available.}
 \citep{goriely2010} or the Thomas-Fermi model \citep[][hereafter MS99]{Myers1999} to estimate the fission rates. 
\cref{fig_fiss} shows the regions where fission processes dominate over other decay modes or neutron capture for the HFB-14 and MS99 fission barriers using either mass model HFB-21 or BSkG2 for the calculation of the neutron capture rates.
The MS99 barriers are usually lower than the predictions of the HFB model, so more nuclei are found to be affected by fission processes when the MS99 barriers are adopted. Note that the MS99 barriers are not always available for very exotic n-rich nuclei close to the neutron-drip line, especially for $Z \geq 106$. In these cases, the HFB-14 barriers are adopted. 
During the freeze-out phase, if we adopt the HFB-14 fission barriers, fission takes place around $Z \simeq 101 -102$ along the abundant $A \simeq 278$ isobar, while the lower MS99 barriers lead to fission already around the $Z \simeq 95 - 97$ isotopes.
The applied mass model impacts the balance between neutron-induced fission and neutron capture through the calculation of the neutron capture rates but also through the position of the neutron drip line since this determines which nuclei are available to fission. This is illustrated in \cref{fig_fiss} where we can see that mass model HFB-21 allows for fission in a large region for $N>230$, while mass model BSkG2 predicts this region to be unstable for neutron emission.

For the fission fragment distributions we adopt either the renewed microscopic Scission Point Yield (SPY) model \citep{Lemaitre2019,Lemaitre2021} or the 2018 version of the semi-empirical GEF model (``GEneral description of Fission observables'')  \citep{Schmidt2016}. Unless otherwise specified, we use the HFB-14 barriers and the SPY fission fragment distributions as default for our calculations.

\subsection{The default nuclear input set}
\label{sec_network_input}

For each r-process calculation, nuclear models for the mass, rates of radiative neutron capture, $\beta$- and $\alpha$-decay, and fission fragment distributions and barriers are required as input. The chosen model combinations that together describe these properties are in the following referred to as an input set.
\cref{tab_nuc_mods} lists the nuclear models considered in this work. 
Unless otherwise specified, our calculations use input set 1 as default in our uncertainty study in \cref{ch_nucuncert}.
For the calculations in \cref{ch_dynweak} which investigate the impact of weak nucleonic reactions in the dynamical ejecta on the r-process nucleosynthesis, we have applied input set 6 with one modification. 
To calculate the theoretical reaction rates the E1 and M1 HFB+QRPA \citep{goriely2004} strength functions have been used instead of the E1 and M1 D1M+QRPA strength functions \citep{Goriely18a} (which are used for all calculations in \cref{ch_nucuncert}). 
As stated in \cref{subsec_ncap}, the microscopic HFB plus combinatorial nuclear level densities \citep{goriely2008} and the KD optical nuclear potential \citep[KD;][]{koning2003} are applied in all calculations.

%
%
%
%
 
A quantity which have not been varied in this work is the model for $\alpha$-decay, where we adapt the model of \citet{Koura2002} for all calculations.


\section[Computer facilities and high-performance computing][Computer facilities and HPC]{Computer facilities and high-performance computing}
\label{sec:HPC}

\epigraph{\itshape ``In FORTRAN, GOD is REAL (unless declared INTEGER).''}{--- \textup{J. Allan Toogood}}

Modern computers have become more powerful, increasing the complexity of the science questions that we can solve.
Instead of simplifying physical problems to be able to solve them analytically, we can now tackle complicated problems by solving them numerically.
The fastest computers in the world, supercomputers, are used in many fields of science and engineering where high-speed computations are required to solve advanced computational problems. 
A subclass of supercomputers are computer clusters which consist of a number of computers set up to work as one single, very large computer. 
Compared to a typical desktop computer, supercomputers deliver a much higher performance; tasks that may take hours, days, weeks, months, years on a standard computer might only take minutes, hours, days, or weeks on a supercomputer. 

The ability to process data and perform complex computations at high speed, either by using supercomputers or by using specially designed numerical methods, are often referred to as high-performance computing (HPC) \citep[e.g.,][]{gorelick2020}. 
At the simplest level, the choice of programming language can be seen as a HPC tool. 
While each programming language has its pros and cons and particular problems for which they perform the best, some languages are objectively faster than others.
For example, low-level compiled languages such as Fortran or C/C++ perform calculations much faster than non-compiled scripting languages such as Python \citep[see for example][]{BenchmLang}. 
However, many tools and techniques exist so that Python can achieve run-time speeds at the level of what low-level languages can obtain \citep{gorelick2020,reitz2016,harris2020,behnel2011}.
In the following we will limit our discussion to tools related to the Python language since it is widely used in scientific research \citep[in fact currently the most popular language in the world][]{TIOBE} and relevant for the present work. 
Despite the slower computation speeds, there are many benefits of using Python \citep{Dubois2007,Millman2011}.
Python is known for a clear syntax so that programs are often shorter than the counterpart written in another language, which some argue makes development faster \citep{Faulk2004,BenchmLang,Dubois2007}. 
It is also designed to be readable\footnote{This of course depend on the programmer.}, and ``close to the problem'' in the sense that the program resembles the (mathematical) problem at hand.
Perhaps the biggest advantage of Python is the large amount of available (scientific) packages \citep[see][for a short historical review]{Dubois2007}, allowing the completion of advanced calculations using only a few simple lines of code. 
A widely used Python package is NumPy \citep{harris2020} which significantly lower the computing time for tasks involving arrays or numbers to speeds comparable with C/C++\footnote{NumPy is in fact built on C code which have been optimized and pre-compiled.} and Fortran.

Another HPC tool which can be used to make Python code run faster, in particular code involving loops, is Cython \citep{behnel2011}. This tool requires the Python code to be compiled, but a significant increase in performance can be achieved without changing the syntax\footnote{Unchanged Python code compiled with Cython runs faster than uncompiled Python code, and even faster code can be acquired by introducing minor changes to the syntax such as adding type declarations.}.
It is also possible to achieve a higher performance by using Cython to call functions written in C from a Python script, or f2py (now a part of NumPy) to use functions or subroutines written in Fortran.
These hybrid HPC tools yield large performance gains per hour spent, in particular in the beginning. This is in contrast to more intricate HPC techniques like for example MPI parallelization which can give a significantly higher performance, but usually at a much larger human cost, i.e., in hours invested (often even before the code can run for the first time). \todo{check the HPC book}

In this work, the computationally heavy tasks (excluding the hydrodynamical simulations), i.e., the r-process calculations, were performed using Fortran, while Python (and NumPy) was used for most of the post-processing, including data analysis and visualisations.
The data analysis, in particular the presentation in \cref{sec_init_cond,sec_analysis} benefited from the data analysis and manipulation package Pandas \citep{mckinney2010} and some machine learning tools in the scikit-learn package \citep{scikit-learn}.
Plots in \cref{ch_dynweak,ch_nucuncert} (and elsewhere) were produced with the Matplotlib \citep{hunter2007} library. 
For the most part, the post-processing did not require high-speed calculations, however, some exceptions were identified in the Python scripts used to analyse the data. 
For example, in the code used to make animations of the r-process flow (described in \cref{sec:nuc_net}), it was found that most of the execution time was spent in one (or a few) of the functions. By using Cython the run-time was lowered by a factor of $\sim10$. 
A similar bottleneck was found in the code used to extract the heat generated by individual nuclei in the ejecta (e.g., see \cref{sec_q}). 
A Fortran subroutine was written (by expanding on previously written code) to solve this.
The f2py package was used to interact with the other Python analysis scripts which significantly reduced the computation times from the order of hours to minutes. 

The trajectories provided by the hydrodynamical models are assumed to be independent from each other, i.e., they do not interact with each other during the expansion, and therefore the r-process network calculation for one trajectory can be performed independently from another. 
From a computational perspective this is an advantage, meaning that we have what is often referred to as a pleasingly parallel problem (also called embarrassingly parallel, perfectly parallel or delightfully parallel).
A pleasingly parallel problem require that little or (as is the case here) no effort is needed to separate the problem into a number of parallel tasks.
In this work, each task (or job) is an r-process calculation which can be sent to one central processing unit (CPU) of the computer to be completed. 
Since we have several thousand trajectories for which we want to calculate the r-process, our tasks are best solved by supercomputers.
The work presented in this thesis have benefited greatly from Zenobe\footnote{This supercomputer was in fact on the list of the top 500 fastest computers in the world between 2014 and 2016 \citep{Zenobe500}.}
, the Belgian Tier-1 supercomputer of Wallonia with a total of 13968 CPU cores (however each user can maximally only use in total 480 cores at the same time and maximally 192 cores per job). 
Some of the r-process calculations were also performed using one of the Wallonia-Brussels supported CECI (Consortium des Équipements de Calcul Intensif) clusters called Hercules2 with 1536 CPU cores, and the desktop computers available at Institut d'Astronomie et d'Astrophysique (IAA) at Université libre de Bruxelles (ULB). Additional early tests were also performed with the now decommissioned Hydra cluster of ULB. 


There are often many users of a supercomputer or cluster, and anyone that wants to use the system has to request the desired resources and queue their jobs. 
This is done through a management script specifying the amount of memory required for the job. For example, Zenobe prefer that each job stay as close as possible to 2625~Mb/core to best exploit the resources. 
Other specifications of the management script include the expected run time (or ``walltime'') and other instructions related to the execution of the job.
For Zenobe the walltime cannot exceed 7~days per job, and 15~days for Hercules. 
Once submitted to the queue, the job scheduler\footnote{Usually \citet{Slurm} or \citet{OpenPBS} is used.} will optimize the queue and decide which jobs to start first. 
The queueing time for a particular job depends on the length of the queue, but more importantly on how the resources you are requesting fit with the resources that become available.
Some systems also prioritize certain types of jobs, i.e., particularly large or long-running jobs. 
Array jobs can be used to simplify the management of many similar jobs, i.e., if many jobs only differ by the input files. The benefit of array jobs is that only one management script is needed for all jobs in the array and the scheduler will do the work of starting the next job(s) in the array when more CPUs become available. 
For the r-process calculations the same executable was used to run the calculations, and the input files for each trajectory were different.
This enabled the submission of many jobs (i.e., as many as the system allows per user) to the queue at once. 

A rough estimate of the computing time used in the present thesis work can be done by assuming that the average trajectory uses about 5~days to complete an r-process calculation (on one CPU).  
This is only a guess based on the fact that some calculations finish within minutes, while others do not finish within the maximum allowed run time of 7~days at the supercomputer.
Across all hydrodynamical models considered in this work there are 
\begin{equation*}
783+1263+4398+1290+150+4150+2116=14\ 150
\end{equation*}
trajectories in total (see \cref{tab:astromods}), where the 150 trajectories correspond to the NS-BH model for which only the subset were included and not all 13175 trajectories. 
This alone implies 
\begin{equation*}
14\ 150 \cdot 5 \cdot 24\ \mathrm{h} = 1\ 698\ 000\ \mathrm{h} 
\end{equation*}
of CPU-time for the full set of trajectories. 
In the study of the nuclear physics uncertainties (\cref{ch_nucuncert}) the nuclear physics input was varied between 15 input sets for all the hydrodynamical models (see \cref{tab:astromods} and \cref{tab_nuc_mods}) by using a subset of
\begin{equation*}
266+256+150+296+177=1145
\end{equation*}
trajectories giving 
\begin{equation*}
15\cdot 1145\cdot 5 \cdot 24\ \mathrm{h}= 2\ 061\ 000\ \mathrm{h}
\end{equation*}
of elapsed CPU-time.
The grand total is then an estimated \textbf{3~755~760 CPU hours} used to acquire the results of this thesis. This measure does not include calculations that failed, were restarted\footnote{For example, all of the 7707 trajectories of the dynamical ejecta presented in \cref{ch_dynweak} were re-calculated once due to an input error.} and tests, or any of the sensitivity calculations discussed in \cref{ch:sensitivity}. 
Therefore, assuming that the applied average time of 5~days used for an r-process calculation to finish is correct, the total estimate of 4 million CPU hours must be a lower limit, and the true number is much larger. 

\todo{can actually check the computing time...}



\chapter[Sensitivity studies and results analysis][Sensitivity studies and analysis]{Sensitivity studies and results analysis}
\label{ch:sensitivity}

\epigraph{\itshape \enquote{I mean, it’s sort of exciting, isn’t it, breaking the rules?} Hermione said.}{--- \textup{J. K. Rowling}, Harry Potter and the Order of the Phoenix}


In order to study the synthesis of the r-process elements, several simplifications and assumptions have to be made, as already discussed in \cref{sec:nuc_net}. This chapter is devoted to testing the impact and validity of some essential assumptions (\cref{sec_sensitivity}) and presents a detailed analysis of the physical observables from the r-process calculations (\cref{sec_analysis}). The analysis includes the selection of trajectory subsets, which will be used in the nuclear uncertainty study in \cref{ch_nucuncert} and the sensitivity tests performed in \cref{sec_sensitivity}.

\section{Results analysis}
\label{sec_analysis}

Within a specific hydrodynamical model, the trajectories experience a range of initial conditions (e.g., \cref{sec_init_cond}). The time evolution of the temperature, density and electron fraction may differ significantly between the trajectories (except for the BH-NS model, which has quite similar ejecta conditions, see \cref{fig_xn0_ye_distr} and the discussion in \cref{subsec_photofis}). 
Thus the resulting r-process abundance distribution for one trajectory may differ significantly from another. 
As illustrated in \cref{sec:HPC}, running the r-process results for a large number of trajectories requires an excessive amount of computing resources. One way to reduce the required CPU time when studying the r-process results with respect to input variations is to select a subset of trajectories that satisfactorily represent the complete set.
The aim of \cref{subsec_cluster} is to motivate the method used to select the trajectory subsets in \cref{sec_sel_subs}. 
Thus, some r-process results will be shown ``in advance'' in \cref{subsec_cluster} in order to justify the trajectory subsets extracted in \cref{sec_sel_subs}.
These trajectory subsets will be applied in the sensitivity tests in \cref{sec_sensitivity} and the study of nuclear uncertainties in \cref{ch_nucuncert}, not in \cref{ch_dynweak}.
In this section, nuclear input set 6 in \cref{tab_nuc_mods} has been used for all calculations, which for the four NS-NS dynamical ejecta models are the same data as presented (with full details) in \cref{ch_dynweak}.


\subsection{Trajectory clusters}
\label{subsec_cluster}

\begin{figure}[tbp]
\begin{center}
\includegraphics[width=\columnwidth]{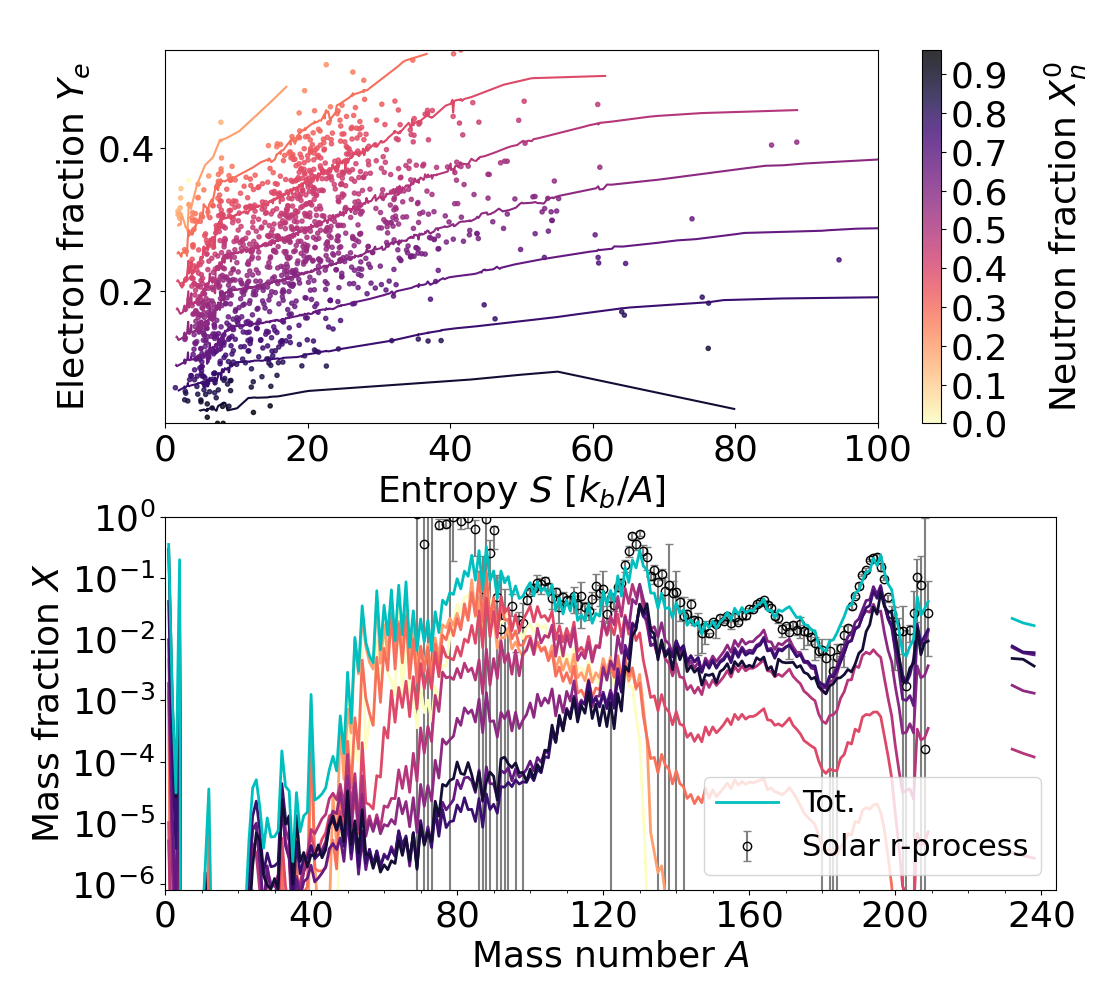}
\caption[The initial electron fraction versus initial entropy and the mass fraction versus mass number when using all trajectories of model SFHo-135-135.]{
The initial electron fraction $Y_e$ versus initial entropy $S$ (top) and the mass fractions $X$ versus mass number $A$ (bottom) when using all 1263 trajectories of model SFHo-135-135. Initial refer here to the values extracted at $\rho_\mathrm{net}$. The trajectories are divided into bins of the initial neutron mass fraction $X_n^0\in[0.,0.2,0.3,0.4,0.5,0.6,0.7,0.8,0.9,1.]$ which are indicated by the contour lines in the top plot and the colour scale. The cyan line in the bottom panel shows the mass-averaged mass fractions when using all trajectories, and the other lines indicate the mass fractions mass-averaged over the trajectories in each $X_n^0$-bin. 
}
\label{fig_Xn0_ye_ss}
\end{center}
\end{figure}

As discussed in \cref{sec_init_cond}, the neutron mass fraction $X_n^0$ at $\rho_\mathrm{net}$ is a suitable quantifier of the r-process efficiency, and in particular, if a strong r-processing with significant production of heavy r-process elements and actinides occur. In the following, we will refer to all values extracted at the network initiation time ($\rho_\mathrm{net}$) as ``initial'' values. 
The top panel of \cref{fig_Xn0_ye_ss} shows the initial electron fraction versus the initial entropy for all trajectories of the SFHo-135-135 model. The contour lines divide the trajectories into nine $X_n^0$ bins, emphasized by the colour scale. In the bottom panel, the corresponding mass fractions versus mass number is shown for the trajectories in each of the nine $X_n^0$-bins, where the mass fraction has been calculated according to \cref{eq:mass_avr} for each bin. The final mass fraction is shown as the cyan line when using the complete set of 1263 trajectories. \cref{fig_Xn0_ye_ss} highlights that the diversity of trajectory conditions also yields a diversity in the r-process results, where the smallest $X_n^0$-values lead to the production of only $A<140$ nuclei, while larger values of $X_n^0$ lead to the production of progressively more heavy r-process elements. 

As the cyan line in \cref{fig_Xn0_ye_ss} shows, it is the conditions of the bulk of the ejecta that dominates the final results (see \cref{fig_xn0_ye_distr} which shows $X_n^0\sim 0.4-0.6$ for the bulk of the SFHo-135-135 material). 
However, even if $X_n^0$ successfully divides the trajectories into groups with similar mass fraction distributions, other physical observables can show significant variations within a group. For example, the trajectory velocity can significantly impact the final abundance distribution. 
Here, we will only use the velocity as an example of how other physical observables than the $X_n^0$ can impact the final results. The complete discussion about the high-velocity component of the dynamical ejecta and how it impacts the r-process results will be postponed to \cref{sec_ang_vc}. 

\begin{figure}[tbp]
\begin{center}
\includegraphics[width=0.8\textwidth]{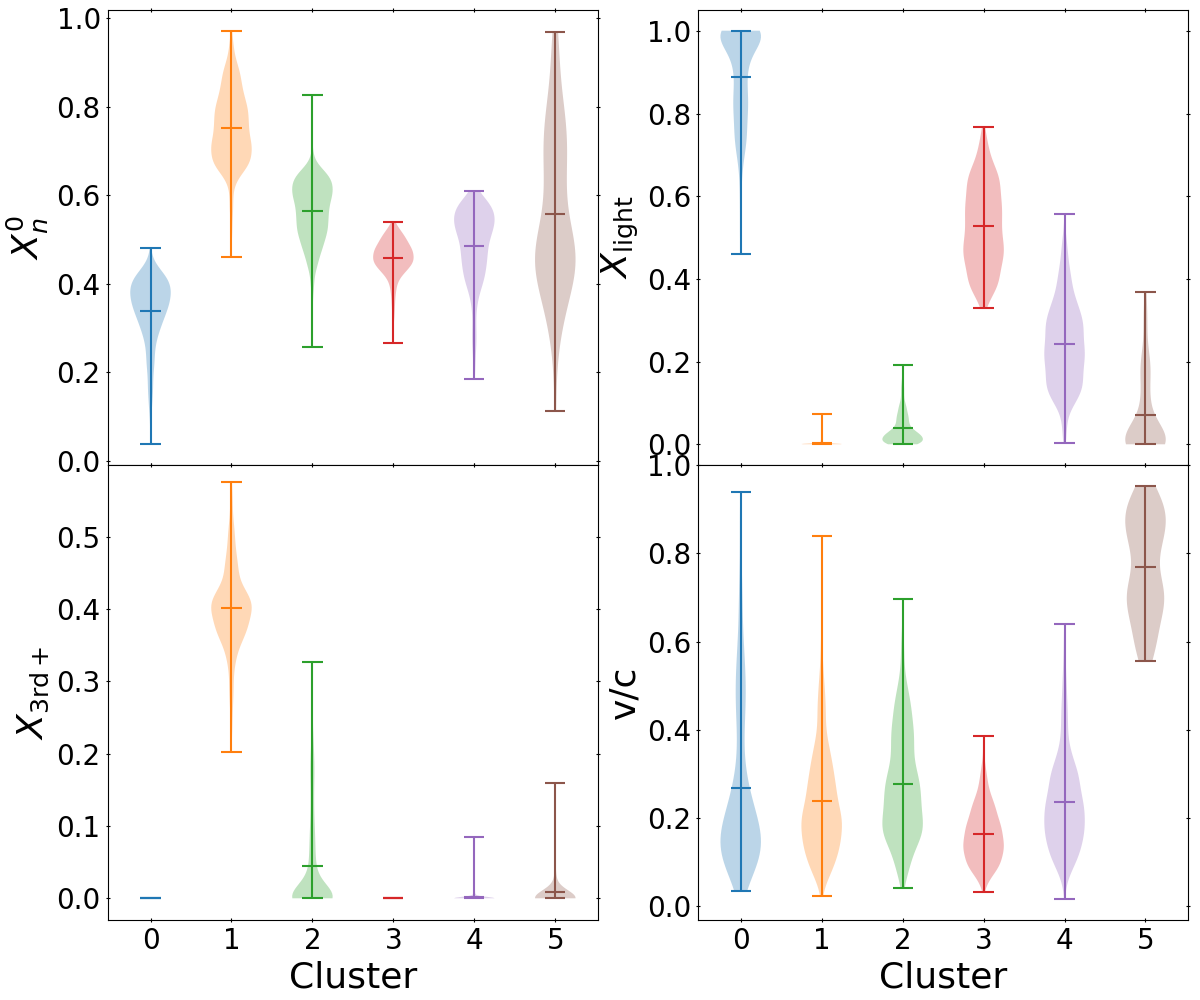}
\caption[A violin plot of the physical observables of model SFHo-125-145 used as input to the clustering algorithm.]{
A violin plot of the physical observables of model SFHo-125-145 (with 4398 trajectories) used as input to the clustering algorithm: the initial neutron mass fraction $X_n^0$ (top left), the sum of the final mass fractions of light elements $X_\mathrm{light}$ with $5<A\leq 95$ (top right), the sum of the final mass fractions of third-peak and heavy r-process nuclei $X_\mathrm{3rd+}$ with $180>A$ (bottom left), and the initial velocity $v/c$ (bottom right). An horizontal bar indicates the mean, minimum and maximum of each cluster distribution. 
}
\label{fig_clu_param}
\end{center}
\end{figure}

\begin{figure}[tbp]
\begin{center}
\includegraphics[width=0.8\textwidth]{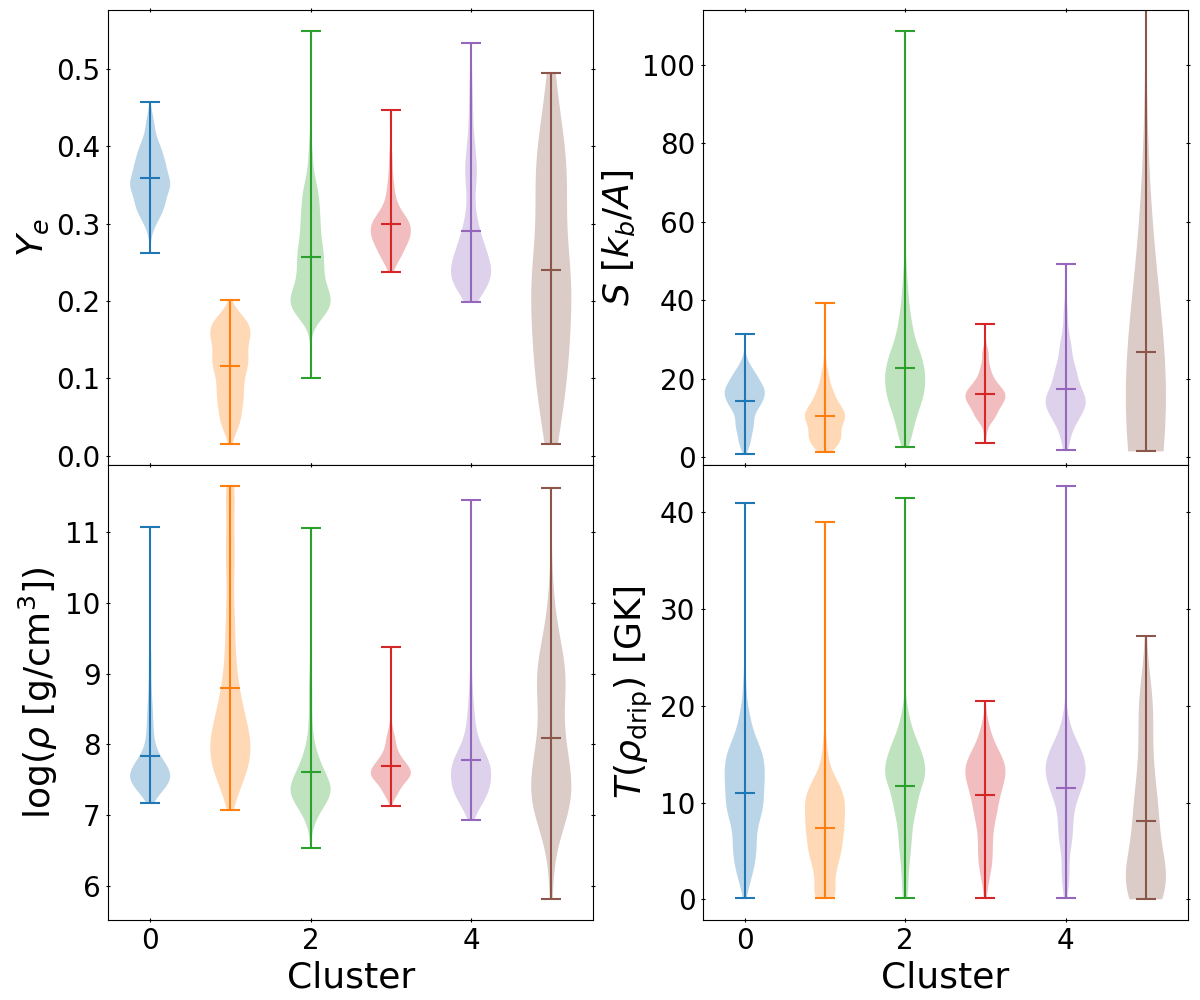}
\caption[Same as \cref{fig_clu_param} for physical observables not used as input to the clustering algorithm.]{
Same as \cref{fig_clu_param} for physical observables not used as input to the clustering algorithm: the initial electron fraction $Y_e$, the initial entropy $S$, the log of the density at the network initiation time, and the temperature at the drip density $\rho_\mathrm{drip}$. 
}
\label{fig_clu_other}
\end{center}
\end{figure}

\begin{figure}[tbp]
\begin{center}
\includegraphics[width=\textwidth]{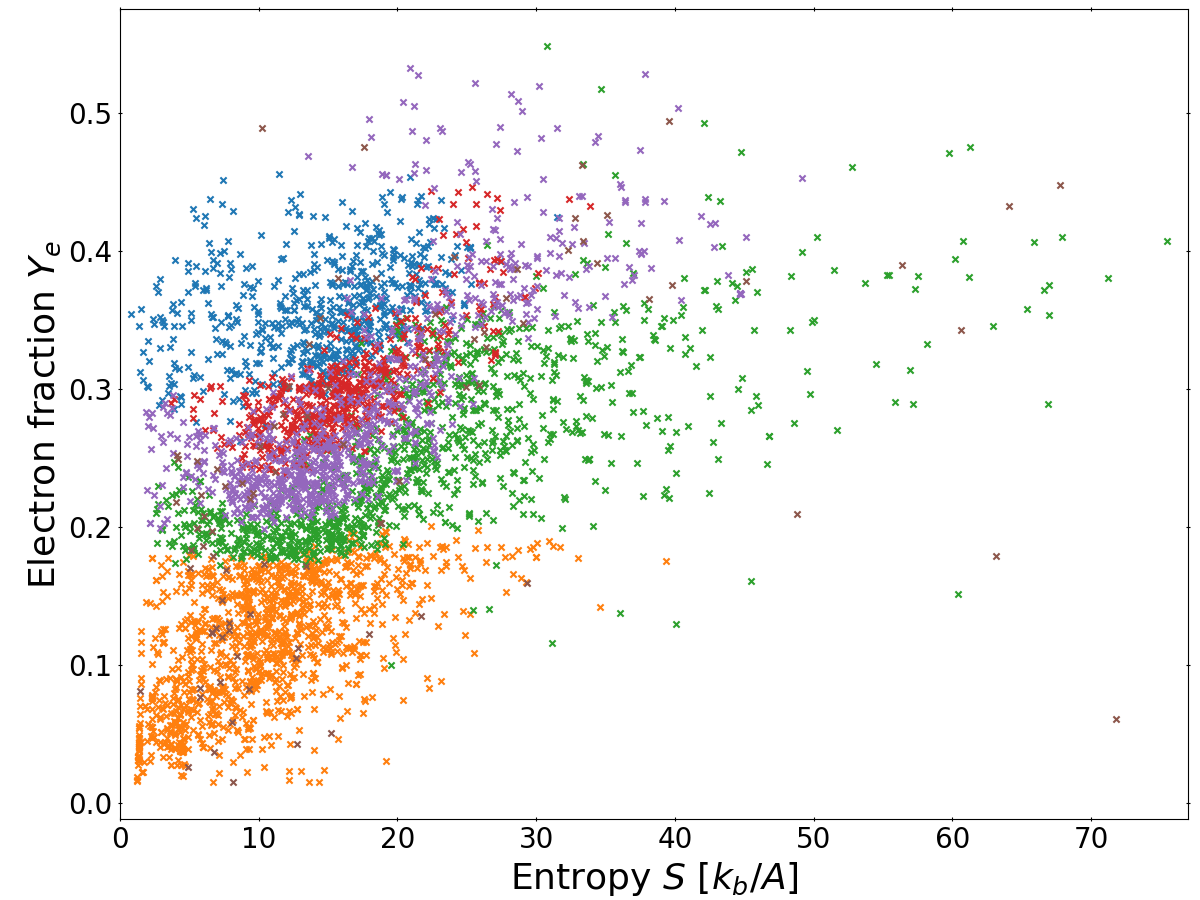}
\caption[A scatter plot of the electron fraction versus entropy for all 4398 trajectories of model SFHo-125-145.]{
A scatter plot of the electron fraction $Y_e$ versus entropy $S$ at $\rho_\mathrm{net}$ for all 4398 trajectories of model SFHo-125-145. The colours indicates the six clusters also shown in \cref{fig_clu_param,fig_clu_rpro}, see the text for more details. Note that four trajectories with large entropies are not shown as they are outside the range of the $x$-axis.
}
\label{fig_clu_S0_Ye}
\end{center}
\end{figure}

\begin{figure}[tbp]
\begin{center}
\includegraphics[width=\textwidth]{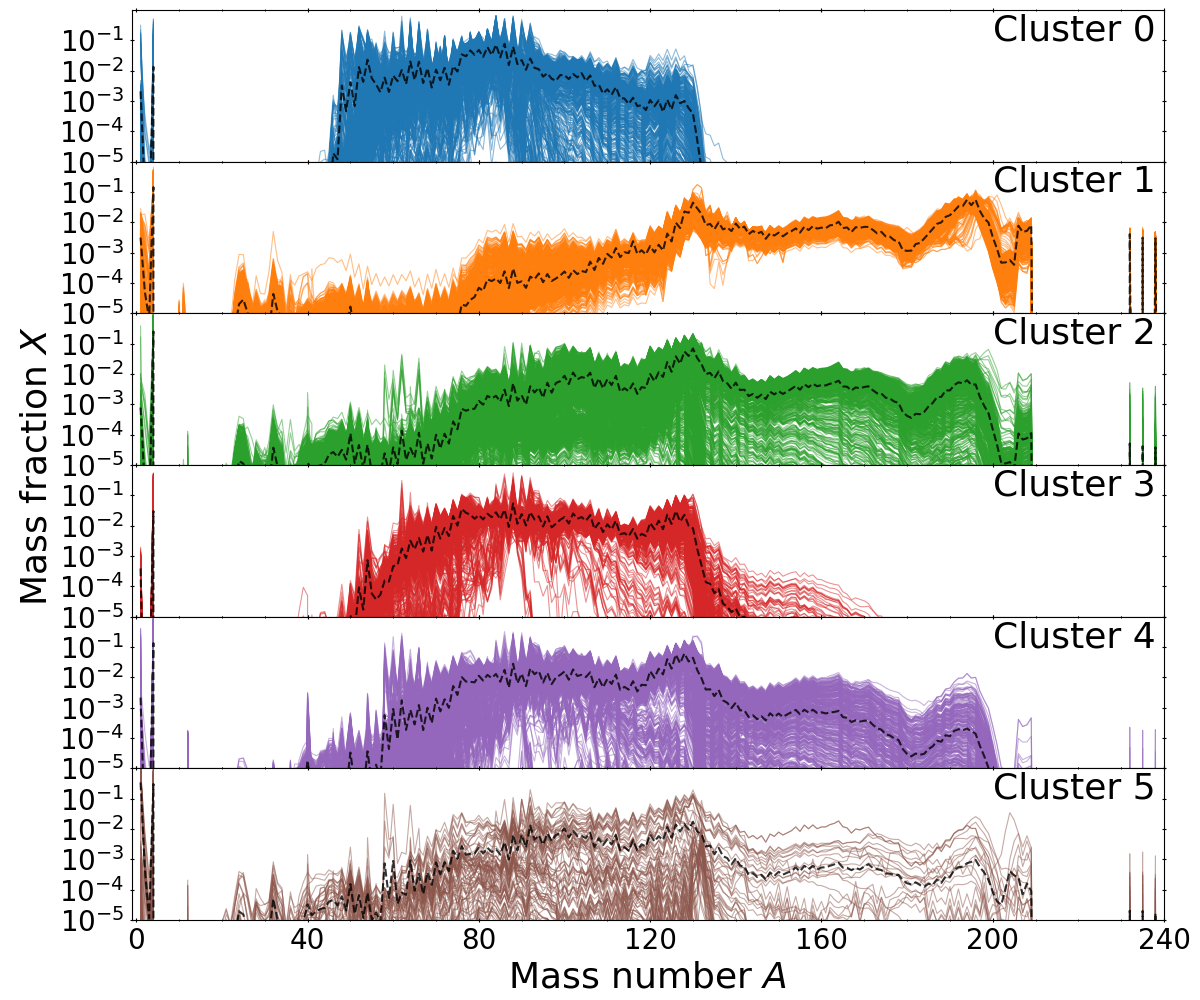}
\caption[The final mass fraction versus mass number for all 4398 trajectories of model SFHo-125-145 divided into six clusters.]{
The final mass fraction $X$ versus mass number $A$ for all 4398 trajectories of model SFHo-125-145 divided into six clusters also shown in \cref{fig_clu_param,fig_clu_S0_Ye}, see the text for more details.
The black dashed lines are the mass-averaged sum of all trajectories within each cluster. 
}
\label{fig_clu_rpro}
\end{center}
\end{figure}

Let us now use an algorithm, or more specifically, a form of machine learning, to automatically sort the trajectories into clusters, i.e., into well-defined bundles of trajectories. In the following discussion, we will show results obtained using the Agglomerative Clustering algorithm of the Python scikit-learn package \citep{scikit-learn}. 
However, other clustering algorithms, such as Gaussian Mixture and K-means, were also tested. 
The Agglomerative clustering algorithm requires the number of clusters to be decided by the user in advance\footnote{Other clustering algorithms can decide the number of clusters based on the input data \citep{scikit-learn}.} and a set of input data to be provided.
The data can have as many dimensions as the user wishes. However, it is easier to interpret the results if the dimension is not too large. 
In our case, each data point corresponds to a trajectory, and each dimension to a (chosen) physical observable. 
Many observables were tested as input to the clustering algorithm to find the best-performing combination in terms of separating the trajectories into groups with similar mass fraction distributions while also highlighting the correlation between an observable and the final r-process results. 
The combination of the following observables yielded the most interesting results: the initial neutron mass fraction $X_n^0$, the sum of the final mass fractions of light elements $X_\mathrm{light}$ with $5<A\leq 95$, the sum of the final mass fractions of third-peak and heavy r-process nuclei $X_\mathrm{3rd+}$ with $180>A$, and the initial velocity $v/c$. 
\cref{fig_clu_param} shows a violin plot of the data used as input to the clustering algorithm using all 4398 trajectories of the SFHo-125-145 model. 
A similar plot of the physical observables not used as input to the clustering algorithm is shown in \cref{fig_clu_other}.
Based on the four observables $X_n^0$, $X_\mathrm{light}$, $X_\mathrm{3rd+}$, and $v/c$, the algorithm automatically, i.e., without further intervention by the user, divided the trajectories into six clusters numbered from 0 to 5. 
The $X_n^0$-distributions in \cref{fig_clu_param} overlap somewhat; however, the mean $X_n^0$ values of clusters 0 to 4 differ significantly.
This is particularly visible in \cref{fig_clu_S0_Ye}, which shows the initial electron fraction $Y_e$ versus entropy $S$ for the same clusters. 
Except for cluster 5, it is interesting that the algorithm has chosen the cluster boundaries in the $Y_e$-$S$ plane in \cref{fig_clu_S0_Ye} at similar locations as some of the $X_n^0$ contour lines in \cref{fig_Xn0_ye_ss}. An apparent division around $Y_e\sim0.2$ is found between clusters 1 and 2, which are the two clusters with the highest $X_n^0$ mean values in \cref{fig_clu_param}. 
Cluster 5 differs from the other clusters since its $X_n^0$ distribution covers the entire range of possible values. As seen in the bottom right panel of \cref{fig_clu_param}, this cluster contains the bulk of the high-velocity trajectories, with a mean close to 0.8~c, while the other clusters have $v\sim 0.2-0.3$~c. Thus, the trajectories of cluster 5 scatter over the entire $Y_e$-$S$ plane in \cref{fig_clu_S0_Ye}. 

Some of the observables are (anti)correlated. For example, $X_\mathrm{3rd+}$ is a measure of strong r-processing and is correlated with $X_n^0$ since neutron-rich initial conditions are required to produce the heaviest r-process elements. 
Conversely, $X_\mathrm{light}$ and $X_\mathrm{3rd+}$ are anti-correlated since a significant production of heavy nuclei occurs at the expense of lighter nuclei (remember that $\sum_A X=1$). 
\cref{fig_clu_rpro} shows the final mass fraction distributions for the five clusters. For each cluster, the mass fractions of the individual trajectories and the mass-averaged distribution are plotted. 
Cluster 1 contains only strong r-processing trajectories since  it has the largest $X_n^0$ and $X_\mathrm{3rd+}$ mean values among the clusters (\cref{fig_clu_param}). Actinides and third-peak nuclei are also produced in clusters 2 and 4 (and 5) but to a lesser extent. 
The opposite is found for cluster 0, which contains almost all trajectories with high $Y_e$-values and low entropies (\cref{fig_clu_S0_Ye}), so only $A<140$ nuclei are synthesized. 

The trajectories in cluster 3 (and some in clusters 2 and 4) are just below the limit of successfully producing the third r-process peak.
Clearly, the limit is not exactly at $Y_e\sim 0.25$ (see the discussion in \cref{sec_init_cond}), but ranging from $0.2$ to even 0.45 for higher initial entropies, which correspond to $X_n^0$-values around 0.3 to 0.5. 
The sorting of the trajectories around this limit is particularly challenging to do manually (i.e., \cref{fig_Xn0_ye_ss}), but also with an automatic clustering algorithm as shown here in \cref{fig_clu_param,fig_clu_S0_Ye,fig_clu_rpro}.

In summary, the large variety of initial conditions for the ejecta within a given hydrodynamical model leads to different final mass fraction distributions for individual trajectories that differ significantly in shape and the amount of heavy r-process nuclei produced. Among the physical observables analyzed here, $X_n^0$ remains the best observable to predict the success of the r-process, except for the high-velocity component of the ejecta, which can, if present, lead to significantly different results (see also \cref{sec_ang_vc}). 
Although only results from NS-NS dynamical ejecta models using the SFHo EoS were shown here (i.e., \cref{fig_Xn0_ye_ss,fig_clu_param,fig_clu_S0_Ye,fig_clu_rpro}), the discussion remains valid for the BH-torus models, and the NS-BH model applied in \cref{ch_nucuncert}. 



 

\subsection{Trajectory subsets}
\label{sec_sel_subs}



The remainder of \cref{sec_analysis} is devoted to selecting a subset of all trajectories for each hydrodynamical model used in the nuclear uncertainty study in \cref{ch_nucuncert}, i.e., all hydrodynamical models using the SFHo EoS (see \cref{sec:astro_mods}).
Note that the results regarding the dynamical ejecta in \cref{ch_dynweak} consider all trajectories. 
Certainly, the ideal approach would be to always use all trajectories for the r-process studies. However, as shown in \cref{sec:HPC}, this quickly becomes an overwhelming task when an extensive set of inputs are varied for a substantial number of trajectories. 
In addition to choosing the subset of trajectories, this section aims to verify that the selected subset of trajectories represents the properties of the complete set adequately, both in terms of the initial conditions and for the final observables (such as abundance distributions and evolution of the decay heat). 
Thus, as a basis for the verification method described below, r-process calculations using nuclear physics input set 6 in \cref{tab_nuc_mods} have been performed for all trajectories of the hydrodynamical models, except NS-BH model SFHo-11-23. 
For this model, a subset of 150 trajectories was chosen randomly without obtaining a ``basis'' set of calculations. 
This is assumed to be a reasonable choice since the initial conditions of this model show little diversity (see \cref{fig_hyd_trajsbhns,fig_distr_t9_rho,fig_xn0_ye_distr}), which then yields very similar r-process results (see the discussion in \cref{subsec_cluster}). 
Therefore, the improvement of running all 13175 trajectories of model SFHo-11-23 would be small for a huge increase in computation time. 

\begin{figure}[tbp]
\begin{center}
\includegraphics[width=\textwidth]{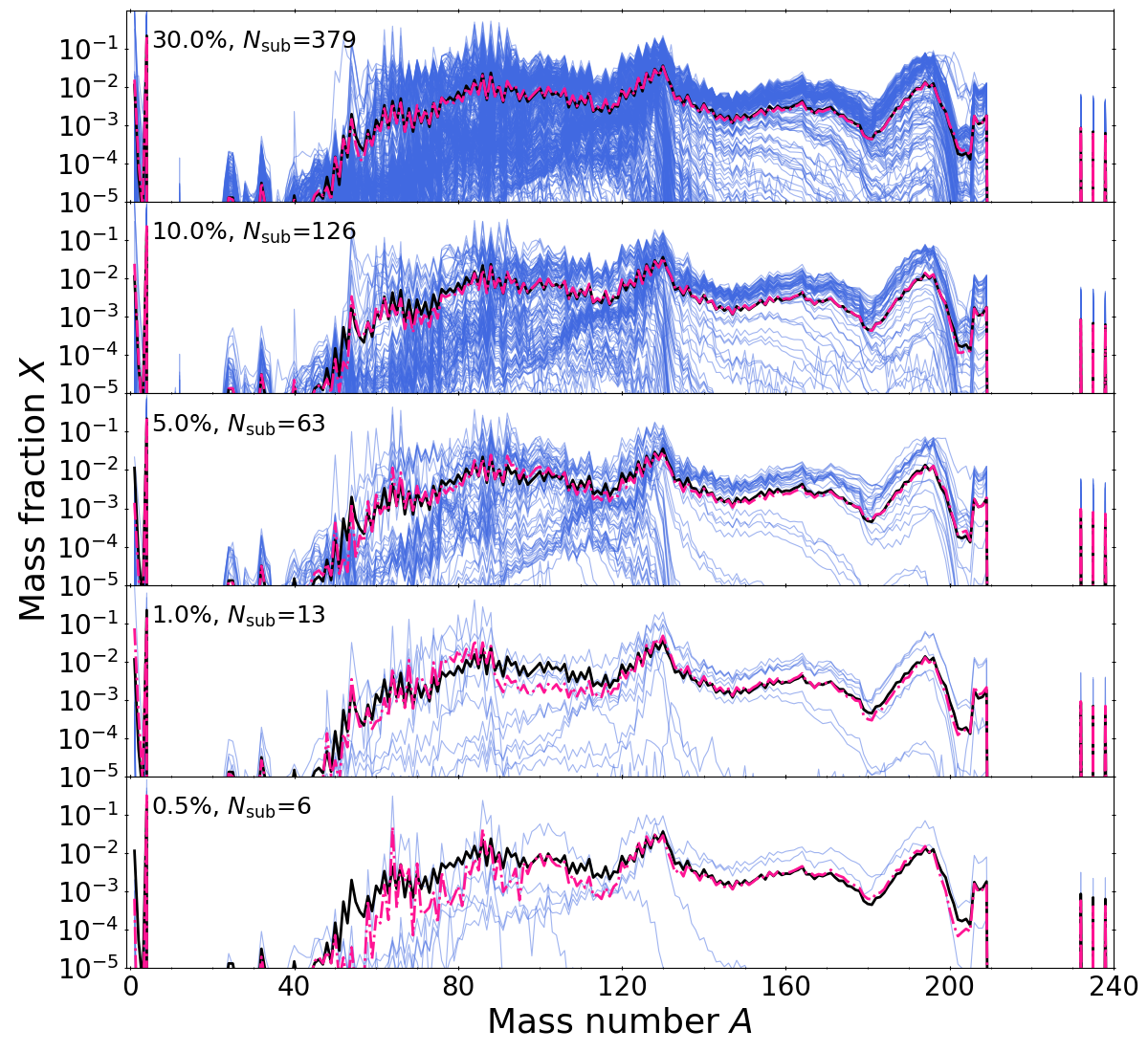}
\caption[Final mass fractions versus mass number for a selection of 30\%, 10\%, 5\%, 1\%, and 0.5\% of the 1263 trajectories for model SFHo-135-135.]{
Final mass fractions $X$ versus mass number $A$ for a selection of 30\%, 10\%, 5\%, 1\%, and 0.5\% of the 1263 trajectories for model SFHo-135-135. In each panel, the blue lines are the $N_\mathrm{sub}$ number of selected trajectories, the pink dashed-dotted line is the mass-averaged sum of the blue lines, and the black dashed lines are the mass-averaged sum using the total number of trajectories. 
Here, we apply the numpy \texttt{random} sampling generator with a constant random seed to select $N_\mathrm{sub}$ trajectories in each panel. 
}
\label{fig_Nsub}
\end{center}
\end{figure}
\begin{figure}[tbp]
\begin{center}
\includegraphics[width=\textwidth]{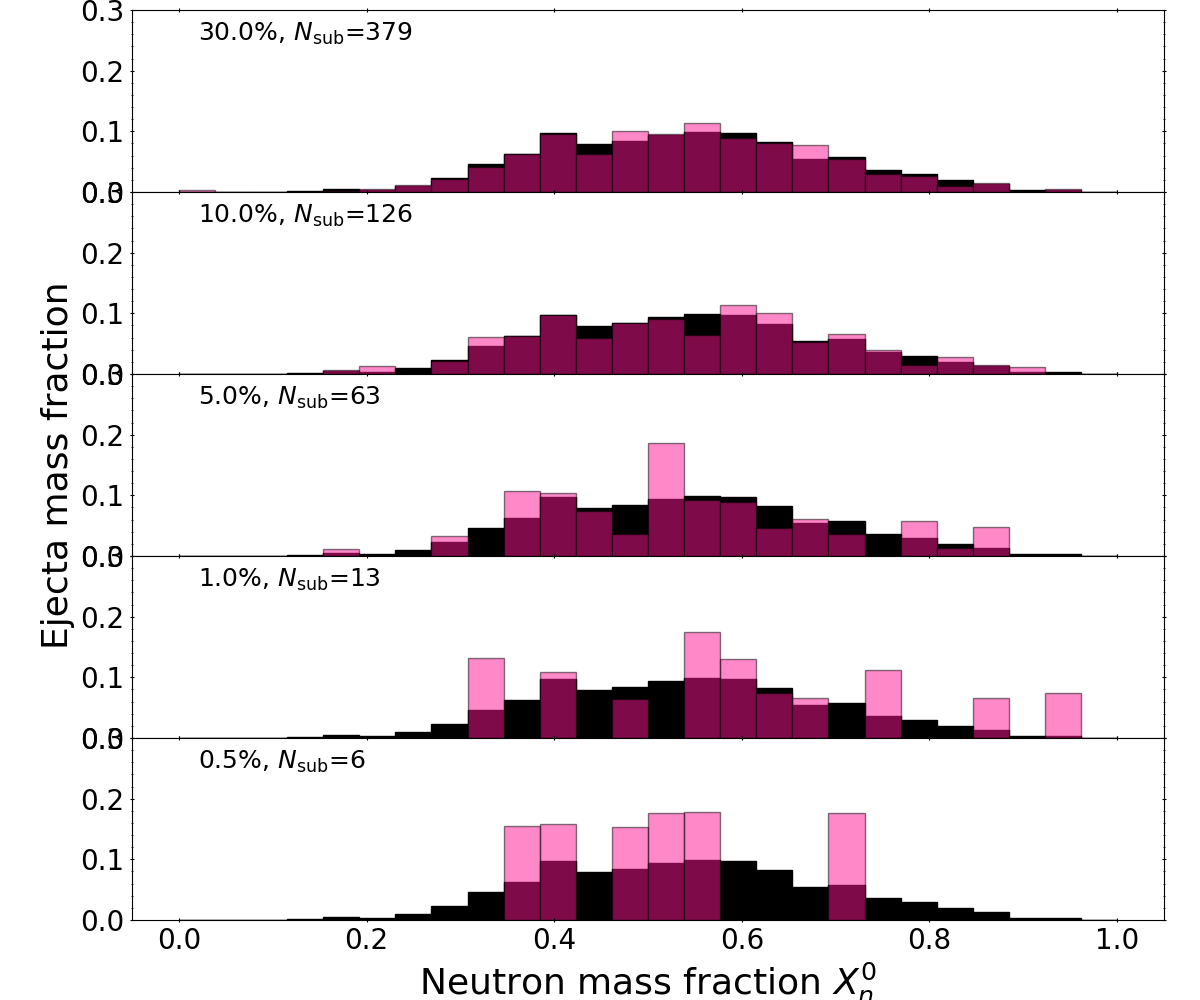}
\caption[A comparison of the $X_n^0$ distribution obtained using all 1263 trajectories of model SFHo-135-135 with the distributions obtained using a selection of 30\%, 10\%, 5\%, 1\%, and 0.5\% of all trajectories.]{
A comparison of the $X_n^0$ distribution obtained using all 1263 trajectories (black) of model SFHo-135-135 with the distributions obtained using a selection of 30\%, 10\%, 5\%, 1\%, and 0.5\% of all trajectories (pink).
The $N_\mathrm{sub}$ trajectories of the pink distributions in each panel are identical to the trajectories plotted (in blue) in the corresponding panel in \cref{fig_Nsub}. Here, the black distributions have been normalized to the total ejected mass of model SFHo-135-135, while each subset distribution has been normalized to the summed mass of the $N_\mathrm{sub}$ trajectories within the distribution. 
}
\label{fig_Nsub_hist}
\end{center}
\end{figure}

Let us consider the most straightforward method for choosing a trajectory subset: the random selection of $N_\mathrm{sub}$ trajectories from the total number of $N$ trajectories. 
\cref{fig_Nsub} shows the final mass-averaged abundance distributions for increasing $N_\mathrm{sub}$ and the corresponding $X_n^0$ distributions are shown in \cref{fig_Nsub_hist}. As expected, the agreement between the pink and black lines (distributions) corresponding to $N_\mathrm{sub}$ and $N$ trajectories, respectively, increase for larger values of $N_\mathrm{sub}$. The mass-averaged abundance distribution using $N_\mathrm{sub}$ trajectories resembles the total distribution already at 30 per cent of the total number of trajectories. This is connected to the similar shape of the $X_n^0$-distribution, as discussed in \cref{subsec_cluster}. 
\cref{fig_Nsub} demonstrates that the total distribution is better reproduced with an increasing $N_\mathrm{sub}$; however, the computing time also escalates for larger $N_\mathrm{sub}$.  
Thus, the subset selection has to balance the trade-off between CPU time and accuracy. 

\begin{figure}[tbp]
\begin{center}
\includegraphics[width=\textwidth]{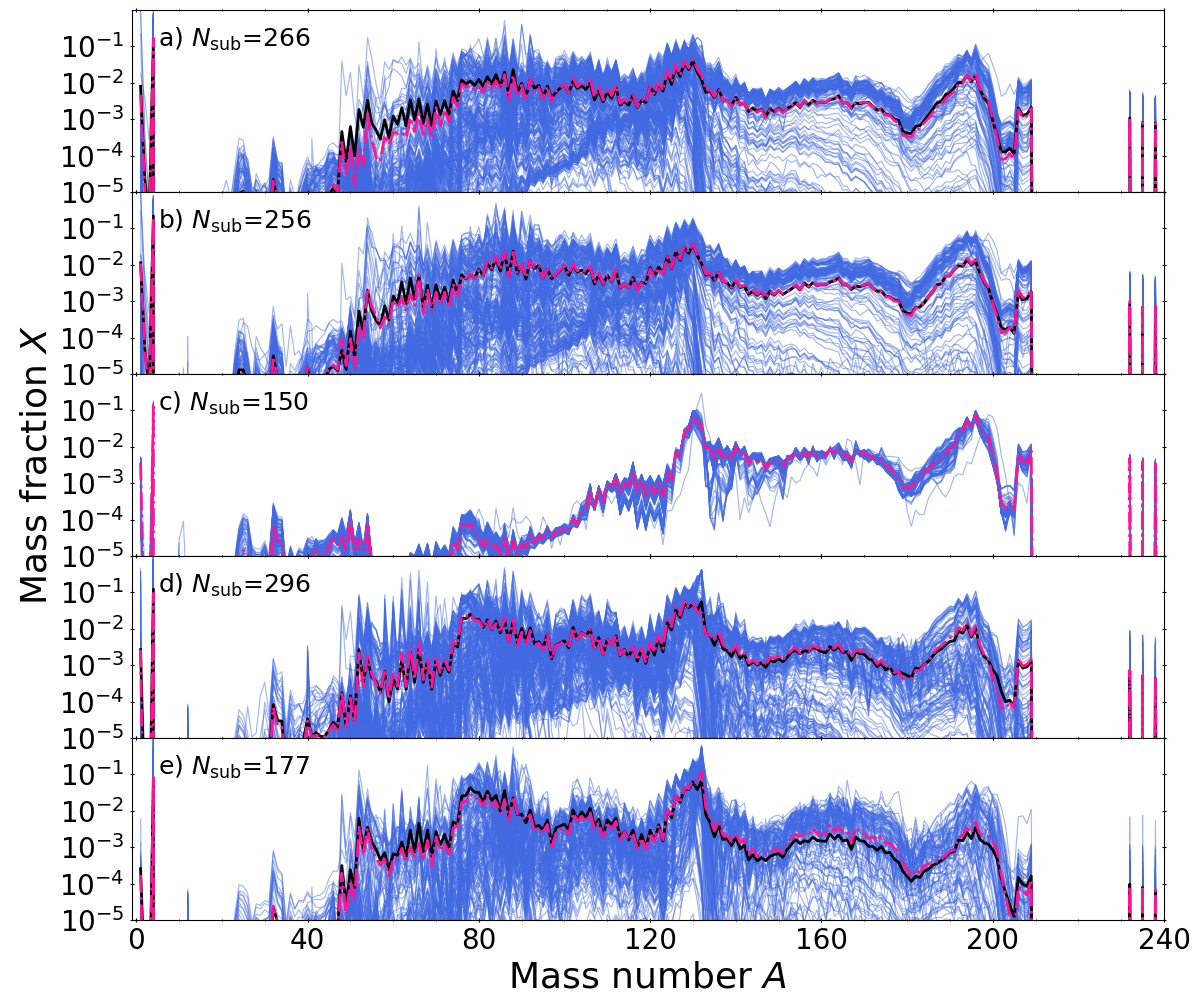}
\caption[Comparison of the final mass fractions for the selected trajectory subset with the complete set of trajectories for all models.]{
Comparison of the final mass fractions $X$ for the selected trajectory subset with the complete set of trajectories for models a) SFHo-125-145, b) SFHo-135-135, c) SFHo-11-23, d) M3A8m1a5, and e) M3A8m3a5-v2 obtained with the ``improved sampling'' method described in the text. The blue lines are the mass fractions of the $N_\mathrm{sub}$ trajectories in the subset, the pink dashed-dotted line is the mass-averaged sum of the blue lines, and the black dashed line is the mass-averaged sum using the total number of trajectories. 
}
\label{fig_subs_rpro}
\end{center}
\end{figure}

\begin{figure}[tbp]
\begin{center}
\includegraphics[width=\textwidth]{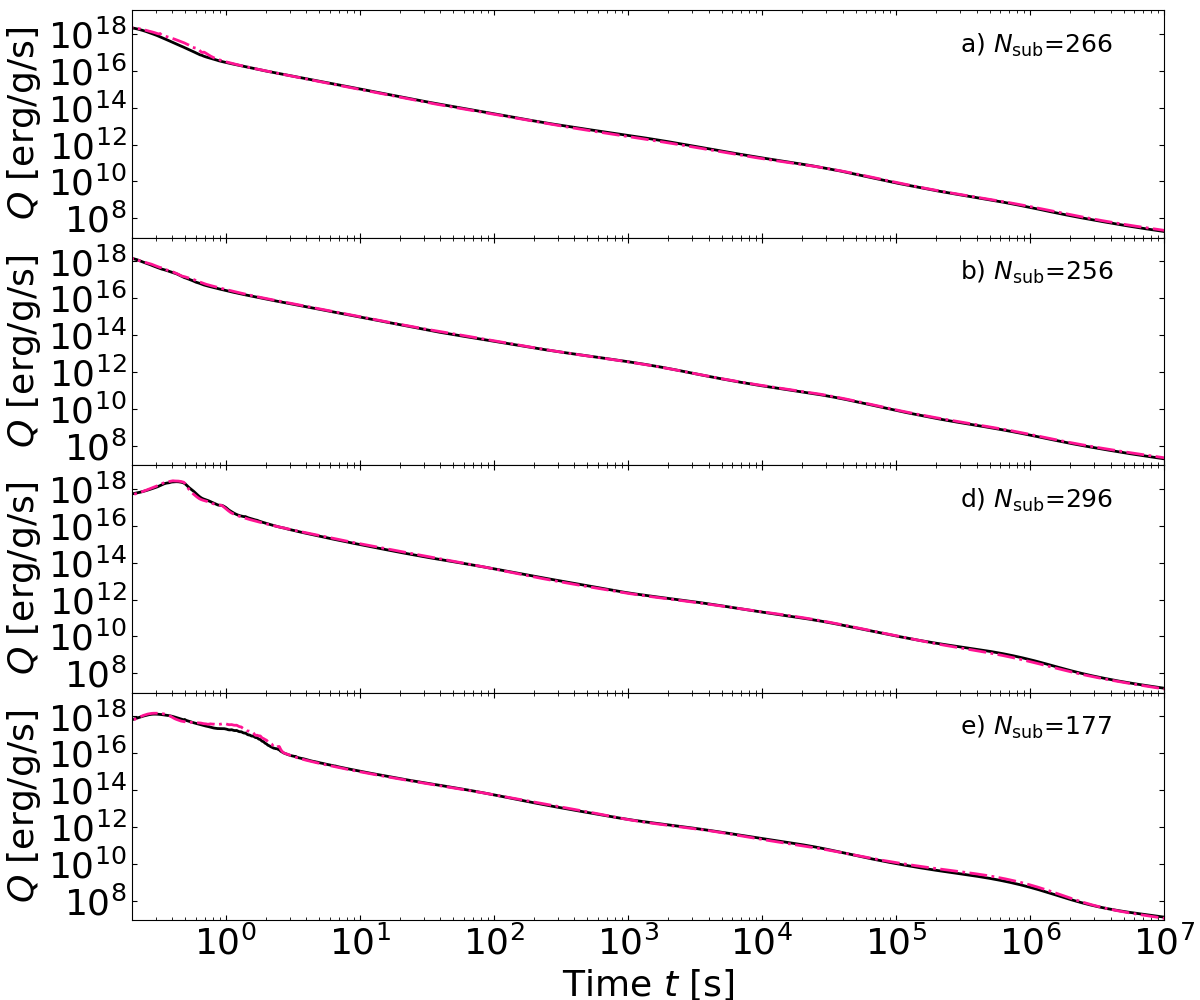}
\caption[Comparison of the heating rate for the selected trajectory subset with the complete set of trajectories for all models.]{
Comparison of the heating rate $Q(t)$ for the selected trajectory subset with the complete set of trajectories for models a) SFHo-125-145, b) SFHo-135-135, d) M3A8m1a5, and e) M3A8m3a5-v2 obtained with the ``improved sampling'' method described in the text. The pink dashed-dotted line is the mass-averaged sum of the $N_\mathrm{sub}$ trajectories in the subset, and the black dashed line is the mass-averaged sum using the total number of trajectories. Model SFHo-11-23 is omitted here since the r-process calculations have only been performed for the subset of 150 trajectories, see the text for details.  
}
\label{fig_subs_Qt}
\end{center}
\end{figure}

\begin{figure}[tbp]
\begin{center}
\includegraphics[width=\textwidth]{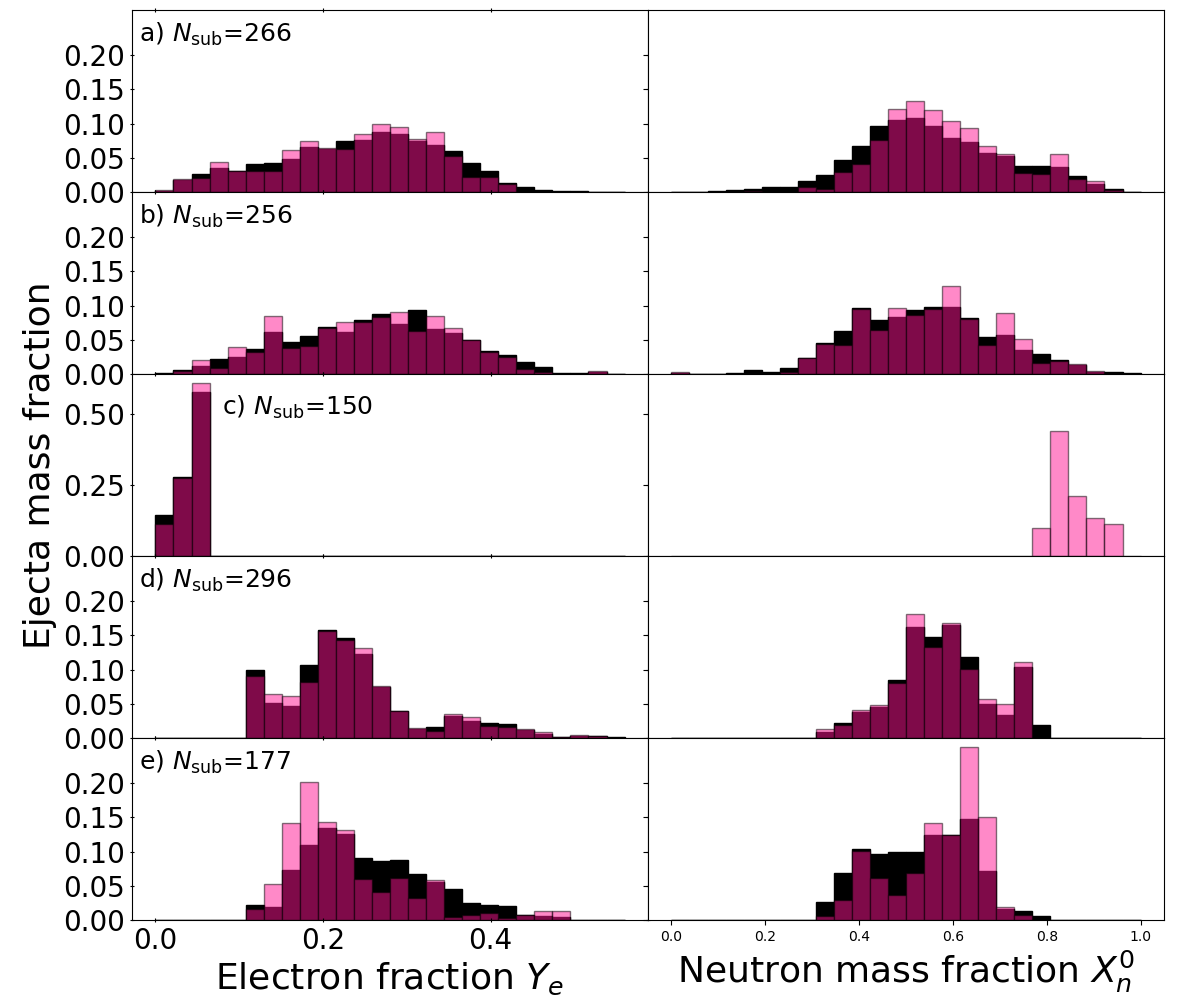}
\caption[Comparison of the electron fraction distributions for the selected trajectory subset with the complete set of trajectories for all models.]{
Comparison of the electron fraction $Y_e$ (left) and neutron mass fraction $X_n^0$ (right) distributions for the selected trajectory subset (pink) with the complete set of trajectories (black) for models a) SFHo-125-145, b) SFHo-135-135, c) SFHo-11-23, d) M3A8m1a5, and e) M3A8m3a5-v2 obtained with the ``improved sampling'' method described in the text.
Here, the black distributions have been normalized to the total ejected mass of each model (listed in \cref{tab:astromods}), while each subset distribution has been normalized to the summed mass of the $N_\mathrm{sub}$ trajectories within the distribution. The $X_n^0$ values for model SFHo-11-23 have only been calculated for the subset.
}
\label{fig_subs_hist}
\end{center}
\end{figure}

The above random selection method works well when the distributions of initial ejecta conditions (i.e., \cref{fig_hyd_trajsbhns,fig_distr_t9_rho,fig_xn0_ye_distr}) are similar in their shape and width. 
The asymmetrical NS-NS model SFHo-125-145 have a larger high-velocity component (see $M_{\rm ej}^{v \ge 0.6c}$ in \cref{tab:astromods}) that affects the final mass fraction distribution significantly, but that also leads to an enhancement in the heating rate (see \cref{sec_ang_vc}). 
Since a small number of trajectories represents this high-velocity component, it is essential that the trajectory selection method samples from this group, which is not necessarily the case when the random selection method is applied. 

Taking the above discussion into account, the trajectory subsets were chosen according to the following ``improved sampling'' method:
\begin{enumerate}
\item Divide the trajectories into 13 $Y_e$ bins\footnote{$Y_e$ was used here instead of $X_n^0$ since it was available for all hydrodynamical models at the time of the subset selection.}, and within each bin, sort the trajectories according to their ejecta mass
\item starting from the trajectories with the largest mass within a $Y_e$ bin, select trajectories until the sample contains at least $P$ per cent of the total bin mass. Repeat for all $Y_e$ bins
\end{enumerate}
If the above sampling from the $Y_e$ distribution did not contain at least a few per cent of $v>0.6$~c trajectories for the NS-NS models, high-velocity trajectories were added iteratively to the subset until a better agreement was found between final mass-averaged mass fractions and heating rates of the subset and complete set of trajectories. 
The per cent $P$ of selected mass from each $Y_e$ bin was varied between 10 to 60 per cent until the best agreement was found for the smallest possible $N_\mathrm{sub}$. See \cref{tab:astromods} for $N_\mathrm{sub}$ and the fraction of mass $M_{\mathrm{sub}}/M_{\mathrm{ej}}$ in each model subset. 
\cref{fig_subs_rpro,fig_subs_Qt} compare the final mass fractions and heating rates for the trajectory subsets and complete set of all trajectories for all hydrodynamical models. 
The $Y_e$ distributions used to select the subsets and the distributions of the subsets are shown in \cref{fig_subs_hist}, where the agreement between the distributions is reasonably good, except for model M3A8m3a5-v2. This model has a peculiar final distribution of mass fractions, as will be discussed in more detail in \cref{ch_nucuncert}. 
When using the algorithm described above to select trajectories from the $Y_e$ distribution (black in \cref{fig_subs_hist}e), the abundance distribution for the heaviest nuclei (including the actinides) is poorly represented. Therefore, additional low-$Y_e$ trajectories were added to the subset manually, i.e., the ``extra'' $Y_e<0.2$ trajectories seen in the pink distribution compared to the black distribution in \cref{fig_subs_hist}e. This ensures that the final abundance distribution is represented accurately for the heaviest r-process elements, including the actinides, at the expense of somewhat misshaping the $Y_e$ distribution. 
By using these subsets, the total number of r-process calculations required for a single input set is reduced by 95 per cent, or if we do not count the over 10~000 trajectories of the NS-BH model SFHo-11-23 by 90 per cent. 



\begin{figure}[tbp]
\begin{center}
\includegraphics[width=\textwidth]{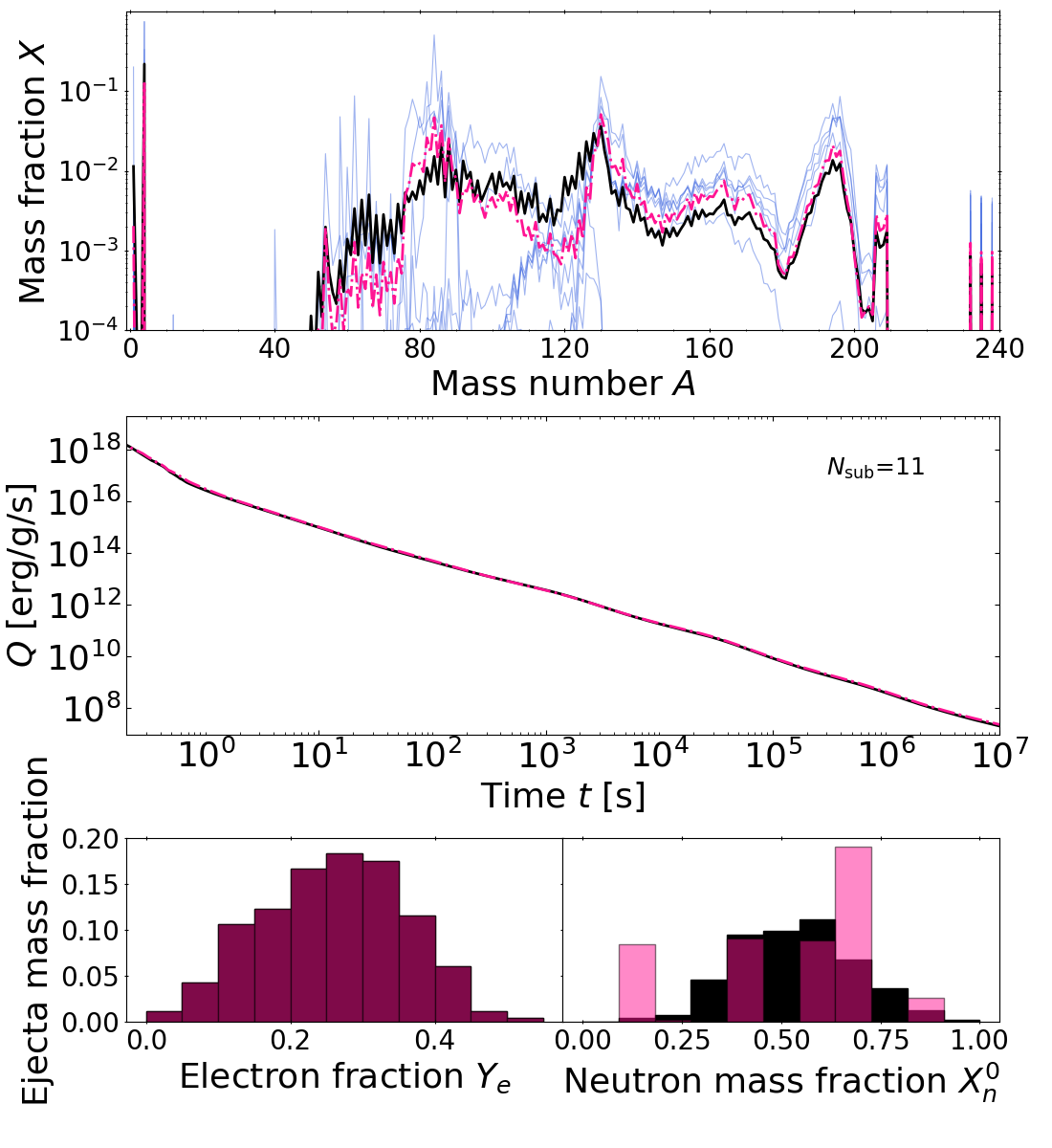}
\caption[An example of using a small number of trajectories ($N_\mathrm{sub}=11$) in a subset for model SFHo-135-135 when using the $Y_e$-distribution to select the subset.]{
An example of using a small number of trajectories ($N_\mathrm{sub}=11$) in a subset for model SFHo-135-135. The top, middle and bottom panels show the same quantities as in \cref{fig_subs_rpro,fig_subs_Qt,fig_subs_hist}, respectively. 
Here, the 11 trajectories have been selected based on their $Y_e$ value so that one trajectory is sampled from each bin of 0.05 evenly spaced from 0 to 0.55.
The subset distribution (pink) has artificially been normalized to follow the $Y_e$-distribution of the total ejecta (black) of model SFHo-135-135.
}
\label{fig_subs_fewtraj}
\end{center}
\end{figure}

\begin{figure}[tbp]
\begin{center}
\includegraphics[width=\textwidth]{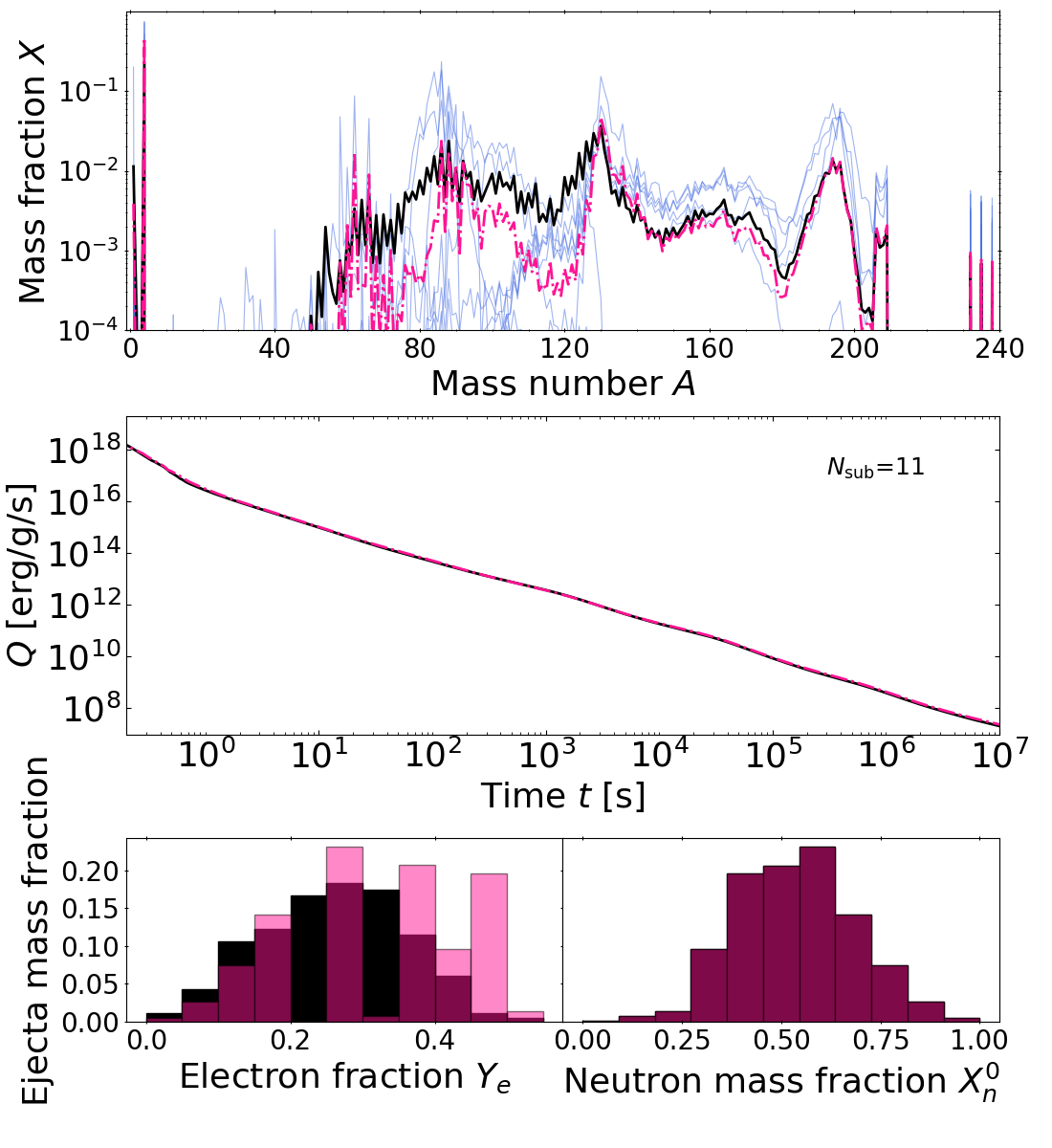}
\caption[Same as \cref{fig_subs_fewtraj} when using the $X_n^0$-distribution to select a small subset.]{
Same as \cref{fig_subs_fewtraj} when selecting 11 trajectories based on their $X_n^0$ value so that one trajectory is sampled from each bin.
}
\label{fig_subs_fewtraj_xn0}
\end{center}
\end{figure}

The subset in \cref{fig_Nsub,fig_Nsub_hist}, using only six trajectories of model SFHo-135-135, performs quite well in terms of representing the abundance distribution given its small size (to a lesser extent for $A<120$ nuclei). 
Let us consider a subset of model SFHo-135-135 extracted with a method similar to the ``improved sampling'' method above but with a significantly smaller $N_\mathrm{sub}$, see \cref{fig_subs_fewtraj}.
With $Y_e$-bins of 0.05, we randomly select one trajectory from each bin between 0 and 0.55, giving $N_\mathrm{sub}=11$. Instead of using the trajectory mass as a weight when doing the mass-averaged sum to calculate the final mass fractions, we use an artificially constructed mass distribution. This distribution is identical to the total ejecta $Y_e$ distribution of model SFHo-135-135, i.e., pink distribution in the lower panel of \cref{fig_subs_fewtraj} is normalized with the same weights as the black distribution. 
For the test shown in \cref{fig_subs_fewtraj}, the heating rate is well represented by the 11 trajectories in the test subset. However, contrary to expectation, the test subset does not represent the final mass fraction distribution any better than the six randomly selected trajectories in \cref{fig_Nsub}. 
Similar results are found when using the same test method to select 11 trajectories based on their $X_n^0$ values and renormalized to the total ejecta $X_n^0$ distribution, as shown in \cref{fig_subs_fewtraj_xn0}.
Thus, the criteria for selecting a subset cannot only be based on a good representation of the $Y_e$ or $X_n^0$ distribution, but the subset also has to be large enough to be statistically significant. 
This is particularly important when we want to study the impact of varying the nuclear physics input on the r-process results. 
\cref{sec_comp2} will also discuss some of the possible caveats of only choosing a single or a few trajectories when studying the r-process.
Interestingly, the pink subsets in \cref{fig_subs_fewtraj,fig_subs_fewtraj_xn0} do not resemble the black total distributions simultaneously. A good correspondence between the pink and black distributions for the $Y_e$ gives a pink subset $X_n^0$ distribution which does not mirror the black total distribution. 
Although $X_n^0$ and $Y_e$ are highly correlated, as discussed in \cref{sec_analysis}, they represent different features of the ejecta, which highlights the importance of selecting a statistically significant subset when studying the r-process.




\section{Sensitivities to assumptions}
\label{sec_sensitivity}

\epigraph{\itshape ``You must learn from the mistakes of others. You can't possibly live long enough to make them all yourself.''}{--- \textup{Samuel Levenson}}

The current section aims to test the reliability of the results presented in \cref{ch_dynweak,ch_nucuncert} with respect to the various assumptions and approximations made for the r-process calculations. 
Unless otherwise specified, the subset of model SFHo-135-135 described in \cref{sec_sel_subs} is used for all sensitivity tests.
This model subset represent a broad distribution of initial conditions, see \cref{fig_xn0_ye_distr} and \cref{sec_init_cond} for more details. 
The nuclear input set 1 in \cref{tab_nuc_mods} has been used as the default for all calculations herein. 

\subsection{Network limits and ODE solver tolerance}
\label{sec_netlimtol}

\begin{figure}[tbp]
\begin{center}
\includegraphics[width=0.98\textwidth]{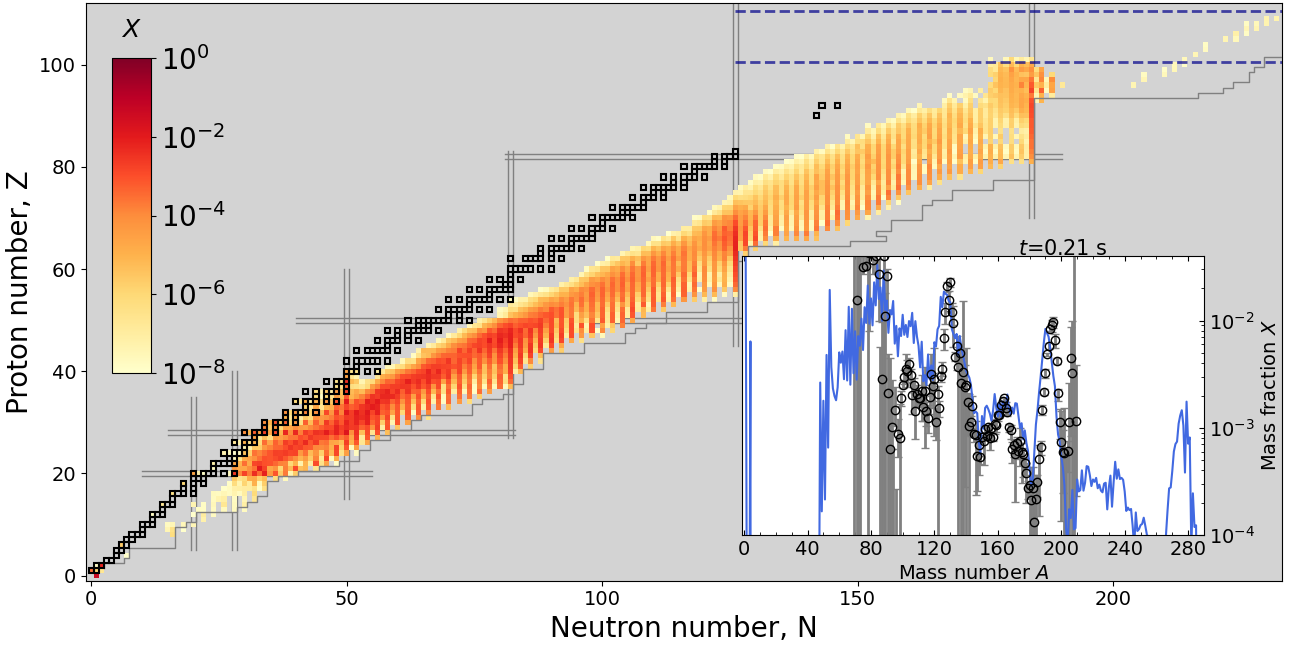}
\includegraphics[width=0.98\textwidth]{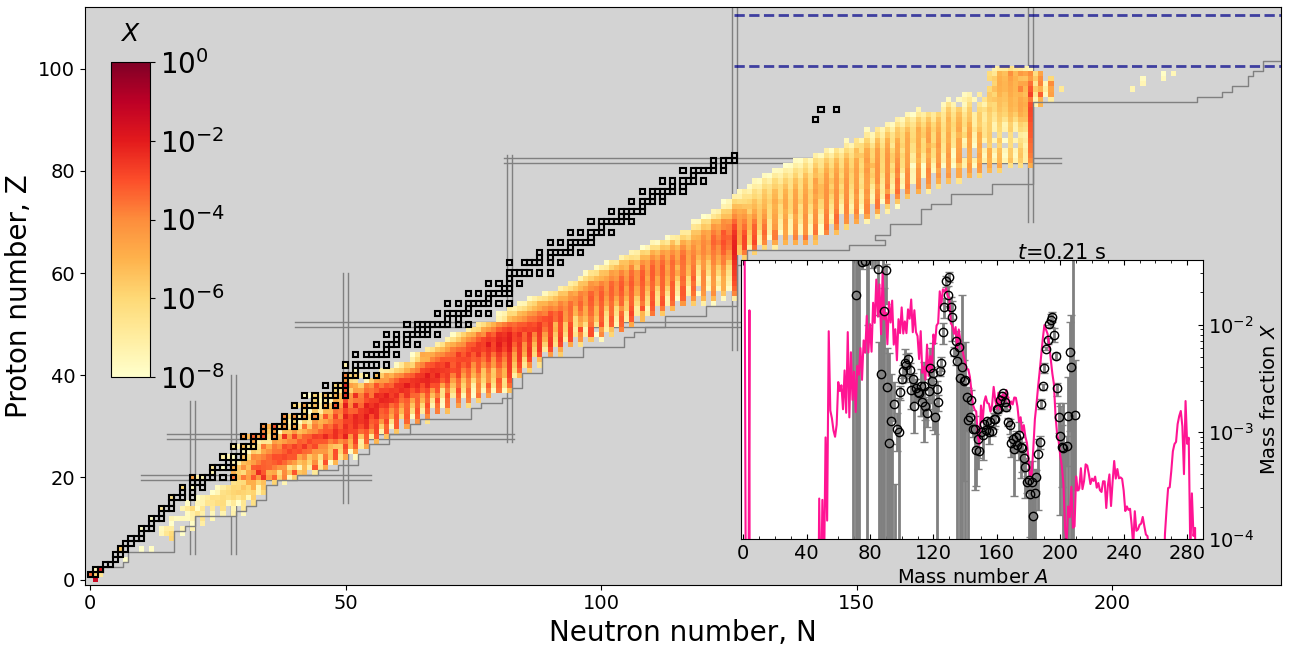}
\caption[The mass fraction (for 256 trajectories) shown in the $(N,Z)$ plane at $\sim0.2$~s when limiting the maximum proton number of the reaction network to $Z=110$ and 100 for model SFHo-135-135.]{
The mass fraction $X$ mass-averaged over all 256 trajectories in the subset of model SFHo-135-135 represented by the colour bar shown in the $(N,Z)$ plane at $t\sim0.2$~s after the network was initiated when limiting the maximum proton number of the reaction network to $Z_\mathrm{max}=110$ (top) and $Z_\mathrm{max}=100$ (bottom). 
The black squares correspond to the stable nuclei, the double solid lines depict the neutron and proton magic numbers, the neutron drip line is shown as a grey line to the right of the valley of stability, and two dark blue dotted lines are indicated at $Z=110$ and 100. 
Note that all mass fractions below $10^{-8}$ are not displayed.
The insert shows the mass fraction $X$ versus mass number $A$ (also at $t\sim0.2$~s) compared to the solar r-process distribution, including error bars, shown in black. 
}
\label{fig_Zmax_absum}
\end{center}
\end{figure}

Two different approximations were introduced to achieve convergence for trajectories that could not finish within the allocated computing time (usually 7~d on the supercomputer). 
These approximations include reducing the maximum $Z$-limit of the reaction network (lowering $Z_\mathrm{max}$) and lowering the tolerances for the ODE solver, which is used to evolve the abundances in time. 
Only \cref{ch_nucuncert}, which studies the nuclear uncertainties on the r-process results, applies these assumptions.
To save computing time (and make sure all trajectories converge), all calculations for the NS-BH model SFHo-11-23 used a lower network limit with $Z_\mathrm{max}=100$ (in contrast to the default of $Z_\mathrm{max}=110$). 
For the four NS-NS models, less than ten trajectories in each variation of the nuclear input sets had convergence problems.
The calculations were restarted on one of the IAA desktop computers, which does not have any time restrictions for the run-time of the programs. Most trajectories finished within a reasonable computing time; however, a few trajectories did not converge even after 35~days, and for these, the approximations discussed below were applied. 

The mass fraction (in the $(N,Z)$ plane) mass-averaged over all 256 trajectories in the SFHo-135-135 subset is shown in \cref{fig_Zmax_absum} for $Z_\mathrm{max}=110$ and 100 at $t\sim0.2$~s after the network was initiated. 
The mass-averaged sum of $X$ is calculated as $X=\sum_j X_j m_j/\sum_j m_j$ for all isotopes in the distribution, where $m_j$ is the ejecta mass of trajectory $j$ (see \cref{eq:mass_avr}).
An animation of this figure can be seen on the \citet{GitHubAnims}.
The chosen time of $t=0.2$~s shortly after freeze-out for most trajectories in the subset, i.e., the nuclei have generally started to decay back to the valley of stability. 
Due to the difficulty of passing the $N=184$ shell, a significant amount of $A\sim280$ nuclei have accumulated (see the inserts in \cref{fig_Zmax_absum}). 
As described in \citet{Lemaitre2021} (see their Fig.~7), this is related to the ``fission bottleneck'' for $A<292$. As seen in \cref{fig_fiss}, fission processes even dominate along the neutron-drip line at $Z\sim 95$ and $N\sim190$, prohibiting the flow towards higher $Z$-values.
The abundances of the heaviest elements also reach up to $Z\sim100$ (and beyond for the $Z_\mathrm{max}=110$ calculation) with mass fractions below $10^{-4}$. 
As discussed in \citet{Lemaitre2021}, there is a ``fission roof'' around $Z\sim100$ for $160<A<300$, which means that after undergoing $\beta$-decay, all $A>160$ nuclei will fission when reaching the roof (for the fission models applied here). Similarly, there is a fission roof at $Z=110$ for $A>320$, which terminates the r-process and hinders the production of super-heavy nuclei (nuclei in the $300<A<320$ region undergo fission between $Z=100$ and 110). 
The above discussion is true for both fission barriers considered herein, i.e., the HFB-14 and MS99 barriers shown in \cref{fig_fiss}.


\begin{figure}[tbp]
\begin{center}
\includegraphics[width=0.98\textwidth]{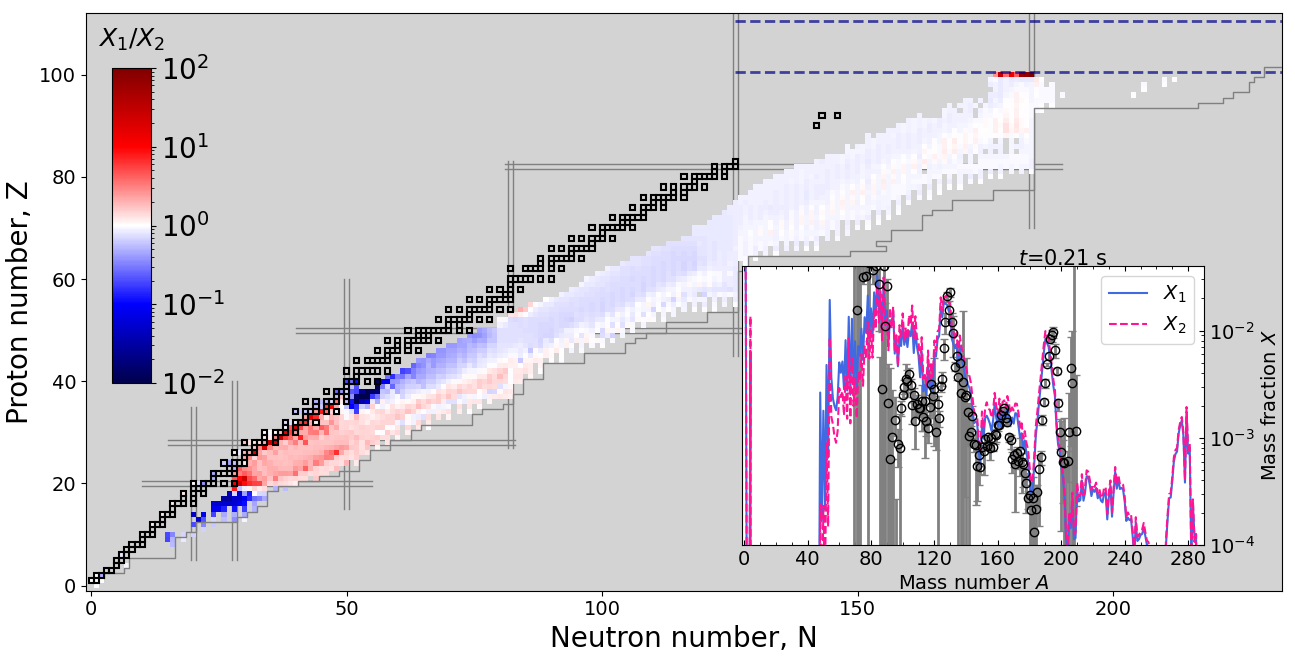}
\caption[Same as \cref{fig_Zmax_absum} for the mass fraction ratio between calculations using a maximum $Z$-value of $Z=110$ and 100.]{
Same as \cref{fig_Zmax_absum} for the ratio between the mass fractions $X_1$ and $X_2$ for the calculations using a maximum $Z$-value of $Z=110$ and 100, respectively.
The ratio $X_1/X_2$ is only calculated when $X_1$ and $X_2$ are larger than $10^{-8}$ (i.e., the values displayed in \cref{fig_Zmax_absum}), and ratios larger or smaller than the maximum or minimum values of the colour bar are represented by red and blue, respectively.
}
\label{fig_Zmax_absumrat}
\end{center}
\end{figure}

\begin{figure}[tbp]
\begin{center}
\includegraphics[width=\textwidth]{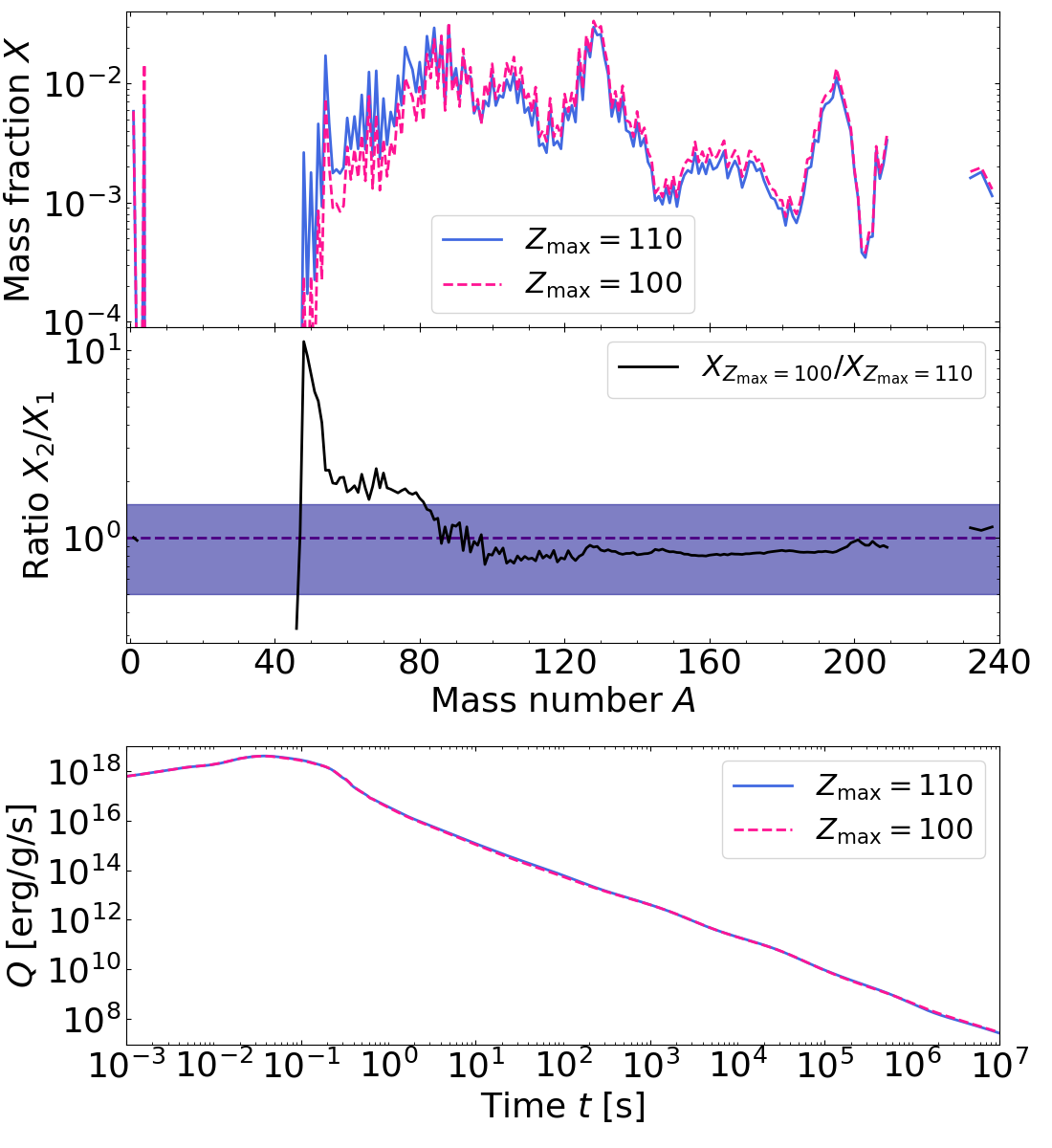}
\caption[The sensitivity of the final mass fraction and heating rate to changes in the maximum $Z$-value of the reaction network.]{
The sensitivity of the final mass fraction and heating rate to changes in the maximum $Z$-value of the network when using the 256 subset trajectories of model SFHo-135-135. The top panel show the mass fraction $X_{Z_\mathrm{max}=110}$ and $X_{Z_\mathrm{max}=100}$, the middle panel shows the mass fraction ratio $X_{Z_\mathrm{max}=100}/X_{Z_\mathrm{max}=110}$, and the bottom panel shows the heating rate for the two calculations with the maximum limit of the network at $Z_\mathrm{max}=110$ and 100. The purple dashed line $y=1$ and a dark purple colour band from $y=0.5$ to 1.5 is shown in the middle panel. The ratio $X_{Z_\mathrm{max}=100}/X_{Z_\mathrm{max}=110}$ is only calculated when the mass fraction is greater than $10^{-6}$. 
}
\label{fig_Zmax_sens}
\end{center}
\end{figure}

\begin{figure}[tbp]
\begin{center}
\includegraphics[width=\textwidth]{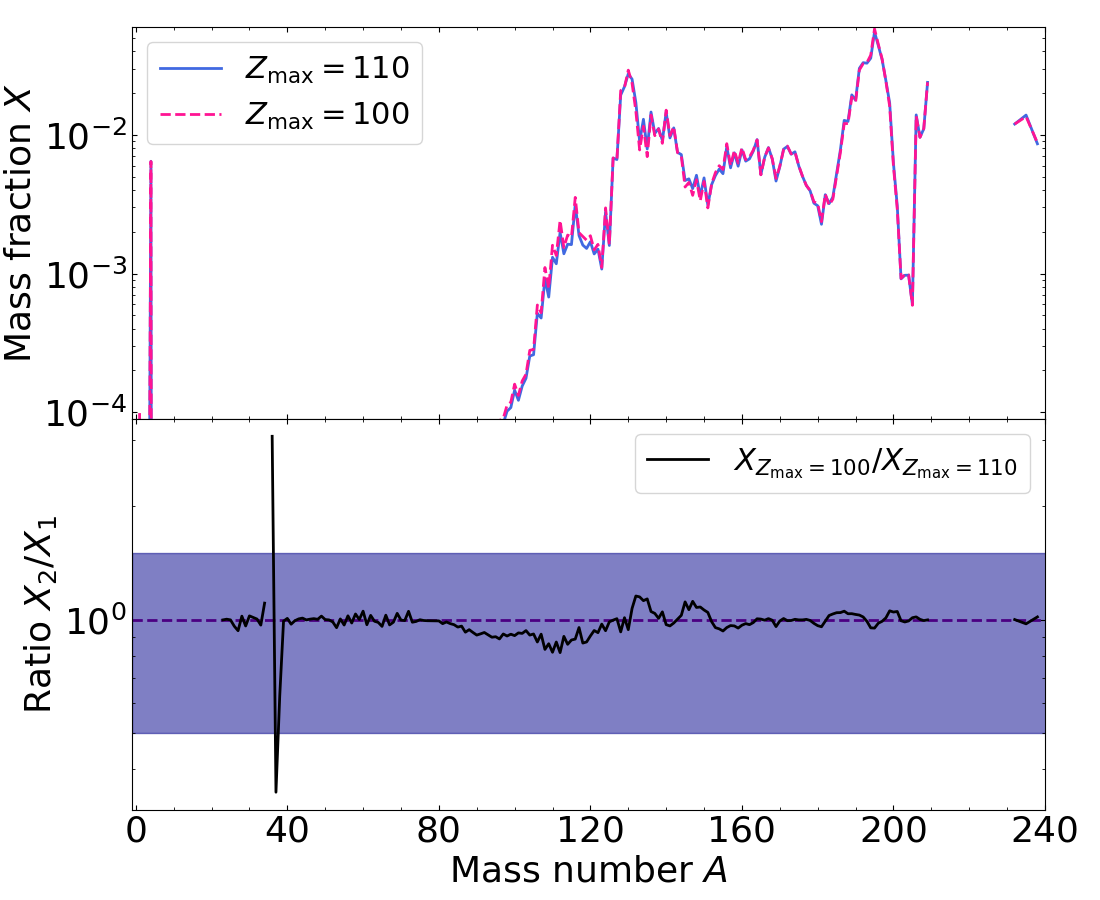}
\caption[Same as \cref{fig_Zmax_sens} for one trajectory with $Y_e=0.05$.]{
Same as the two top panels of \cref{fig_Zmax_sens} for one trajectory with $Y_e=0.05$.
The ratio $X_{Z_\mathrm{max}=100}/X_{Z_\mathrm{max}=110}$ is only calculated when the mass fraction $X$ is greater than $10^{-8}$. 
}
\label{fig_Zmax_sens1traj}
\end{center}
\end{figure}

The ratio between the mass fractions of the two calculations using $Z_\mathrm{max}=110$ and 100 at $t=0.2$~s for the 256 subset trajectories of model SFHo-135-135 (\cref{fig_Zmax_absum}) is displayed in \cref{fig_Zmax_absumrat}. Some deviations between the calculations are visible at this time, particularly for $A<130$ and $Z=100$ nuclei.
These differences even out by the end of the simulation (at $t\sim1$~yr), as shown in \cref{fig_Zmax_sens}, where a comparison between the mass fraction versus mass number and the evolution of the heating rate is shown for both calculations. 
Note that although our r-process calculations only run up to 1~yr, all nuclei have been instantaneously decayed back to the stable nuclei, except for the long-lived actinides, as described in \cref{sec:nuc_net}.
For the mass fraction, the ratio between the two calculations is within a factor of 10 for all mass numbers and less than 2 for $A>80$. Similarly, the heating rate when using $Z_\mathrm{max}=100$ is within a factor of two up and down compared to the heating rate obtained when using $Z_\mathrm{max}=110$ for all time steps. 
Thus, \cref{fig_Zmax_sens} confirms that the final abundances are not significantly affected by limiting the network to $Z_\mathrm{max}=100$ for the subset of 256 trajectories of model SFHo-135-135, particularly for the $A>80$ nuclei. This subset includes trajectories with a wide distribution of $Y_e$-values (see \cref{sec_sel_subs}), which is representative of the $Y_e$ distributions of the NS-NS merger models. 
The NS-BH model generally has much lower $Y_e$-values than the NS-NS dynamical ejecta (see \cref{fig_xn0_ye_distr}). Therefore, it is interesting to investigate the impact of changing the network's maximum $Z$-limit for a trajectory that undergoes strong r-processing and fission recycling.
The final mass fraction for one trajectory with $Y_e=0.05$ when using $Z_\mathrm{max}=110$ and 100 is shown in \cref{fig_Zmax_sens1traj} (an animation of the abundance evolution for this trajectory can be found on the \citet{GitHubAnims}). 
For this trajectory, the impact of reducing the maximum $Z$-limit of the network is smaller than for all trajectories in the subset (\cref{fig_Zmax_sens}) and insignificant for $A>50$ nuclei. Similar results are found for other low-$Y_e$ trajectories. 
The low impact of $Z_\mathrm{max}$ on the results can be explained by the fact that even for very neutron-rich conditions, almost no abundance manages to flow past the ``fission bottleneck'' around $A<292$ \citep[see][for more details]{Lemaitre2021}.



As discussed in \cref{sec:ab_evo}, the ODE solver \texttt{DVODE} takes the molar fractions \textit{\textbf{Y}} and $\textit{\textbf{f}}(\textit{\textbf{Y}},t)$ (the RHS of \cref{eq:ab_evo}) at the time $t_\mathrm{in}$, and calculates the new abundances at the next time step $t_\mathrm{out}$. In between $t_\mathrm{in}$ and $t_\mathrm{out}$ \texttt{DVODE} will take many smaller time steps $h$ with a length determined internally in \texttt{DVODE} by an algorithm \citep{Brown1989}. 
The user controls the relative and absolute error in \textit{\textbf{Y}} allowed for each time step through the input parameters \texttt{rtol} and \texttt{atol}, respectively, which in turn regulate the size of the time step $h$. Setting smaller tolerances is equivalent to demanding higher resolution for the calculations, which will generally require an increase in the number of time steps taken by \texttt{DVODE} and, therefore, also escalate the computational demand for the calculations. 
In the following, we will compare r-process results obtained using three sets of the tolerances \texttt{rtol} and \texttt{atol}:
\begin{description}
\item[Default:] \texttt{rtol}$=10^{-3}$, \texttt{atol}$=10^{-11}$
\item[High resolution:] \texttt{rtol}$=10^{-4}$, \texttt{atol}$=10^{-12}$
\item[Extreme resolution:] \texttt{rtol}$=10^{-5}$, \texttt{atol}$=10^{-12}$
\end{description}
The latter ``extreme resolution'' set will only be discussed for a single trajectory.

\begin{figure}[tbp]
\begin{center}
\includegraphics[width=\textwidth]{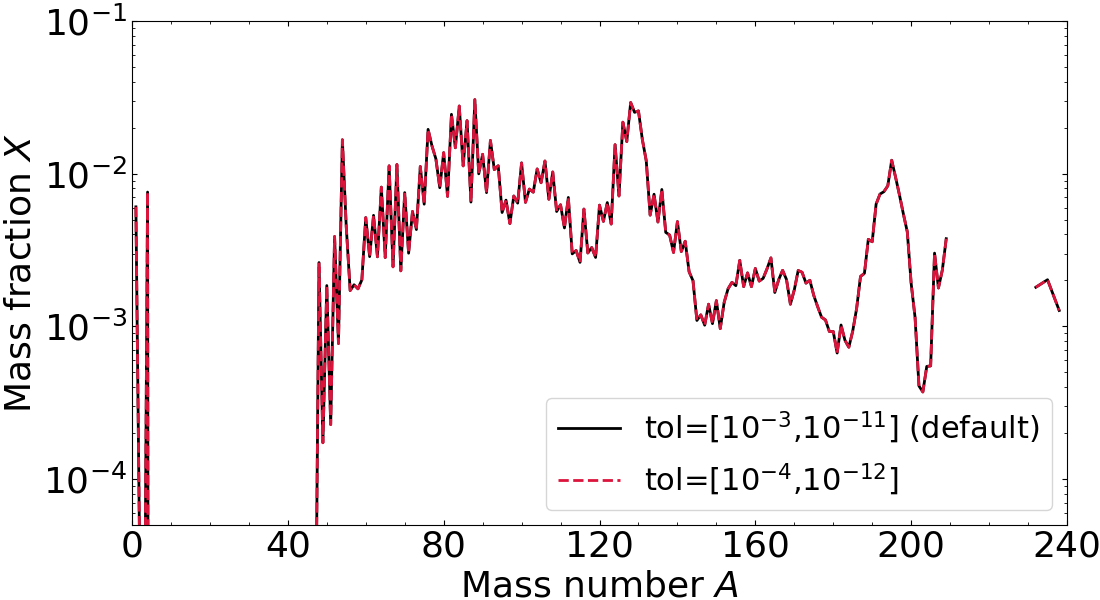}
\caption[The sensitivity of the mass fraction to changes in the tolerances used in the ODE solver.]{
The sensitivity of the final mass fraction to changes in the relative and absolute tolerances (tol=[\texttt{rtol},\texttt{atol}]) used in the ODE solver which evolves the reaction network. The mass fraction shown here sums over the abundances from all 256 trajectories in the subset of model SFHo-135-135.
}
\label{fig_tol_sens}
\end{center}
\end{figure}
\begin{figure}[tbp]
\begin{center}
\includegraphics[width=\textwidth]{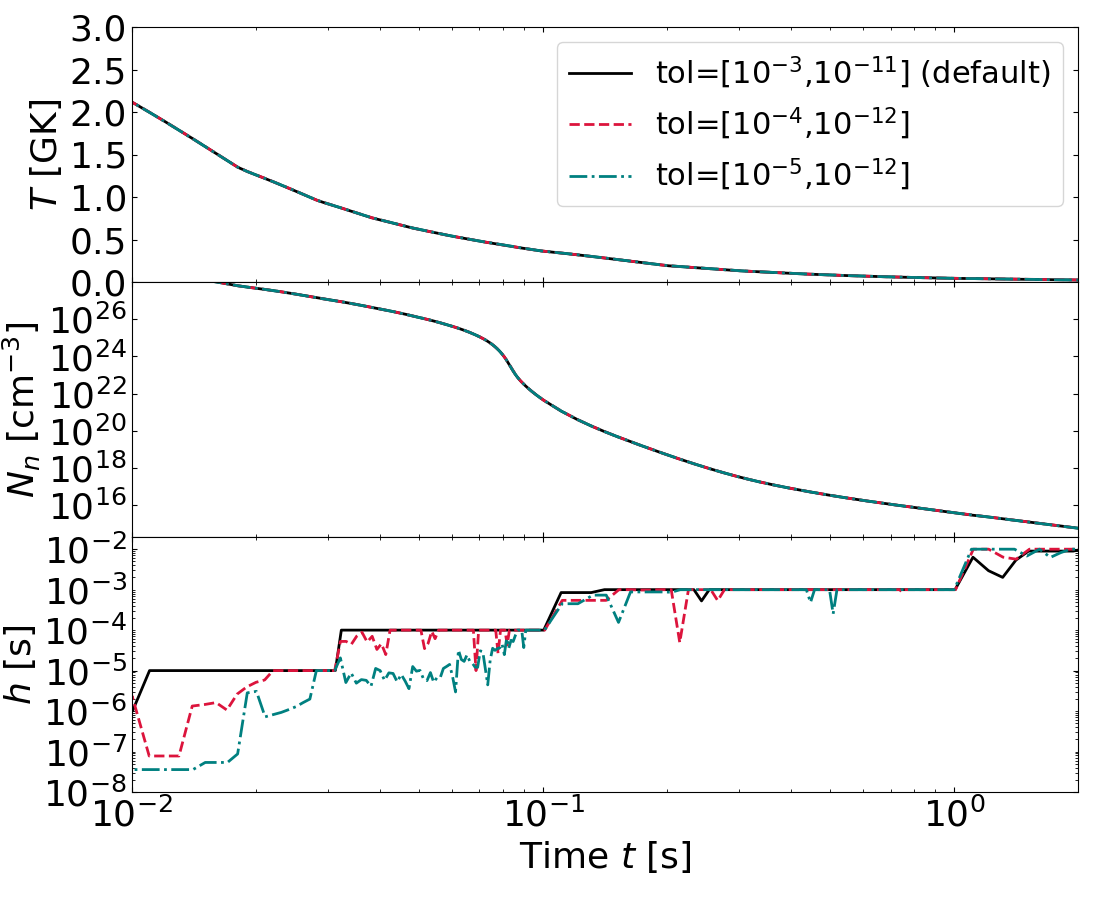}
\caption[The sensitivity of the temperature, neutron density and time step evolution to changes in the tolerances used in the ODE solver for one trajectory.]{
The sensitivity of the temperature $T$ (top), neutron density $N_n$ (middle) and time step $h$ evolution (bottom) to changes in the relative and absolute tolerances (tol=[\texttt{rtol},\texttt{atol}]) used in the ODE solver which evolves the reaction network for one trajectory. 
}
\label{fig_tol_sens1traj}
\end{center}
\end{figure}

\cref{fig_tol_sens} compares the mass fraction distributions obtained when applying the default and high-resolution tolerances for the 256 trajectories in the subset of model SFHo-135-135. The ratio between the default and high-resolution set is close to one for all mass numbers; thus, it is evident that the choice of tolerances does not impact the final abundances. Similar results are also found for the temperature and heating rate evolution.
The time evolution of the temperature, neutron density and time step is shown in \cref{fig_tol_sens1traj} for a typical trajectory around $\sim0.1$~s. We can see that the tolerance does not significantly affect the evolution of the temperature or neutron density; however, the timestep $h$ taken by \texttt{DVODE} varies, and is generally smaller when low tolerances are used. 
During periods with rapid changes in the physical variables, such as the one seen for the neutron density around 0.1~s, the changes in the abundances have to be resolved with sufficiently high resolution. In the example shown in \cref{fig_tol_sens1traj} (and also for all trajectories in the subset), the default resolution (i.e., the highest tolerances) yields results with satisfactory precision. Thus, the benefit of decreasing the tolerances from the default is negligible. At the same time, the computing time almost doubles for a calculation with high-resolution tolerances and is up to 20 times longer for an extreme-resolution simulation. 


\subsection{Late time evolution}
\label{sec_extrap_sens}

After the hydrodynamical simulation ends, the thermodynamical quantities are extrapolated according to a homologous expansion, assuming that the density of the ejecta decreases proportionally to $1/t^3$. 
Two calculations with a faster and slower expansion of the ejecta have been performed to test the sensitivity to this assumption. In summary, we have:
\begin{equation}
\begin{aligned}
\rho \propto \left\{
  \begin{array}{ll}
    t^{-3}, & \text{default} , \\
    t^{-2.7}, & \text{slow expansion}, \\
    t^{-3.3}, & \text{fast expansion},
  \end{array}
  \right.
\end{aligned}
\end{equation}
for the time-extrapolation method.

\begin{figure}[tbp]
\begin{center}
\includegraphics[width=\textwidth]{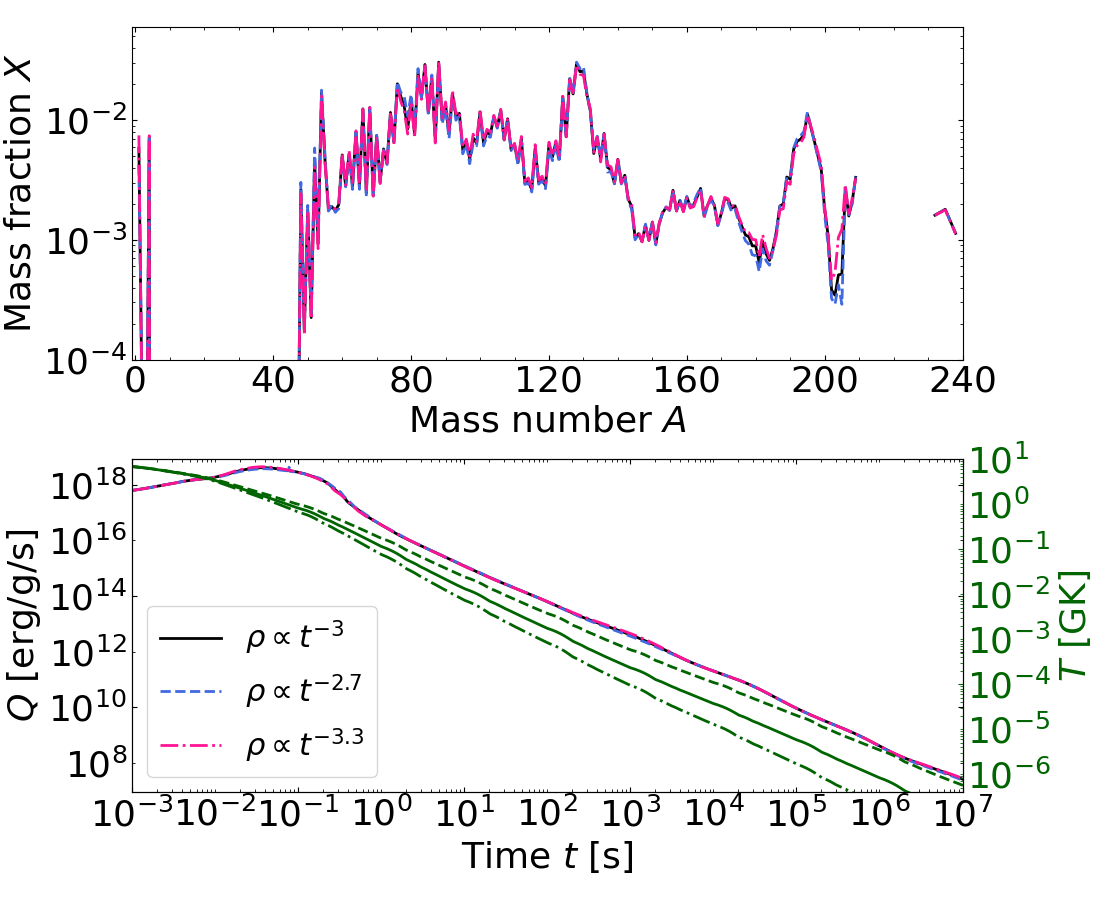}
\caption[The mass fraction, heating rate and temperature sensitivity to the time-extrapolation method for the thermodynamic quantities.]{
The sensitivity of the final mass fraction $X$, heating rate $Q$ and temperature $T$ to the time-extrapolation method for the thermodynamic quantities, i.e., the density $\rho$, using the default assumption (solid black line) compared to assuming a faster (blue dashed line) and slower (pink dash-dotted line) expansion. The temperature is shown in the same plot as the heating rate (sharing the $x$-axis and $y$-axis on the right) with green lines using the same line styles as for $Q$ and $X$ to indicate the assumed expansion model.
}
\label{fig_extrap_sens}
\end{center}
\end{figure}

The final mass fractions versus mass number are shown in \cref{fig_extrap_sens}, together with the time evolution of the heating rate and temperature for the 256 trajectories in the subset of model SFHo-135-135.
The mass fractions obtained when using the fast or slow expansions are within a factor of two larger or smaller than the default calculation when we only consider nuclei with mass fractions larger than $10^{-8}$. For $A>50$ nuclei, the differences between the three calculations are insignificant, except for some variations at the base of the third r-process peak $A\sim205$. 
Similar results are found for the heating rate when varying the time-extrapolation method; the results stay within a maximum 10 per cent change with respect to the default calculation. 
As expected, the assumed expansion model impacts the temperature evolution, where the temperature is five times larger (smaller) than the default calculation for the slow (fast) expansion.

\begin{figure}[tbp]
\begin{center}
\includegraphics[width=\textwidth]{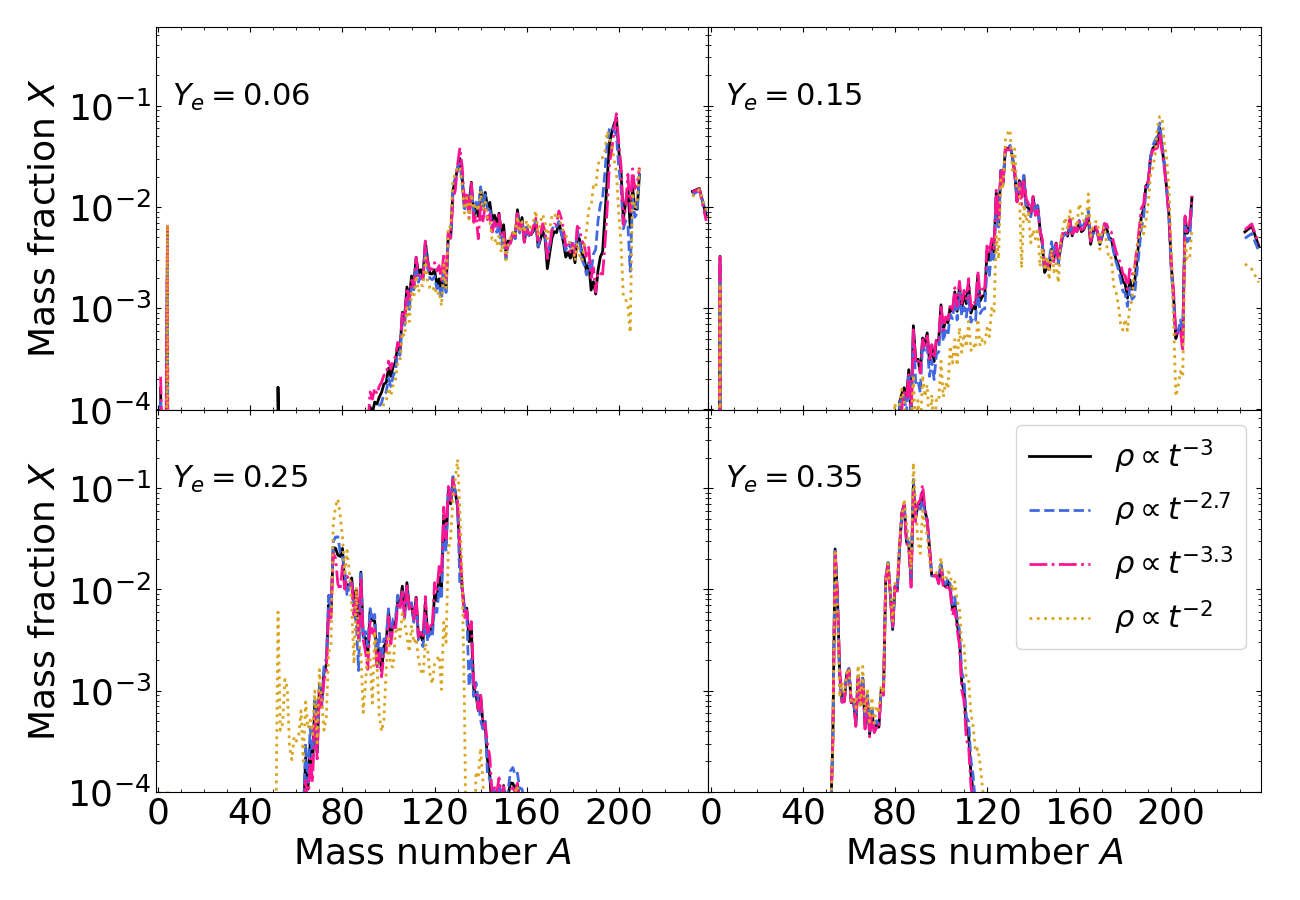}
\caption[The sensitivity of the mass fraction to the time-extrapolation method for four trajectories.]{
Same as \cref{fig_extrap_sens} for four trajectories in addition to results obtained when assuming a very slow expansion $\rho\propto t^{-2}$ (yellow dotted line). The $Y_e$-value at the network initiation time ($\rho_\mathrm{net}$) is shown for each trajectory. 
}
\label{fig_extrap_sens4}
\end{center}
\end{figure}

Similar to \cref{fig_extrap_sens}, \cref{fig_extrap_sens4} shows the impact of varying the expansion model on the mass fraction for four individual trajectories with $Y_e$-values ranging from 0.06 to 0.35. The sensitivity of the final mass fractions to the expansion model is quite independent of the initial $Y_e$ value at $\rho_\mathrm{net}$. A larger impact is found when assuming a very slow expansion, i.e., $\rho\propto t^{-2}$ (yellow dotted line in \cref{fig_extrap_sens4}). The changes with respect to the default calculation are smaller than a factor of two for most $A>80$ nuclei, while changes up to a few orders of magnitude are found for the extreme case $\rho\propto t^{-2}$.

\subsection{Calculating heating rates}
\label{sec_sensheat}

As discussed in \cref{sec:heat}, two methods can be applied to calculate the radioactive heating rate from the freshly synthesized r-process elements. 
The first method (\cref{eq:Q_m1}) sums up the radioactive heat generated by each decay mode, while the second method (\cref{eq:Q_m2}) uses abundance changes of the species in the network and their corresponding binding energies to calculate the heat released. 
The heating rate affects the temperature evolution, which in turn affects the abundance evolution, and therefore possible differences between the methods could affect the final r-process results. 

\begin{figure}[tbp]
\begin{center}
\includegraphics[width=\textwidth]{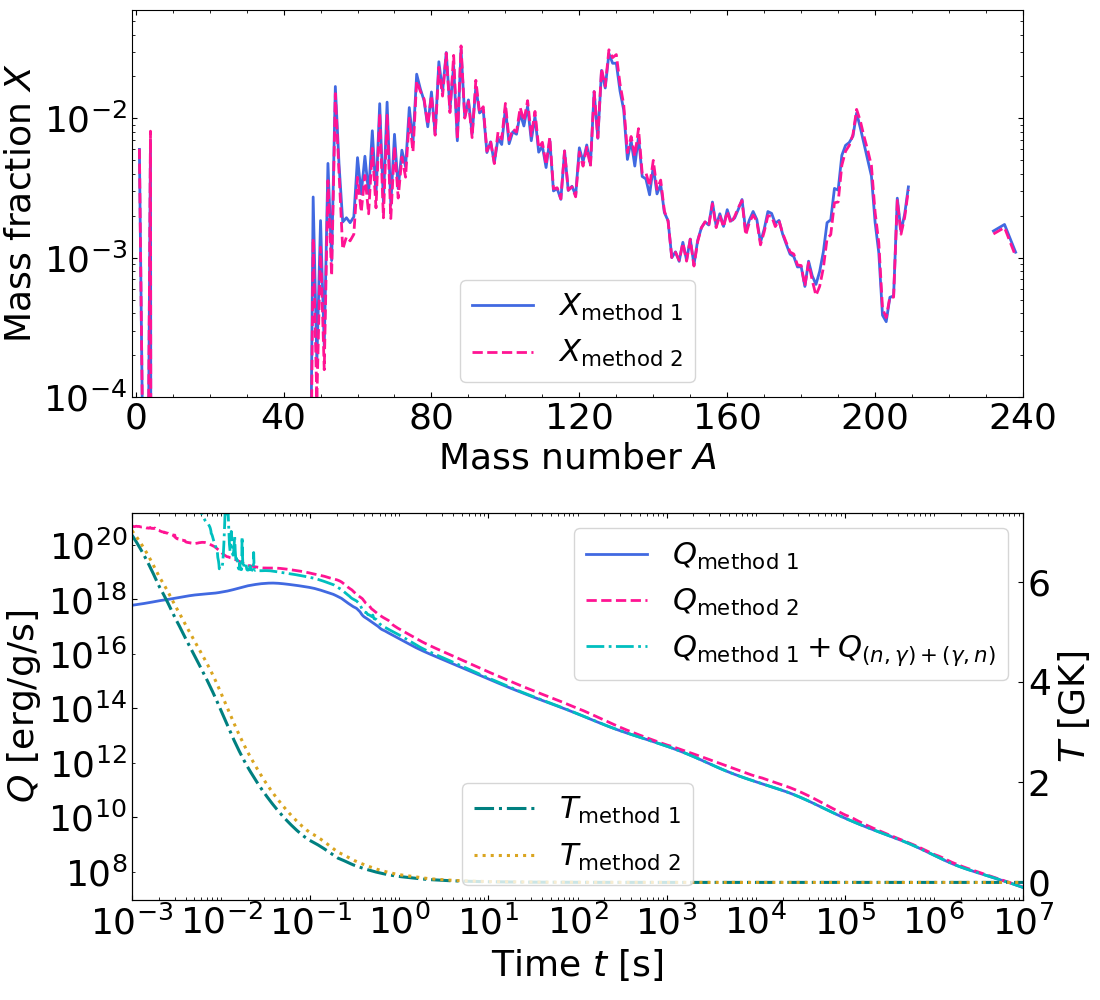}
\caption[Comparison of the mass fraction, heating rate and temperature evolution when the heating rates have been calculated with two different methods.]{
The sensitivity of the final mass fractions (top), heating rate (bottom left $y$-axis) and temperature evolution (bottom right $y$-axis) to the method used to calculate the radioactive heating rate. Here, method 1 and 2 refers to using \cref{eq:Q_m1} and \cref{eq:Q_m2}, respectively, to calculate the heating rate.
The bottom panel also shows the heating rate for method 1 plus the heat generated by $(n,\gamma)$ and $(\gamma,n)$ reactions $Q_\mathrm{method~1} + Q_{(n,\gamma)+(\gamma,n)}$.
}
\label{fig_Qsens}
\end{center}
\end{figure}

The subset of 256 trajectories of model SFHo-135-135 has been used to run r-process calculations to test the sensitivity of the r-process results to both methods used to calculate the heating rate. 
\cref{fig_Qsens} shows a comparison of the results obtained when \cref{eq:Q_m1} (method 1) and \cref{eq:Q_m2} (method 2) are used to calculate the heating rate. The largest differences in the heating rate are found at early times. However, for $t>0.02$~s, the factor between methods 2 and 1 is less than 5 and less than 2 after $t\sim 5$~s. 
By default, method 1 does not include the heat generated by $(n,\gamma)$ and $(\gamma,n)$ reactions, which are important at early times. The cyan dash-dotted line in \cref{fig_Qsens} displays the heat of method 1 plus the heat from $(n,\gamma)$ and $(\gamma,n)$ reactions ($Q_\mathrm{method~1} + Q_{(n,\gamma)+(\gamma,n)}$), where we can see that the heating rate of method 1 is more similar to that of method 2 if these reactions are included. 
Some fluctuations at $\sim0.1$~s stemming from the $(n,\gamma)$ and $(\gamma,n)$ contribution can be seen for $Q_\mathrm{method~1} + Q_{(n,\gamma)+(\gamma,n)}$. Since the nucleosynthesis calculations start from a NSE distribution (see \cref{sec_init_cond}), the network needs some ``relaxation'' time before the $(n,\gamma)$ and $(\gamma,n)$ reactions equilibrate, which is why their contribution is ignored at early times. At late times $Q_{(n,\gamma)+(\gamma,n)}$ becomes insignificant compared to the heat generated by the other decay modes. 
Method 2 
does not have any instabilities at early times. However, it is impossible to separate the heat stemming from the separate decay modes and reactions when using this method. 

The temperature of method 1 starts to decrease slightly earlier than for method 2. However, the temperatures of both methods decrease at a similar rate, and the temperature of method 2 is at most 30\% higher than method 1.
Minor differences are found between the two methods in the abundance distribution. 
The factor between the mass fractions for both methods is less than 2 for all nuclei with $X>10^{-8}$. 

\subsection{Temperature evolution}
\label{sec_Tsens}

For each time step taken by the nuclear reaction network, the abundances but also the temperatures are updated when considering the SPH simulations (see \cref{sec:nuc_net}). 
Before the network initiation time, the temperature evolution is provided by the hydrodynamical model. For the secular ejecta (i.e., the BH-torus) models, the temperature evolution from the hydrodynamical simulation is also provided during the nucleosynthesis calculations. 
Thus, the possible re-heating of the ejecta from the radioactive decay of nuclei (see \cref{sec:heat}) is only included in the dynamical ejecta models, i.e., for the NS-NS and NS-BH mergers.
\citet{Just2015} tested the impact of ignoring the radioactive heat on the r-process results and found it to be small for the BH-torus models, see in particular their Figs.~17 and 18. 

The temperature evolution provided by the hydrodynamical simulations has significant uncertainties. 
In previous works \citep[e.g.,][]{goriely2011,Bauswein2013,goriely2013} based on the same SPH code as our NS-NS and NS-BH merger models, the SPH temperature evolution was ``post-processed'' to avoid shocks with extreme temperature changes. 
However, the most recent works and the models applied here \citep[e.g.,][]{goriely2015a,ardevol-pulpillo2019,kullmann2021} have not included such post-processing for the temperature evolution.
The NS-NS dynamical ejecta models include weak nucleonic reactions with the ILEAS framework. These neutrino reactions are temperature dependent and can, therefore, not be uncoupled from the temperature evolution of the hydrodynamical  simulation. Thus, for consistency, the nucleosynthesis calculations should follow the SPH temperature evolution. 

Three methods have been implemented to test the sensitivity of the nucleosynthesis calculations to uncertainties in the temperature evolution, where the temperature $T_i$ at timestep $i$ is evolved according to:
\begin{equation}
\begin{aligned}
T_i= \left\{
  \begin{array}{ll}
    T_{i-1} + \mathrm{d}T, & \text{default step} , \\
    T_{i-1} + 3\mathrm{d}T, & \text{large step}, \\
    T_{i-1} + \mathrm{d}T/1.3, & \text{small step},
  \end{array}
  \right.
\end{aligned}
\end{equation}
Here, the method with a large (small) temperature step d$T$ artificially increases (decreases) the change in temperature from the previous time step $i-1$ to the next $i$.

\begin{figure}[tbp]
\begin{center}
\includegraphics[width=\textwidth]{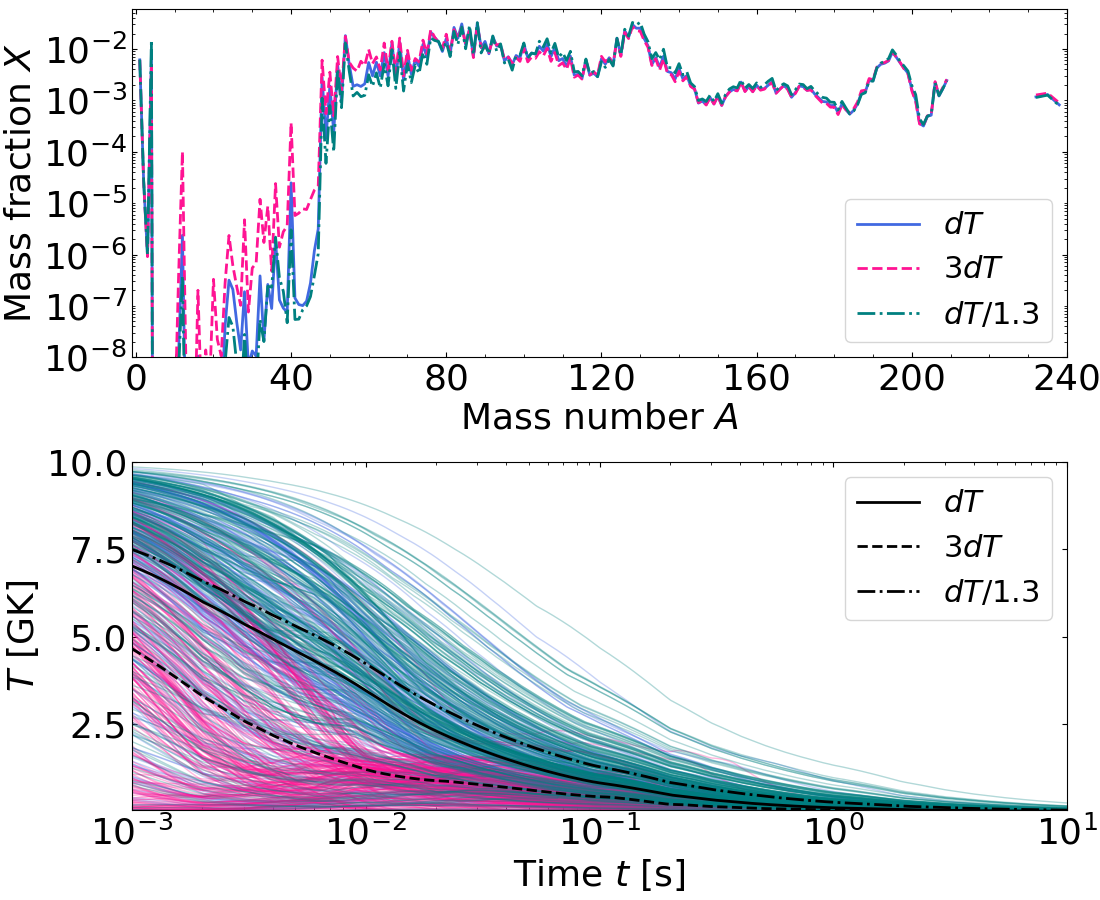}
\caption[The sensitivity of the r-process results to the temperature evolution.]{
The sensitivity of the r-process results to the temperature evolution $T$ when using 256 trajectories of the SFHo-135-135 subset: final mass fractions versus mass number (top) and temperature versus time (bottom). 
All 256 trajectories in the SFHo-135-135 subset have been used to obtain the results. 
Each coloured line in the bottom panel represents the temperature evolution of a trajectory, using the same colours as in the top panel, where the mass-averaged values of $X$ are shown. 
The black lines are the mass-averaged temperatures for all trajectories using the same line styles as for $X$.
Note that the x-axis in the bottom plot starts from $t=10^{-3}$~s.
}
\label{fig_T_sens}
\end{center}
\end{figure}
\begin{figure}[tbp]
\begin{center}
\includegraphics[width=\textwidth]{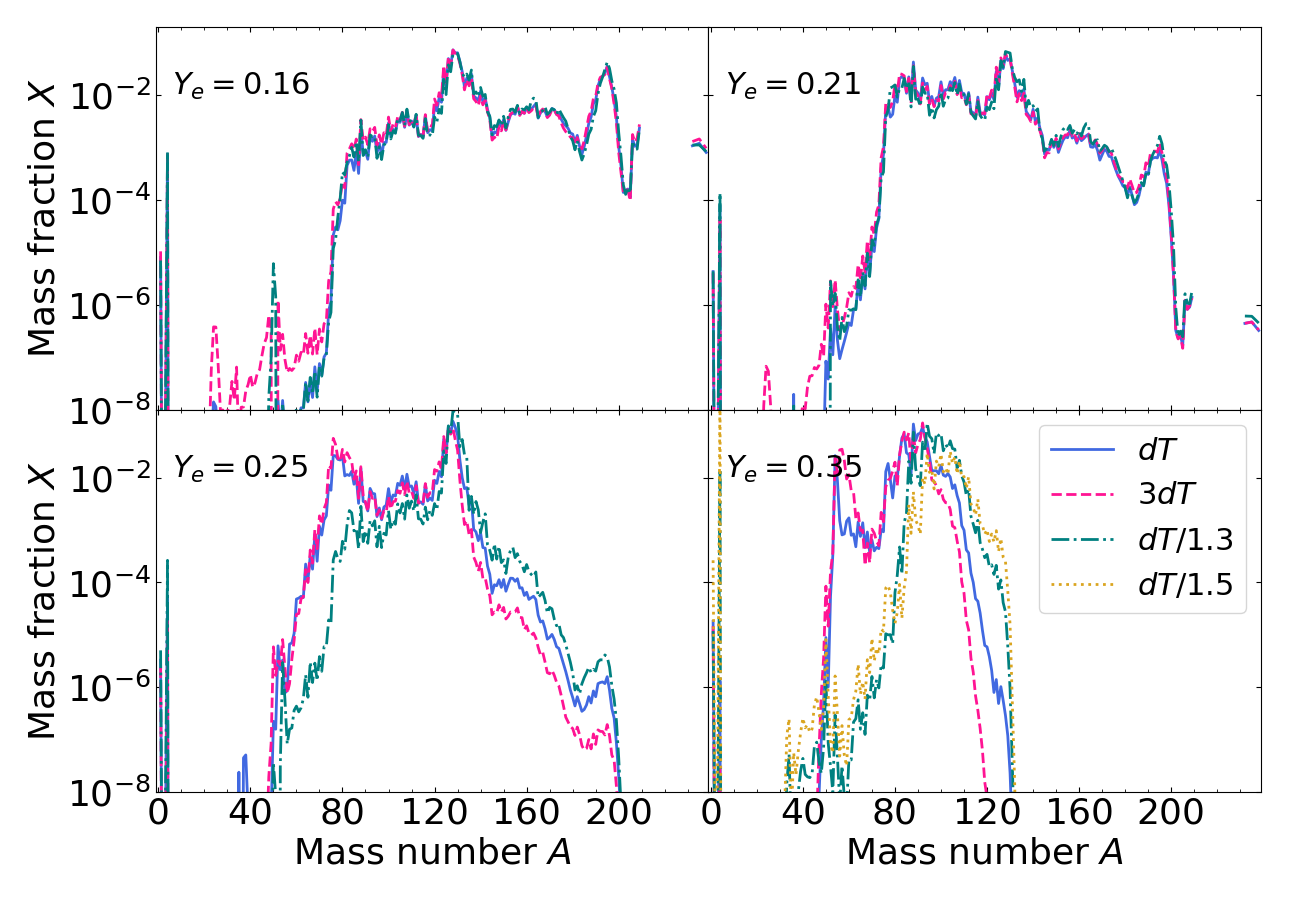}
\caption[The sensitivity of the mass fraction to the temperature evolution for four trajectories.]{
Same as the top panel in \cref{fig_T_sens} for four trajectories of various $Y_e$-values at the network initiation time. 
}
\label{fig_T_sens_4trajs}
\end{center}
\end{figure}

\cref{fig_T_sens} shows the sensitivity of the r-process results to the temperature, where the final mass fractions are shown in the top panel and the temperature evolution in the bottom panel. 
We can see that the modification of the temperature steps d$T$ significantly impacts the temperature evolution.
For the mass fractions, minor differences are found for the mass fraction of the heavier $A>80$ nuclei. Considerable deviations are found for $A<50$ nuclei, however, except for protons and $\alpha$-particles, the abundances for these nuclei are small (generally $X<10^{-4}$). 
The final mass fraction for four trajectories with $Y_e\in[0.16,0.21,0.25,0.35]$ are shown in \cref{fig_T_sens_4trajs}.
The two trajectories with $Y_e<0.25$ are not significantly impacted by the three methods for the temperature timestep if only nuclei with $X>10^{-4}$ are considered. However, for the $Y_e=0.25$ trajectory, the d$T/1.3$ calculation starts to deviate from the default, and even more for the $Y_e=0.35$ trajectory.
An extreme case with d$T/1.5$ is shown for the $Y_e=0.35$ trajectory, which gives a similar mass fraction to the d$T/1.3$ case. The first r-process peak ($A\sim90$) is shifted towards higher mass numbers compared to the default calculation for these two calculations. The first r-process peak is also impacted  for the $Y_e=0.25$ trajectory when using d$T/1.3$. 
When a small temperature step is taken, the temperature decreases more slowly (\cref{fig_T_sens}), which in particular impacts the high-$Y_e$ trajectories and the first r-process peak. Conversely, when the temperatures stay higher for a longer time during the r-process calculations, the $(n,\gamma)$ and $(\gamma,n)$ reactions will stay in semi-equilibrium for a longer time, restricting the amount of material that can reach further than the first r-process peak. 

\subsection{Inclusion of photo-induced fission and (n,2n) reactions}
\label{subsec_photofis}


\begin{figure}[tbp]
\begin{center}
\includegraphics[width=\textwidth]{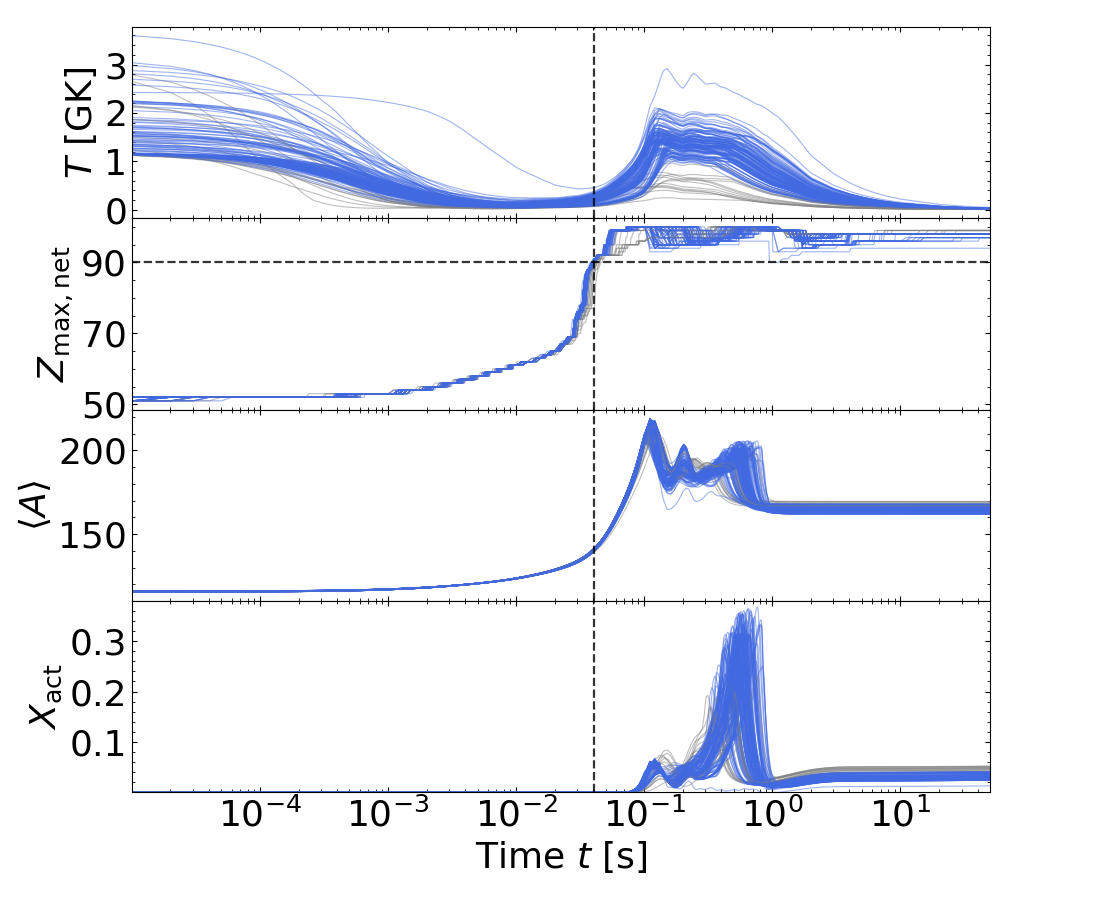}
\caption[Temperature, $Z_\mathrm{max,net}$, $\langle A \rangle$ and $X_\mathrm{act}$ evolution during nucleosynthesis of the 150 trajectories of BH-NS merger model.]{
Time evolution of the 150 trajectories of BH-NS merger model SFHo-11-23 during the nucleosynthesis, starting from the network initiation time.
From the top: the temperature, maximum $Z$ in the network $Z_\mathrm{max,net}$, mean mass number in the network $\langle A \rangle$, and actinide mass fraction $X_\mathrm{act}$ defined as the sum of the mass fractions of $90\leq Z\leq 103$ nuclei. 
All trajectories with $T>1$~GK when $Z_\mathrm{max,net}>90$ are in blue, while the remaining trajectories are in grey. A vertical dotted black line marks the time where most trajectories reach $Z_\mathrm{max,net}>90$, and a horizontal line is plotted at $Z=90$.
}
\label{fig_photofis_trajs}
\end{center}
\end{figure}
\begin{figure}[tbp]
\begin{center}
\includegraphics[width=\textwidth]{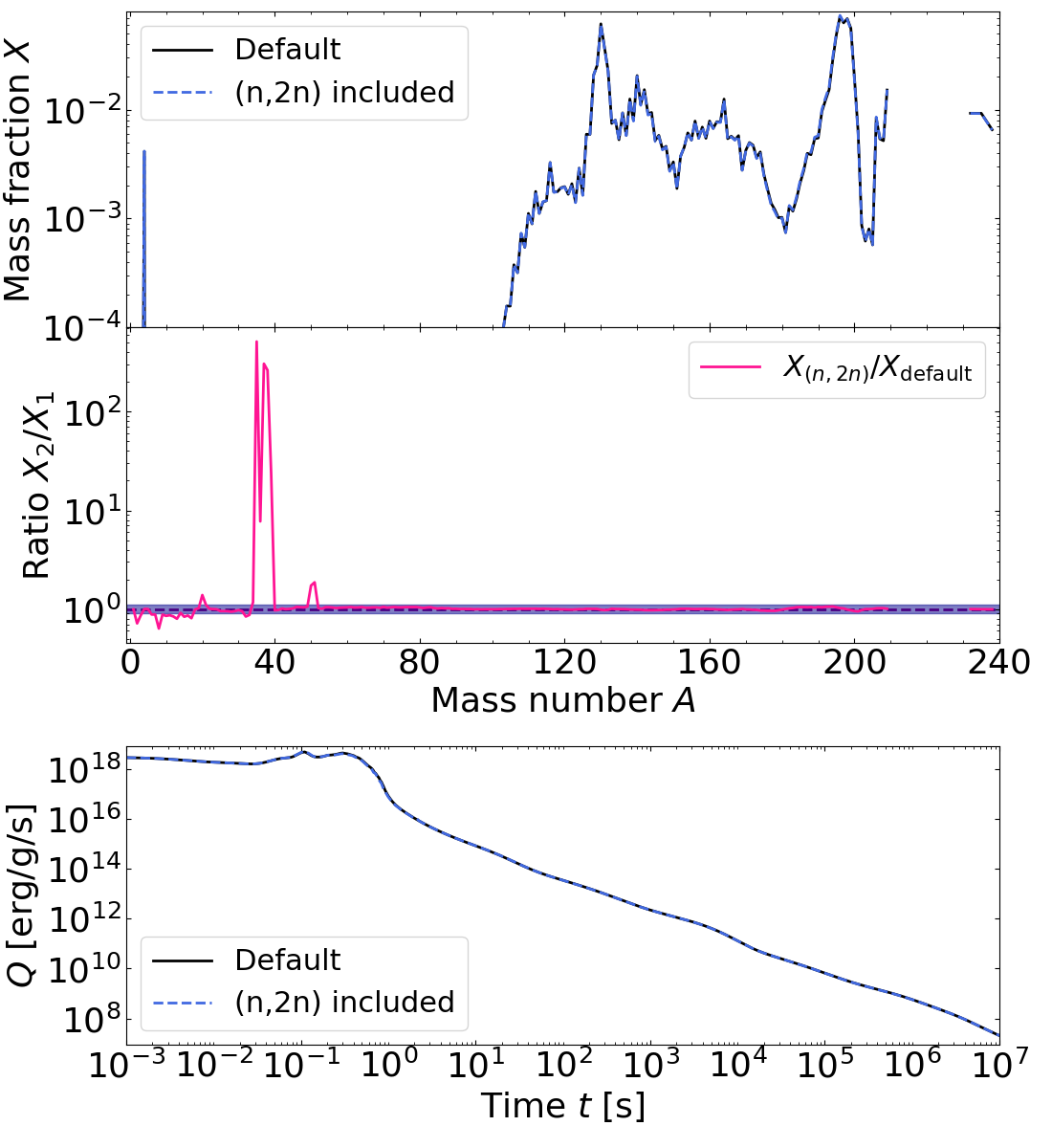}
\caption[The final mass fractions and heating rate with and without the inclusion of (n,2n) reactions for NS-BH model SFHo-11-23.]{
Comparison of the r-process results when ignoring or including (n,2n) reactions for NS-BH model SFHo-11-23: the final mass fractions $X$ (top), the ratio of the mass fractions with and without (n,2n) reactions (middle), and the heating rate $Q$ (bottom).
Similar to \cref{fig_Zmax_sens}, the ratio $X_{\mathrm{(n,2n)}}/X_{\mathrm{default}}$ is only calculated when the mass fractions are greater than $10^{-6}$. The purple dashed line $y=1$ and a dark purple colour band from $y=0.5$ to 1.5 are shown in the middle panel. 
}
\label{fig_n2n_sens}
\end{center}
\end{figure}

By default, our r-process calculations ignore photo-fission and (n,2n) reactions since they are assumed to be unimportant for temperatures typically below 1~GK. 
For the four NS-NS models considered here, the temperature is never above 1~GK when the r-process flow reaches $Z>90$, and the ejecta is only re-heated to moderate temperatures by radioactive decays. 
A different situation is found for the BH-torus remnant models, where most trajectories experience high temperatures and re-heating during the r-process nucleosynthesis, see \cref{fig_hyd_wm}\footnote{Note that the first timestep in this figure is at $\rho_\mathrm{net}$.}.
The trajectories are generally re-heated to temperatures around 6 to 8~GK, where photodisintegration reactions start to play a major role, breaking up the recently synthesized nuclei into lighter nuclei. 
However, very few trajectories reach the fissile $Z>90$ region due to the generally higher $Y_e$-values for the BH-torus models compared to the dynamical ejecta (see \cref{fig_xn0_ye_distr}).
The top panel of \cref{fig_photofis_trajs} shows the temperature evolution of NS-BH dynamical ejecta model SFHo-11-23 starting from $\rho_\mathrm{net}$ for the 150 trajectories of the subset. Although all trajectories start with low temperatures ($T<5$~GK), they all experience re-heating. The trajectories that are re-heated to temperatures above 1~GK are shown in blue in \cref{fig_photofis_trajs}. 
We can measure if the network has reached the fissile region by defining a time-dependent quantity $Z_\mathrm{max,net}(t)$ that contains the largest $Z$-value among all nuclei with a molar fraction $Y$ larger than $10^{-8}$.
Thus, the temperature $T$ at the time $t$ when $Z_\mathrm{max,net}>90$  determines if photo-fission could play a significant role in the r-process calculations or not. 
The evolution of $Z_\mathrm{max,net}$ is shown in the second panel of \cref{fig_photofis_trajs}, where we can see that the majority of the trajectories reach high temperatures after $Z_\mathrm{max,net}>90$ (see the dotted lines).
In addition, the mean mass number $\langle A \rangle$ in the network and the actinide mass fractions are displayed in \cref{fig_photofis_trajs}. 
As seen in \cref{fig_xn0_ye_distr}, the NS-BH model SFHo-11-23 has a narrow $Y_e$-distribution where all values are below 0.1. Thus, a large amount of heavy r-process elements and actinides is expected to be formed in all trajectories, which is what we see in \cref{fig_photofis_trajs}. 
For the BH-torus models considered here this is not the case, where only a few trajectories produce $Z>90$ nuclei, and $Z_\mathrm{max,net}$ is only larger than 90 for a short time due to the high temperatures (which destroy the heaviest nuclei). Thus the impact of photo-induced fission is expected to be small in the secular ejecta models. 

Similar to photofission, $(n,2n)$ reactions are also expected to play a role at elevated temperatures \citep{goriely2008}. 
\cref{fig_n2n_sens} shows the impact of including $(n,2n)$ (and the reverse $(2n,n)$) reactions on the r-process results when using 150 trajectories of the NS-BH model SFHo-11-23. 
Except for nuclei in the $A=34$ to 40 region, the impact of including (n,2n) reactions is insignificant on the abundance distribution. The impact is particularly small for $A>52$ nuclei, where the ratio $X_{\mathrm{(n,2n)}}/X_{\mathrm{default}}$ stays within 1.07 and 0.95. 
Even smaller differences are found when comparing the default results for 150 trajectories to those obtained when including photofission.

\subsection{Summary of sensitivity studies}

In \cref{sec_netlimtol,sec_extrap_sens,sec_sensheat,sec_Tsens,subsec_photofis}, the sensitivity of the r-process results to several assumptions and approximations have been studied: the $Z$-limit of the reaction network, the ODE solver tolerances, the late time evolution extrapolation of thermodynamical variables, the method used to calculate the heating rates, the temperature evolution, and the inclusion of photo-induced fission and (n,2n) reactions. Except for the latter section, where 150 trajectories from the BH-NS model SFHo-11-23 were applied, the subset of NS-NS model SFHo-135-135 with 256 trajectories was used for all sensitivity tests.
Nuclear physics input set 1 in \cref{tab_nuc_mods} was applied in all calculations as default for models SFHo-135-135 and SFHo-11-23. 

Generally, our results obtained with the default input configuration are robust with respect to the assumptions tested here. Globally, the shape of the r-process abundance distribution is maintained for all sensitivity variations, and the deviations are smaller than the nuclear physics uncertainties, which will be presented in \cref{ch_nucuncert}. 
The largest deviations are found in the mass fractions of light nuclei with $A\lesssim 50$, in particular for changes in the temperature evolution (\cref{sec_Tsens}), and if (n,2n) reactions are included (\cref{subsec_photofis}). 
However, the impact on the mass fractions of $A>80$ nuclei is of most interest to the r-process studies performed in the following \cref{ch_dynweak,ch_nucuncert} and was found to be insignificant for all sensitivity variations studied here. 





\chapter[Dynamical ejecta with nucleonic weak processes][Dynamical ejecta]{Dynamical ejecta with nucleonic weak processes}
\label{ch_dynweak}


\epigraph{\itshape ``Be naive enough to start the PhD and stubborn enough to finish.''}{--- \textup{Unknown}}

Four hydrodynamical simulations of the NS-NS dynamical ejecta have been applied as the basis for the nucleosynthesis calculations (see \cref{subsec_dynsim}).
In the following, the ``ILEAS'' models refer to these four models, which include weak nucleonic reactions in the ejecta, and involve two NS-NS systems with equal and unequal mass for the NSs, varying between two different EoSs, namely the DD2 and SFHo EoSs, in the simulations. 
Particular attention to the angular and velocity dependence of the r-process abundance distributions (presented in \cref{sec_nuc}) and heating rates (presented in \cref{sec_q}) is given in \cref{sec_ang_vc}. 
The `no neutrino' case introduced in \cref{sec_init_cond} differ from the corresponding ILEAS model mainly by the $Y_e$-value adopted at the network initiation time (\cref{fig_distr_t9_rho,fig_xn0_ye_distr,fig_yedist_cases}), and is used to study if a more accurate description of neutrino interactions can significantly affect the r-process yields, heating rates and kilonova light curve (presented in \cref{sec_kilonova}).
In \cref{sect_sens}, the sensitivity of the results to the electron fraction is investigated further by `artificially' increasing and decreasing the electron fraction at the network initiation time by 0.05 or 0.1. 
A comparison to other works is given in \cref{sect_comp}, and \cref{sect_concl} summarizes and concludes the study.
The following section results have been adapted from \citet{kullmann2021}.

\todo{[...make landscape table work and longtable]}

\section{Nucleosynthesis}
\label{sec_nuc}

\begin{figure}[tbp]
\includegraphics[width=\columnwidth]{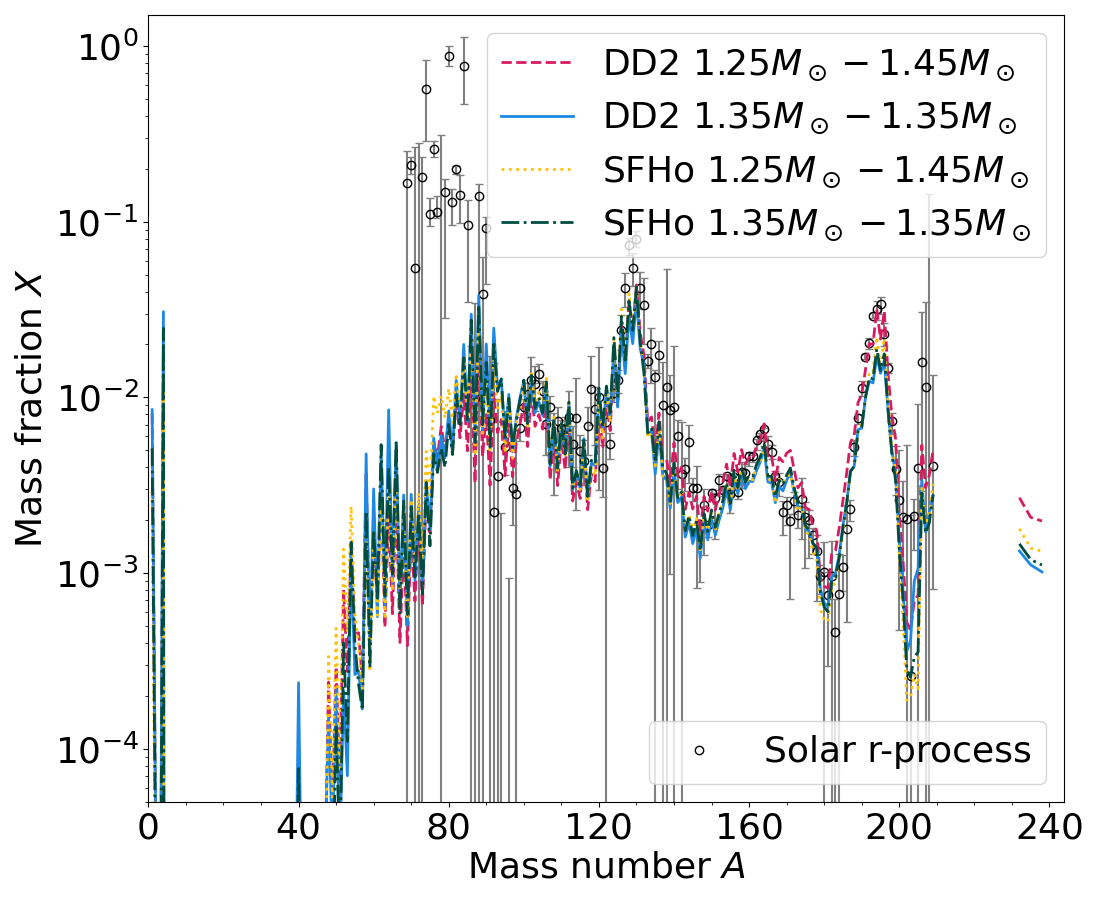}
  \caption[Final mass fractions of the material ejected as a function of the atomic mass for the four NS-NS merger models.]{
  Final mass fractions of the material ejected as a function of the atomic mass $A$ for the DD2 1.25--1.45~\Msun,  DD2 1.35--1.35~\Msun, SFHo 1.25--1.45~\Msun\ and  SFHo 1.35--1.35~\Msun\ NS-NS merger models.  The solar system r-abundance distribution  (open circles) from \citet{goriely1999} is shown for comparison and arbitrarily normalized to the DD2 asymmetric model at the third r-process peak ($A\simeq 195$).}
\label{fig_rpro_aa}
\end{figure}

\cref{fig_rpro_aa} shows the final isotopic abundance distributions obtained if we adopt the initial ILEAS $Y_e$ distributions of \cref{fig_xn0_ye_distr} for the four hydrodynamical merger models. Given the similar and relatively wide initial $Y_e$ distributions, the resulting abundance distributions are almost identical and reproduce rather well the solar system r-abundance distribution above $A\ga 90$. For all models, we have efficient r-process nucleosynthesis with the production of lanthanides, second- and third-peak nuclei. The lanthanide plus actinide mass fraction $X_{LA}$, the relative amount of r-process nuclei $x_{A>69}$ and of third-r-process peak nuclei $x_{A>183}$, (i.e., with $A>183$) are summarized in \cref{tab:rpro_regions} for each model. In particular, the ejecta of all four systems consists of 88 up to 95\% of $A>69$ r-process material with lanthanides plus actinides ranging between 11 and 15\% in mass. In all four models, the third r-process peak is rather well produced and includes between 11 to 15\% of the total mass. The DD2-125-145 model has a relatively larger production of the heaviest r-process elements, as indicated by a larger value of $x_{A>183}$, which reaches about 30\% in the equatorial region.

\begin{figure}[tbp]
\includegraphics[width=\columnwidth]{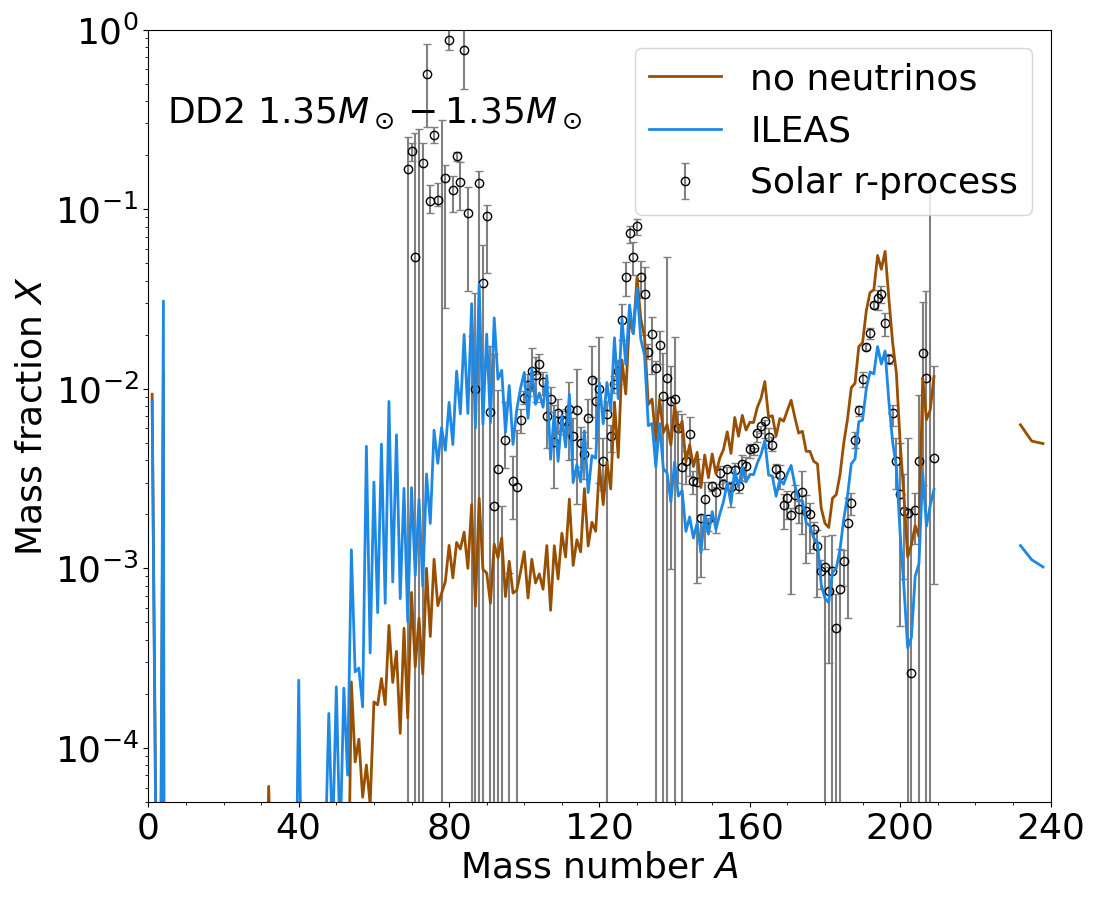}
  \caption[Same as \cref{fig_rpro_aa} for the ``no neutrino'' case.]{
  Mass fractions of the $2\times 10^{-3}$~\Msun\ of material
    ejected in the 1.35--1.35~\Msun\ NS-NS merger model with DD2 EoS as a function of the atomic mass $A$ for the two cases studied here, i.e., with (ILEAS) or without (no neutrinos) weak nucleonic interactions.  The solar system abundance distribution is normalized as in \cref{fig_rpro_aa}. }
\label{fig_rpro_aa_cases}
\end{figure}

For the DD2-135-135 model, \cref{fig_rpro_aa_cases} shows the final isotopic abundance distributions of the case without neutrinos compared to the case where neutrino interactions are included.
If we assume the initial $Y_e$ distribution to be unaffected by weak interactions (\cref{fig_yedist_cases}a), the resulting distribution is characteristic of what has been obtained by most of the calculations neglecting neutrino absorption, i.e., the production of $A \ga 130-140$ is considerably enhanced due to the dominance of $Y_e<0.1$ trajectories and an efficient fission recycling. The production of $A\simeq 130$ nuclei in the second r-process peak is linked to the non-negligible presence of $Y_e> 0.15 $ trajectories (see \cref{fig_yedist_cases} and the discussion in \cref{sec_init_cond}). The ``no neutrino'' case is found to be composed of 2.1 (4.7) times more lanthanides (actinides) and a significantly more pronounced third r-process peak (\cref{tab:rpro_regions}).
\todo{at which point should I talk about recoil and weak magnetism corrections?}

\begin{figure}[tbp]
\includegraphics[width=0.85\columnwidth]{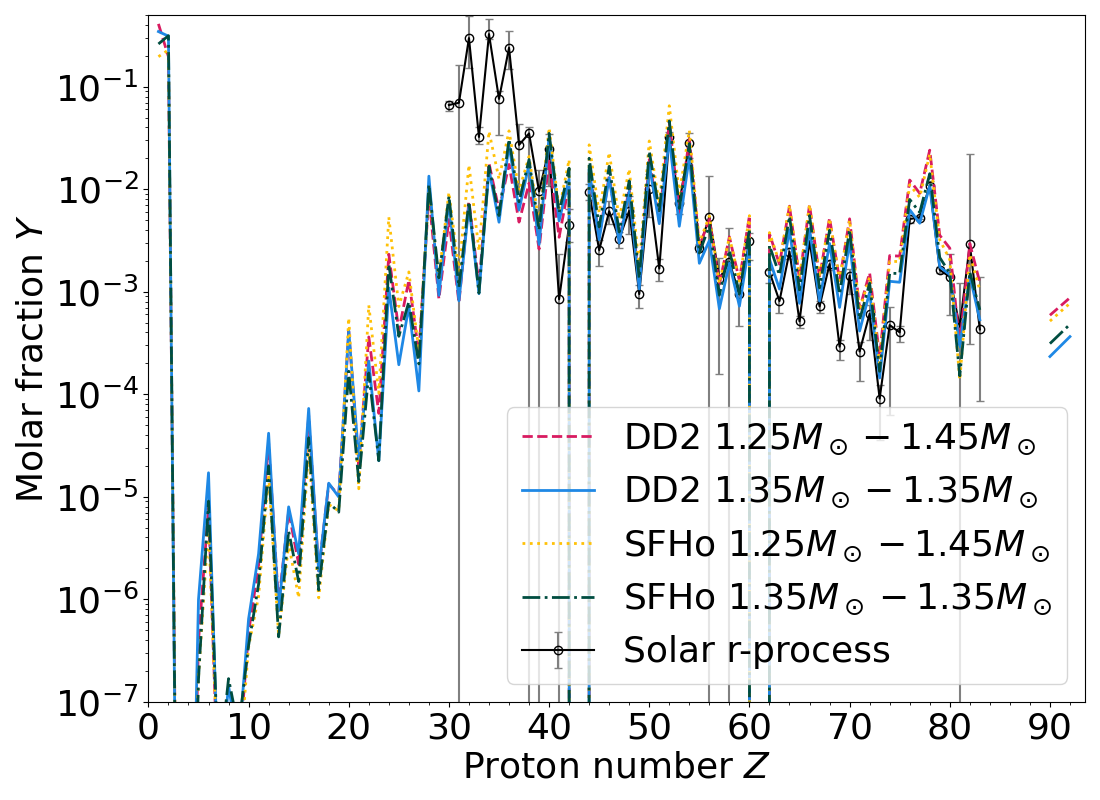}
  \caption[Same as \cref{fig_rpro_aa} for the elemental abundance distributions.]{
  Same as \cref{fig_rpro_aa} for the elemental abundance distributions given by the molar fractions. The solar distribution is normalized to the DD2-125-145 prediction of the third r-process peak.}
\label{fig_rpro_zz}
\end{figure} 
\begin{figure}[tbp]
\includegraphics[width=0.85\columnwidth]{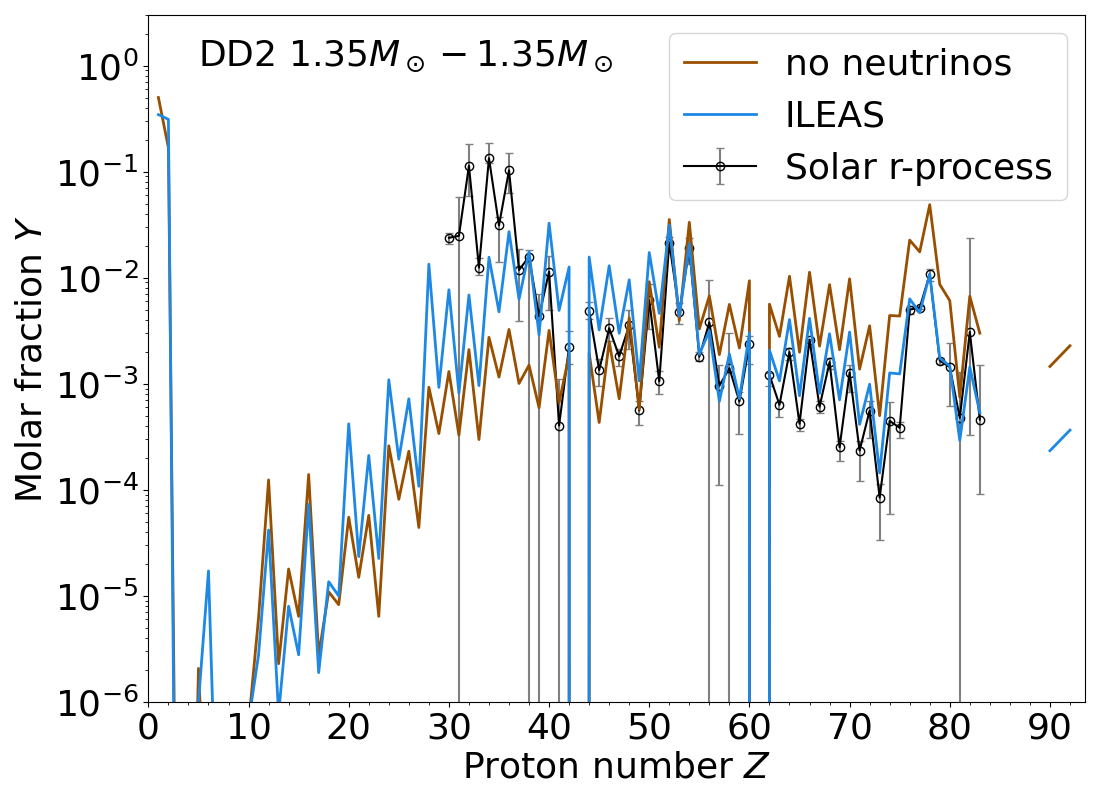}
  \caption[Same as \cref{fig_rpro_aa_cases} for the elemental abundance distributions.]{
  Same as \cref{fig_rpro_aa_cases} for the elemental abundance distributions given by the molar fractions. The solar distribution is normalized as in \cref{fig_rpro_zz}.}
\label{fig_rpro_zz_cases}
\end{figure} 

The final elemental abundance distributions obtained from the four hydrodynamical models including weak processes are shown in \cref{fig_rpro_zz}. As for the isotopic distributions, there are only minor differences between the four elemental distributions. In particular,  the production of actinides is larger for the two asymmetric merger models. However, the $^{232}$Th to $^{238}$U ratio remains rather constant and equal to 1.35--1.39 for all four models (\cref{tab:rpro_regions}), a property of particular interest to cosmochronometry, see \cref{sec:cosmo}. 

The elemental distributions of the DD2-135-135 cases with and without neutrinos are presented in \cref{fig_rpro_zz_cases}.  We can see that a somewhat different prediction is obtained when including weak processes, in particular, a significantly smaller amount of $Z \ga 50$ elements is produced.  
However, although the actinide production for the ILEAS case is significantly smaller compared to the reference neutrino-less simulation, the elemental ratio Th/U remains rather constant.  
For the ILEAS case, the Ni to Zr region dominates the ejecta,  and the production of Sr ($Z=38$) is 14 times larger compared to the case without neutrinos.  Sr is of special interest after its identification by \citet{watson2019} in the AT2017gfo spectrum.  Such a different elemental distribution impacts the observed kilonova light curve,  as discussed in \cref{sec_kilonova} \citep[see also][]{just2021}.

\begin{landscape}
\begin{table}[p]
\centering
\caption[Properties of the NS-NS merger dynamical ejecta in three angular regions.]{Summary of the ejecta properties for the five models considered in the present study, including the contribution stemming from three different angular regions, namely the polar ($0^\circ < |90^\circ - \theta| \leq 30^\circ$), middle ($30^\circ < |90^\circ - \theta| \leq 60^\circ$) or equatorial ($60^\circ < |90^\circ - \theta| \leq 90^\circ$)  regions abbreviated as Pol., Mid. and Eq., respectively. 
These properties correspond to the total ejected mass $M_{\rm ej}$, the mass of the high-velocity ejecta $M_{\rm ej}^{v \ge 0.6c}$, the mean velocity $\langle v/c \rangle$, the mean $Y_e$ at $\rho=\rho_{\rm net}$, the neutron mass fraction $X_n$ at $t=20$~s, the lanthanide plus actinide mass fraction $X_{LA}$ (i.e., with respect to the total ejecta), the relative amount nuclei $x$ with respect to the mass in each region $x_{A>69}$ and $x_{A>183}$ for r-process nuclei and third-r-process peak nuclei, respectively, and the $^{232}$Th to $^{238}$U ratio. 
The masses $M$ are given in solar masses \Msun, and the first four columns are identical to the values given in \cref{tab:astromods} for the total ejecta (for these quantities, the entries for the `no neutrino' case are identical to the entries of model DD2-135-135). 
}
\begin{tabular}{llccccccccc}
\hline \hline
Model & Reg. & $M_{\rm ej}$ & $M_{\rm ej}^{v \ge 0.6c}$ & $\langle v/c \rangle$ & $\langle Y_e\rangle$ & $X_n^{t=20s}$ & $X_{LA}$ & $x_{A>69}$ & $x_{A>183}$ & Th/U \\
 & & [$10^{-3}$] & [$10^{-4}$] & & & [$10^{-3}$] & & & & \\
\hline
 DD2-125-145 & Tot.  & 3.20 & 1.77  & 0.25  & 0.22 & 7.57  & 0.152  & 0.90  & 0.15  & 1.39 \\
             & Eq.   & 2.20 & 0.91  & 0.24  & 0.19 & 1.60  & 0.126  & 0.97  & 0.30  & 1.40 \\
             & Mid.  & 0.78 & 0.62  & 0.24  & 0.26 & 4.22  & 0.022  & 0.95  & 0.11  & 1.35 \\
             & Pol.  & 0.22 & 0.23  & 0.27  & 0.33 & 1.76  & 0.004  & 0.78  & 0.05  & 1.30 \\
\hline
 DD2-135-135 & Tot.  & 1.99 & 0.87  & 0.25  & 0.27 & 7.21  & 0.107  & 0.88  & 0.11  & 1.35 \\
             & Eq.   & 1.35 & 0.46  & 0.25  & 0.25 & 4.71  & 0.086  & 0.95  & 0.17  & 1.36 \\
             & Mid.  & 0.45 & 0.23  & 0.24  & 0.29 & 1.72  & 0.017  & 0.93  & 0.09  & 1.30 \\
             & Pol.  & 0.20 & 0.18  & 0.27  & 0.34 & 0.78  & 0.003  & 0.75  & 0.07  & 1.33 \\
\hline
 DD2-135-135 & Tot.  & -    & -     & -     & 0.13  & 9.17  & 0.246  & 0.97  & 0.43  & 1.31 \\
 no neutrino & Eq.   & -    & -     & -     & 0.11  & 5.51  & 0.174  & 0.98  & 0.48  & 1.31 \\
             & Mid.  & -    & -     & -     & 0.14  & 2.83  & 0.048  & 0.97  & 0.42  & 1.29 \\
             & Pol.  & -    & -     & -     & 0.18  & 0.83  & 0.024  & 0.96  & 0.40  & 1.28 \\
\hline \hline
\end{tabular}
\label{tab:rpro_regions}
\end{table}

\begin{table}
\centering
\begin{tabular}{llccccccccc}
\hline \hline
Model & Reg. & $M_{\rm ej}$ & $M_{\rm ej}^{v \ge 0.6c}$ & $\langle v/c \rangle$ & $\langle Y_e\rangle$ & $X_n^{t=20s}$ & $X_{LA}$ & $x_{A>69}$ & $x_{A>183}$ & Th/U \\
 & & [$10^{-3}$] & [$10^{-4}$] & & & [$10^{-3}$] & & & & \\
\hline
SFHo-125-145 & Tot.  & 8.67 & 2.56  & 0.24  & 0.24 & 2.75  & 0.114  & 0.95  & 0.12  & 1.38 \\
             & Eq.   & 5.05 & 1.46  & 0.24  & 0.22 & 0.32  & 0.079  & 0.97  & 0.20  & 1.38 \\
             & Mid.  & 2.89 & 0.79  & 0.24  & 0.27 & 1.22  & 0.030  & 0.95  & 0.11  & 1.38 \\
             & Pol.  & 0.73 & 0.31  & 0.26  & 0.30 & 1.21  & 0.005  & 0.92  & 0.05  & 1.34 \\
\hline
SFHo-135-135 & Tot.  & 3.31 & 1.53  & 0.29  & 0.26 & 4.76  & 0.115  & 0.92  & 0.11  & 1.35 \\
             & Eq.   & 2.17 & 0.93  & 0.31  & 0.24 & 0.46  & 0.088  & 0.95  & 0.18  & 1.36 \\
             & Mid.  & 0.86 & 0.51  & 0.25  & 0.29 & 2.54  & 0.020  & 0.93  & 0.08  & 1.34 \\
             & Pol.  & 0.28 & 0.09  & 0.26  & 0.30 & 1.76  & 0.007  & 0.89  & 0.07  & 1.26 \\
\hline \hline
\end{tabular}
\end{table}
\end{landscape}

\section{Radioactive decay heat}
\label{sec_q}

\begin{figure}[tbp]
\includegraphics[width=\columnwidth]{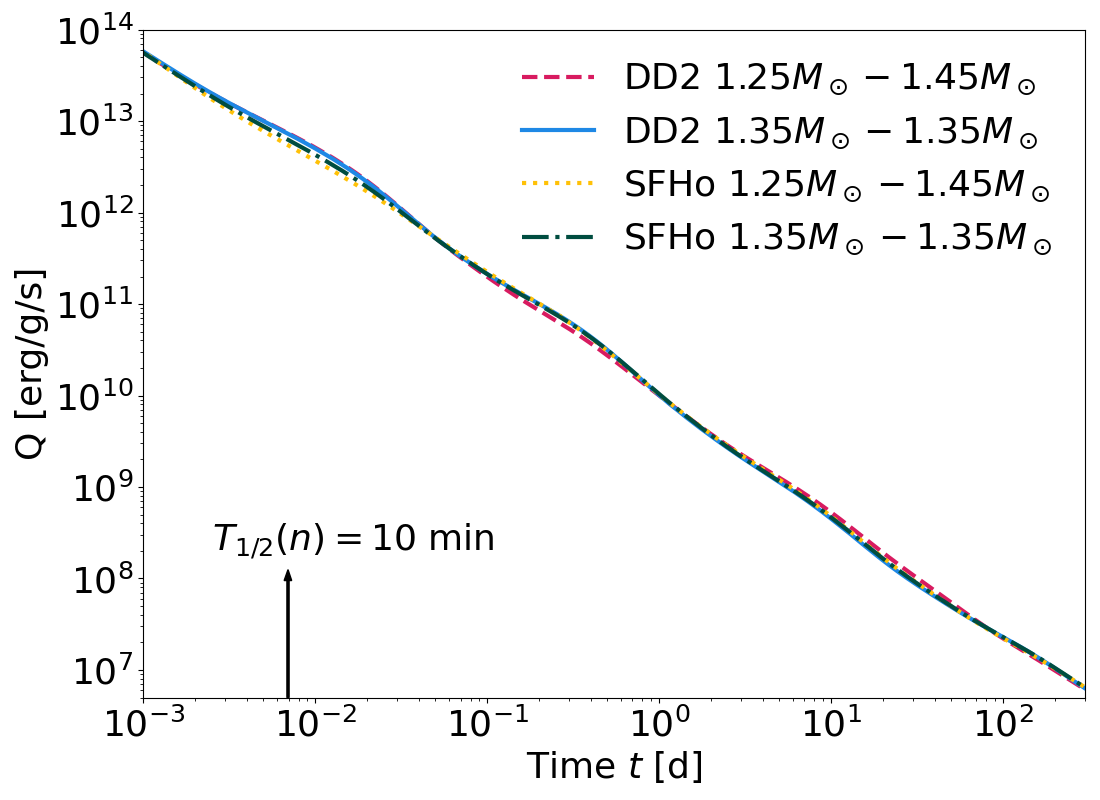}
  \caption[Time evolution of the radioactive heating rate for the four NS-NS merger models including weak interactions. ]{
  Time evolution of the radioactive heating rate $Q$  for the four hydrodynamical models including weak interactions. The arrow at $t=10$~min corresponds to the decay half-life of the neutron.}
\label{fig_Q_t}
\end{figure} 

\begin{figure}[tbp]
\includegraphics[width=\columnwidth]{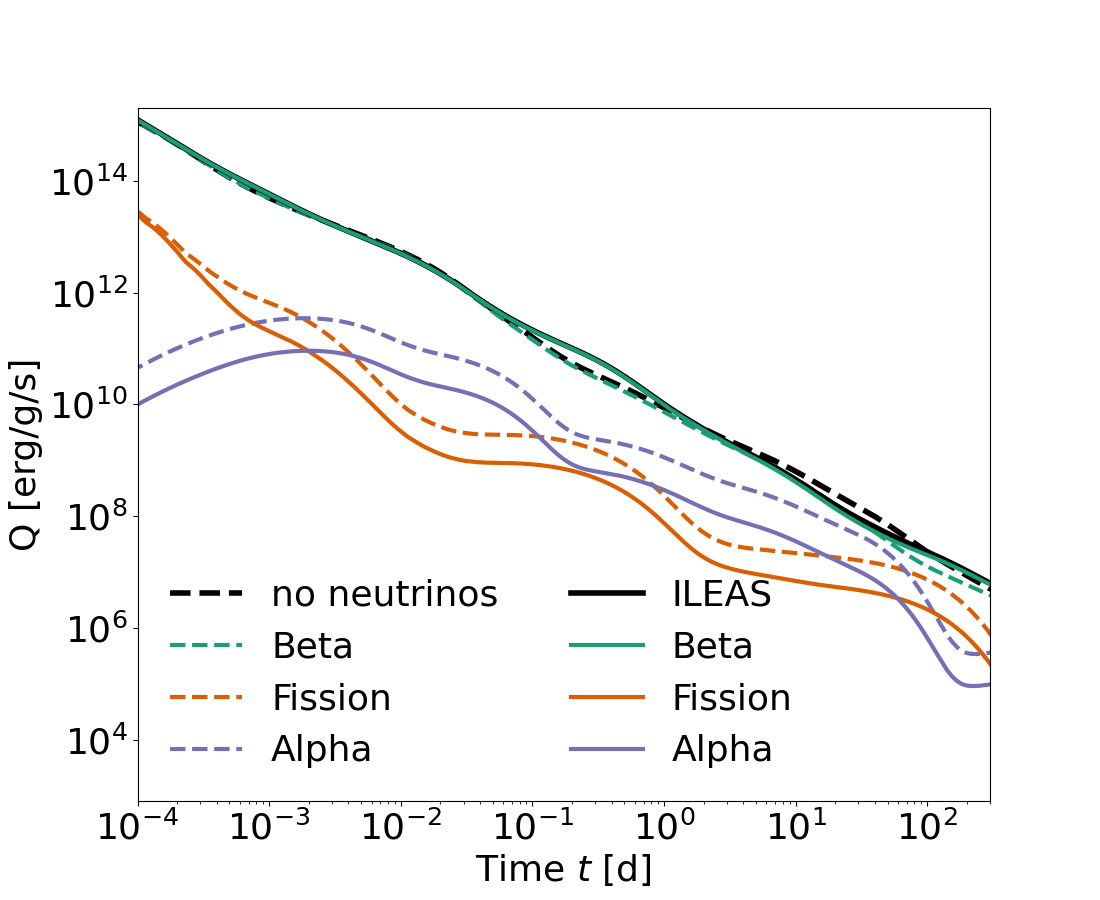}
  \caption[Same as \cref{fig_Q_t} for the ``no neutrino'' case. ]{
  Same as \cref{fig_Q_t} for the DD2-135-135 model with (solid lines) and without (dashed lines) weak nucleonic interactions. The total heating rate is shown in black, the $\beta$-, $\alpha$- and fission contributions in green, orange and purple colours, respectively. }
\label{fig_Q_t_cases}
\end{figure} 

\begin{figure}[tbp]
\includegraphics[width=\columnwidth]{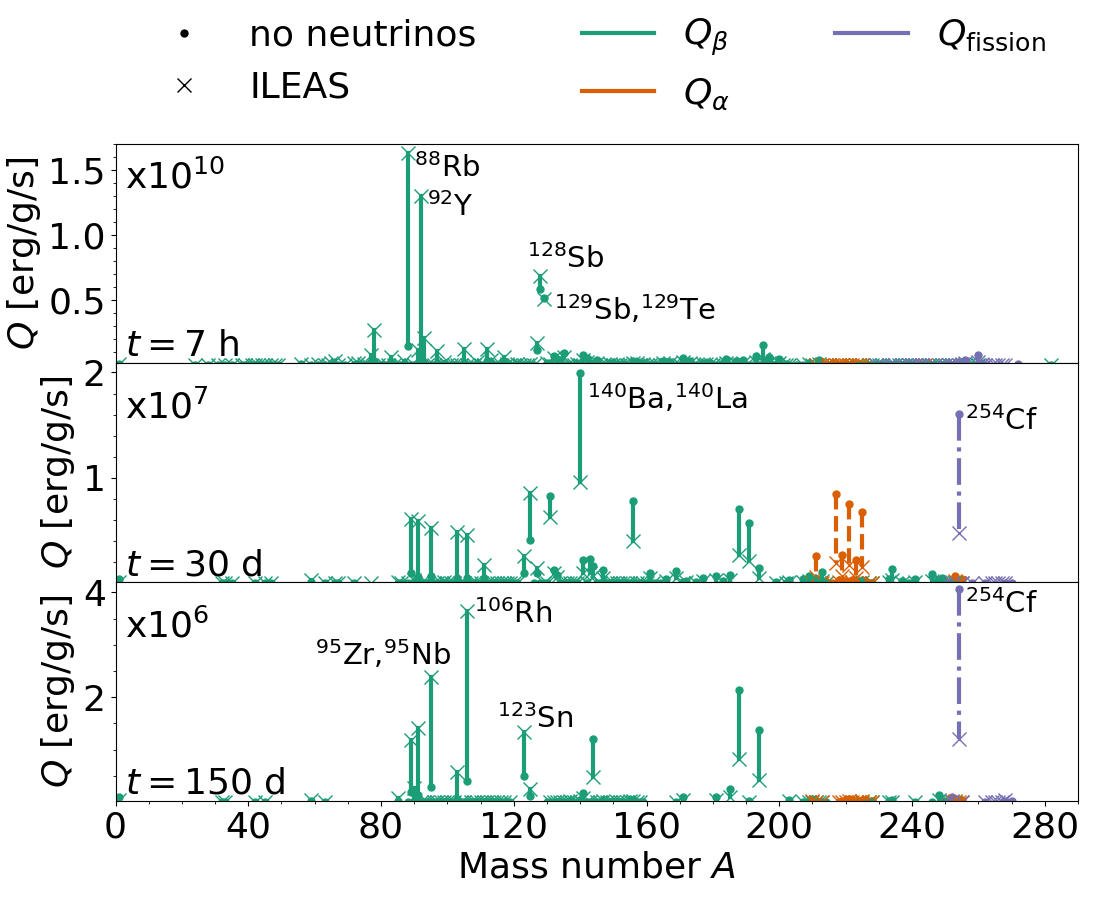}
  \caption[Contribution to the radioactive heating rate stemming from individual nuclear decays versus mass number  at $t=7$~h, 30~d and 150~d.]{
  Contribution to the radioactive heating rate $Q$ stemming from individual nuclear decays versus mass number $A$ at $t=7$~h, 30~d and 150~d in the top, middle and bottom panels, respectively. The no-neutrinos model is represented by a dot, the ILEAS model by a cross, and a line connects both predictions. The colours represent the decay modes as in \cref{fig_Q_t_cases}.  }
\label{fig_dQ}
\end{figure} 

The energy released by radioactive decay is estimated consistently within the same nucleosynthesis network for each of the models considered. No thermalization efficiency is included at this stage. Only the contribution from neutrino energy loss in each of the $\beta$-decays is removed, as described in \cref{sec:heat}, so that what we denote as radioactive heating rate $Q$ is the total energy per time per unit mass carried by $\alpha$, $\beta$, and $\gamma$ particles as well as fission products.  
\cref{fig_Q_t} shows the time evolution of the radioactive heating rate $Q$ for the four hydrodynamical models including weak interactions. 
As indicated by the arrow, a bump corresponding to the ejection of free neutrons that decay after $t=10$~min is visible. 
In particular, the models using the DD2 EoS have the largest ejection of free neutrons with a mass fraction of 0.7\% at late time $t\simeq 20$~s (see  \cref{tab:rpro_regions}). 
The absence of neutrino interactions would provide even larger quantities, typically a $\sim 30$\% increase of ejected free neutrons. 
For both SFHo models, the ejected amount of free neutrons is smaller, corresponding to 0.3\% and 0.5\% of the ejected material for the SFHo-125-145 and SFHo-135-135 models, respectively. 
In these cases, lower ejection velocities are found compared to the DD2 models, and therefore, a less noticeable feature linked to neutron decay is observed.
The decay of free neutrons may power a precursor signal of the kilonova,  as discussed in \citet{metzger2015}. The inclusion of neutrino interaction may, however, decrease its strength, though it is essentially linked to the high-speed ejection of a few neutron-rich mass elements.  

In general, the heating rate is dominated by a large number of neutron-rich $A \la 200$ nuclei that $\beta$-decay towards the valley of stability. Even after several days, the main contribution to the heating rate comes from a few $\beta$-unstable isotopes with half-lives on the order of days. However, at late times ($t>10$~d), the contribution from $\alpha$-decay and spontaneous fission of trans-Pb species starts to become significant (\cref{fig_Q_t_cases}). 
More specifically, the $\alpha$-decay chains starting from $^{222}$Rn, $^{223-224}$Ra and, in particular, $^{225}$Ac significantly contribute to the heating rate at time $t>10$~d. In addition, the $\alpha$-decay of $^{253}$Es and $^{255}$Es produces a substantial amount of heat, however,  several orders of magnitude less than the four aforementioned $\alpha$-decay chains since their daughter nuclei quickly $\beta$-decay to long-lived Cf. 
Among the fissioning species, $^{254}$Cf is the only notable isotope; in fact, it has one of the largest heating rates across all decay modes and isotopes for $t>10$~d. 
The production of trans-Pb is often taken as  the signature of r-processing beyond the heaviest stable elements \citep{wanajo2018,wu2019,zhu2018}.
Among our four hydrodynamical models, the asymmetric DD2 model shows the most pronounced contribution to the heating rate by trans-Pb species for  $t > 10$~d, as also shown by the large production of Th and U in \cref{fig_rpro_aa}. 

In \cref{fig_Q_t_cases}, the heating rate of our two models with and without neutrino interactions are compared, highlighting the relative contribution from the main decay modes $\beta$, $\alpha$ and fission.  In general, the time evolution of the heating rate due to $\alpha$-decay and spontaneous fission follow the same trend for both models, where the heating rate of the no-neutrinos model is up to one order of magnitude higher. For the total heating rate, the main differences occur around $t\sim7$~h, 30~d, and after 150~d (see \cref{fig_dQ} for the heating rate versus mass number at these three time points). 
At 7~h after the merger,  the discrepancies are caused by an excess of heat for the ILEAS model mainly generated by the $\beta$-decay of  $^{88}$Kr-$^{88}$Rb and $^{92}$Y (top panel \cref{fig_dQ}).  
This can be seen in \cref{fig_rpro_zz_cases} where an excess amount of elements in the $33 \le Z \le\ 39$ region is produced in the simulation including neutrinos, and is responsible for the large radioactive heat through $\beta$-decays at times between 0.1 and 1 day (\cref{fig_Q_t_cases}).  The same effect can be seen after $t>60$~d, where in the `no-neutrino' $\beta$-decay heating rate starts to decrease, while this does not occur within the ILEAS model due to heat generated by $A<130$ nuclei with longer half-lives. 
After $t>10$~d, the signature of trans-actinides production starts to differ significantly. The ILEAS model is still dominated by $\beta$-decays of $A \la 200$ species, however, for the no-neutrinos model, the contribution from the $A=222-225$ $\alpha$-decay chains and spontaneous fission of $^{254}$Cf becomes important.  In addition, the case without weak interactions has a larger production of elements with $A>130$ (see \cref{fig_rpro_zz_cases}), leading to more heat generated by $\beta$-decays, in particular by $^{140}$Ba and $^{140}$La  (see middle panel \cref{fig_dQ}).
Later at $t=150$~d (bottom panel \cref{fig_dQ}), the major trans-actinide contributor is $^{254}$Cf, with a heating rate about 4 times larger in the no-neutrinos model, while the $\beta$-decay of $A<130$ nuclides powers the heating rate in the ILEAS model.

\section{Sensitivity to the electron fraction}
\label{sect_sens}

\begin{figure}[tbp]
\includegraphics[width=\columnwidth]{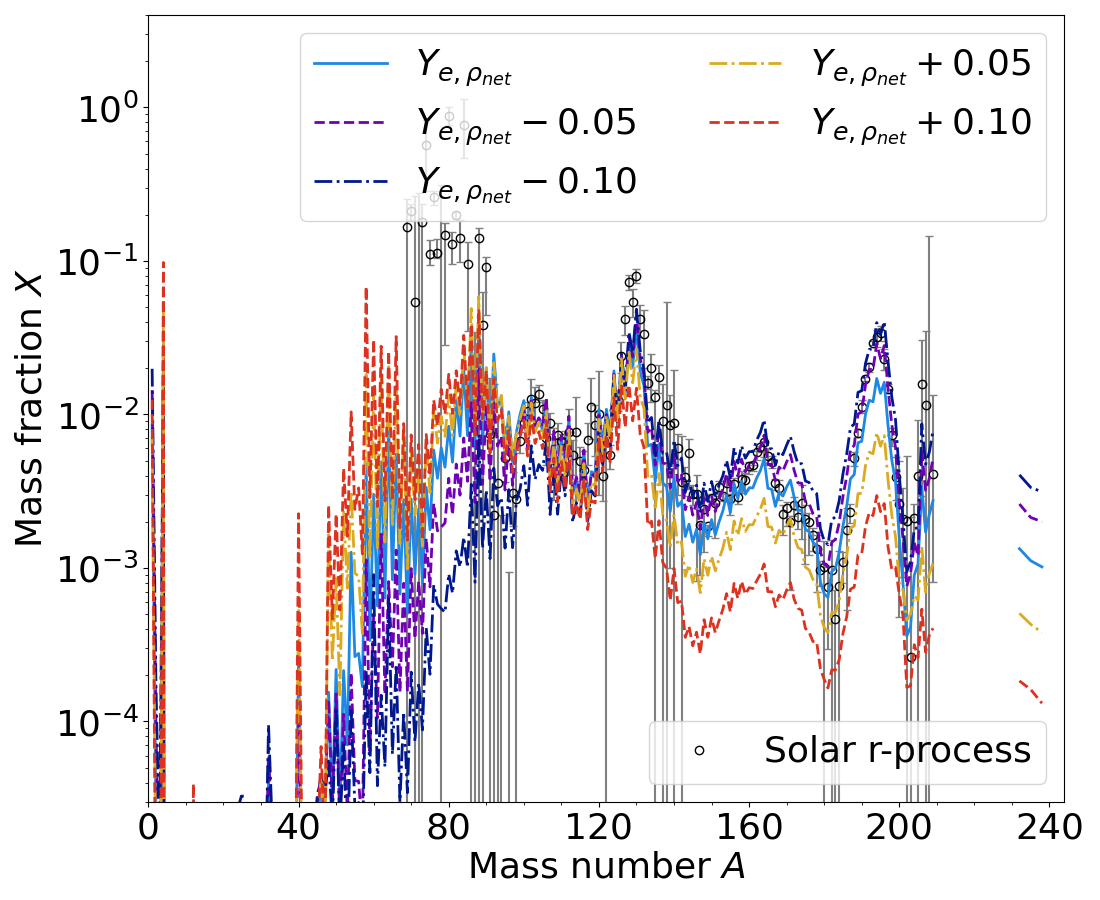}
  \caption[Final abundance distribution for the DD2-135-135 model assuming $Y_e\pm 0.05$ or $\pm 0.10$.]{
  Final abundance distribution in the dynamical ejecta of the DD2-135-135 model assuming the initial $Y_e$ is systematically modified with respect to ILEAS prediction by $\pm 0.05$ or $\pm 0.10$. The solar system abundance distribution is normalized as in \cref{fig_rpro_aa}.}
\label{fig_rpro_dye}
\end{figure}

\begin{figure}[tbp]
\includegraphics[width=\columnwidth]{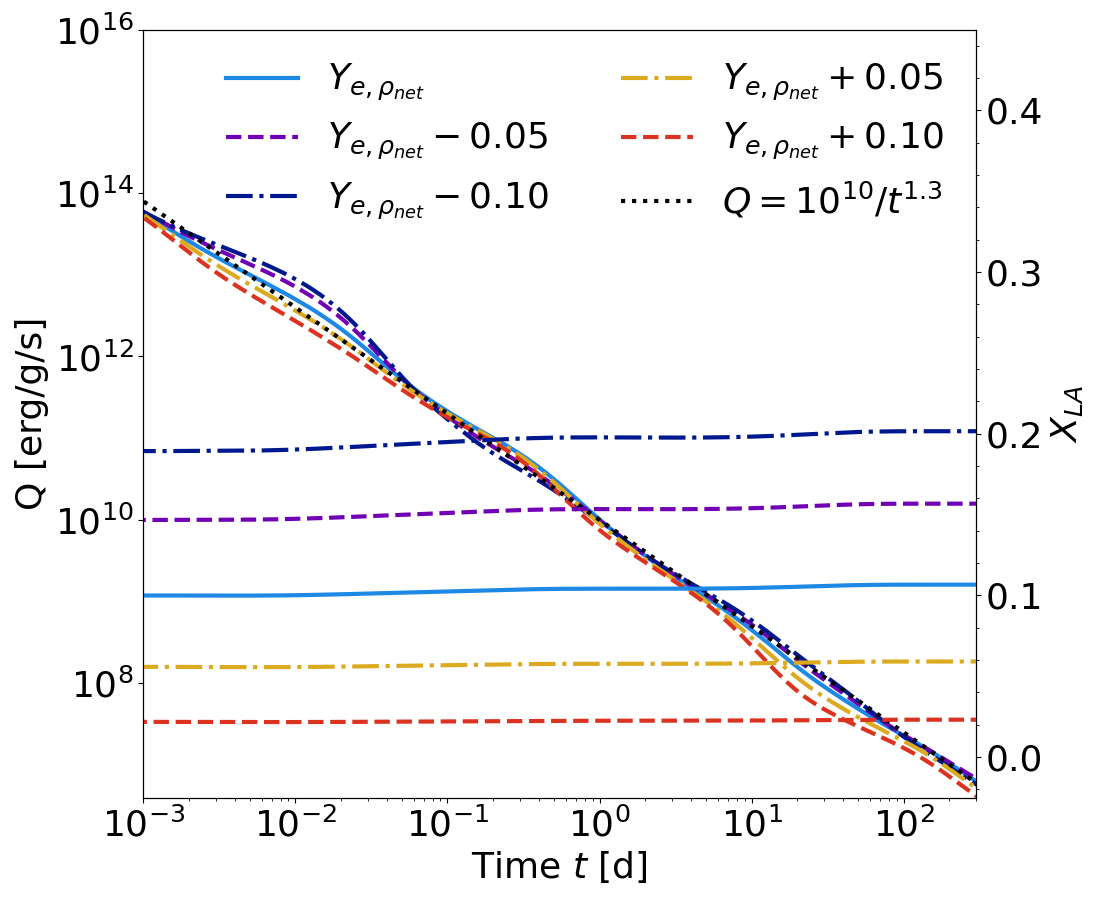}
  \caption[Time evolution of the radioactive heating rate and the lanthanide+actinide mass fraction assuming $Y_e\pm 0.05$ or $\pm 0.10$.]{
  Time evolution of the radioactive heating rate $Q$ and the lanthanide+actinide mass fraction $X_{LA}$ for the DD2-135-135 model assuming the initial $Y_e$ is systematically modified with respect to ILEAS prediction by $\pm 0.05$ or $\pm 0.10$. The black dotted line corresponds to the approximation $Q_0[{\rm erg/g/s}]=10^{10}~(t[{\rm d}])^{-1.3}$ (\cref{eq:Q_0}).}
\label{fig_Q_dye}
\end{figure}

As discussed in \cref{sec_intNS}, NS merger simulations still suffer from numerous uncertainties connected to, among others, the neutrino transport and the EoS. Those uncertainties inevitably affect the initial conditions for the nucleosynthesis calculations within the models described in \cref{subsec_dynsim}. In particular, the initial $Y_e$ at the time of $\rho=\rho_\mathrm{net}$ may differ by $\pm 0.05$ or even $\pm 0.10$ with respect to the present modelling. To test the sensitivity with respect to the many uncertainties affecting still the hydrodynamical simulations, we have considered extreme cases in which the  $Y_e$ at the time of $\rho=\rho_\mathrm{net}$ is systematically increased or decreased by 0.05 or even 0.10. We show in \cref{fig_rpro_dye,fig_Q_dye} the impact of such different initial conditions on the final composition of the ejected material as well as the time evolution of the radioactive heating rate and the lanthanide+actinide mass fraction $X_{LA}$ for the DD2-135-135 model.

As seen in \cref{fig_rpro_dye}, a systematic increase of $Y_e$ by 0.1 is not sufficient to give rise to the production of the first r-process peak but instead favours the synthesis of the Fe, Ni or Zn  even-$N$ isotopes in the $A=58-66$ region, as well as $^{88}$Sr and $^{84,86}$Kr. Consequently, a significant reduction in the production of the second and third r-process peaks, as well as lanthanides and actinides, can be observed (\cref{fig_Q_dye}). 
Note, however, that even in the least neutron-rich case where $Y_e$ is increased by 0.10, the mass fraction of $X_{LA}\simeq 0.02$ remains not negligible.
The radioactive decay heat $Q$ is found to follow rather well the empirical approximation $Q_0[{\rm erg/g/s}]=10^{10}~t[{\rm d}]^{-1.3}$ (e.g., \cref{eq:Q_0}) in all the five cases shown in \cref{fig_Q_dye}.  The differences seen in the various $Y_e$ cases in \cref{fig_Q_dye} can be explained by the same effects as those found in the  comparison between the cases with and without weak interactions described in \cref{sec_q}.
In the most neutron-rich case ($Y_e-0.10$), the $Q$ enhancement around $t=10^{-2}$~d and $t>10$~d is due to the decay of free neutrons and actinides, respectively. 
For the least neutron-rich case, the trans-actinides signal at $t>10$~d completely disappears, and the heating rate is essentially powered by $\beta$-decay of $A\leq 140$ nuclei, except for the spontaneous fission of $^{254}$Cf. 

The highest $Y_e+0.10$ case with the lowest $X_{LA}$ gives rise to a lower heating rate $Q$ with respect to the most neutron-rich case, in contrast to what is obtained by \citet{martin2015} (see their Fig.13) who found,  relative to $Q_0$, a heating rate enhancement by a factor of 2.5 at $t\sim 4$~h for their neutrino-driven wind producing the light r-process elements. Our heating rate is also significantly lower than the value found by \citet{perego2017} (see in particular their Eq. 2, and \citet{just2021} for a detailed comparison with their heating rates) for $Y_e \ga 0.25$ ejecta. 

\section[Angular and velocity dependence of the ejecta][Angular and velocity dependence]{Angular and velocity dependence of the ejecta}
\label{sec_ang_vc}

\begin{figure}[tbp]
\includegraphics[width=\columnwidth]{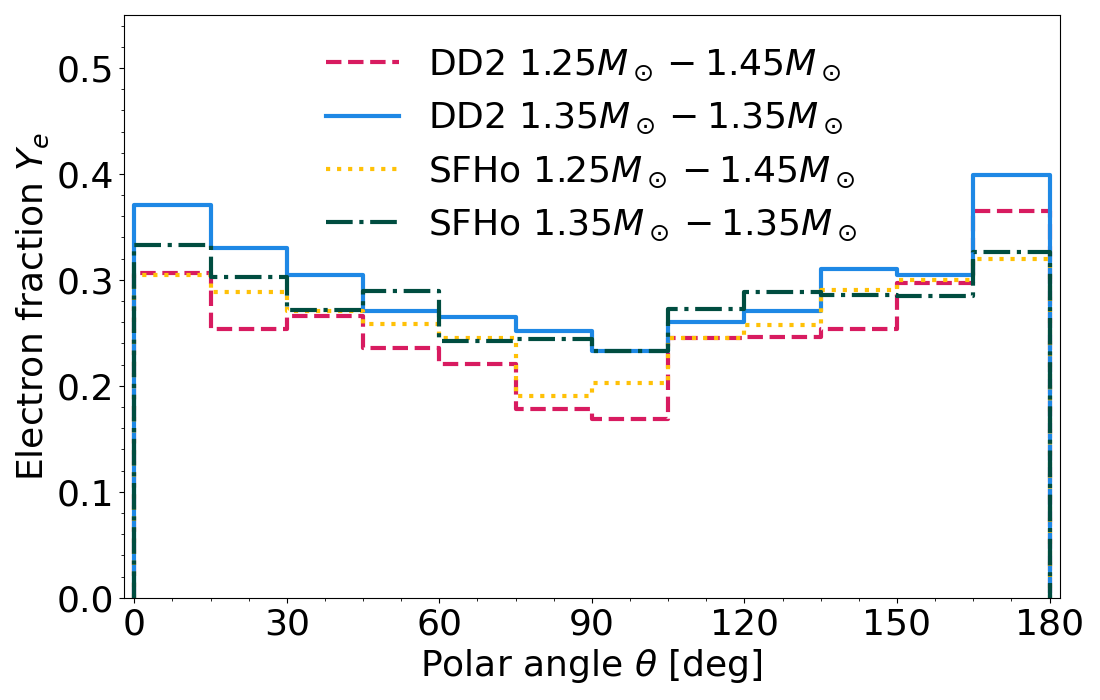}
  \caption[Average electron fraction per polar angle bin for the four NS-NS merger models. ]{
  Average electron fraction per polar angle bin $d\theta$
   at $\rho=\rho_\mathrm{net}$ 
  as predicted by the ILEAS hydrodynamical simulation for the four NS-NS merger models: 
  DD2-125-145, DD2-135-135, SFHo-125-145 and SFHo-135-135. 
  }
\label{fig_yedist_ang}
\end{figure} 
\begin{figure}[tbp]
\includegraphics[width=\columnwidth]{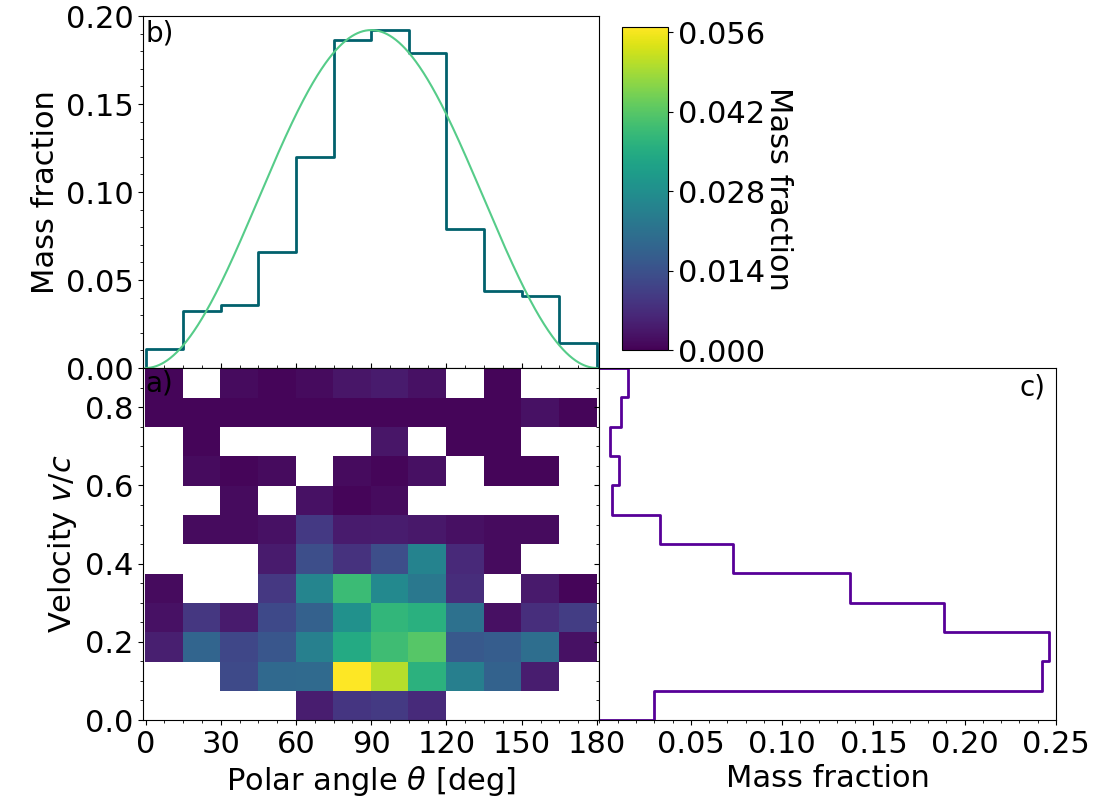}
  \caption[Mass distribution of the ejecta in the two-dimensional plane of the polar 
  angle versus velocity at the end of the DD2-135-135 NS-NS hydrodynamical simulation.]{
  Mass distribution of the ejecta a) in the two-dimensional plane of the polar 
  angle versus velocity at the end of the DD2-135-135 NS-NS hydrodynamical simulation. b) and c) panels represent the one-dimensional projections of the integrated ejected masses on the $\theta$ and $v/c$ axes, respectively. In panel a), the white
regions depict the absence of any  trajectory. The green line in panel b) corresponds to the $\rm{sin}^2(\theta)$ function, arbitrarily normalized.
  }
\label{fig_dM2d}
\end{figure} 
%

The three-dimensional implementation of the weak nucleonic reactions (\cref{eq:betareac}) within the ILEAS framework of the hydrodynamical simulations allows us to study the angle and velocity distributions of the ejecta and nucleosynthesis yields. 
As discussed in \cref{subsec_eject} and also shown by previous studies \citep[see, e.g.,][]{goriely2011,sekiguchi2015}, two major mechanisms during the merger phase are responsible for the dynamical mass ejection, namely tidal stripping preferentially in the orbital plane and the more isotropic mass ejection by shock compression at the NS contact interface.
Most high-$Y_e$ mass elements originate from the collision interface (blue, light blue and white dots in \cref{fig_snapshot}). 
In the equatorial plane, mostly lower-$Y_e$ material is ejected (red and purple dots in \cref{fig_snapshot}), whereas regions at higher polar angles contain a mix of low-$Y_e$ and high-$Y_e$ ejecta.
Indeed, \cref{fig_yedist_ang} displays the electron fraction distribution of the ejected matter as a function of polar angle $\theta$ for the four merger models. It can be seen that the material at the poles is significantly less neutron-rich, in particular for the DD2-125-145, SFHo-125-145 and DD2-135-135 models. It is consequently of particular interest for observation (see \citet{just2021} and \cref{sec_kilonova}) to study in more detail the composition and decay heating rate obtained as a function of the polar angle, but also as a function of the expansion velocity. Some ejecta properties in the different angular regions, namely the polar, middle or equatorial regions, are given in \cref{tab:rpro_regions}.
The following section presents a detailed analysis of the angle and velocity distribution of the ejected material and the resulting nucleosynthesis for the DD2-135-135 model.
If the other dynamical models deviate from its trend, a special mention will be made.

\cref{fig_dM2d}a illustrates the mass distribution of the ejecta in the two-dimensional plane of the polar angle $\theta$ and  velocity $v/c$.  Panels b) and c) of \cref{fig_dM2d} give the corresponding projection of the integrated mass fraction on the $\theta$ and $v/c$ axes, respectively.  \cref{fig_dM2d}b shows that most of the material is ejected in the equatorial region ($60^\circ < \theta \leq 120^\circ$) of the NSs merger plane. The angular distribution is seen to be roughly following a $\rm{sin}^2(\theta)$ function, as pointed out by \citet{perego2017}. However, compared to the $\rm{sin}^2(\theta)$ distribution, the simulations tend to produce a pattern that is slightly more peaked towards $\theta=90^\circ$.

In \cref{fig_dM2d}c, we see that most of the ejected material has a velocity below $0.3c$ and that the high-velocity tail ($v/c>0.6$) contains only a minor fraction of the total ejected mass (between 1-5\% of the total mass, see also \cref{tab:rpro_regions}) and is represented by 63 (22) and 37 (13) particles for the DD2 and SFHo asymmetric (symmetric) models, respectively. These fast-moving particles contain over 99\% of the $X_n^{t=20\mathrm{s}}$ mass.
The high-velocity ejecta is not isotropic but more concentrated in the equatorial plane, though they are distributed over all angles, especially for the DD2 EoS models (\cref{fig_dM2d}a).

\begin{figure}[tbp]
\includegraphics[width=\columnwidth]{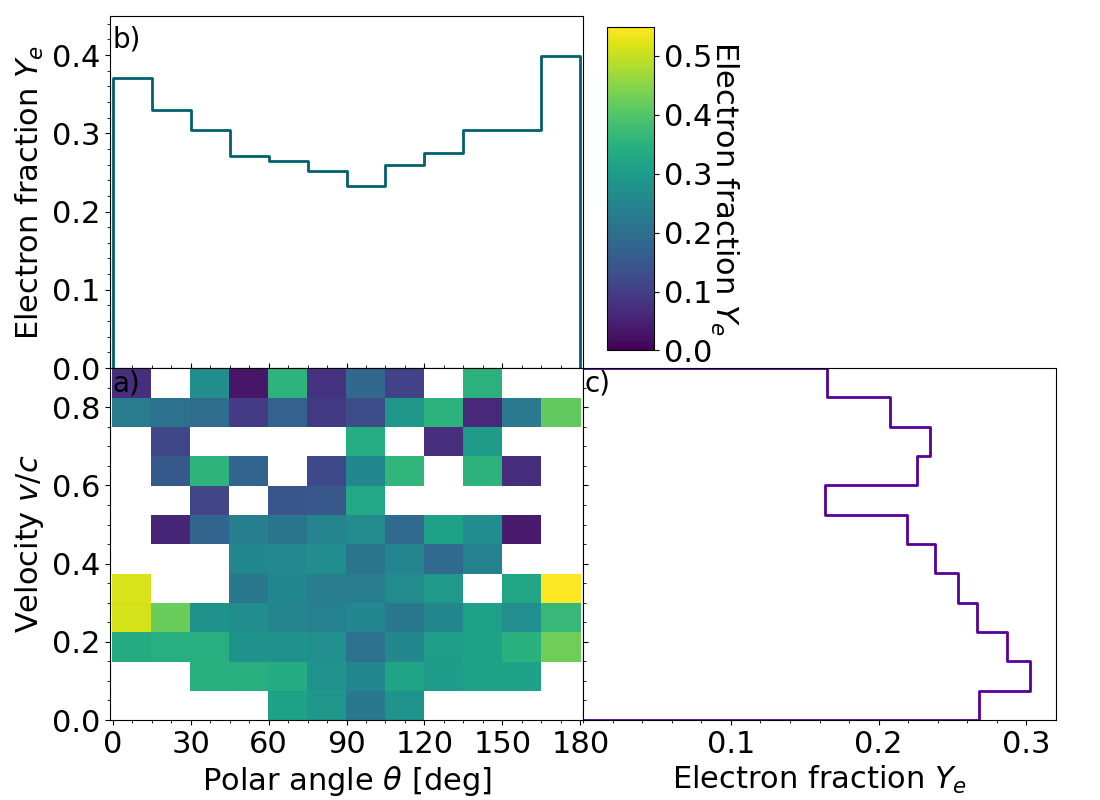}
  \caption[Same as \cref{fig_dM2d} for the electron fraction distribution.]{
  Same as \cref{fig_dM2d} for the electron fraction distribution at $\rho=\rho_\mathrm{net}$ 
  as predicted by the ILEAS hydrodynamical simulation. b) and c) panels represent the mass-averaged $Y_e$ along the $\theta$ and $v/c$ axes, respectively.
  }
\label{fig_ye2d}
\end{figure} 
\begin{figure}[tbp]
\includegraphics[width=\columnwidth]{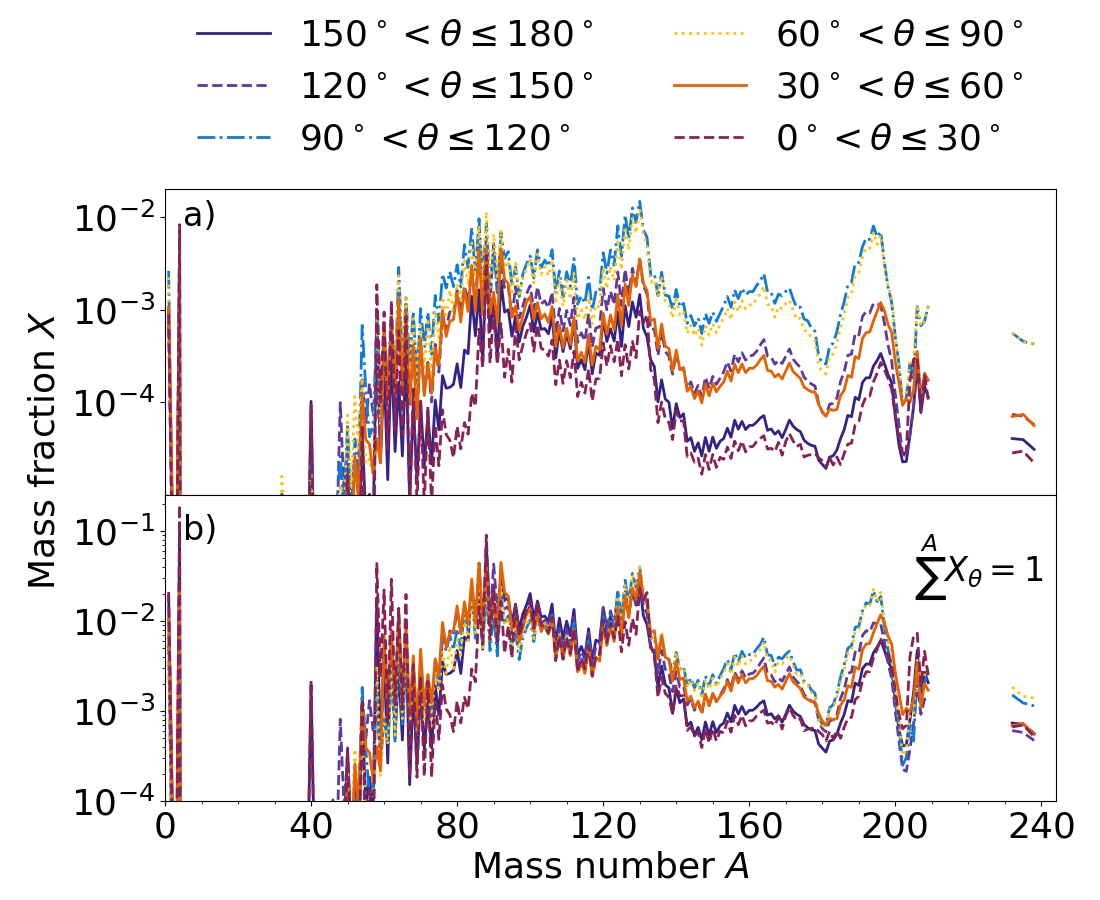}
  \caption[Final mass fractions of the material ejected as a function of the atomic mass at six angle bins for the DD2-135-135 model. ]{
  Final mass fractions of the material ejected as a function of the atomic mass $A$
  divided into six angle bins for the DD2-135-135 model.  a) Shows the ejected mass fraction $X_\theta$ 
   relative to the total ejected mass, while in b) the mass fractions are renormalized so that the sum over the atomic
     mass numbers is equal to one for each angle bin.}
\label{fig_rpro_aa_angs}
\end{figure} 

\begin{figure}[tbp]
\includegraphics[width=\columnwidth]{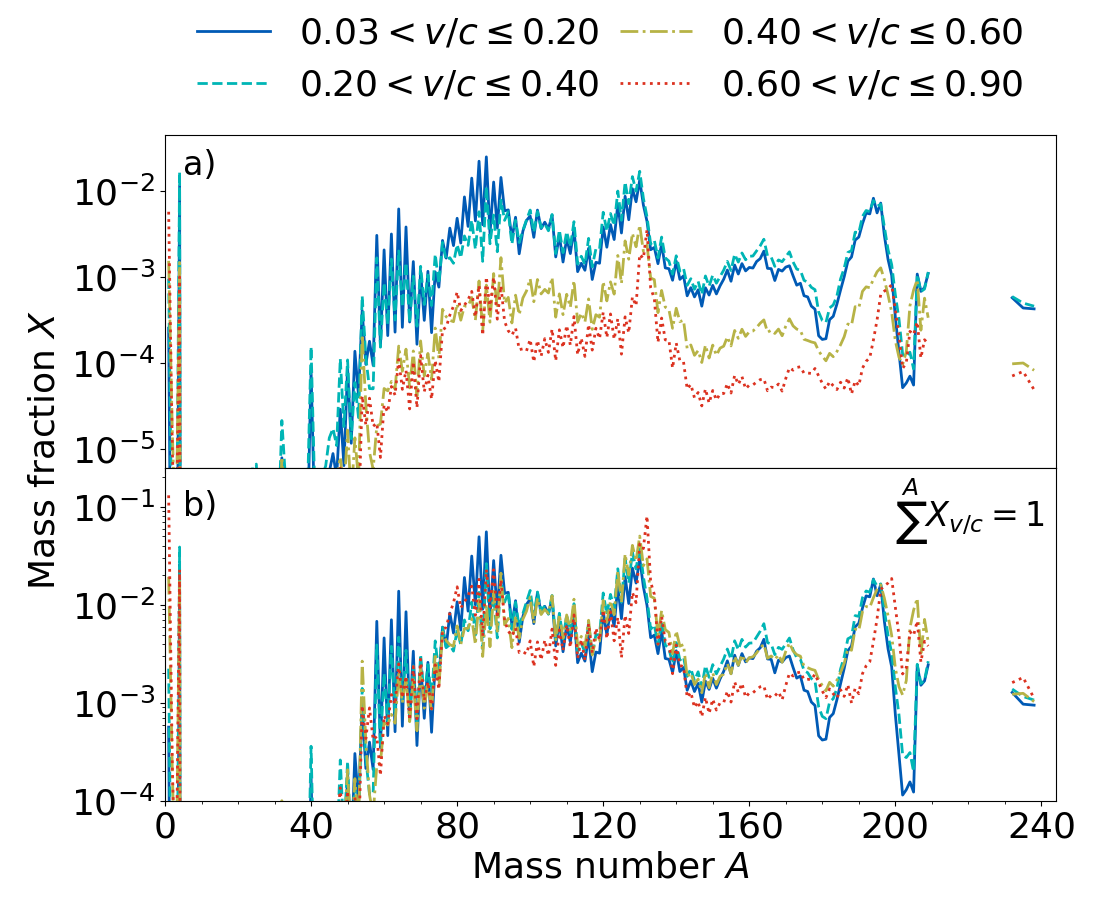}
  \caption{
  Same as \cref{fig_rpro_aa_angs} for four velocity bins. }
\label{fig_rpro_aa_vc}
\end{figure} 

\begin{figure}[tbp]
\includegraphics[width=\columnwidth]{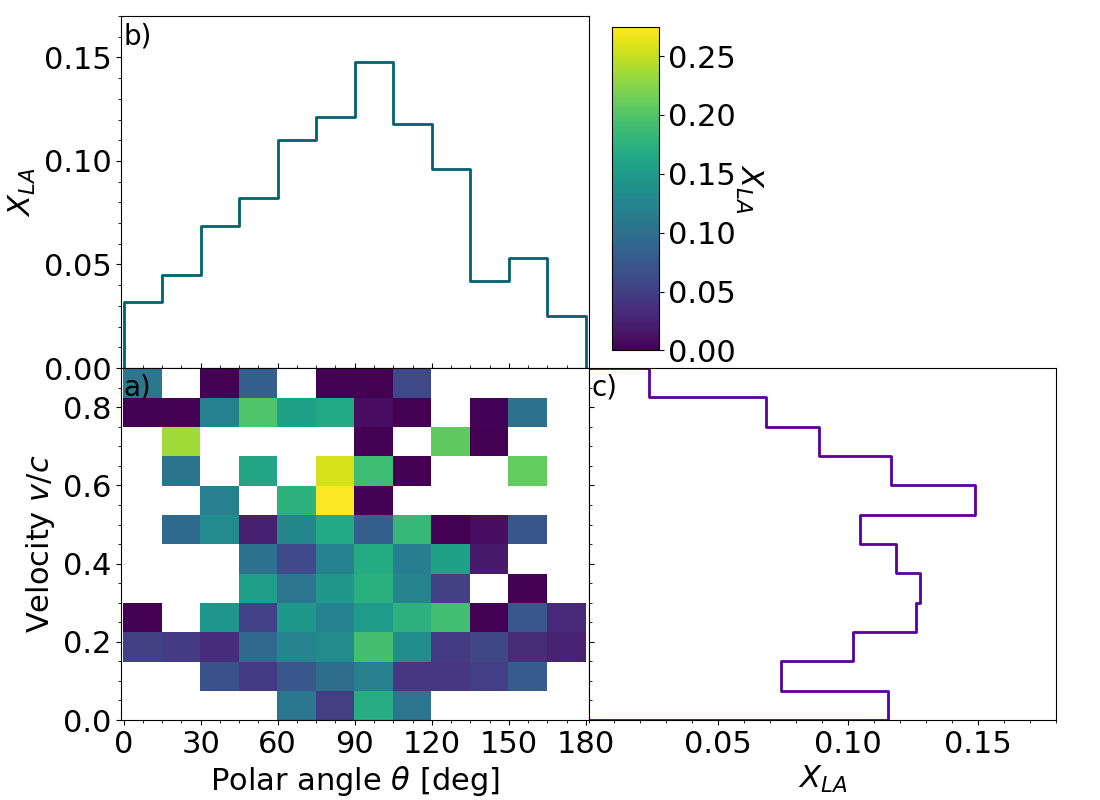}
  \caption[Same as \cref{fig_dM2d} for the ejected mass fraction of lanthanides and actinides.]{
  Same as \cref{fig_dM2d} for the ejected mass fraction of lanthanides and actinides $X_{LA}$. 
  }
\label{fig_Xla2d}
\end{figure} 
\begin{figure}[tbp]
\includegraphics[width=\columnwidth]{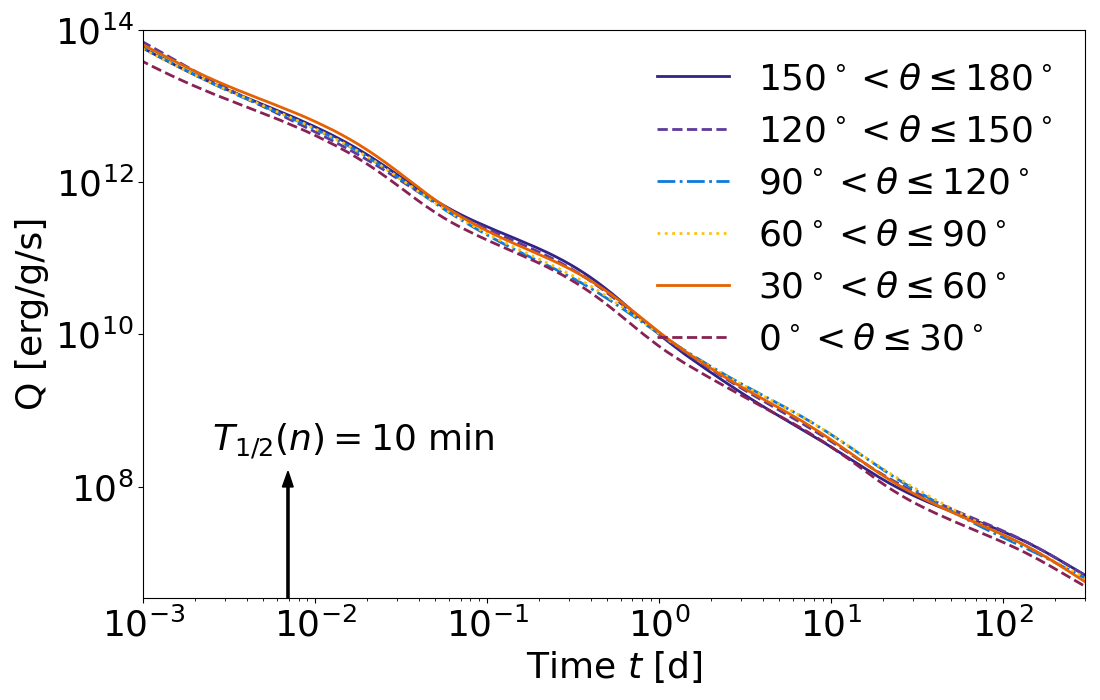}
  \caption{
  Same as \cref{fig_Q_t} for the heating rate for the DD2-135-135 model in the six angle bins.   }
\label{fig_Q_t_angles}
\end{figure} 

\begin{figure}[tbp]
\includegraphics[width=\columnwidth]{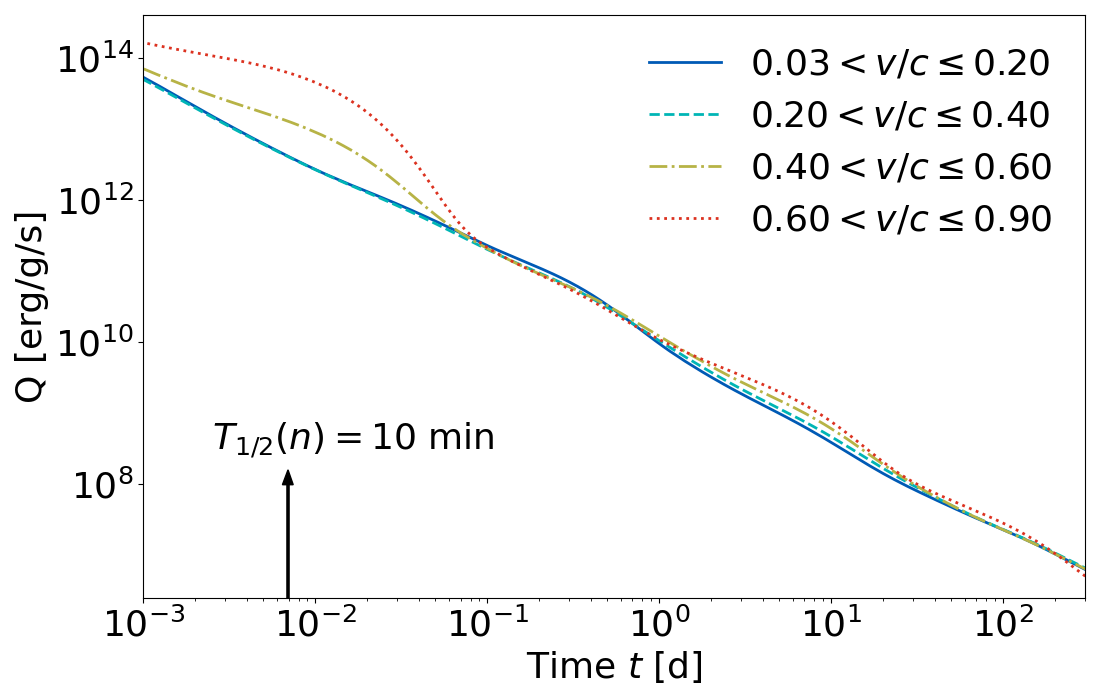}
  \caption{
  Same as \cref{fig_Q_t_angles} for the four velocity bins of the DD2-135-135 model.  }
\label{fig_Q_t_vc}
\end{figure} 


In general, the further away from the equatorial plane the mass element, the larger its minimum velocity. In particular,  the minimum velocity in the extreme polar regions ($\theta \simeq 0^\circ$ and $\theta \simeq 180^\circ$) is larger than $0.2c$, while $v/c\simeq 0.03$ is the lower limit in the equatorial plane.
\cref{fig_ye2d}a shows the electron fraction distribution in the $\theta-v/c$ plane, and panel b) and c) of \cref{fig_ye2d} show the projection of the average $Y_e$ on the $\theta$ and $v/c$ axes, respectively. The ejecta in the equatorial regions is the most neutron-rich.  As seen in \cref{fig_ye2d}c, most of the high-$Y_e$ mass elements have velocities in the range $\sim 0.1-0.4c$.  

In \cref{fig_rpro_aa_angs}, the ejecta composition in the polar,  middle and equatorial regions are shown as a function of mass number for the DD2-135-135 model. \cref{fig_rpro_aa_angs}a emphasizes the differences in ejected yields in the separate angle bins, with a particularly small contribution from the polar regions. \cref{fig_rpro_aa_angs}b compares the nucleosynthesis results of the angle regions renormalized such that in that region $\sum_A X_A=1$, showing that the production of $A>140$ nuclei is proportionally smaller in the polar regions compared to the other angle regions.
As seen in \cref{fig_yedist_ang}, the material in the middle and polar regions has a significantly higher $Y_e$ and is therefore expected to yield a weaker r-process.  However, since the material ejected in both the polar  and middle regions is less massive (see \cref{tab:astromods}),  the global effect of the high-$Y_e$ trajectories remains small. In the case of the symmetric SFHo system, a more isotropic distribution of $Y_e$ is found (\cref{fig_yedist_ang}) so that a somewhat similar pattern is observed in all directions. In this case, like in the others, the mass ejected along the poles remains relatively low, i.e., smaller than typically 10\% of the total ejected mass (\cref{tab:rpro_regions}).

The velocity dependence of the isotopic composition  is given in \cref{fig_rpro_aa_vc} where panel a) shows the yields relative to the total ejecta mass, and b)  the renormalized mass fractions, so that the sum over the atomic mass for each velocity bin is 1.  The mass elements with the highest velocities ($v\ge 0.6c$) have a small contribution to the overall production of all isotopes.  Comparing only the shape of the abundance distributions in \cref{fig_rpro_aa_vc}b, those $v/c>0.6$ trajectories have their second and third r-process peaks shifted to higher mass numbers and relatively low content in lanthanides with respect to the slow ejecta. This specific distribution is related to the fact that for the fast ejecta, not all neutrons are captured, and the final composition results from the competition between neutron captures and $\beta$-decays during the fast expansion \citep{goriely2016b}.

The distribution of ejected lanthanides and actinides, $X_{LA}$,  in the $\theta-v/c$ plane is presented in \cref{fig_Xla2d}a for the DD2-135-135 model, where b) and c) show the projection of the average $X_{LA}$ fraction on the $\theta$- and $v/c$ axes, respectively. Both \cref{fig_rpro_aa_angs}a and \cref{fig_Xla2d}b show that the production of lanthanides and actinides are the largest in the equatorial plane (see also \cref{tab:rpro_regions}). However, as displayed in \cref{fig_Xla2d}c, the production of $X_{LA}$ does not seem to have a large dependence on the velocity for $v/c<0.6$ while the production falls rapidly for the fastest escaping mass elements, as mentioned above. 

The time evolution of the heating rate in six different angle bins is displayed in \cref{fig_Q_t_angles} for the DD2-135-135 model. The shape of the curves follows more or less the same $t^{-1/3}$ dependence (e.g., see \cref{eq:Q_0}) and are rather similar for the different viewing angles. Some variations are found for the polar directions due to a relatively different composition, as shown in \cref{fig_rpro_aa_angs}.
On the contrary,  the velocity dependence of the heating rate displayed in \cref{fig_Q_t_vc} differs for the highest escape velocities, especially at specific times.  In particular, for the high-velocity ejecta ($v/c>0.6$), the decay of free neutrons at $t\sim 10$~min creates a large bump in the heating rate, and the decay of largely produced $^{132}$Te and $^{132}$I (see \cref{fig_rpro_aa_vc}) is responsible for the smaller bump at $t\simeq 10$~d.
In \cref{fig_rpro_aa_vc}b, a relative enhancement in the actinide production is also observed for the largest velocity bin ($v/c>0.6$), compared to the low velocities, which leads to an increase in the heating rate for $t \ga 30$~d due to the decay of $^{254}$Cf. 
To a smaller extent, the same features as for the high-velocity ejecta can be seen in the second largest velocity bin ($0.4 < v/c<0.6$) and are explained by the same effects. A somewhat similar evolution of the heating rate is found for the SFHo-135-135 model and the asymmetric models. We note that the high-velocity tail is discussed as a possible source of the late-time synchrotron emission observed in GW170817 \citep{hajela2021,nedora2021}.

\section{Kilonova}
\label{sec_kilonova}

\begin{figure}[pbt]
\centering
\includegraphics[width=\textwidth]{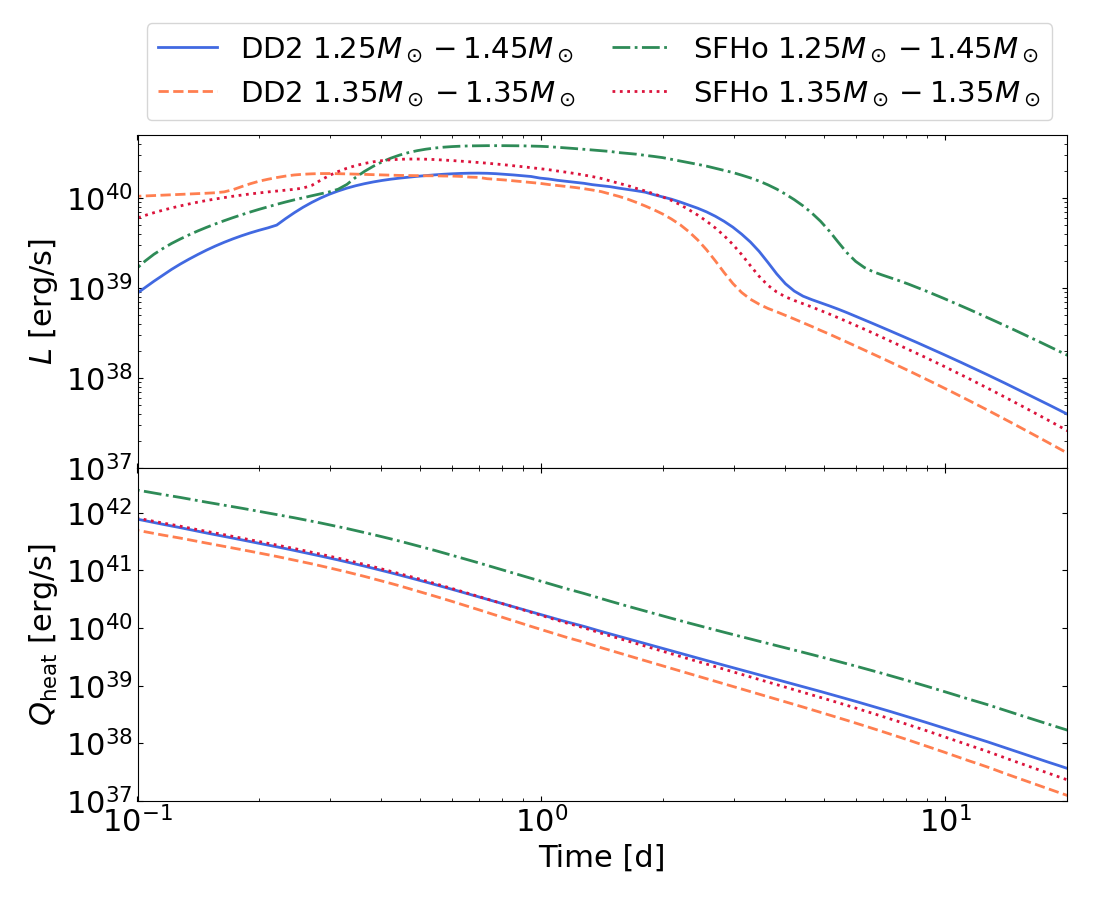}
  \caption[Some kilonova properties based on the nucleosynthesis yields of the four NS-NS merger systems presented in \cref{sec_nuc}.]{
  Some kilonova properties based on the nucleosynthesis yields of the four NS-NS merger systems presented in \cref{sec_nuc}: the bolometric luminosity $L$ (top) and thermalized heating rate $Q_\mathrm{heat}$ (bottom). The light curves and thermalized heating rates shown here are identical to those shown in the top right and middle panels of Fig.~7 in \citet{just2021}, respectively.
  }
\label{fig_kilo}
\end{figure} 
\begin{figure}[pbt]
\centering
\includegraphics[width=\textwidth]{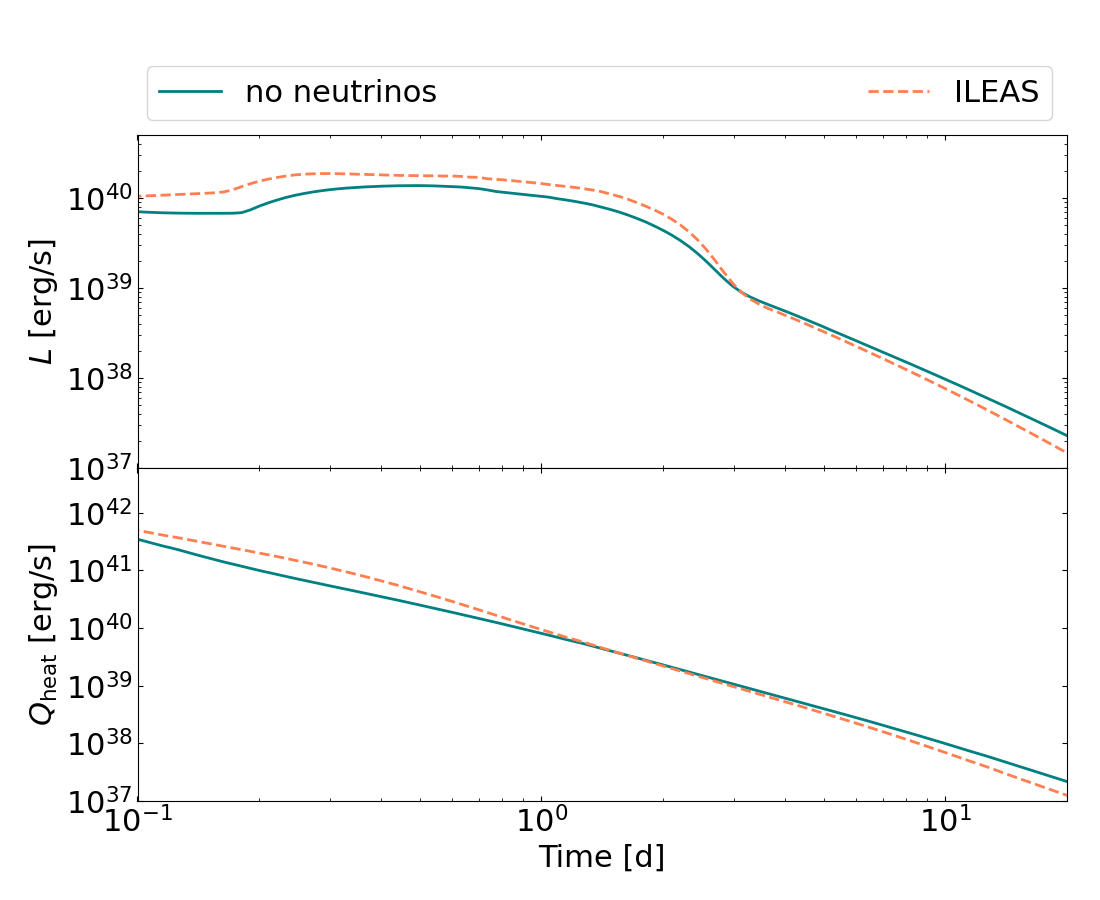}
  \caption{
 Similar to \cref{fig_kilo} for the `no neutrino' case compared to the original ILEAS calculation.}
\label{fig_kilnonu}
\end{figure} 

The kilonova light curves obtained using the model discussed in \cref{sec_kilo} \citep[see also][]{just2021} and based on the r-process calculations presented in \cref{sec_nuc,sec_q} are shown in the top panel of \cref{fig_kilo}, and the thermalized heating rate (including the heat carried away by neutrinos) $Q_\mathrm{heat}$ in the lower panel.
The kilonova typically peaks after 0.7 to 1.5~d with luminosities between $3-7\cdot 10^{40}$~erg/s. 
Since the composition, i.e., $X_\mathrm{LA}$, is rather similar for the four regular models, the observed differences of peak times and luminosities between these models are a result mainly of the different ejecta masses (see \cref{tab:rpro_regions}).
A softer EoS or higher binary-mass asymmetry lead to a longer and brighter signal.

Neglecting all neutrino interactions after both NSs first touched for the symmetric model with DD2 EoS reduces the average electron fraction from 0.27 to 0.13 and increases the lanthanide fractions by a factor of about 2 (see \cref{tab:rpro_regions}). The `no neutrino' case also moves the radioactive heating power effectively from times $t<1$~d to $t>1$~d (\cref{fig_Q_t_cases}). The main consequences for the kilonova shown in \cref{fig_kilnonu} is an extended peak (by $\sim15$ per cent) and a luminosity reduction (by $\sim40$ per cent). These changes are comparable to the changes obtained when reducing $Y_e$ by 0.1 globally for all trajectories as done in \cref{sect_sens}; see the corresponding light curves in \citet{just2021}. Many more details on the kilonova properties, including, in particular, the light curves, can be found in \citet{just2021}.

\section{Comparison to other studies}
\label{sect_comp}

In recent years a variety of publications have focused on the dynamically ejected material from NS-NS mergers \citep[e.g.,][]{wanajo2014, Palenzuela2015,lehner2016,sekiguchi2015,foucart2016,foucart2020,bovard2017,radice2018a,nedora2021a,martin2018}. However, only a subset of the simulations available in the literature includes neutrino absorption and in addition performs r-process nucleosynthesis network calculations. In the following, we highlight the most relevant differences between our simulations and those of others that also use approximately the same NS mass configurations and EoSs as we do.

The works of \citet{radice2018a} and more recently \citet{nedora2021a} perform fully GR hydrodynamical calculations with neutrino absorption.
Their calculations are performed with the grid-based, adaptive mesh code WhiskyTHC \citep{radice2014} in contrast to our simulations based on the SPH code \citep{oechslin2007,Bauswein2013} described in \cref{subsec_dynsim}. 
In the optically thick regions, they also implement a leakage scheme for the weak interactions, but, in contrast to our ILEAS scheme, without ensuring that the neutrino fluxes obey the diffusion-law and without conserving the lepton number in neutrino-trapping regions. In order to describe net neutrino absorption in optically thin regions, they evolve free-streaming neutrinos using a one-moment (``M0'') scheme, whereas our ILEAS scheme employs a ray-tracing algorithm. Lacking, to our knowledge, comparisons with reference solutions or with other approximate schemes, we are unable to assess the accuracy of their M0 scheme. We note that while ILEAS has not been benchmarked by Boltzmann solutions in a binary NS merger environment, it showed excellent agreement with accurate neutrino transport solutions for the case of a proto-neutron star formed in a core-collapse supernova \citep{ardevol-pulpillo2019}.

For the r-process nucleosynthesis, \citet{radice2018a} and \citet{nedora2021a} use a mapping technique to calculate the ejecta composition, i.e., the expansion history of every trajectory is parametrized depending on its density, entropy, electron fraction and velocity at a fixed instant of time (when crossing the sphere of 443~km). On this basis, the composition is deduced from the corresponding pre-computed r-process calculation of \citet{lippuner2015}.  In our simulations, we perform one r-process calculation per unbound SPH mass element, following its detailed expansion history. 
A subset of models by \citet{radice2018a}, as well as \citet{nedora2021a}, employs general-relativistic large eddy simulations (GRLES) calibrated to mimic viscosity due to magnetohydrodynamic turbulence. Since the GRLES simulations of \citet{radice2018a} do not include neutrino absorption, we only include the calculations without GRLES in the following comparison.  
Note that some complementary information about models discussed in \citet{radice2018a} is provided in  \citet{perego2017}.

Numerical simulations in full general relativity have also been performed by \citet{sekiguchi2015,sekiguchi2016} with a leakage-plus-M1 scheme for neutrino cooling and absorption using finite-volume methods on a nested grid. The corresponding nucleosynthesis calculations are reported in \citet{wanajo2014}. 
When calculating the r-process composition, \citet{wanajo2014} only include ejecta from the orbital plane and choose a representative trajectory from each $Y_e$-bin from 0.09 to 0.44, giving a total of 35 trajectories.

The recent study by \citet{foucart2020} applied a sophisticated Monte-Carlo scheme for neutrino transport to directly solve the transport equations in low-density regions for a 1.27 -- 1.58~\Msun\ NS-NS merger system. The simulations ran for 5~ms on a Cartesian grid, using high-order finite volume shock-capturing methods with the GR SpEC code, and did not include nucleosynthesis calculations. We note here that the total mass of the binary system in \citet{foucart2020} is 0.15~\Msun\ larger than the total mass of the models discussed above, and in addition, a different mass configuration was used. 

As is common in grid-based codes, the NSs in the simulations of \citet{radice2018a,nedora2021a,sekiguchi2015,foucart2020} are placed in a low-density artificial atmosphere. If not checked carefully,  the atmosphere material may suppress mass ejection or decelerate parts of the ejecta \citep{sekiguchi2016}. This problem is not present in SPH-based simulations where no artificial atmosphere needs to be used. In general, it is assuring that very different numerical schemes, i.e., particle-based and grid-based hydrodynamics methods, yield ejecta properties in the same ballpark noting that, in addition, we employ the conformal flatness approximation, whereas the aforementioned studies consider a fully relativistic evolution. Note that the differences due to the use of the conformal flatness condition are likely to be small since calculations employing this approximation reproduce post-merger gravitational-wave frequencies of full-general relativity simulations quantitatively very well, implying that our simulations capture well the gravitational field and dynamics of the bulk matter \citep[e.g.,][]{bauswein2012b}. Also, compared to the range of literature data on, for instance, total ejecta masses, our simulations do not seem to feature some systematic offset from fully relativistic simulations.
We emphasize that direct quantitative comparisons with other works remain difficult and should be taken with a grain of salt, as discussed below, in particular when dealing with the total ejected mass. 

We first consider the ejecta masses summarized in \cref{tab:comp}. 
When comparing our results to previously published studies, it should be kept in mind, as already discussed in \cref{subsec_dynsim}, that estimating the exact amount of ejected mass remains a difficult task since it depends on the extraction time as well as on the criterion adopted to count gravitationally unbound material and the radius or volume at which this criterion is evaluated. While we use the $\varepsilon_{\rm stationary}>0$ criterion (see \cref{subsec_dynsim}), \citet{radice2018a} and  \citet{sekiguchi2015} use a more restrictive so-called geodesic criterion which defines a mass element as unbound if its kinetic energy only is larger than the gravitational one (or, more precisely, if the time component of the fluid four velocity $u_t$ is smaller than $-1$). 
In this case, we expect the ejecta to be less massive than determined by applying our criterion.  
The way this criterion is applied also differs between published works. 
While we apply it at a given time $t_{\rm ej}$ in a given volume (with $r> r_{\rm ej}=100$~km), \citet{radice2018a} apply the criterion at a sphere of a given radius (443~km) and time integrates the fluxes at this radius to obtain the ejecta mass. 
\citet{sekiguchi2015,sekiguchi2016} followed our approach, except that
they use $r_{\rm ej}=0$\footnote{One can see in \cref{fig_mejec}, however, that our ejecta masses would be very similar if we had used $r_{\rm ej}=0$.} and ran their simulations for a longer time.
\citet{foucart2020} apply a specific version of the Bernoulli criterion to identify unbound ejecta, namely version B of \citet{Foucart2021}, where a variety of ejection criteria are compared and discussed. 
(Note that most merger simulations ---in line with our work--- have not followed the potential ejecta to very large distances, in which case their gravitationally unbound state could not be unambiguously diagnosed.)  For all these reasons, the comparison of the quoted ejected masses should be taken with care.

\begin{table}
\centering
\caption[Comparison of dynamical ejecta properties with other studies. ]{Summary of the compared ejecta properties from \citet{radice2018a}, \citet{nedora2021a}, \citet{sekiguchi2015}, \citet{sekiguchi2016}, \citet{foucart2020}, abbreviated as R18, N21, S15, S16, F20, respectively, and this work. All systems have the same total mass $M_1+M_2=2.7$~\Msun, except for F20 where it is 0.15~\Msun\ larger. The symmetric mergers have $q=M_1/M_2=1$ for the NSs star masses $M_1$ and $M_2$, while the asymmetric have $q=0.86$, except for F20 where $q=0.80$. }
\begin{tabular}{llcccc}
\hline \hline 
 Ref. & EoS & $q$ & $M_{\rm ej}$         &  $M_{\rm ej}^{v \ge 0.6c}$ & $\langle Y_e\rangle$ \\ 
       &  &         & [$10^{-3}$~\Msun] & [$10^{-5}$~\Msun] &  \\ 
\hline
R18 & DD2 & 1 & 1.4 & 0.2 & 0.23 \\ 
N21 & DD2 & 1 & 1.1 & - & 0.25 \\ 
S15 & DD2 & 1 & 2.1 & - & 0.29 \\ 
This work & DD2 & 1 & 2.0 & 8.7 & 0.27 \\ 
 \hline
 R18 & SFHo & 1 & 4.2 & 3.3 & 0.22 \\ 
 N21 & SFHo & 1 & 2.8 & - & 0.23 \\ 
S15 & SFHo & 1 & 11 & - & 0.31 \\ 
This work & SFHo & 1 & 3.3 & 15.3 & 0.26\\ 
 \hline
F20 & DD2 & 0.80 & 8.3 & - & 0.13 \\
S16 & DD2 & 0.86 & 5 & - & 0.20 \\
This work & DD2 & 0.86 & 3.2 & 17.7 & 0.22\\ 
 \hline
S16 & SFHo & 0.86 & 11 & - & 0.18 \\ 
This work & SFHo & 0.86 & 8.7 & 25.6 & 0.24\\ 
\hline \hline 
\end{tabular} 
\label{tab:comp}
\end{table}

This being said, our dynamically ejected mass for the symmetric 1.35--1.35~\Msun\ system with the DD2 EoS is found to be 40-80\% larger than the one obtained by \citet{radice2018a} and recently \citet{nedora2021a}, while, with the SFHo EoS, it is  20\% lower than the values found by \citet{radice2018a} and up to 20\% larger than those of \citet{nedora2021a}.  Compared to the results of \citet{sekiguchi2015}, our ejected mass is found to be similar for the DD2 EoS but 3 times lower for the SFHo EoS. 
When comparing the symmetric and asymmetric merger simulations with the SFHo EoS in \citet{sekiguchi2016}, they yield the same dynamical ejecta mass. At the same time, we observe about three times more ejected mass in our asymmetric case.  For the DD2 EoS, their ejected mass is over two times larger for the asymmetric merger,  while our ejected mass only increases by 60\%. For the angular distribution of the ejected mass, \citet{perego2017} (see their Fig.~1) report a $\sin^2\theta$ dependence for the SFHo symmetric model including neutrino heating,  which corresponds well with our SFHo EoS results (see \cref{fig_dM2d} for the DD2 EoS).
The seemingly large ejected mass of \citet{foucart2020} for the asymmetric DD2 EoS could also be due to the larger total mass or, the smaller mass ratio $q$ of their merger system.

Looking now at the mean electron fractions (cf. \cref{tab:comp} and \cref{fig_xn0_ye_distr}), \citet{radice2018a} and \citet{nedora2021a} found with their M0 scheme for the symmetric systems using the DD2 (SFHo) EoSs,  values that are lower compared to our mean $Y_e$ by 0.04 and 0.02 (0.04 and 0.03), respectively, and \citet{sekiguchi2015} with their leakage-plus-M1 scheme found larger values, namely by 0.02 (0.05). 
However, the $Y_e$ distributions in \citet{radice2018a}, \citet{nedora2021a}, \citet{sekiguchi2015}  and our in \cref{fig_xn0_ye_distr} are quite similar since they all show a wide $Y_e$ range spanning from about 0.4 or 0.5 down to at least 0.05. 
In \citet{sekiguchi2016}, they also report the $Y_e$ values for asymmetric mergers and find a decrease of the average value relative to symmetric mergers by 0.09 (0.13) compared to 0.05 (0.02) in our case for the DD2 (SFHo) EoS. 
The mean $Y_e$ of 0.13 reported by \citet{foucart2020} for their asymmetric DD2 system is relatively low compared to the other simulations discussed here. While the $Y_e$ distribution of \citet{foucart2020} spans a similar range as in \cref{fig_xn0_ye_distr}a,  their distribution peaks at the lowest $Y_e$-value around $\sim 0.05$ (compared to $\sim 0.2$ in our calculations). It remains, however,  difficult to draw conclusions here on the impact of the neutrino treatment since the discrepancy between their $\langle Y_e \rangle$ and ours can also be due  to the different NS masses and, in particular, their lower mass ratio leading to a larger fraction of low-$Y_e$ tidal ejecta. 
As can be observed in Fig.~1 of \citet{perego2017} for the SFHo symmetric merger, the hydrodynamical simulations of \citet{radice2018a} including neutrino interactions produce larger $Y_e$ values in the polar regions compared to the equatorial plane, which can also be seen in our \cref{fig_yedist_ang}. However,  as pointed out in \cref{sec_ang_vc} (see \cref{fig_ye2d}), the mass of the polar ejecta is smaller than the equatorial ejecta, which is also what \citet{perego2017} report. 

When comparing our abundance distribution in \cref{fig_rpro_aa} with the simulation of \citet{radice2018a} (see their Fig.~20), some differences can be observed. 
In particular, a different abundance distribution is obtained for $A>90$ nuclei that probably finds its origin in the different nuclear data used as input to the nuclear reaction network (e.g., fission fragment distributions) adopted to estimate the nucleosynthesis yields. However, our isotopic composition is characterized by a systematically smaller production of $A \le 80$ nuclei despite our higher mean electron fraction. The DD2-135-135 (SFHo-135-135) model is characterised by $\langle Y_e\rangle$=0.27 (0.26) versus 0.23 (0.22) for the equivalent model in \citet{radice2018a}. It remains unclear whether this discrepancy is connected to the nuclear network solver, or the electron fractions, or other trajectory properties that may be differently predicted by \citet{radice2018a}. The mapping technique applied by \citet{radice2018a} and \citet{nedora2021a} to calculate the ejecta nucleosynthesis and taken from the parametrized r-process calculation of \citet{lippuner2015} is also likely to contribute to such differences. The updated calculations of \citet{nedora2021a} show similar nucleosynthesis results as \citet{radice2018a}.

The nucleosynthesis results reported in \citet{wanajo2014} should be compared to our equatorial composition for the SFHo EoS. As seen in \cref{fig_rpro_aa_angs} and \cref{fig_rpro_aa}, the equatorial ejecta dominates the shape of the final abundance distribution for the DD2 EoS, and the same is also true for the SFHo EoS.  Similar to our results, \citet{wanajo2014} report a distribution covering the range of mass numbers from 240 down to $\sim 80$, which is a consequence of the wide $Y_e$ distribution when including neutrino absorption.  \citet{wanajo2014} predict a larger production of $A\sim 80-120$ nuclei compared to our composition but follow the solar r-process distribution fairly well for $A>120$. Differences in the predicted composition may also originate from the use of different $\beta$-decay rates and fission fragment distributions~\citep{goriely2015,Lemaitre2021,kullmann2022}, see also \cref{ch_nucuncert} which discuss variations to the nuclear input.

As mentioned in \cref{sec_ang_vc}, the high-velocity ($v/c>0.6$) tail in our simulations contains only a minor fraction of the total ejected mass. Compared to previous simulations \citep{Bauswein2013} based on the same SPH code ignoring neutrino emission and absorption, the inclusion of neutrinos does not give rise to an increase of the total ejecta mass. However, it does enhance the mass of ejecta with $v/c>0.6$ by a factor of about two. 
For comparison, \citet{radice2018a} (see their Table 2) find that simulations with the M0 scheme compared to a leakage scheme not accounting for neutrino absorptions tend to increase the total ejected mass as well as the mass of the fast ejecta significantly. Still, in the case of the 1.35-1.35~\Msun\ models, they predict masses of fast ejecta with $M_{\rm ej}^{v \ge 0.6c}=1.7\times 10^{-6}$ and $3.3\times 10^{-5}$~\Msun\ for the DD2 and SFHo EoSs, respectively, which are about 50 and 5 times lower than those given in \cref{tab:comp}. Similarly, the recent calculations of \citet{nedora2021} (see their Table 1) give masses about 90 and 5 times lower than our results for the fast ejecta of the DD2 and SFHo EoSs, respectively.
These differences are likely to be connected to limitations in the affordable numerical resolution in both our SPH models as well as the grid-based models of \citet{radice2018a}, such that details of the numerical methods have a strong impact. This also includes a possible influence of the aforementioned atmosphere treatment in grid-based codes and potentially how ejecta properties are extracted, i.e., on a fixed sphere in \citet{radice2018a} compared to Lagrangian tracers in SPH. Future better-resolved simulations will have to reduce this uncertainty. 

\section{Summary and Conclusions}
\label{sect_concl}

In this chapter, we have studied r-process nucleosynthesis and the subsequent radioactive heating rate and kilonova produced by material dynamically ejected from four NS-NS merger models including a proper description of neutrino interactions. 
In contrast to more ``conventional'' leakage schemes,  the ILEAS framework coupled to the hydrodynamical calculations follows the detailed evolution of the electron fraction and its changes due to weak interactions through a neutrino-equilibration treatment in the high-density regions and an absorption module in semitransparent conditions \citep{ardevol-pulpillo2019}.  This allows us to study the impact of the neutrino interactions with nucleons on the composition and heating rate of the dynamical ejecta of NS mergers. In addition, the kilonova light curve resulting from the r-process nucleosynthesis calculations is briefly discussed.
Our four merger models include two EoSs, namely DD2 and SFHo. For each EoS, we have included symmetric and asymmetric merger cases, i.e.,  systems of 1.35-1.35~\Msun\ and 1.25-1.45~\Msun\ NSs, which all lead to the formation of a longer-lived neutron star surrounded by an accretion torus as a remnant. 


The main conclusions from our study can be summarized as follows:
\begin{itemize}
\item Proper inclusion of neutrino interactions yields dynamical ejecta with a solar-like composition down to $A \simeq 90$ compared to 140 without neutrino interactions. This includes, in particular, a significant enrichment in Sr. This conclusion is rather similar to the one drawn initially by \citet{wanajo2014}, though the final abundance distribution obtained in our case with different nuclear physics inputs is found to be in much closer agreement with the solar system distribution. 

\item The composition of the ejected matter and as the corresponding heating rate are found to be rather independent of the system mass asymmetry and the EoS. Despite differences in the initial electron fraction distributions, somewhat similar nucleosynthesis results are found in the four hydrodynamical models. This approximate degeneracy in abundance pattern and heating rates can be favourable for extracting the ejecta properties (such as masses, opacities, and expansion velocities) from kilonova observations.  
However, our study has two clear limitations. First, only the dynamical ejecta are taken into account (see \cref{subsec_combej} for the combined dynamical and secular ejecta). The HMNS or BH-torus wind ejecta may not follow the same trend with EoS and mass ratio as the dynamical ejecta. Second, here we only consider delayed-collapse cases in which a NS remnant survives for a longer period of time (around 10~ms) after the merger. Cases of prompt BH formation may exhibit different nucleosynthesis signatures, albeit the ejecta in these cases is typically much less massive \citep[e.g.,][]{Bauswein2013}. 

\item Uncertainties in the initial electron fraction impact not only the composition and, in particular, the amount of lanthanides and actinides ejected but also the heating rate, mainly at 10~min, 7~h or 30~d after merger through the $\beta$-decay of residual nuclei. Spontaneous fission of $^{254}$Cf or $A=222-225$ $\alpha$-decay chains contribute significantly less to the heating rate when weak nucleonic processes are taken into account.

\item  The nucleonic weak processes have the largest impact on the composition of the material ejected in the polar regions but also affect the r-process efficiency in the equatorial plane.

\item Even with nucleonic weak processes, the fast-expanding layers are found to include a non-negligible amount of free neutrons that should leave an observable signature in the light curve and may contribute to late-time X-ray emission \citep{hajela2021}.

\item The kilonova light curve, based on nucleosynthesis calculations using the complete set of trajectories from the four NS-NS dynamical ejecta models with regular treatment of weak interactions, peaks around $3-7\cdot 10^{40}$~erg/s. These models are quite similar in their input to the kilonova model as they have ejecta masses between 0.002 and 0.009~\Msun, average velocities of $\approx0.25$~c, and lanthanide mass fractions of $\approx0.1$.
Thus, the  observed differences in peak times and luminosities for the ILEAS models are primarily due to the differences in ejecta masses. However, a softer EoS and a more asymmetric binary produce slightly more luminous transients with longer durations.

\item When including weak processes, the early kilonova light curve tends to be brighter as a result of the reduced opacity (due to lower lanthanide mass fractions) and enhanced heating rate (due to higher abundances of elements between Ca and Sn). On the other hand, at late times ($t\ga 10$~d), the light curve tends to be dimmer because of a reduced abundance of trans-Pb nuclei.

\end{itemize}

In addition to the important role of weak nucleonic processes on the final composition of the ejecta, the final abundance distribution and decay heat are known to be affected by nuclear uncertainties. Despite numerous works dedicated to the impact of nuclear physics uncertainties in this context, the proper inclusion of neutrino interactions may change the conclusion of such sensitivity studies. In particular, as already shown by \citet{Lemaitre2021}, the role of fission may be significantly reduced if neutrino interactions are included and found to affect the initial neutron richness.  
Finally, before drawing conclusions on the contribution of NS mergers to the Galactic enrichment, the outflow generated during the evolution of the merger remnant must be consistently included in the simulations and the discussion.
Both of the aforementioned points, i.e., the inclusion of weak interactions and the impact of nuclear uncertainties on the r-process, as well as the combination of the dynamical and secular ejecta, will be studied in detail in the following \cref{ch_nucuncert}. 



\chapter{Nuclear uncertainties} 
\label{ch_nucuncert}

\epigraph{\itshape \enquote{And now for something completely different.}}{--- \textup{Monty Python's Flying Circus}}  

The r-process calculations in this section are based on hydrodynamical models of both the dynamical and secular ejecta. The dynamical ejecta models include two NS-NS models also studied in \cref{ch_dynweak} (the models using the SFHo EoS) and a NS-BH model (see \cref{subsec_dynsim}). 
The two BH-torus remnant systems described in \cref{subsec_BH-t_sim} are combined with the dynamical ejecta into an approximate model for the total ejecta.
The nuclear uncertainties have been consistently propagated into the r-process calculations, including abundance distributions, heating rates and cosmochronometric age estimates, by varying the nuclear input between the 15 input sets discussed in \cref{sec_network_input}.
For each nuclear input set, all quantities (except those presented in \cref{fig_rpro_onetraj}) are mass averaged over $\sim$200-300 representative trajectories for each hydrodynamical model (see \cref{sec:astro_mods,sec_sel_subs} and \cref{tab:astromods}). 
This work is based on the study of \citet{kullmann2022}.

First, we show the r-process nucleosynthesis distributions and heating rates obtained using the astrophysical models for the dynamical (\cref{sec_dyn_unce}) and BH-torus (\cref{sec_resbh_torus}) ejecta separately before we discuss the results for the combined ejecta (\cref{subsec_combej}) and the cosmochronometric age estimates for six metal-poor r-process-enhanced stars (\cref{sec:cosmo}).
\cref{sec_comp2} compares this section's work to similar studies, and \cref{sect_concl_p2} summarizes and concludes the study.

\section{Dynamical ejecta}
\label{sec_dyn_unce}

\subsection{NS+NS merger}

\begin{figure}[tbp]
\begin{center}
\includegraphics[width=\columnwidth]{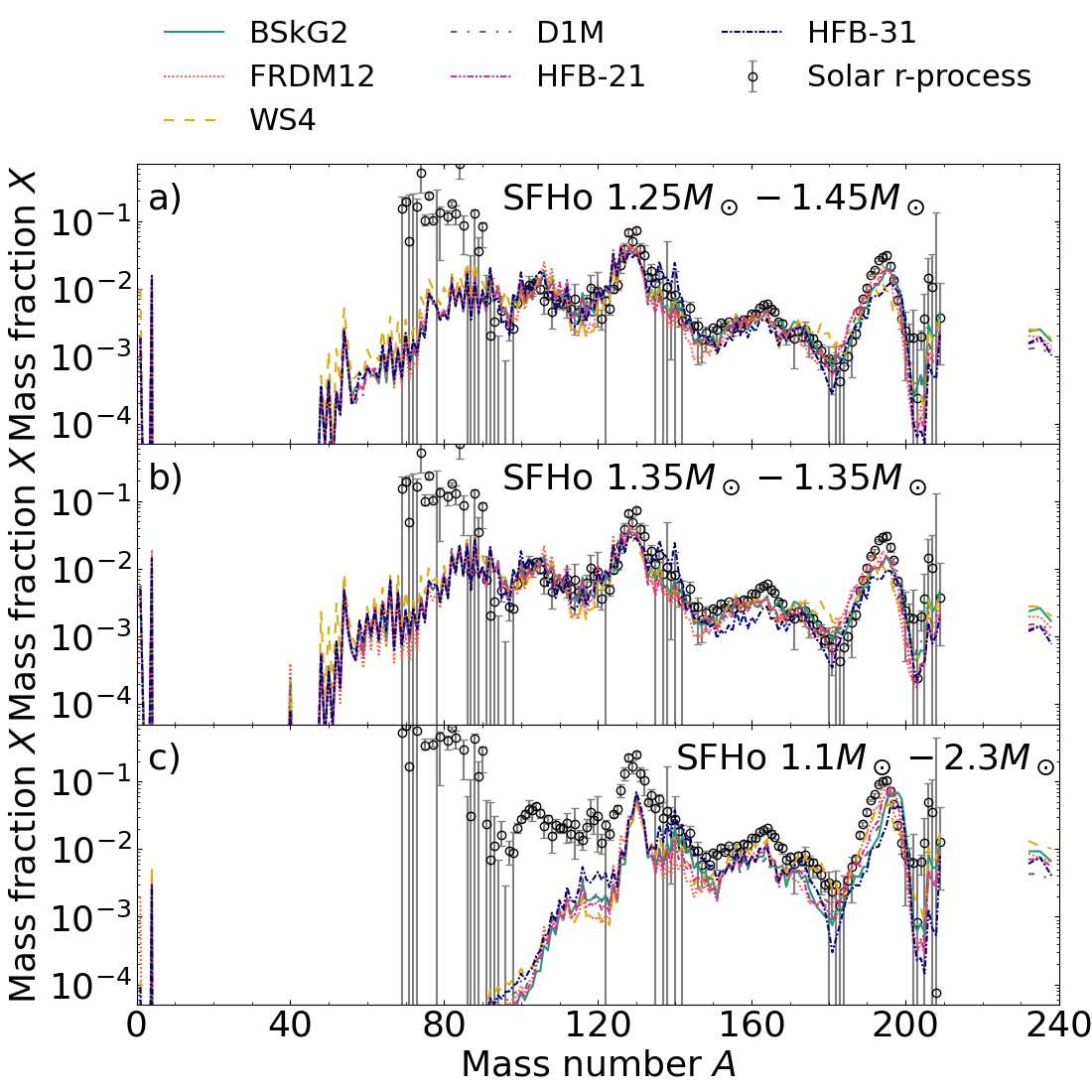}
\caption[Final mass fractions as a function of the atomic mass for three dynamical ejecta NS--NS and NS--BH models when varying the applied mass model.]{
Final mass fractions of stable nuclei (and long-lived Th and U) of the material ejected as a function of the atomic mass $A$ for the three NS--NS and NS--BH dynamical ejecta merger models: a) SFHo-125-145, b) SFHo-135-135, and c) SFHo-11-23 when varying the applied mass model (i.e., input sets 1-6 in \cref{tab_nuc_mods}).  See the text for references and details about the different mass models. 
The solar system r-abundance distribution (open circles) \citep{goriely1999} is shown for comparison and arbitrarily normalized at the third r-process peak using input set 1 in \cref{tab_nuc_mods} of model SFHo-135-135 in a) and b), and SFHo-11-23 in c).
}
\label{fig_rpro_u1-7}
\end{center}
\end{figure}

In \cref{fig_rpro_u1-7}(a,b) we show the r-process abundance distributions for the SFHo symmetric and asymmetric merger models, respectively, computed with six different mass models (see \cref{tab_nuc_mods}). 
Except for two mass models, HFB-31 around $A\sim 130$ and WS4 around $A\sim 180$, the other mass models agree globally well and follow the solar system distribution for $A>90$.  
All models produce a significant amount of actinides, ranging from 0.2-0.7\% of the total abundance. Local differences, in particular around the third r-process peak, can be observed in relation to the prediction of the strength of the $N=126$ closed shell.

\begin{figure}[tbp]
\begin{center}
\includegraphics[width=\columnwidth]{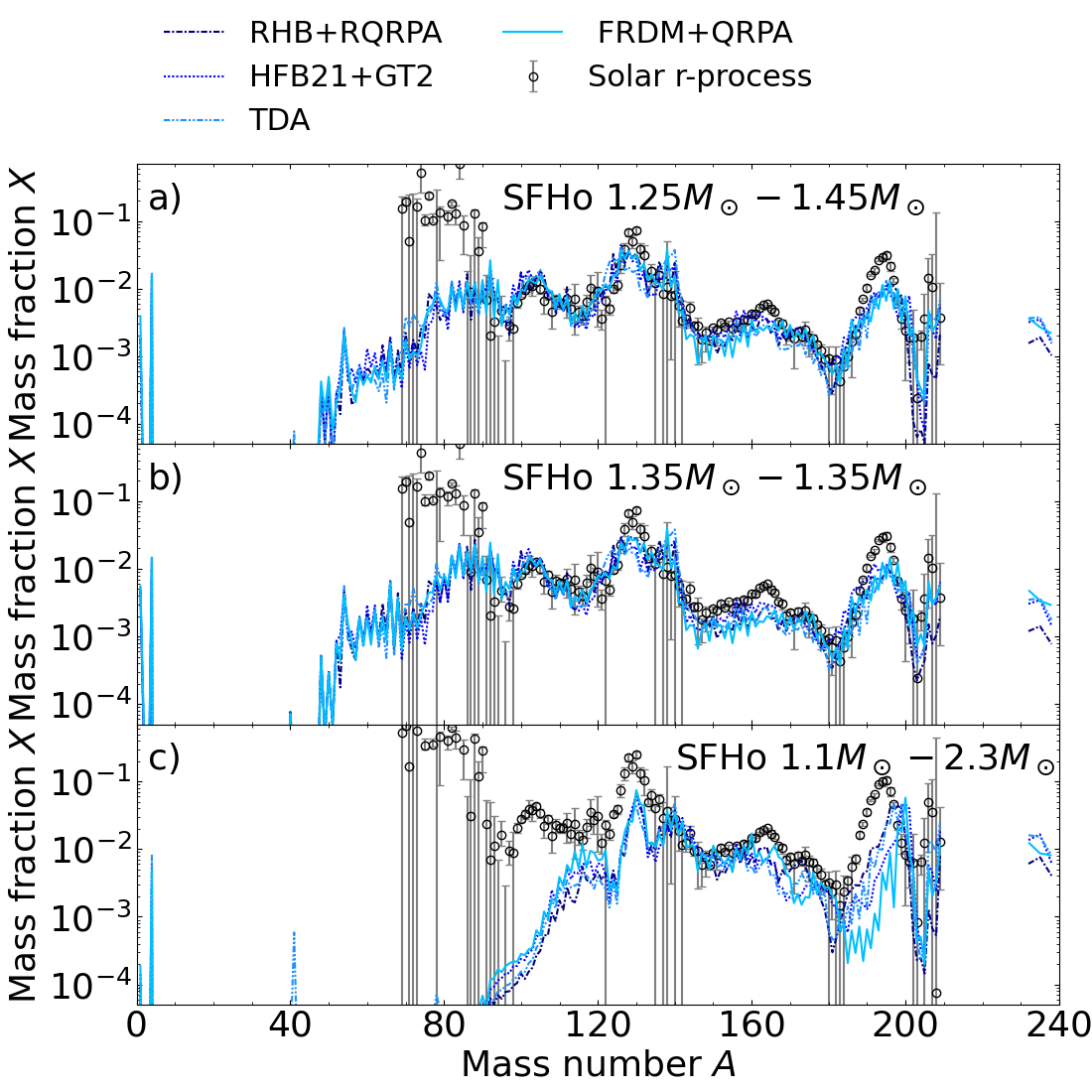}
\caption[Same as \cref{fig_rpro_u1-7} when varying the four models for the $\beta$-decay rates, but using the same HFB-31 mass model. ]{
Same as \cref{fig_rpro_u1-7} when varying the four models for the $\beta$-decay rates, but using the same HFB-31 mass model (i.e., input sets 6-9 in \cref{tab_nuc_mods}). See the text for references and details about the different models. 
The solar system abundance distribution is normalized as in \cref{fig_rpro_u1-7}.
}
\label{fig_rpro_u7-10}
\end{center}
\end{figure}

The calculations based on the HFB-31 mass model (\cref{fig_rpro_u7-10}), shows an `extra'' peak around $A\sim 130-140$. 
This over-abundance stems from a significantly stronger odd-even effect predicted around $N=90$ for HFB-31 compared to the other mass models considered.
Similar to what can be observed at closed neutron shells,  the larger $S_{2n}$ values at $N=90$ lead to    abundance accumulating during the r-process simulation at this isotone, which is later transformed into the $A\sim 130-140$ peak after neutron freeze-out.

\begin{figure}[tbp]
\begin{center}
\includegraphics[width=\columnwidth]{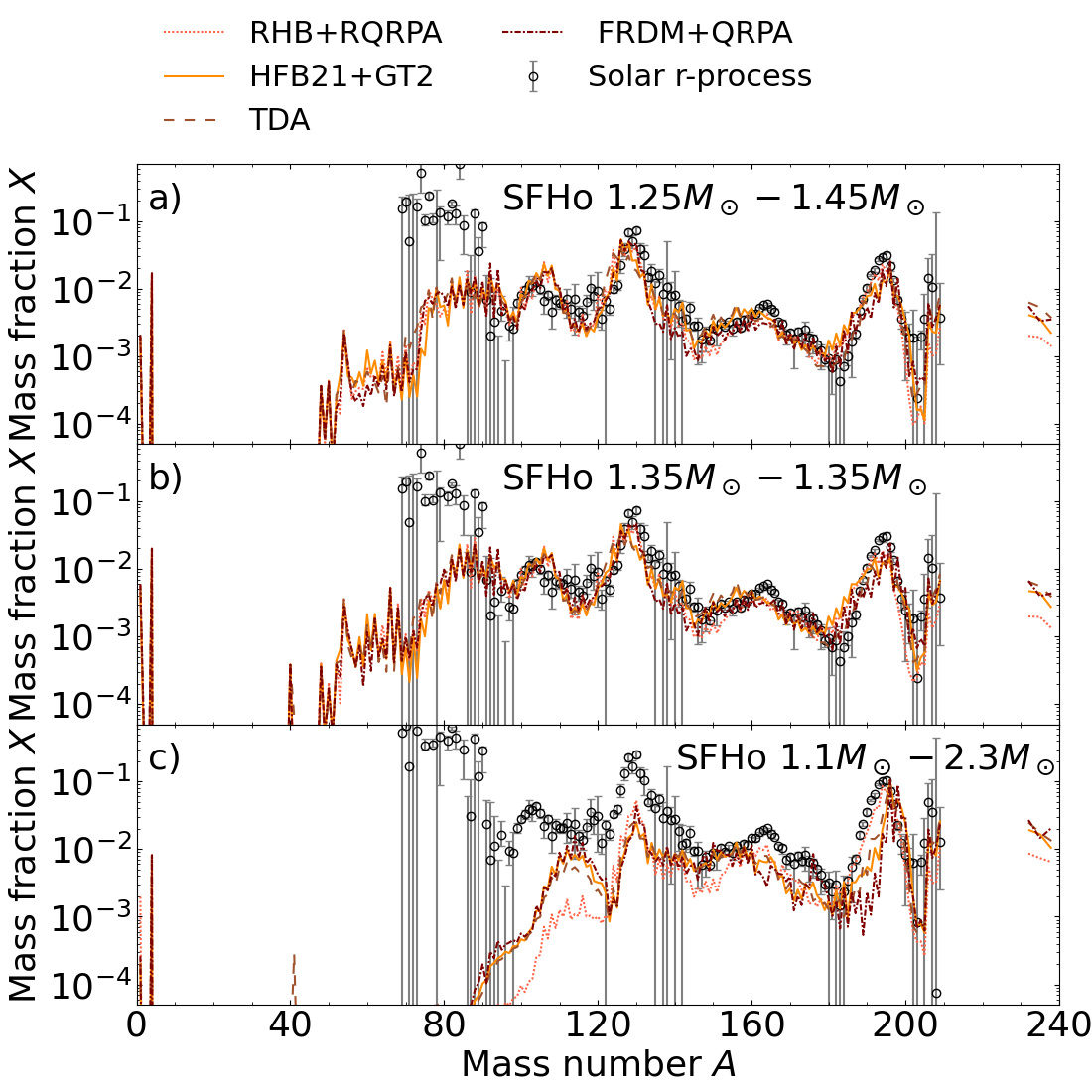}
\caption[Same as \cref{fig_rpro_u1-7} when varying the four models for the $\beta$-decay rates, but using the same FRDM12 mass model.]{
Same as \cref{fig_rpro_u1-7} when varying the four models for the $\beta$-decay rates, but using the same FRDM12 mass model (i.e., input sets 2 and 10-12 in  \cref{tab_nuc_mods}). See the text for references and details about the different models. 
The solar system abundance distribution is normalized as in \cref{fig_rpro_u1-7}.
}
\label{fig_rpro_u3-13}
\end{center}
\end{figure}

Similarly, Figs.~\ref{fig_rpro_u7-10}(a,b) and \ref{fig_rpro_u3-13}(a,b) give the abundance distributions calculated using four different models for the $\beta$-decay rates applying the two mass models HFB-31 and FRDM12, respectively (see \cref{tab_nuc_mods}).  Compared to the mass models, the differences are larger between the $\beta$-decay models, particularly for $A>130$.  
The actinide production is larger when applying the FRDM12 masses (Fig.~\ref{fig_rpro_u3-13}), ranging from $1.3$ to $1.7$\%, compared to using HFB-31 masses, which result in $0.4$ to $1.1$\% of actinides.\todo{The above sentence is unclear (ref. report)}

\begin{figure}[tbp]
\begin{center}
\includegraphics[width=\columnwidth]{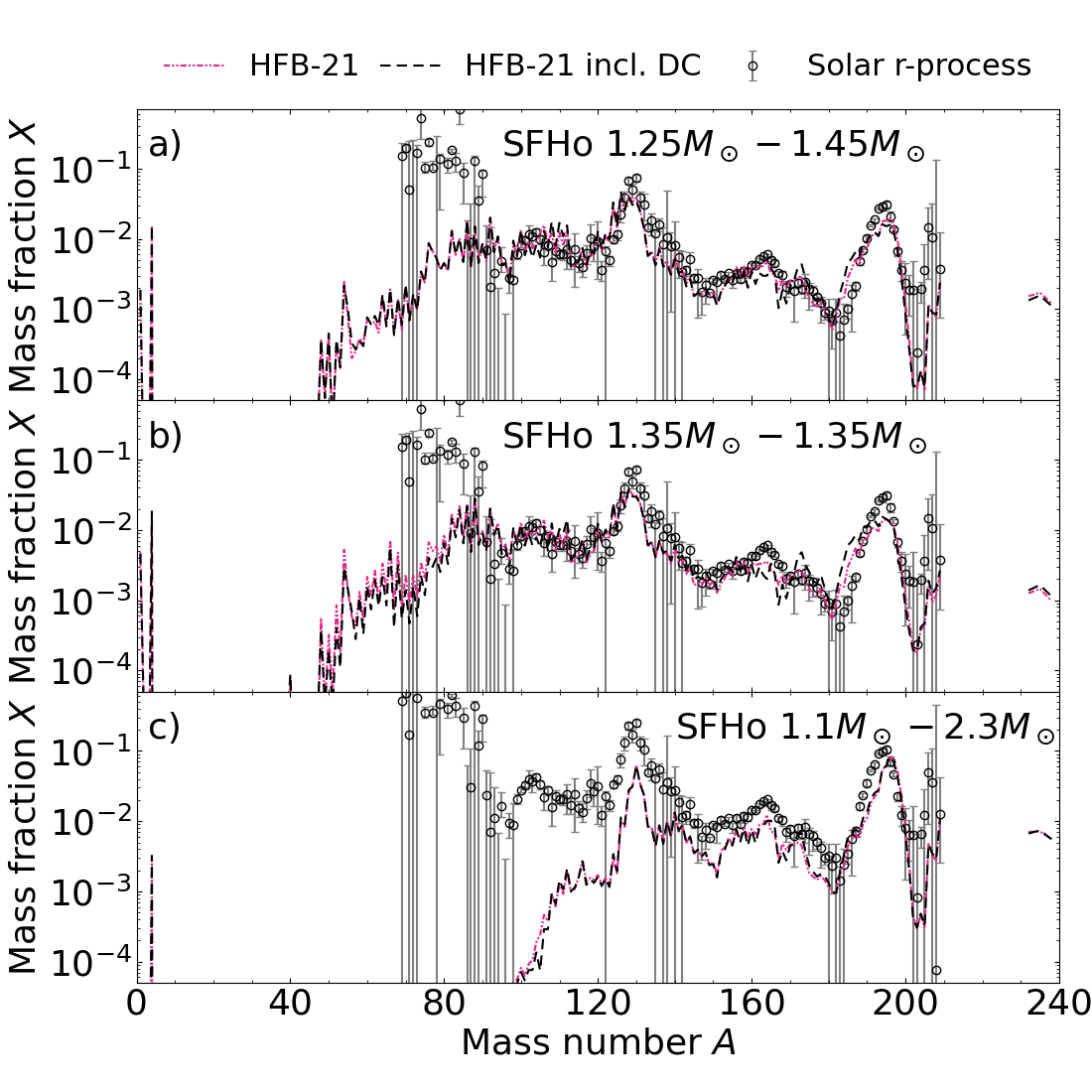}
\caption[Same as \cref{fig_rpro_u1-7} when including the DC component for the radiative neutron capture rates, but using the same mass model, the HFB-31 model.]{
Same as \cref{fig_rpro_u1-7} when including the DC component for the radiative neutron capture rates, but using the same mass model, the HFB-21 model (i.e., input sets 5 and 15 in  \cref{tab_nuc_mods}). See the text for references and details about the different models. 
The solar system abundance distribution is normalized as in \cref{fig_rpro_u1-7}.
}
\label{fig_rpro_uDC}
\end{center}
\end{figure}

\begin{figure}[tbp]
\begin{center}
\includegraphics[width=\columnwidth]{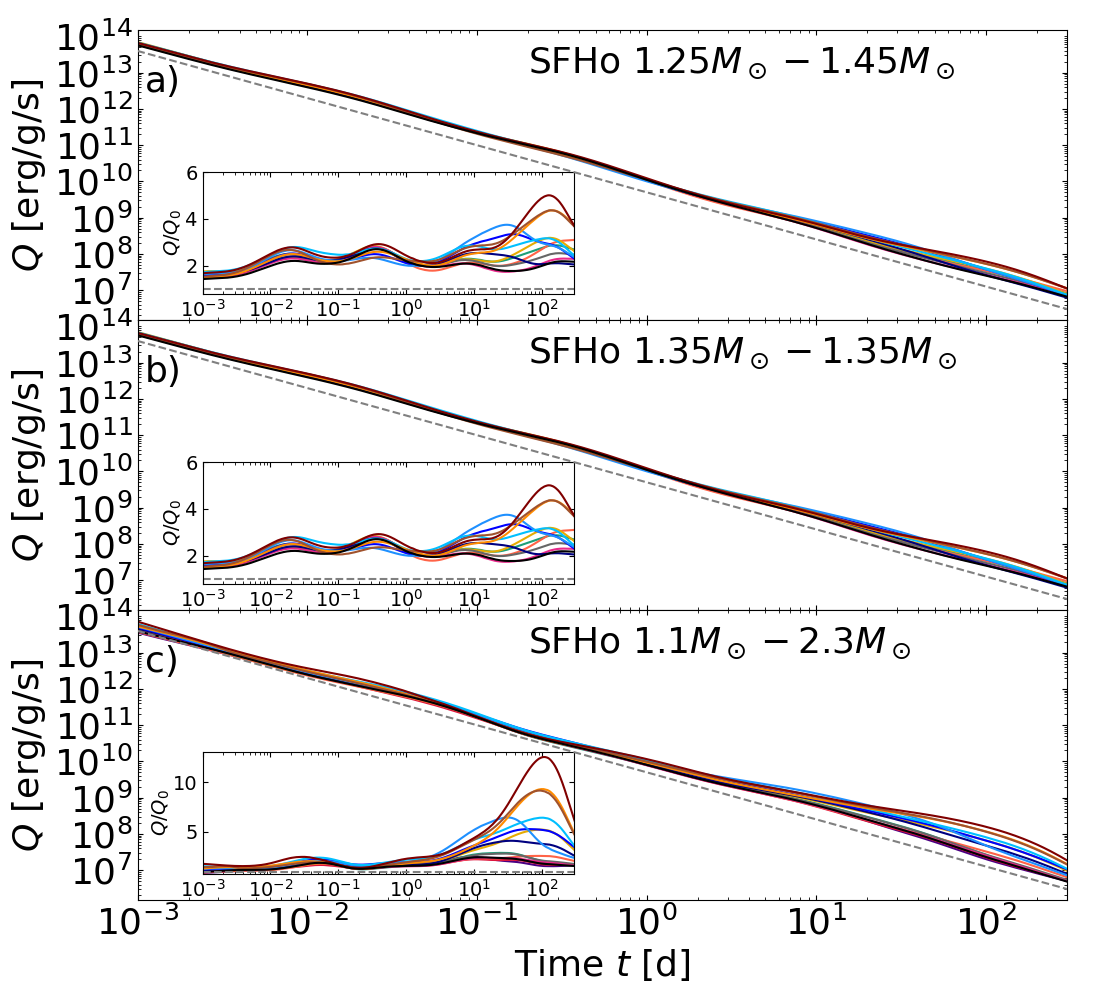}
\caption[Time evolution of the radioactive heating rate for the NS–NS and NS-BH dynamical ejecta merger models when varying the nuclear input between all input sets listed in \cref{tab_nuc_mods}.]{
Time evolution of the radioactive heating rate $Q$ (before thermalization) for the NS–NS and NS-BH dynamical ejecta merger models: a) SFHo-125-145, b) SFHo-135-135, and c) SFHo-11-23 when varying the nuclear input between all models listed in  \cref{tab_nuc_mods}.  
The grey dashed line corresponds to the global trend followed by $Q_0=10^{10}[t/1~\mathrm{day}]^{-1.3}$~erg/g/s (\cref{eq:Q_0}). The ratio $Q/Q_0$ is also shown as an insert, where the grey dashed line indicates $y=1$.
See \cref{fig_rpro_u1-7,fig_rpro_u7-10,fig_rpro_u3-13,fig_rpro_uDC} for the colour labels for each nuclear input combination. \todo{[put inset figs on same y-scale as the others]}
}
\label{fig_Qt_all_dyn}
\end{center}
\end{figure}

\cref{fig_rpro_uDC} compares the difference between the abundance distributions when the radiative neutron capture rates are estimated with and without the DC component for hydrodynamical models SFHo-125-145 and SFHo-135-135. The most considerable differences between both sets of rates are found around $A\sim150$, where the model including the DC component has a peak structure (double lanthanide peak), and in particular, for model SFHo-125-145, the third r-process peak is also broader when the DC is included.

As described in \cref{sec:heat}, the heating rate is defined as the radioactive energy release rate per unit of mass due to $\beta$-decay, $\alpha$-decay, and fission, not including the energy lost into neutrino emission. 
\cref{fig_Qt_all_dyn}(a-b) shows the time evolution of the heating rate ($Q$) for all combinations of nuclear models (see \cref{tab_nuc_mods}) for the dynamical ejecta of the NS-NS systems. 
All nuclear models predict roughly the same r-process heating.
After $t>30$ days, the contribution to the heating rate from fission differs between the nuclear input cases, where the largest heating is found when using mass model FRDM12 and either FRDM+QRPA, HFB21+GT2 or TDA $\beta$-decay rates. 

\subsection{NS-BH merger}

\begin{figure}[tbp]
\begin{center}
\includegraphics[width=\columnwidth]{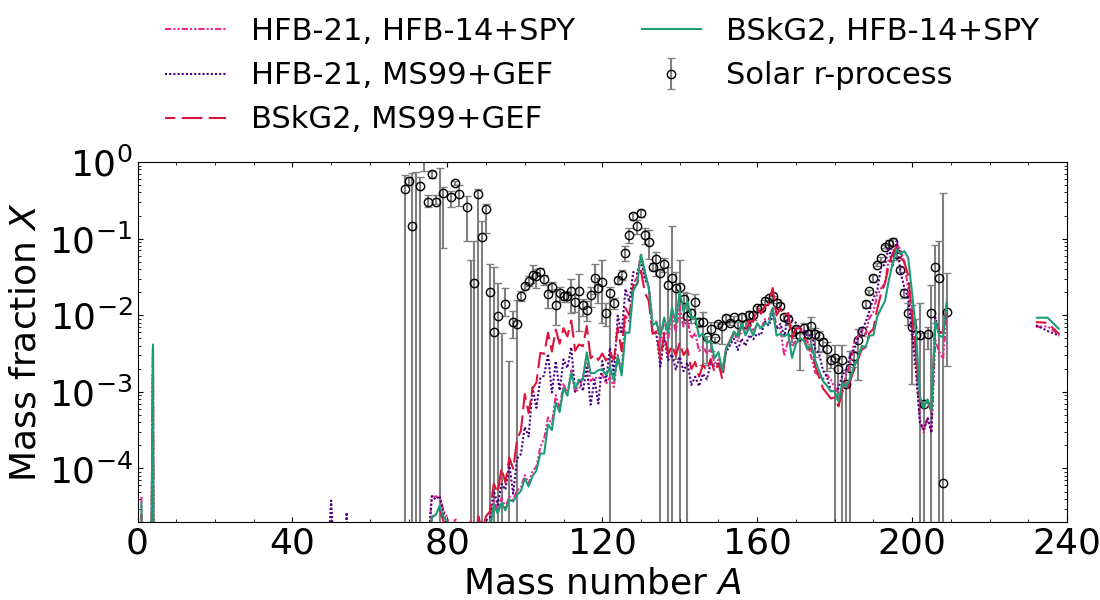}
\caption[Same as \cref{fig_rpro_u1-7}c when varying the fission barriers and fragment distributions, using mass model HFB-31 or BSkG2.]{
Same as \cref{fig_rpro_u1-7}c when varying the fission barriers and fragment distributions, using mass model HFB-31 or BSkG2 (i.e., input sets 1, 5, 13 and 14 in  \cref{tab_nuc_mods}) for the material ejected by the NS-BH merger model SFHo-11-23.  See the text for references and details about the different nuclear models. 
The solar system abundance distribution is normalized as in \cref{fig_rpro_u1-7}c.
}
\label{fig_rpro_ufiss}
\end{center}
\end{figure}

\cref{fig_rpro_u1-7}c displays the composition estimated when varying the six mass models for the NS-BH merger model SFHo-11-23. We can see that compared to the dynamical ejecta of the NS-NS merger systems (panels a and b), the NS-BH merger ejecta undergoes a strong r-process and follows the solar system r-process distribution for $A\ga 140$ (compared to $A \ga 90$ for NS-NS systems). The astrophysical conditions lead to a weak second r-process peak and an additional structure around $A\sim140$ due to fission processes (see below); however, the strength of this structure varies between the mass models and is found to be particularly strong for mass model HFB-31.

When varying the models for the $\beta$-decay rates in \cref{fig_rpro_u7-10}c (using mass model HFB-31), the structure around $A\sim 140$ is present with the same strength, which was not always the case for the NS-NS merger models. \todo{[there are large deviations everywhere not just for 160?]} 
However, significant deviations are found for $A>160$ in \cref{fig_rpro_u3-13}c, which shows the impact of the $\beta$-decay models when using mass model FRDM12. 

The material ejected from the NS-BH merger model has initial conditions that favour a larger production of fissile material compared with the NS-NS merger models considered here and are therefore well suited to study the impact of varying the fission properties.
In \cref{fig_Qt_all_dyn}c, we can see that the impact of fission on the heating rate at $t>30$d is more prominent than in the NS-NS merger models.  

\cref{fig_rpro_ufiss} shows that fission barriers and fragment distributions significantly affect the predicted abundance distributions (and proportionally more than varying the mass model). For example, all models predict about the same amount of actinides (2--2.5\%); however, BSkG2+MS99+GEF (brown line) predicts a sharper peak at $A\sim 165$ than the other models. The asymmetric nature of the fission fragment distribution for HFB-14+SPY, particularly around $A\simeq 278$ isobars, is also found to impact the final abundance distribution in the $A\simeq 140$ region, as already pointed out by \citet{goriely2013}.


\section{BH-torus ejecta}
\label{sec_resbh_torus}

\begin{figure}[tbp]
\begin{center}
\includegraphics[width=\columnwidth]{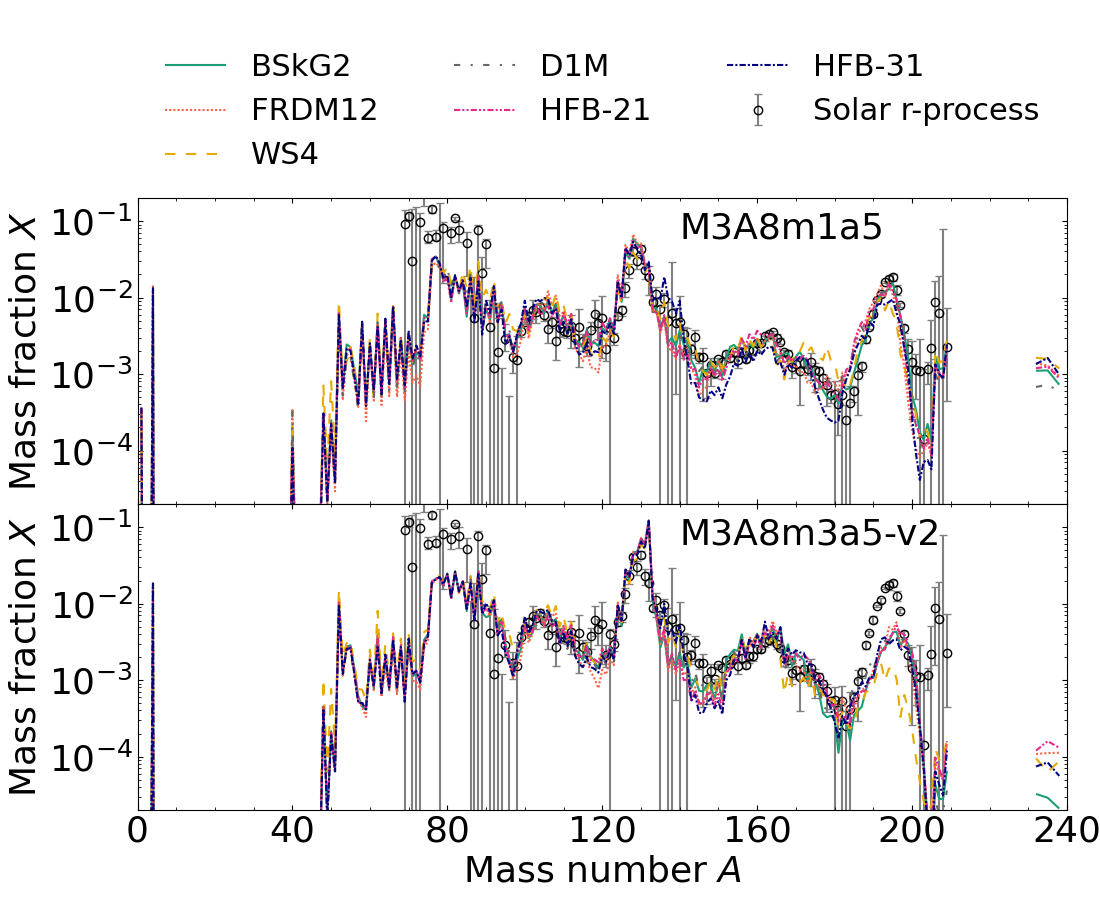}
\caption[Final mass fractions of the material ejected as a function of the atomic mass for two BH-torus models when varying between six mass models.]{
Final mass fractions of the material ejected as a function of the atomic mass $A$ for the two BH-torus models (M3A8m1a5 and M3A8m3a5-v2) when varying between six mass models (i.e., input sets 1-6 in  \cref{tab_nuc_mods}). See the text for references and details about the different models. 
The solar system r-abundance distribution (open circles) from \citet{goriely1999} is shown for comparison and arbitrarily normalized at the third r-process peak of model M3A8m1a5.
}
\label{fig_rprowind_u1-7}
\end{center}
\end{figure}

\begin{figure}[tbp]
\begin{center}
\includegraphics[width=\columnwidth]{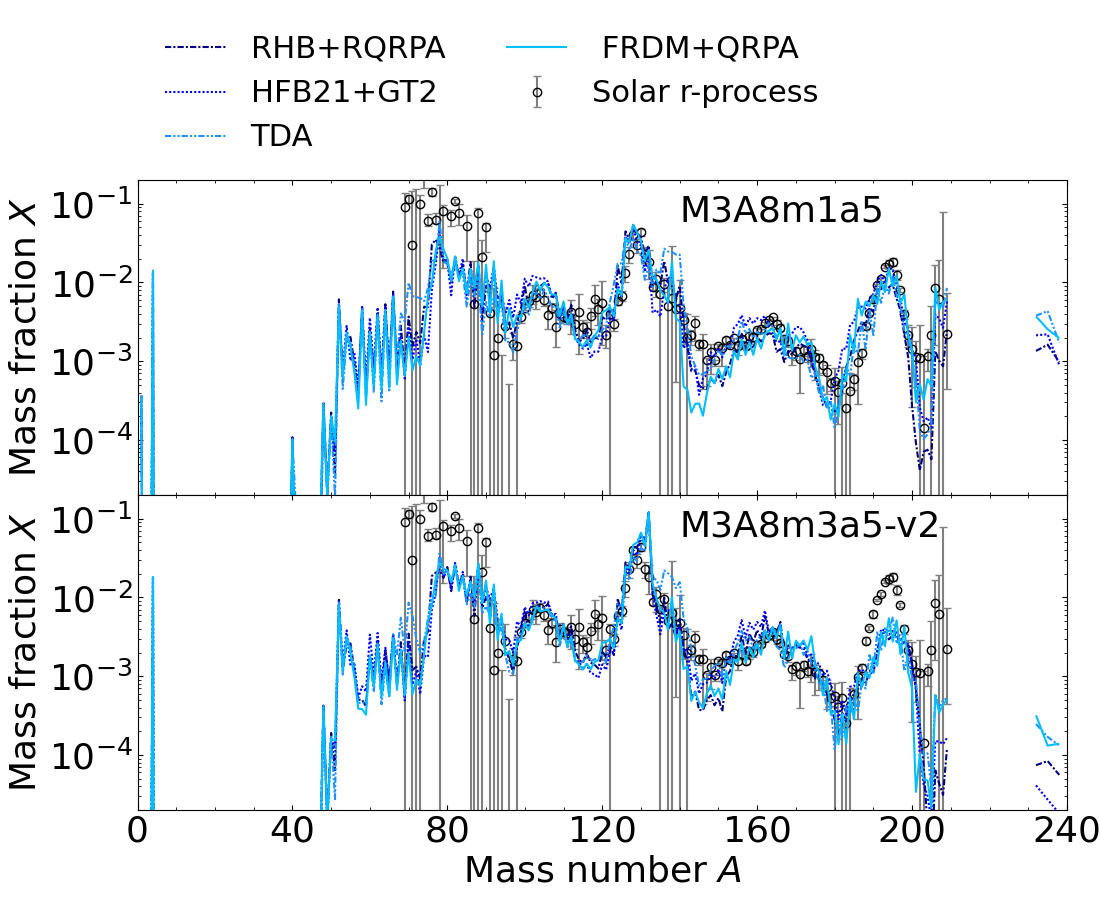}
\caption[Same as \cref{fig_rprowind_u1-7} when varying the models for the $\beta$-decay rates, but using the same HFB-31 mass model.]{
Same as \cref{fig_rprowind_u1-7} when varying the models for the $\beta$-decay rates, but using the same HFB-31 mass model (i.e., input sets 6-9 in  \cref{tab_nuc_mods}). See the text for references and details about the different models. 
The solar system abundance distribution is normalized as in \cref{fig_rprowind_u1-7}.
}
\label{fig_rprowind_u7-10}
\end{center}
\end{figure}

\begin{figure}[tbp]
\begin{center}
\includegraphics[width=\columnwidth]{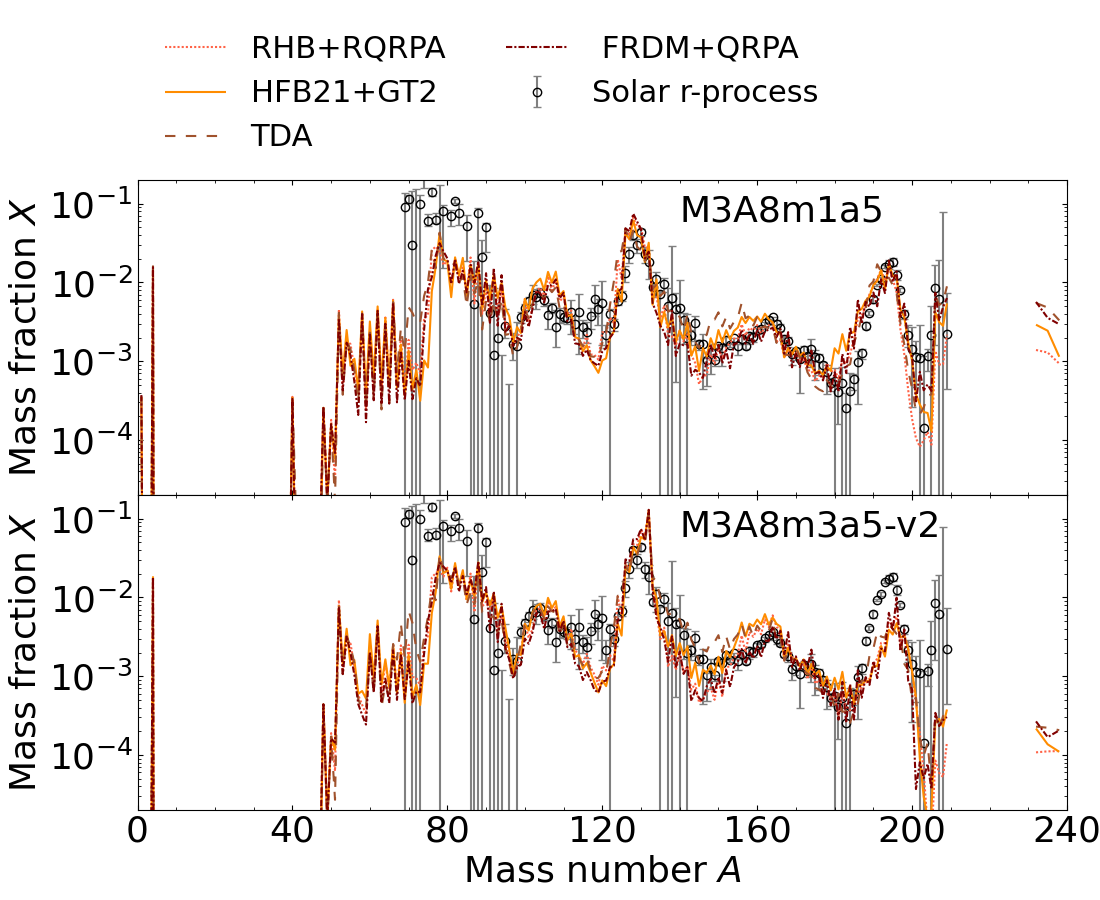}
\caption[Same as \cref{fig_rprowind_u1-7} when varying the models for the $\beta$-decay rates, but using the same FRDM12 mass model]{
Same as \cref{fig_rprowind_u1-7} when varying the models for the $\beta$-decay rates, but using the same FRDM12 mass model (i.e., input sets 2 and 10-12 in  \cref{tab_nuc_mods}). See the text for references and details about the different models. 
The solar system abundance distribution is normalized as in \cref{fig_rprowind_u1-7}.
}
\label{fig_rprowind_u3-13}
\end{center}
\end{figure}

\begin{figure}[tbp]
\begin{center}
\includegraphics[width=\columnwidth]{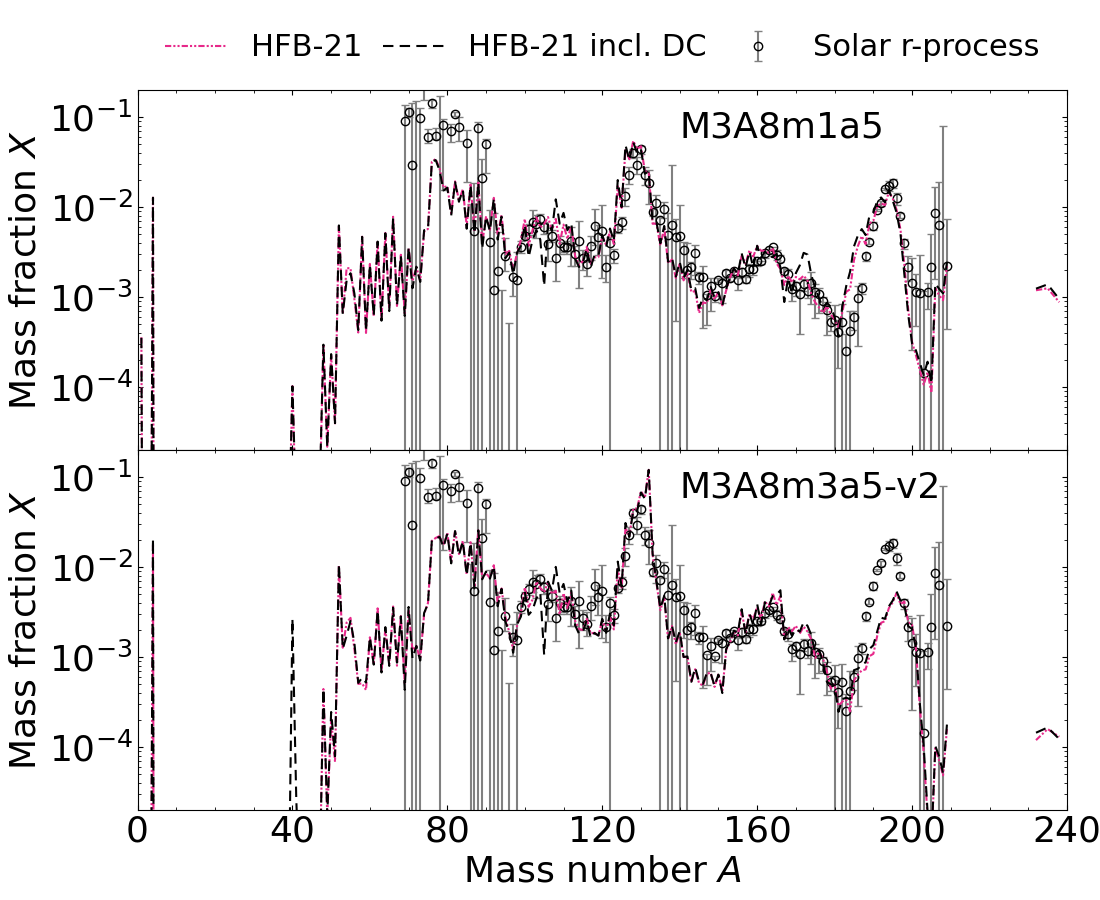}
\caption[Same as \cref{fig_rprowind_u1-7} when including the DC component for the radiative neutron capture rates, but using the same mass model, the HFB-31 model.]{
Same as \cref{fig_rprowind_u1-7} when including the DC component for the radiative neutron capture rates, but using the same mass model, the HFB-31 model (i.e., input sets 5 and 15 in  \cref{tab_nuc_mods}). See the text for references and details about the different models. 
The solar system abundance distribution is normalized as in \cref{fig_rprowind_u1-7}.
}
\label{fig_rprowind_DC}
\end{center}
\end{figure}

\begin{figure}[tbp]
\begin{center}
\includegraphics[width=\columnwidth]{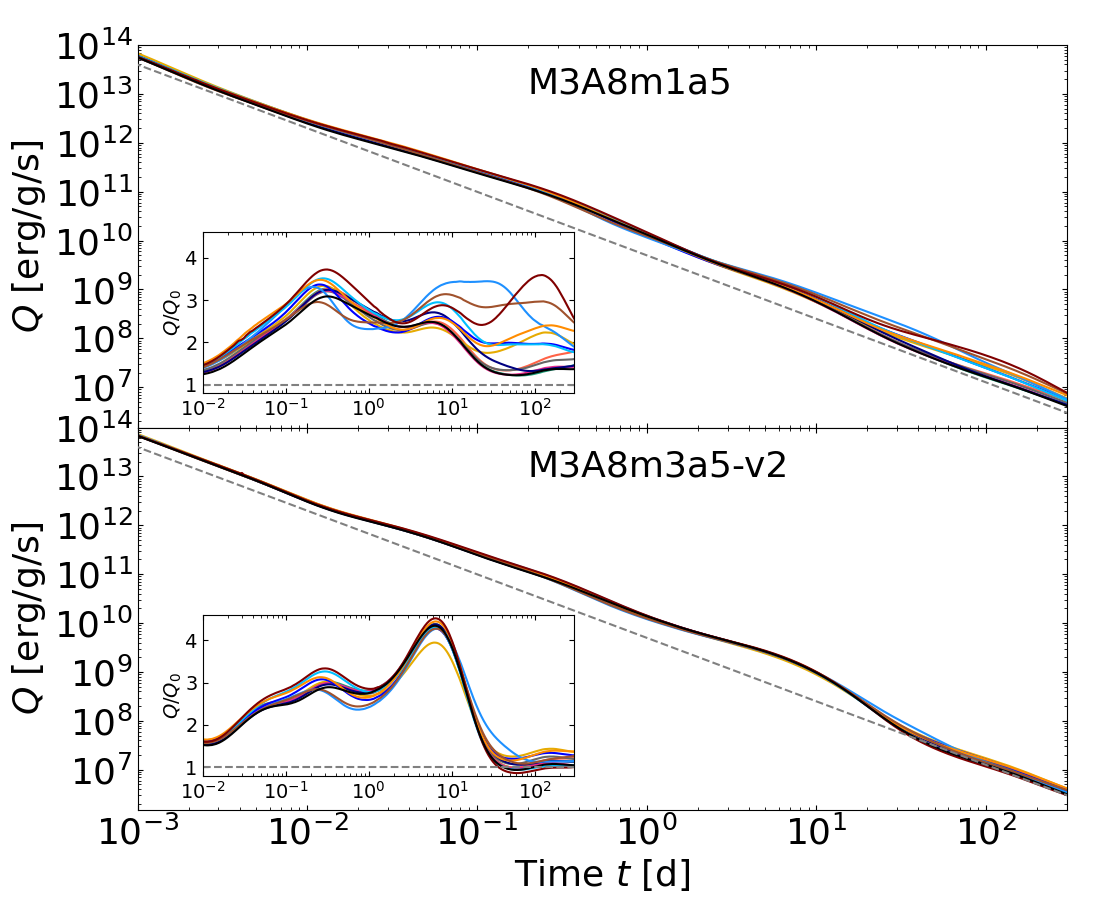}
\caption[Time evolution of the radioactive heating rate for two hydrodynamical BH-torus models when varying the nuclear input between all input sets in \cref{tab_nuc_mods}.]{
Time evolution of the radioactive heating rate $Q$ (before thermalization) for the two hydrodynamical BH-torus models M3A8m1a5 and M3A8m3a5-v2 when varying the nuclear input between all models listed in  \cref{tab_nuc_mods}.
The grey dashed line corresponds to the approximation $Q_0=10^{10}[t/1~\mathrm{day}]^{-1.3}$~erg/g/s (\cref{eq:Q_0}). The ratio $Q/Q_0$ is also shown as an insert, where the grey dashed line indicates $y=1$.
See \cref{fig_rprowind_u1-7,fig_rprowind_u7-10,fig_rprowind_u3-13,fig_rprowind_DC} for the colour labels for each nuclear input combination.     
}
\label{fig_Qt_all_wind}
\end{center}
\end{figure}
The material ejected in the post-merger phase from the torus surrounding the BH has quite different conditions than the dynamical ejecta (see \cref{fig_xn0_ye_distr}). Therefore, the r-process distribution differs in its shape, as seen in \cref{fig_rprowind_u1-7,fig_rprowind_u7-10,fig_rprowind_u3-13} for the various mass and $\beta$-decay models.  In general, the disk wind models produce more nuclei in the $A\sim80$ region, and in particular, model M3A8m3a5-v2 produces less of third peak elements and actinides, which is also where the most considerable differences between the mass models can be seen.  
The shape of the peak around $A\sim132$ also differs,  where a very narrow structure can be seen for the M3A8m3a5-v2 model\footnote{This feature is also observed in \citet{Wu2016}. It is related to trajectories, in which material falls back towards the torus and gets re-heated before it ultimately becomes ejected.}, compared to the wider peak for model M3A8m1a5 and the dynamical ejecta.

\cref{fig_rprowind_u1-7,fig_rprowind_u7-10,fig_rprowind_u3-13} show the r-process results when varying the four models for the $\beta$-decay rates and the two mass models HFB-31 and FRDM12, respectively.  Just as for the dynamical ejecta,  the HFB-31 mass model combined with the TDA or GT2 $\beta$-decay models overproduce abundances for nuclei around $A\sim 130-140$. 

The time evolution of the heating rate is shown in \cref{fig_Qt_all_wind} for all the 15 nuclear models. We can see that the heating rate varies less for different nuclear models in the BH-torus ejecta than in the dynamical ejecta (\cref{fig_Qt_all_dyn}).
The nuclear mass models only start to differ significantly at $t>10$ d for model M3A8m1a5 due to the more significant production of heavy and fissile r-process elements. Model M3A8m3a5-v2 peaks at around 0.2 and 10~d due to the $\beta$-decays of $A\sim80-90$ and $\sim140$ nuclei, respectively, and has a negligible contribution from fission and $\alpha$-decay at late times ($t>20$~d).  

\section[Combining dynamical and BH-torus ejecta][Combining the ejecta]{Combining dynamical and BH-torus ejecta}
\label{subsec_combej}

\begin{landscape}
\begin{figure*}
\begin{center}
\includegraphics[width=1.3\textwidth]{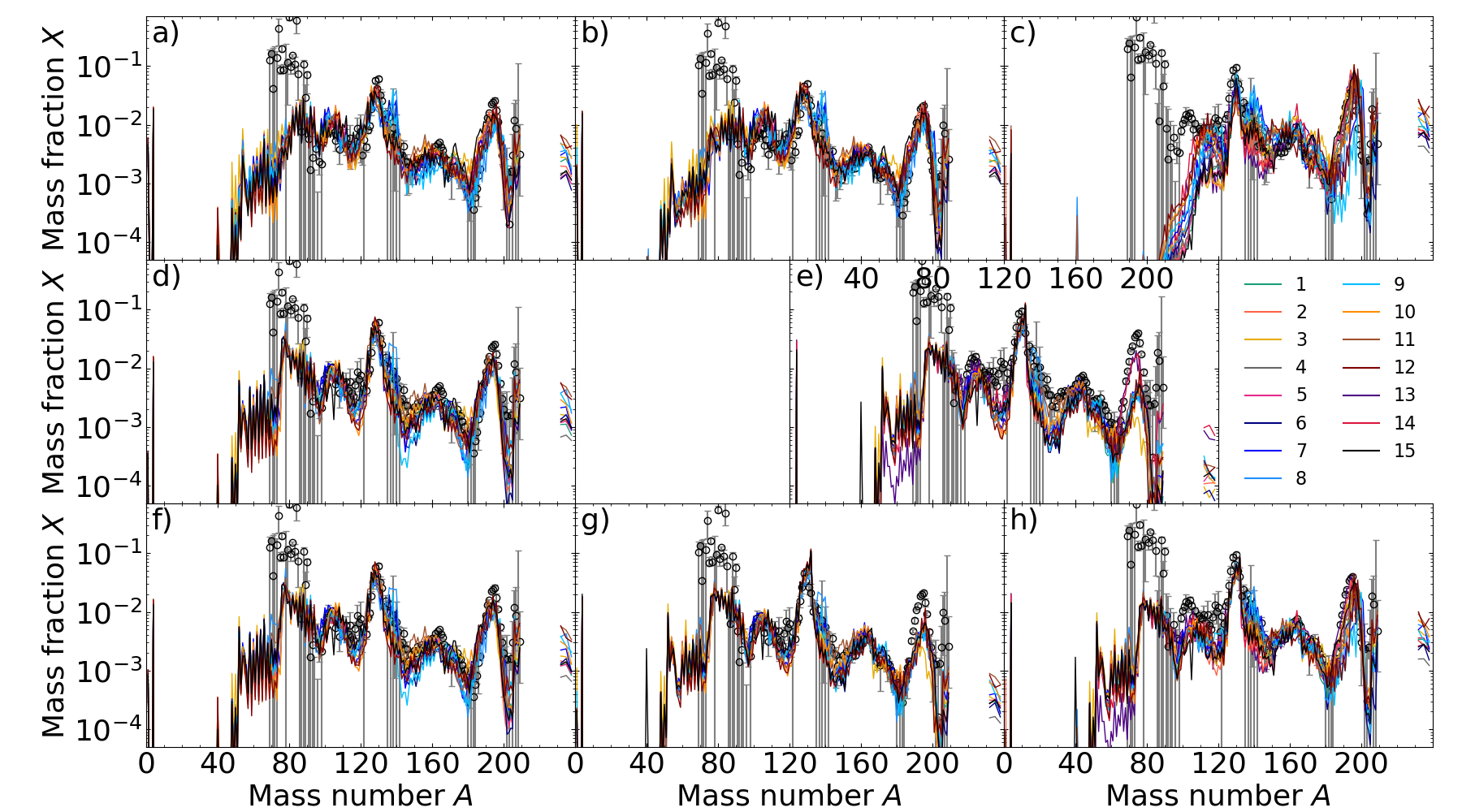}
\caption[Final mass fractions of the material ejected as a function of the atomic mass when varying all nuclear models listed in \cref{tab_nuc_mods} for each hydrodynamical model and the combined models for the total ejecta.]{
Final mass fractions of the material ejected as a function of the atomic mass $A$ when varying all nuclear models listed in \cref{tab_nuc_mods} for each hydrodynamical model: a) SFHo-135-135, b) SFHo-125-145, c) SFHo-11-23, d) M3A8m1a5 and e) M3A8m3a5-v2. The total ejecta (bottom panels) are calculated by summing the dynamical ejecta (top panels) with the BH-torus ejecta (middle panels), weighted by their respective ejected masses. For example panel f) is calculated as $X_\mathrm{f}=(X_\mathrm{a} M_\mathrm{a}+X_\mathrm{d} M_\mathrm{d})/(M_\mathrm{a}+M_\mathrm{d})$, and similarly panel g) by combining b) and e), and h) by combining c) and e). 
The colour scheme for the nuclear models is consistent throughout the thesis. 
The solar system r-abundance distribution (open circles) from \citet{goriely1999} is shown for comparison and arbitrarily normalized to the combined ejecta models (i.e., panel a, d use the normalisation from f, panel b from g, and panel c and e from h). \todo{[lines cant be distinguished very well smaller line width?]}
}
\label{fig_rpro_comb_all}
\end{center}
\end{figure*}
\end{landscape}

\begin{figure}[tbp]
\begin{center}
\includegraphics[width=\columnwidth]{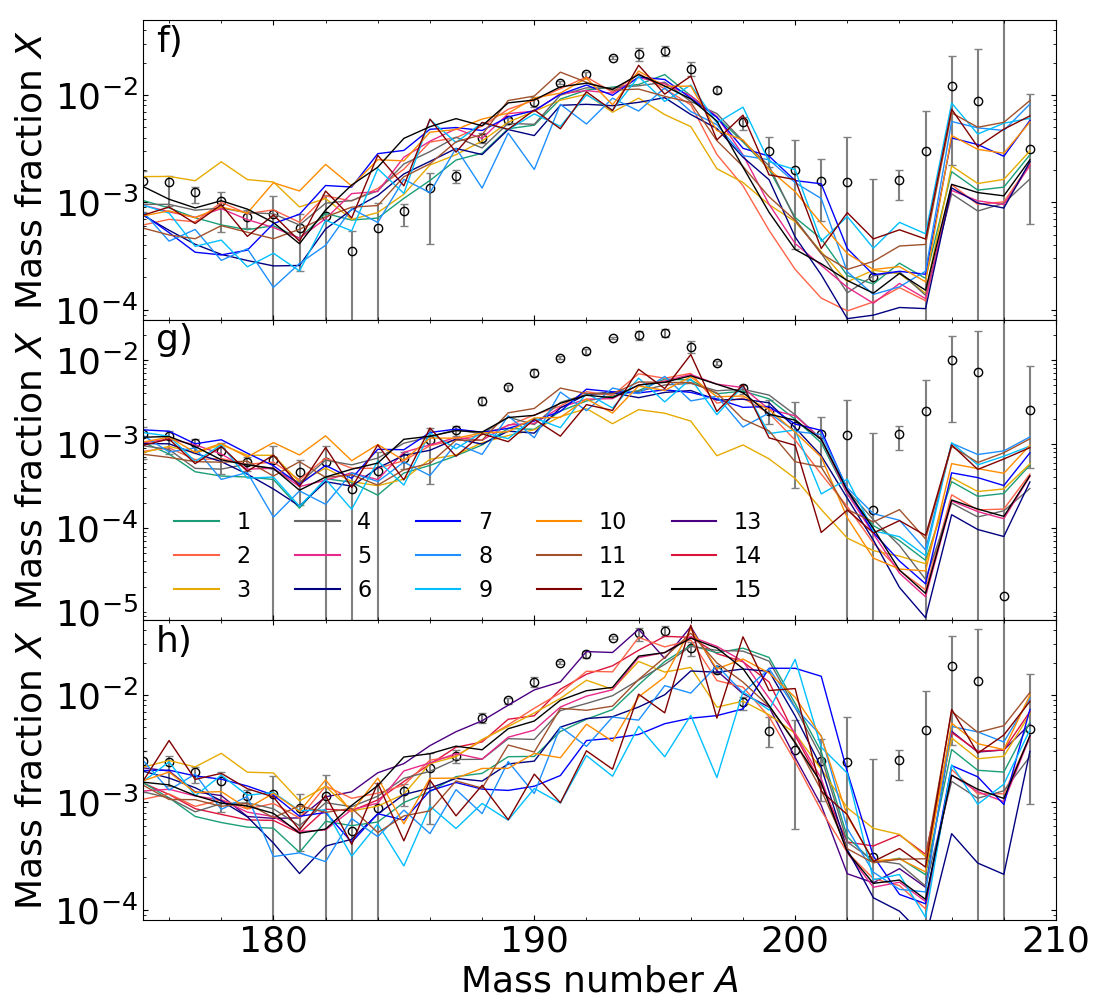}
\caption[Same as \cref{fig_rpro_comb_all} for the combined ejecta models zoomed in on the third r-process peak.]{
Same as \cref{fig_rpro_comb_all} panel f), g) and h) for the combined ejecta models zoomed in on the third r-process peak.}
\label{fig_rpro_comb_zoom}
\end{center}
\end{figure}

\begin{landscape}
\begin{figure*}
\begin{center}
\includegraphics[width=1.45\textwidth]{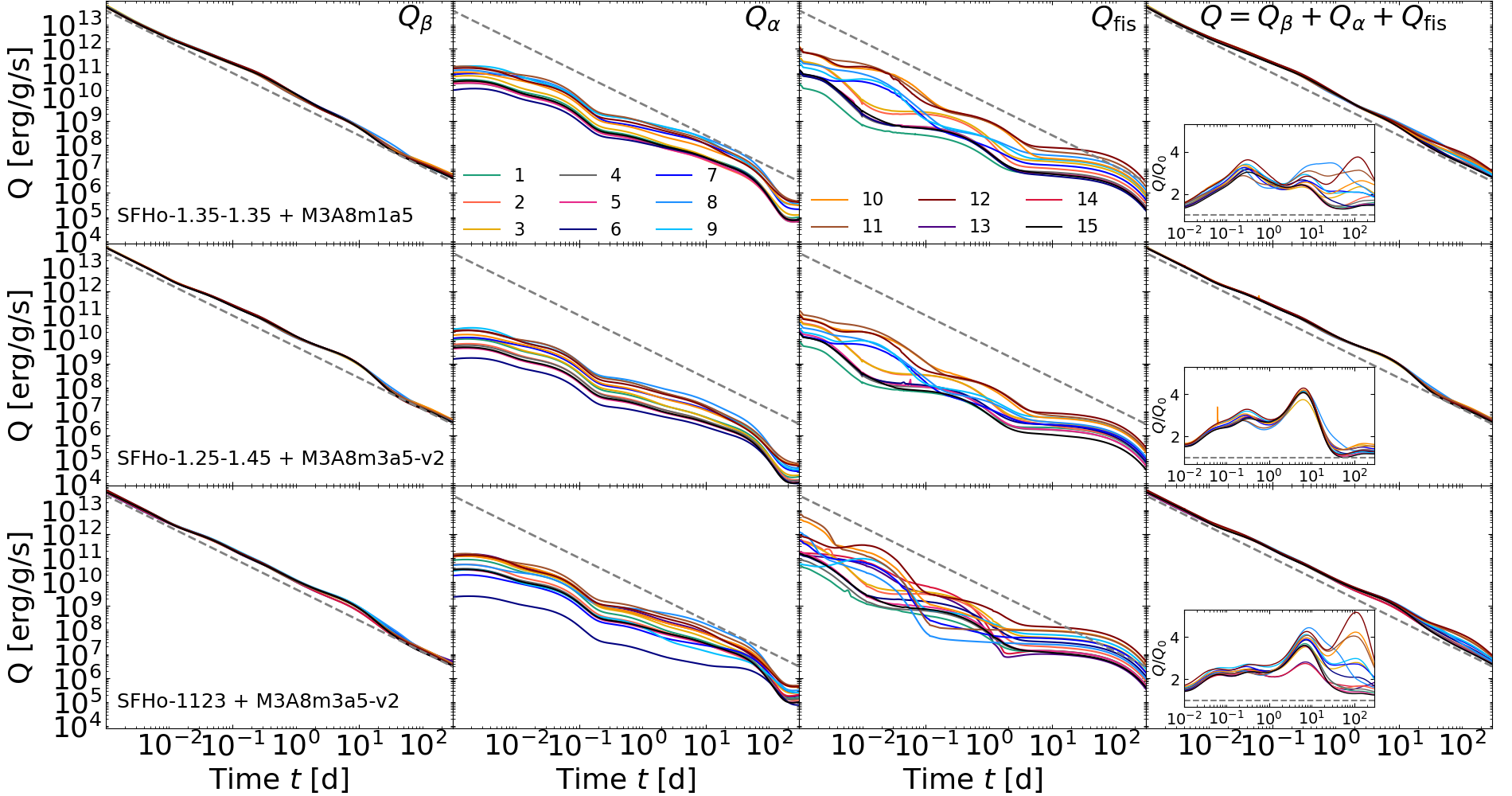}
\caption[Time evolution of the radioactive heating rate for the combined dynamical and secular ejecta when varying the nuclear input between all models listed in \cref{tab_nuc_mods}.]{
Time evolution of the radioactive heating rate $Q$ (before thermalization) for the combined dynamical and secular ejecta for model SFHo-135-135 plus M3A8m1a5 (top), SFHo-125-145 plus M3A8m3a5-v2 (middle) and SFHo-11-23 plus M3A8m3a5-v2 (bottom) when varying the nuclear input between all models listed in  \cref{tab_nuc_mods}.
The columns displays the heating generated by $\beta$-decay ($Q_\beta$), $\alpha$-decay ($Q_\alpha$), fission ($Q_\mathrm{fis}$) and the total heat from all or the decay modes ($Q=Q_\beta + Q_\alpha + Q_\mathrm{fis}$).
The ratio $Q/Q_0$ (e.g., see \cref{eq:Q_0}) is also shown as an insert in the right column, where the grey dashed line indicates $y=1$.
See \cref{fig_rprowind_u1-7,fig_rprowind_u7-10,fig_rprowind_u3-13,fig_rprowind_DC} for the colour labels for each nuclear input combination. \todo{[do in paper too, move this fig earlier]}
}
\label{fig_Qmodes}
\end{center}
\end{figure*}
\end{landscape}

\begin{landscape}
\begin{figure}[tbp]
\begin{center}
\includegraphics[width=1.45\textwidth]{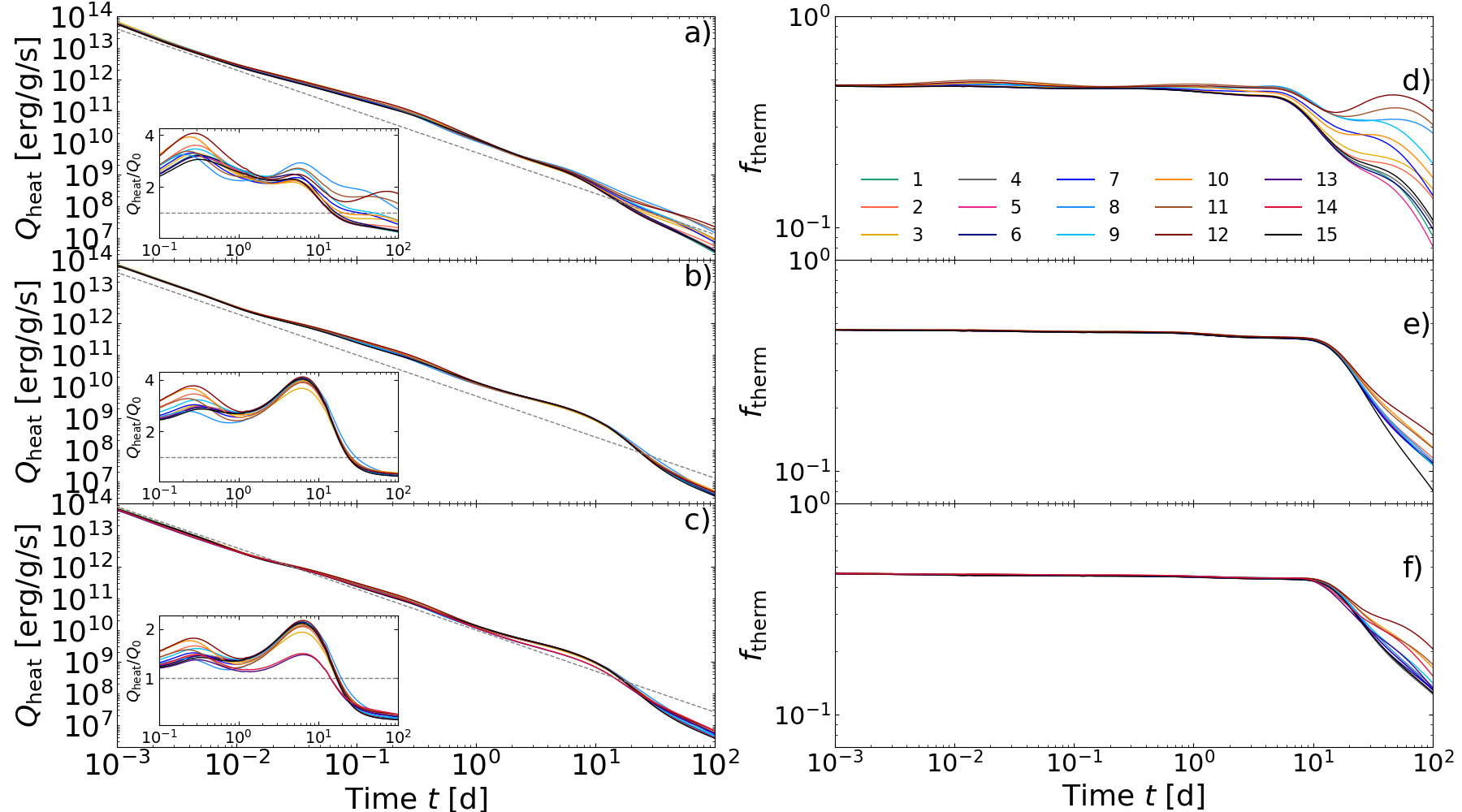}
\caption[Time evolution of the effective heating rate and the total thermalization efficiency for the combined dynamical and secular ejecta models when varying the nuclear input between all models listed in  \cref{tab_nuc_mods}.]{
Time evolution of the effective heating rate $Q_\mathrm{heat}$ (left column) and the total thermalization efficiency (right column) as defined in \cref{eq:Qheat} for the combined dynamical and secular ejecta for model SFHo-135-135 plus M3A8m1a5 (top), SFHo-125-145 plus M3A8m3a5-v2 (middle) and SFHo-11-23 plus M3A8m3a5-v2 (bottom) when varying the nuclear input between all models listed in  \cref{tab_nuc_mods}. 
The ratio $Q_\mathrm{heat}/Q_0$ (e.g., see \cref{eq:Q_0}) is also shown as an insert for $t>0.1$~d in the left panels, where the grey dashed line indicates $y=1$.
See for example \cref{fig_rprowind_u1-7,fig_rprowind_u7-10,fig_rprowind_u3-13,fig_rprowind_DC} for the colour labels for each nuclear input combination. 
}
\label{fig_Qheat}
\end{center}
\end{figure}
\end{landscape}

The sensitivity to the nuclear physics input when we combine the material from the dynamical ejecta with the post-merger BH-torus ejecta are displayed in \cref{fig_rpro_comb_all,fig_rpro_comb_zoom,fig_Qmodes,fig_Qheat}. Since the ejected mass of the BH-torus systems dominates over the mass ejected from the dynamical component for the NS-NS systems (see \cref{tab:astromods}), the total r-process abundance distribution (and its uncertainty due to the nuclear physics input) in \cref{fig_rpro_comb_all}(f-g) mostly resembles that of the disk ejecta. However, the dynamically ejected mass from the NS-BH merger system (model SFHo-11-23) is of the same order of magnitude as the one ejected from the BH-torus model M3A8m3a5-v2. In this case, the total ejecta given in \cref{fig_rpro_comb_all}h shows a composition where both components play a significant role, which resembles the solar system distribution fairly well for $A>90$. See \cref{fig_rpro_comb_zoom} for a zoom in on the third r-process peak for the combined models shown in \cref{fig_rpro_comb_all}.

The radioactive heat generated from the separate decay modes $Q_\beta$, $Q_\alpha$ and $Q_\mathrm{fis}$ and the total heating rate $Q$ are displayed in \cref{fig_Qmodes}. 
The heating rate is generally dominated by heat stemming from $\beta$-decays. However, at $t>10$~d, the contributions from fission and $\alpha$-decay become significant.  Overall, we observe small relative variations ($\lesssim2$) for the $\beta$-decay heat, whereas the heating rates produced by $\alpha$-decay and fission can vary by up to 1-2 orders of magnitude.

The impact of varying the nuclear physics input on the shape and magnitude of $Q_\alpha$ is directly linked to the number of heavy r-process elements produced and the detailed decay path taken back to the stable or long-lived nuclei, while it is not affected by the involved $\alpha$-decay rates since these are experimentally known.
Model SFHo-125-145 plus M3A8m3a5-v2 generates fewer trans-Pb species than models SFHo-135-135 plus M3A8m1a5 and SFHo-11-23 plus M3A8m3a5-v2,  and therefore $Q_\alpha$ has an insignificant impact on the total heating rate for this model. 
All of the nuclear input sets generate heat from the $\alpha$-decay chain starting from $^{224}$Ra, however, only a few input sets have non-negligible contributions from other decay chains\footnote{See the supplemental material of \citet{wu2019} for a complete list of the most important $\alpha$-decay chains contributing to the heating rate and also the discussion in \cref{sec_q} for the dynamical ejecta models applied in this work.}.
In particular, for input sets 9, 10, 12 and 13 that apply the TDA or FRDM+QRPA $\beta$-decay models (see \cref{tab_nuc_mods}), the heat generated by the $\alpha$-decay chains starting from $^{223}$Ra, $^{225}$Ac or $^{221}$Fr leads to a large increase in $Q_\alpha$. 
This is related to the faster $\beta$-decay rates for the most neutron deficient $Z>83$ nuclei (see \cref{fig_betadiff})
for models TDA and FRDM+QRPA, but also to the larger production of heavy r-process elements for these models. 

The spontaneous fission of $^{254}$Cf completely dominates the $Q_\mathrm{fis}$ curve at late times; no other fissile nuclei generate any considerable heat.
Therefore, the impact on $Q_\mathrm{fis}$ arising when varying the nuclear physics input is related to the ability of a particular input set to produce the heaviest elements, which is not only sensitive to the various fission models but all of the nuclear physics properties considered in this work. 

At $t\sim 10$~d, another enhancement can be seen in the heating rate, which is particularly strong for the combined ejecta models involving the BH-torus model M3A8m3a5-v2 (i.e., the two bottom panels of  \cref{fig_Qmodes}). This structure in the heating rate is related to the $\beta$-decay of $^{132}$I, which forms the over-abundance peak at $A=132$ seen in \cref{fig_rprowind_u1-7}, as discussed earlier. \todo{[explain why $^{132}$I is so important here, is it because it is populated at late times? since half live is 2h]}

\cref{fig_Qheat} shows the effective heating rate $Q_\mathrm{heat}$ as well as the total thermalization efficiency $f_\mathrm{therm}$ (see the definition in \cref{sec_kilo} and \cref{eq:Qheat}) relevant for kilonova models. Note that $f_\mathrm{therm}$ starts well below 1 due to the energy lost to neutrinos (contained in $Q_\nu$).
\todo{from Oli paper part 2: The average thermalization efficiencies for all investigated models as functions of time are displayed in Fig. 1. Almost perfect thermalization (Qheat ≈ Q) prevails until about t ∼ 0.5−1 d, whereafter fth declines roughly as ∝ t−1 . Lower ejecta masses or faster expansion velocities accelerate the decay of fth. For completeness, we also plot the fraction f neu ≡ Qneu/(Q + Qneu ) of energy that is carried away by neutrinos, which amounts to about ≈ 50 − 60 per cent in our models.}
For the astrophysical models studied here, we can see that the biggest impact on the thermalization efficiency comes when varying the $\beta$-decay rates and fission properties. 
More specifically, a significant enhancement is found when adopting the FRDM+QRPA, TDA or HFB21+GT2 models (input sets 7-12) compared to the RHB+RQRPA rates, in particular for model SFHo-135-135 plus M3A8m1a5.  
As discussed above, varying the $\beta$-decay model leads to variations in $Q_\alpha$ and $Q_\mathrm{fis}$ (see \cref{fig_Qmodes}) which in turn contributes to the $f_\mathrm{therm}$ variations seen for $t>10$~d in \cref{fig_Qheat}. 
Varying the fission properties also impacts the thermalization efficiency, however, to a lesser extent than varying the $\beta$-decay model.  
The masses have a secondary impact on the thermalization efficiency, though results with FRDM (input set 2) and WS4 (input set 3) mass models deviate significantly from the other models (input sets 1, 4-6). 
The deviations between the nuclear physics input sets found in $Q_\mathrm{heat}$ and $f_\mathrm{therm}$, in particular at late times, are expected to impact the kilonova light curve of the total ejecta if included in a model such as the one presented in \cref{sec_kilonova} for the dynamical ejecta.
\todo{[that whole paragraph above is hard to follow since since figures does not show model names, will it be better when have updated figs? (ref. report)]}

\cref{tab_rpro} summarizes several r-process properties, 
including the mass fraction of strontium, lanthanides plus actinides ($X_\mathrm{LA}$), $A>69$ and $A>183$ nuclei, and the heating rate at 1~d for all models studied herein. The lanthanide and actinide content of the ejecta ($X_\mathrm{LA}$ and $X_{A>183}$) and the heating rates ($Q$) are particularly relevant for kilonova modelling, which is outside the scope of this work but listed here for completeness. 
\todo{referee points out that dont discuss the other values in this table. also update to same sentence as here.}

\begin{landscape}
\begin{table*}
\centering
\caption[Summary of several r-process quantities and their variations arising from the nuclear physics uncertainties for all seven hydrodynamical models.]{ 
Summary of several r-process quantities and their variations arising from the nuclear physics uncertainties for all seven hydrodynamical models.
The strontium mass fraction $X_\mathrm{Sr}$, lanthanide plus actinide mass fraction $X_\mathrm{LA}$, the mass fractions of r-process nuclei ($A>69$) and third-r-process peak nuclei ($A>183$), and the radioactive heating rate before thermalization $Q$ at $t=1$~d for the three dynamical, two BH-torus and the two combined ejecta models considered in the present study.
For each quantity, the minimum (min), maximum (max), mean and standard deviation from the mean ($\sigma$) arising from the variation of the nuclear physics input sets in \cref{tab_nuc_mods} are shown.
\todo{[this table should be formatted to fit the page} \todo{paper 2: should reference and move this table earlier (now only in conclusions??)]}
}
 \begin{tabular}{lcccccc}
\hline
\hline
              & & $X_\mathrm{Sr}$ & $X_\mathrm{LA}$ & $X_{A>69}$ & $X_{A>183}$ & $Q(t=1\mathrm{d})$  \\
              & &          &          &            &             & [$10^{10}$ erg/g/s] \\
\hline
SFHo-125-145 & min  & 0.01 & 0.09 & 0.95 & 0.11 & 0.99 \\
     & max & 0.04 & 0.14 & 0.97 & 0.17 & 1.22 \\
     & mean      & 0.02 & 0.12 & 0.97 & 0.14 & 1.11 \\
     & $\sigma $ & 0.005 & 0.015 & 0.005 & 0.021 & 0.054 \\
\hline
SFHo-135-135 & min  & 0.03 & 0.09 & 0.93 & 0.10 & 1.00 \\
     & max & 0.05 & 0.14 & 0.95 & 0.17 & 1.21 \\
     & mean      & 0.04 & 0.11 & 0.95 & 0.14 & 1.10 \\
     & $\sigma $ & 0.004 & 0.015 & 0.006 & 0.021 & 0.047 \\
\hline
   SFHo-11-23 & min  & $4\cdot10^{-5}$ & 0.21 & 0.99 & 0.19 & 0.78 \\
     & max & $3\cdot10^{-4}$ & 0.38 & 1.00 & 0.52 & 1.15 \\
     & mean      & $8\cdot10^{-5}$ & 0.30 & 0.99 & 0.43 & 0.91 \\
     & $\sigma $ & $6\cdot10^{-5}$ & 0.056 & 0.002 & 0.100 & 0.112 \\
\hline
\hline 
 \end{tabular} 
\label{tab_rpro}
\end{table*}

\begin{table*}
\centering
 \begin{tabular}{lcccccc}
\hline
\hline
              & & $X_\mathrm{Sr}$ & $X_\mathrm{LA}$ & $X_{A>69}$ & $X_{A>183}$ & $Q(t=1\mathrm{d})$  \\
              & &          &          &            &             & [$10^{10}$ erg/g/s] \\
\hline
 M3A8m3a5-v2 & min  & 0.01 & 0.06 & 0.93 & 0.01 & 1.15 \\
     & max & 0.04 & 0.10 & 0.96 & 0.13 & 1.42 \\
     & mean      & 0.03 & 0.08 & 0.95 & 0.05 & 1.32 \\
     & $\sigma $ & 0.008 & 0.012 & 0.007 & 0.032 & 0.084 \\
\hline
    M3A8m1a5 & min  & 0.01 & 0.06 & 0.94 & 0.09 & 1.15 \\
     & max & 0.03 & 0.11 & 0.95 & 0.15 & 1.48 \\
     & mean      & 0.02 & 0.08 & 0.95 & 0.12 & 1.32 \\
     & $\sigma $ & 0.005 & 0.015 & 0.004 & 0.021 & 0.083 \\
\hline
SFHo-125-145 & min  & 0.02 & 0.07 & 0.93 & 0.02 & 1.18 \\
+ M3A8m3a5-v2 & max & 0.04 & 0.11 & 0.95 & 0.05 & 1.40 \\
     & mean      & 0.03 & 0.08 & 0.95 & 0.05 & 1.32 \\
     & $\sigma $ & 0.005 & 0.013 & 0.005 & 0.007 & 0.060 \\
\hline
SFHo-135-135 & min  & 0.02 & 0.07 & 0.94 & 0.09 & 1.14 \\
  + M3A8m1a5 & max & 0.03 & 0.12 & 0.95 & 0.15 & 1.45 \\
     & mean      & 0.02 & 0.08 & 0.95 & 0.12 & 1.29 \\
     & $\sigma $ & 0.005 & 0.015 & 0.004 & 0.020 & 0.077 \\
\hline
   SFHo-11-23 & min  & 0.01 & 0.12 & 0.95 & 0.09 & 1.06 \\
+ M3A8m3a5-v2 & max & 0.02 & 0.20 & 0.98 & 0.27 & 1.32 \\
     & mean      & 0.02 & 0.16 & 0.96 & 0.19 & 1.17 \\
     & $\sigma $ & 0.005 & 0.025 & 0.005 & 0.046 & 0.066 \\
\hline
\hline
 \end{tabular} 
\label{tab_rpro2}
\end{table*}
\begin{table}[p]
\centering
\caption[Age estimates of six metal-poor r-process-enhanced stars based on the Th/U cosmochronometry, including uncertainty estimates arising from varying the nuclear physics input.]{ 
Age estimates of six metal-poor r-process-enhanced stars based on the Th/U cosmochronometry (see \cref{eq_cosmo}). The observational ratios, $(\mathrm{Th/U})_\mathrm{obs}$, and their corresponding uncertainty are from 
\citet{sneden2003c} (S03), \citet{Hill2017} (H17), \citet{Mello2013} (M13), \citet{frebel2007} (F07), \citet{holmbeck2018} (H18) and \citet{Placco2017} (P17). 
The $(\mathrm{Th/U})_\mathrm{r}$ ratios are taken directly from our r-process calculations using the three astrophysical models for the combined ejecta: SFHo-135-135 + M3A8m1a5 (comb1), SFHo-125-145 + M3A8m1a5 (comb2) and SFHo-11-23 + M3A8m3a5-v2 (comb3). 
The minimum, maximum and mean of the Th/U ratio, as well as the standard deviation from the mean ($\sigma_{\log(\mathrm{Th/U})_\mathrm{r}}$) and the stellar age ($\sigma_{t^*_\mathrm{r}}$) arising from varying the nuclear input sets listed in \cref{tab_nuc_mods} are given for the three astrophysical models. 
\todo{fix format of this table, the text should not be directly next to the header right?}
}
 \begin{tabular}{lccccccccc}
\hline \hline
 & $\log(\mathrm{Th/U})_\mathrm{obs}$ & $\sigma_{t^*,\mathrm{obs}}$ & ref. & comb1 & comb2 & comb3 \\
\hline
$ \log (\mathrm{Th/U})_\mathrm{r,min}$ & - & - & - & 0.04 & 0.07 & 0.08 \\
$ \log (\mathrm{Th/U})_\mathrm{r,max}$ & - & - & - & 0.31 & 0.37 & 0.34 \\
$ \log (\mathrm{Th/U})_\mathrm{r,mean}$ & - & - & - & 0.17 & 0.21 & 0.18 \\
$\sigma_{\log(\mathrm{Th/U})_\mathrm{r}}$ & - & - & - &0.09 & 0.10 & 0.08 \\
$\sigma_{t^*_\mathrm{r}}$ & - & - & - &2.0 Gyr & 2.1 Gyr & 1.7 Gyr \\
\hline
CS22892-052 & 0.73 $\pm 0.22$ & 4.9 Gyr & S03 & 12.1 Gyr & 11.2 Gyr & 12.1 Gyr \\
CS29497-004 & 1.04 $\pm 0.33$ & 7.2 Gyr & H17 & 18.9 Gyr & 18.0 Gyr & 18.8 Gyr \\
CS31082-001 & 0.94 $\pm 0.21$ & 4.7 Gyr & M13 & 16.7 Gyr & 15.8 Gyr & 16.6 Gyr \\
HE1523-0901 & 0.86 $\pm 0.13$ & 2.8 Gyr & F07 & 14.9 Gyr & 14.1 Gyr & 14.9 Gyr \\
 J0954+5246 & 0.82 $\pm 0.22$ & 4.9 Gyr & H18 & 14.1 Gyr & 13.2 Gyr & 14.0 Gyr \\
 J2038-0023 & 0.90 $\pm 0.20$ & 4.4 Gyr & P17 & 15.8 Gyr & 14.9 Gyr & 15.8 Gyr \\
\hline \hline
 \end{tabular} 
\label{tab_cosmo}
\end{table}
\end{landscape}

\section{Impact on cosmochronometers}
\label{sec:cosmo}

\todo{[confirm that Th/U are calculated using Y not X in paper too (footnote)]}

Based on the sensitivity analysis performed above, we can estimate the impact of the nuclear uncertainties on the production of the Th and U cosmochronometers and the age of specific metal-poor stars for which the surface abundances of Th and U have been determined.
\cref{tab_cosmo} lists the mean estimated ages ($t^*$) for six metal-poor stars, assuming they have been initially polluted by the combined ejecta corresponding to our models for the NS-NS or NS-BH merger.  
The age is calculated by using \cref{eq_cosmo} with the r-process abundance ratio\footnote{We use the molar fractions to calculate the Th/U ratios, not the mass fractions shown in the above figures.} of (Th/U)$_\mathrm{r}$ consistently obtained from our three combined ejecta models and the 15 different nuclear inputs and the observed abundance ratios (Th/U)$_\mathrm{obs}$ from the literature (see \cref{tab_cosmo} for the references).  
In addition, the standard deviation ($\sigma$) from the mean arising when varying the nuclear physics input is listed for the (Th/U)$_\mathrm{r}$ ratio and stellar ages $t^*_r$.
Note that $\sigma_{t^*_\mathrm{r}}$ is identical for all stars within each hydrodynamical model since a change in the observed Th/U ratio only leads to a shift in the estimated age.
If the references do not provide the uncertainty of $\log(\mathrm{Th/U})_\mathrm{obs}$, we calculate it as the square root of the quadratic sum of the individual, observational Th and U uncertainties given and propagate it to the age estimates ($\sigma_{t^*_\mathrm{obs}}$).
The average ages obtained for all different nuclear physics inputs vary from 11.2 to 18.9 Gyr depending on the star and astrophysical model applied. 
The theoretical Th/U ratio for combined models SFHo-135-135+M3A8m1a5 and SFHo-11-23+M3A8m3a5-v2 are similar, leading to almost identical age estimates, while model SFHo-125-145+M3A8m3a5-v2 has a larger Th/U ratio giving in general smaller age estimates. It is also model SFHo-125-145+M3A8m3a5-v2 which has the largest spread in the calculated values when the nuclear physics inputs are varied. For example, the age of star CS22892-052 ranges between 7.9 and 14.4~Gyr for the minimum and maximum ages, respectively. It is always input set 7 (which applies mass model HFB-31 and $\beta$-decay model HFB21+GT2, see \cref{tab_nuc_mods}) that gives rise to the minimum age estimate. The maximum age is found with input sets 5, 4 and 15 (using mass models HFB-21 or D1M) for models SFHo-135-135+M3A8m1a5, SFHo-125-145+M3A8m3a5-v2 and SFHo-11-23+M3A8m3a5-v2, respectively.
The observational uncertainties for the Th/U ratios and the age estimates ($\sigma_{t^*,\mathrm{obs}}$) are also listed in \cref{tab_cosmo}. We calculate the observational uncertainty of the Th/U ratios as the square root of the quadratic sum of the Th and U abundance uncertainties provided by the references listed in \cref{tab_cosmo} and propagate the observational uncertainty of the ratios to the age estimate ($\sigma_{t^*,\mathrm{obs}}$).
We can see that the observational uncertainties are, in general, significantly larger than the uncertainties stemming from the nuclear physics input ($\sigma_{t^*_\mathrm{r}}$).



\section{Comparison with other works}
\label{sec_comp2}

Many r-process studies in the last decade have focused on the impact of the nuclear physics uncertainties on the r-process yields \citep{Caballero2014,Mendoza-Temis2015,Eichler2015,goriely2015,Martin2016,Liddick2016,Mumpower2016,Nishimura2016,Bliss2017,Denissenkov2018,Barnes2021, Vassh2019,Nikas2020,sprouse2020a,McKay2020,Giuliani2020,Zhu2021,Barnes2021}.
Often, such studies aim to identify specific nuclei or regions in the nuclear chart where the r-process has the largest sensitivity to the experimentally unknown nuclear properties. If such nuclei are identified, they can be targeted by experimental campaigns as long as they are within reach of the given facility \citep[e.g.,][]{Surman18}.
Another aim of r-process sensitivity studies is to estimate the magnitude of the nuclear uncertainties so that they can be compared to other sources of uncertainty, like those arising from hydrodynamical modelling. This is particularly important for applications like cosmochronometry, galactic chemical evolution or kilonova models, which require r-process yields from a given site or ejecta component as input.
A large range of astrophysical conditions has been applied in various r-process sensitivity studies in the literature, making detailed comparisons between results difficult. In the following we will, in a quantitative way, compare our results to other sensitivity studies that apply the same or similar nuclear models and ejecta components as we do and as summarized in \cref{tab_comp}.

\begin{landscape}
\begin{figure*}
\begin{center}
\includegraphics[width=1.4\textwidth]{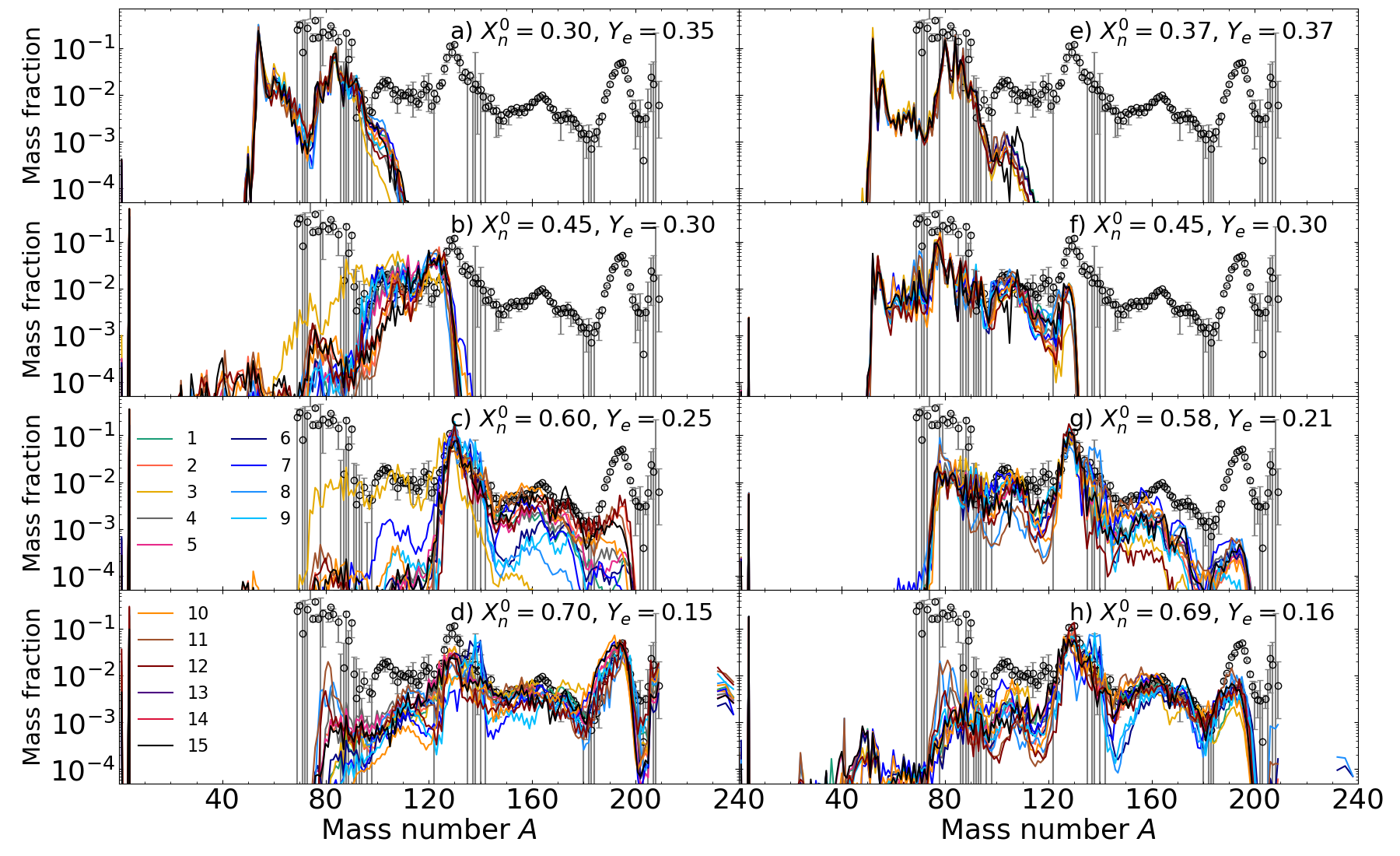}
\caption[Abundance distributions obtained in eight individual trajectories with different initial neutron mass fractions using all nuclear input sets in \cref{tab_nuc_mods}.]{
Abundance distributions obtained in eight individual trajectories with different initial neutron mass fractions (or initial $Y_e$) from model SFHo-135-135 (left column) and M3A8m1a5 (right column) using all of the nuclear input sets listed in \cref{tab_nuc_mods}. 
See for example \cref{fig_rprowind_u1-7,fig_rprowind_u7-10,fig_rprowind_u3-13,fig_rprowind_DC} for the colour labels for each nuclear input combination.
}
\label{fig_rpro_onetraj}
\end{center}
\end{figure*}
\end{landscape}

Since the r-process is very sensitive to the conditions of the environment, it is essential to keep in mind which assumptions and simplifications have been applied for the nucleosynthesis calculations. 
Basically, three approaches to modelling the r-process conditions exist for a given astrophysical site or ejecta component: 1) parametrized trajectories which typically only depend on the initial electron fraction, entropy and expansion time scale, or 2) a single or a few ($<10$) trajectories extracted from a hydrodynamical simulation, or 3) the complete set (or a significant sample) of trajectories representing all the mass elements ejected, as given by a hydrodynamical simulation.\todo{[ref. had comment about this?]}
Although the first two methods are simple idealisations, their obvious benefit is that the required computing time is significantly shorter than the third method, which may require up to several thousand r-process calculations depending on the hydrodynamical model. 
The central assumption adopted when using methods 1) or 2) is that one trajectory (or a few), or parameters characterizing the trajectories, can be chosen in such a way as to represent the conditions of all ejected material.
This assumption is justified only for ejecta components which show minor variations between the trajectories in terms of the evolution of quantities such as electron fraction, temperature and density. 
However, this is not what most hydrodynamical simulations predict \citep{wanajo2014,Mendoza-Temis2015,Just2015,foucart2016,goriely2016b,radice2018a,Siegel2018,ardevol-pulpillo2019}. 
\todo{update in paper too}
\cref{fig_rpro_onetraj} displays the r-process abundance distribution for four individual trajectories from models SFHo-135-135 and M3A8m1a5 with $Y_e$ values ranging from 0.15 to 0.37 using all of the nuclear input sets. 
For the four trajectories shown here, the impact of the nuclear physics uncertainties strongly depends on the adopted trajectory and its neutron richness. For example, the predicted amount of heavy r-process nuclei  ($A\gtrsim 90$) can easily vary by a factor of 100 and more between the adopted nuclear physics input sets when a single or a few trajectories are considered. However, when the r-process abundance distribution is calculated from an ensemble of trajectories representing a range of conditions, the uncertainties due to the nuclear physics input shrink significantly, to at most a factor of about 20 for some specific individual $A>90$ nuclei (see \cref{fig_rpro_comb_all} which is based on $\sim200-300$ hydrodynamical trajectories). 
The comparison of \cref{fig_rpro_onetraj,fig_rpro_comb_all}\footnote{Similar trends can also be seen in Figs.~2a and b in \citet{Zhu2021}.} shows that uncertainty studies based on a single or a few trajectories may artificially exacerbate the impact of nuclear physics uncertainties, particularly in cases of $Y_e\sim0.2-0.3$ close to the threshold of lanthanide production (see also the discussion in \cref{subsec_cluster}). Therefore, one should be careful when drawing conclusions regarding the total r-process yields of a given site or ejecta component from single-trajectory studies.
See also the discussion in \cref{sec_sel_subs} about selecting statistically significant trajectory subsets. 

Another critical point when comparing sensitivity analyses concerns how nuclear uncertainties are propagated to nucleosynthesis calculations. A popular technique used to propagate the nuclear uncertainties to the final r-process results is to increase or decrease, for example, the nuclear mass or the neutron capture rate of a single nucleus by a given factor \citep[e.g., see ][who applied factors of 5, 10, 50 or 100 to the $(n,\gamma)$ or $(\alpha,n)$ rates]{Surman18,Bliss2017}. 

\begin{landscape}
\begin{table*}
\centering
\caption[Comparison of ejecta properties with previous works and the method applied in their nuclear physics sensitivity studies.]{
Comparison with previous works: summary of the adopted ejecta conditions, number of trajectories used ($N_\mathrm{traj}$), the origin of the trajectories (either hydrodynamical simulations or parametrized trajectories), the method used to vary the nuclear physics input (systematic or MC) and an indication of whether the study includes what we refer to as ``advisable'' models, see \cref{sec_network_input} or \cref{sec_comp2} for details. 
The ejecta components refer to either the dynamical ejecta of NS-NS or NS-BH mergers (dyn.) or secular wind ejecta in BH-torus remnant systems (wind). 
The following references have been included in our comparison: \citet{Barnes2021,Zhu2021} (B21\& Z21), \citet{Mendoza-Temis2015} (MT15), \citet{Eichler2015} (E15), \citet{Vassh2019} (V19), \citet{Giuliani2020} (G20), \citet{Caballero2014} (C14),\citet{marketin2016} (M16), \citet{Nikas2020} (N20) and \citet{Mumpower2016} (M16). Note that the work of M16 summarizes the works of several MC studies.
}
\begin{tabular}{cccccc}
\hline 
Ref. & Ejecta comp. & $N_{\mathrm{traj}}$ & Traj. origin & Var. method & ``advisable'' models \\ 
\hline 
This work & wind, dyn. & 150-296 & hydro & systematic & yes \\ 
B21\& Z21 & wind & 1 & param. & systematic & only some \\ 
MT15 & dyn. & 528 & hydro & systematic & yes \\ 
E15 & dyn. & 30 & hydro & systematic & only some \\ 
V19 & dyn. & 1 (\& 30) & hydro & systematic & only some \\ 
G20 & wind, dyn. & 1 & hydro \& param. & systematic & yes \\ 
C14 & dyn. & 1 & hydro & systematic & yes \\
M16 & dyn. & 1 & hydro & systematic & yes \\
N20 & wind, dyn. & 1 & hydro & MC & only some \\ 
M16 & wind, dyn. & 1 & hydro \& param. & MC & yes \\ 
\hline 
\end{tabular} 
\label{tab_comp}
\end{table*}
\end{landscape}

\noindent 
Then, for each variation or change of, for example, the nuclear mass of a given nucleus, the r-process abundances are re-calculated and compared to a base calculation with fixed nuclear (and astrophysical) input. After multiple variations, i.e., r-process calculations, the nuclei for which the abundances are the most sensitive with respect to changes in a given nuclear property are revealed.
This technique can be seen as a variant of the MC method \citep[see][for details about various implementations]{Mumpower2016,Rauscher2020} since it applies random variations of quantities, such as mass, $\beta$-decay half-lives, neutron capture rates or $\beta$-delayed neutron emission, pulled from a distribution which represents the assumed nuclear uncertainty of the property investigated.
When considering uncertainties arising from (theoretical) nuclear physics inputs such as, for example, the nuclear mass or neutron capture rates, there are two sources of uncertainty, namely statistical and systematic errors. The latter is often referred to as model errors. 
MC studies adopt a given nuclear model for the baseline calculation (i.e., they have to choose the mean of the uncertainty distribution) and consider uncorrelated variations around that baseline. Therefore, by design, MC methods and similar techniques can only probe uncorrelated statistical errors.
The claim that the independent and random variations of the nuclear properties of individual nuclei can probe all nuclear uncertainties, including the systematic model uncertainties \citep[][and references therein]{Mumpower2016}, would only be correct if the nuclear properties addressed by the MC studies were uncorrelated, which they are not. As discussed in \cref{sec_nucin}, theoretical physical models considered to estimate all nuclear inputs of relevance for the r-process calculations are responsible for the nuclear correlations between various nuclei involved as well as between different properties of interest. 
For the neutron-rich nuclei relevant to the r-process, for which no experimental information is available, the statistical errors due to variations of model parameters have been shown to be much smaller than the systematic (or model) uncertainties 
\citep[e.g., see][who discuss the extrapolation uncertainties of mass models]{Goriely2014a}. Therefore, the magnitude of the statistical errors, i.e., the width of the uncertainty distributions often assumed in MC studies, may be significantly overestimated, leading to uncertainty bands spanning several orders of magnitude, as shown in Figs.~10 and 11 of \citet{Mumpower2016} or Fig.~13 of \citet{Nikas2020}. With such large uncertainty bands\footnote{
Note that such uncertainty bands can easily be misinterpreted since they only display the uncorrelated uncertainty ranges for individual nuclei and can therefore not be used to interpret the uncertainty between neighbouring nuclei. 
For example, suppose nucleus $A$ has the maximum (or minimum) abundance. In that case, it is incorrect to assume that the possible abundance range (and therefore the uncertainty) of nucleus $A+1$ is as large as the width of the band.  
This problem can be avoided by displaying the individual distributions that gave rise to the uncertainty range together with the band.
}, it is difficult to obtain detailed information about the overall r-process distribution (i.e.,  the shape of the r-process peaks and the ratio between them are hidden within the bands).
This is not the case when systematically and coherently propagating nuclear uncertainties based on global nuclear models (e.g., \cref{fig_rpro_comb_all}).
\todo{referee doesnt like ``coherently'' in this sentence}


\todo{referee doesnt like ``inadvisable'' wording}
As discussed in \cref{sec_nucin}, not all available global nuclear models are suitable for astrophysical applications. For example, mass models that perform poorly in terms of the rms value with respect to available experimental data should not be included when probing the nuclear uncertainties of the r-process. Works that use what we subjectively deem as ``not advisable'' models (see also \cref{sec_nucin}) are indicated in \cref{tab_comp} to caution the reader that these works might achieve excessively large uncertainty bands in their r-process sensitivity studies. For each study, \cref{tab_comp} lists the type of ejecta conditions adopted, the number of and origin of the trajectories used, as well as the method applied to vary the nuclear physics input.
In particular, the works of \citet{Barnes2021} and \citet{Zhu2021} used single parametrized trajectories 
for a given $Y_e$ value, fixed expansion time scale and fixed entropy to mimic disk wind conditions. 
They applied an extensive set of the existing mass models in their input variations, which led to a large uncertainty band spanning up to a factor of 100, in particular for the low-$Y_e$ case shown in Fig.~2 in \citet{Zhu2021}. However, if we exclude the nuclear mass models (such as DZ, UNEDF1, SLy4, TF and ETFSI) that we judge to be ``inadvisable'', their results get closer to ours in terms of the magnitude of the uncertainty stemming from the nuclear mass models. Similarly, for the heating rate shown in Fig.~3 of \citet{Zhu2021}, the factor between the maximum and minimum values can be as large as 1000 for the low-$Y_e$ case if all their applied mass models are included but less than a factor of 100 if only WS3, HFB-22, HFB-27 and FRDM12 mass models are considered. 
Similar results are found in Fig.~1 of \citet{Barnes2021} for the nuclear uncertainties on the r-process abundance distribution and heating rates.

Considering now studies using a larger set of trajectories, \citet{Eichler2015} considered 30 trajectories based on a Newtonian hydrodynamical simulation of the low-$Y_e$ dynamical ejecta for their r-process calculations, and \citet{Mendoza-Temis2015} considered 528 low-$Y_e$ relativistic SPH trajectories. Both studied the impact of varying between a few mass models rather similar to those considered here. 
Their results are shown in Figs.~5 and 6 in \citet{Eichler2015} and \citet{Mendoza-Temis2015}, respectively, are similar to ours (see in particular \cref{fig_xn0_ye_distr}g corresponding to the somewhat similar low-$Y_e$ conditions) in that they predict relatively small abundance variations. 
This is related to the similar nuclear physics input adopted and the fact that they used an ensemble of trajectories instead of single trajectories (see \cref{sec_sel_subs}). \todo{referee has problems with this sentence}

\citet{Eichler2015} also investigated the impact of four fission fragment distribution models on the final r-process abundance distribution. However, they did not implement the SPY or GEF distributions, as we do, but the empirical models of \citet{Kodama1975} and \citet{Panov2001,Panov2008} and the statistical approach by the ABLA07 code \citep{Kelic2008,Kelic2009}. Fig.~4b of \citet{Eichler2015} shows an impact in the $A\sim140-150$ region comparable to the one observed in our \cref{fig_rpro_ufiss} between the GEF and SPY models. 
Similarly, Fig.~8 in \citet{Vassh2019} based on a single $Y_e<0.02$ dynamical ejecta trajectory, as well as the upper and middle panels of Fig.~3 in \citet{Giuliani2020} based on a similar low-$Y_e$ single trajectory, show a significant impact around $A\sim150$ when varying between different models for the fission fragment distributions. 
The largest divergence can be found between an oversimplified symmetric ($A_\mathrm{frag} = \frac{1}{2}A_\mathrm{fission}$) relation for the fragments and the GEF fission fragment distributions in \citet{Vassh2019}, or the HFB-14 and FRDM+TF fission barriers in \citet{Giuliani2020}, both of which are on the same order of magnitude as the deviations we observe between SPY and GEF. 
Figs.~8 and 9 of \citet{Zhu2021} show the impact of assuming symmetric fragments versus a double-peaked Gaussian fission fragment distribution on the heating rates. In order to compare trajectories with the same $Y_e$ and the same nuclear mass model, cases 3 and 4 (blue lines) should be compared, where the relative contribution to the heating rate from $\alpha$-decays and fission changes between the cases, leading to a discrepancy of about a factor of 2, which also impacts the final kilonova light curve shown in their Fig.~11.
Similarly, in Figs~11 and 17 in \citet{Vassh2019}, a three-order-of-magnitude discrepancy in the heating generated by neutron-induced fission is found when varying between symmetric distributions and the GEF fission fragment distributions.
In summary,  the r-process studies that varied the fission properties in their nuclear physics input showed a significant impact on the shape of the r-process abundance distribution. 
In particular large deviations were found when the simplistic symmetric fission fragment distributions were applied, and consequently, such approximations should be avoided in astrophysical applications. 
For the works mentioned here, the magnitude and shape of the r-process abundance distribution are generally found to be more robust with respect to changes in the nuclear physics input when a more extensive set of trajectories are considered, and the input is limited to ``advisable'' models. 
%
%
%
%
%
%
%

Most r-process sensitivity studies have focused on the variation of mass models and neutron capture rates; however, a few have also estimated the impact of varying the theoretical $\beta$-decay rates. In particular, \citet{Eichler2015} implemented the RHB+RQRPA and FRDM+QRPA rates also applied in our work. 
Their Figs. 14a and b show that the $\beta$-decay rates can impact the shape of the third r-process peak in a way comparable to what we find in our NS-BH scenario (in \cref{fig_rpro_u3-13,fig_rpro_u7-10}). The RHB+RQRPA rates reproduce the solar distribution for the low-$A$ end of the peak, while the FRDM+QRPA rates cause deviations from the solar distribution and give rise to a narrower peak. The same trends can also be observed in the lower panel of Fig.~14 in \citet{marketin2016}, Fig.~2 of \citet{Caballero2014} and Fig.~16 of \citet{Vassh2019}.

While our approach to estimating the nuclear uncertainties on the r-process yields is only based on theoretical models that have good accuracy and extrapolation reliability (see \cref{sec_nucin} for more details),  some limitations are present. First of all, due to the limited number of available global nuclear physics models, most of the input sets applied in our r-process calculations are not consistent in terms of, for example, the nuclear mass model and the masses used to calculate the $Q_\beta$-values in the $\beta$-decay models, or the masses used to calculate the fission properties. 
These inconsistencies need to be further improved in future nuclear modelling. 
In addition, as described in \cref{sec_sel_subs}, the $Y_e$ distribution of the trajectory subset for model M3A8m3a5-v2 does not perfectly match that of the complete trajectory set. 
Thus, the method used to select the subset of model M3A8m3a5-v2 should be improved in future studies.

\todo{referee wants to cite sprouse2020a}



\section{Summary and Conclusions}
\label{sect_concl_p2}

In this chapter, we report an extensive study of the uncertainties related to the theoretical nuclear models, which are used as input to r-process nuclear network calculations, and propagate them consistently to the r-process nucleosynthesis to estimate their impact on the abundance distribution and the subsequent radioactive heating rate. 
Our r-process calculations are based on detailed hydrodynamical models, which describe the time evolution of the dynamical and post-merger ejecta, including neutrino interactions and viscosity. The hydrodynamical models do not cover all possible evolutionary paths but include three representative cases of the NS-NS or NS-BH systems, and two subsequent models for the BH-torus system which is expected to form after the merger. These two ejecta components are then combined into the total ejecta. For the NS-NS merger system, we use one system with equal-mass NSs, SFHo-135-135, and one asymmetrical model, SFHo-125-145, which are combined with BH-torus models M3A8m1a5 and M3A8m3a5-v2, respectively. Additionally, we employ one model to describe the dynamical ejecta of a NS-BH merger, model SFHo-11-23, which we combine with model M3A8m3a5-v2 (see \cref{tab:astromods}).
With this setup, we study the impact of varying the nuclear physics input on the r-process nucleosynthesis, the heating rate and thermalization efficiency, as well as the predicted age of six metal-poor stars by Th/U cosmochronometry. 
We vary six nuclear mass models, two frameworks for calculating the radiative neutron capture rates, four $\beta$-decay models and two sets of fission barriers and fragment distributions, as detailed in \cref{tab_nuc_mods}. 
Only global mass models that have an rms value smaller than 0.8~MeV are included in this work. 

For each nuclear input set, $\sim200$-300 representative trajectories of the complete set of $\sim1000-13000$ trajectories (depending on the hydrodynamical model applied) are used to run the r-process nuclear network calculations (see \cref{sec:nuc_net} for more details). This way, our nucleosynthesis trajectories consistently sample a wide range of thermodynamical conditions encountered at different ejection angles and merger phases.
This is in contrast to other sensitivity studies, which often only used one single trajectory, or very few trajectories, to represent entire ejecta components (see \cref{sec_comp2} and \cref{tab_comp}).

The main conclusions from our study can be summarized as follows:
\begin{itemize}
\item The uncertainties connected to the nuclear physics input have a minor impact on the global shape of the r-process abundance distribution for the astrophysical scenarios studied here. Our combined ejecta models (consisting of the dynamical NS-NS or NS-BH merger plus the secular BH-torus ejecta) reproduce the solar system distribution well for $A>90$ and yield a significant amount of Th and U, irrespective of the adopted nuclear physics model. In addition,  only small changes in the amount of heavy r-process elements ($X_{A>69}$) and lanthanides plus actinides ($X_\mathrm{LA}$) are found (see \cref{tab_rpro}) for all model variations.
However, when studying the detailed distribution shape, a variation between different nuclear physics models can have a significant (local) impact (see \cref{fig_rpro_comb_zoom}). 
When fission plays an important role (i.e., for the NS-BH case), the most considerable impact on the abundance distribution and the heating rate is found in connection with the fission barrier height and fragment distribution. 
\item The position, shape and width of the r-process peaks vary with the nuclear physics input used but also with the ejecta model. In particular, when mass model HFB-31 is applied, an additional structure at $A\sim 130-140$ appears. For BH-torus model M3A8m3a5-v2, a very narrow peak around $A\sim 132$ is formed, which is not observed for the other BH-torus or dynamical models. For the high-$X_n^0$ (e.g., low-$Y_e$) conditions in the NS-BH merger, the width of the third r-process peak is very sensitive to the adopted $\beta$-decay rates, where the RHB+RQRPA rates of \citet{marketin2016} lead to better agreement with the solar r-process distribution, as also found in other studies (see \cref{sec_comp2}).
\item While globally, all nuclear models give rise to a relatively similar ejecta composition, deviations up to a factor $\sim20$ can be found for some specific individual abundances of elements with $A>90$, in particular around the second and third r-process peaks (the latter being most prominent for the NS-BH merger scenario). The factor between the largest and smallest yields of actinides ranges between 5 and 7. 
 \item The radioactive heating rate before thermalization is found to be relatively insensitive (within a factor of $\lesssim2$) to variations of the nuclear physics input at early times ($t \la 10$~d). However, more significant deviations are found in particular related to the contribution from fission in cases where the heaviest elements are produced (i.e., for the NS-BH case at $t \ga 5$~d).
The thermalization efficiency introduces an additional spread between different nuclear physics models, particularly at late times when alpha-decay and spontaneous fission become important.
\item We find a similar order of magnitude for the uncertainties due to the nuclear physics input when we compare our results to the works of others, which also systematically varied the nuclear input between global models similar to the ones applied here. 
A larger impact, up to a factor of 100 on individual $A>90$ r-process nuclei and 1000 on the heating rates, can be found when applying global models which, we do not recommend for astrophysical applications due to their poor performance in accurately reproducing measured data \citep{Barnes2021,Zhu2021}, or when considering MC-type methods \citep[e.g.,][and references therein]{Mumpower2016}.
\item For the astrophysical scenarios studied here, the nuclear physics uncertainties are typically small compared to variations related to the different ejecta components.
Moreover, they are also small (or at most similar) compared to the variations encountered when using different disk masses BH masses, and BH spins in BH-torus simulations \citep[e.g.,][]{Wu2016,Just2021c} or different EoSs in NS-NS merger simulations \citep[see \cref{ch_dynweak} and also][]{kullmann2021,sekiguchi2015,radice2016}. This circumstance strengthens the credibility of nucleosynthesis analyses based on hydrodynamical models of these systems, and it lends further support to the notion that NS mergers can be significant sources of r-process elements in the Universe.
\item For the stellar Th/U cosmochronometry age estimates, the uncertainties connected to the nuclear physics input are still larger than the variations associated with changing the hydrodynamical model (for the cases studied here). The mean age inferred by our method for the six metal-poor stars varies between $11\pm 2$ and $19\pm2$ Gyr for stars CS22892-052 and CS29497-004, respectively.
So far, the observational uncertainties (on the order of $\sim$3-7 Gyr) remain significantly larger than the nuclear physics ones.
\end{itemize}

It is difficult to pin down a single nucleus (or a few nuclei) having a dominant impact on the nuclear r-process uncertainties. First, when changing the nuclear input from one model to another, many nuclei are affected in a systematic way (in contrast to other types of sensitivity studies, which neglect systematic correlations and change, for example, the neutron capture rate of individual reactions randomly within a given range). 
Second, because the r-process conditions can vary between the astrophysical scenarios, the significance of a nuclear process can also vary, which would change the sensitivity resulting from a specific nuclear mass region.
Therefore, continued efforts on the experimental and theoretical side have to systematically improve the amount of data available and the description of nuclear structure, reactions and the radioactive decay of neutron-rich nuclei in the foreseeable future.
In particular, a future aim should be to develop fully consistent models for all nuclear physics properties (e.g., masses, $\beta$-decay, fission) required as input for r-process calculations.
Similarly, the push towards improving hydrodynamical simulations of the NS-NS or NS-BH merging systems covering the dynamical as well as post-merging phases will significantly improve our understanding of the conditions for the r-process.
In particular, questions regarding the mass, velocity, entropy, and neutron-richness of the various ejecta components need to be further resolved.
With continuous advancements on the nuclear and astrophysical side, systematic studies of the nuclear physics impact will continue to be essential to quantify the uncertainties of r-process yields and their corresponding production of decay heat.

\todo{[referee comment: Reconsider Zmax for fission study? not really after ch. 3]}



\chapter{Conclusions and outlook}
\label{ch:concl}


\epigraph{\itshape \enquote{I'm wondering what to read next.} Matilda said. \enquote{I've finished all the children's books.}}{--- \textup{Roald Dahl}, Matilda}

This thesis has focused on some of the remaining challenges of modelling r-process nucleosynthesis in NS mergers. 
The main goals of this work were to 1) study the impact of weak nucleonic interactions on the dynamical ejecta of NS mergers and 2) explore the sensitivity of the r-process results with respect to theoretical nuclear physics models used as input in the nucleosynthesis calculations. 

To achieve this, r-process calculations based on hydrodynamical simulations of several thousands of trajectories (i.e., unbound mass elements) were performed. 
The hydrodynamical models include simulations of four NS-NS mergers and one NS-BH merger producing the dynamical ejecta and two BH-torus post-merger systems, combined into three approximate models for the total ejecta (\cref{sec:astro_mods}). 
The r-process calculations were performed in a post-processing step, following the time evolution of the thermodynamic properties predicted by the hydrodynamical simulations (\cref{sec:nuc_net}). A reaction network evolved the abundances in time through a set of coupled ODEs. Solving these equations is a time-consuming task requiring excessive amounts of CPU hours, particularly when, as done herein, running calculations for large numbers of trajectories (\cref{sec:HPC}). 
The network applies experimental nuclear physics data as input when available and otherwise theoretical predictions of the relevant nuclear decay and reaction properties (\cref{sec_nucin}). 
With this setup, the final r-process abundance yields and the evolution of the nuclear heating stemming from nuclear decays were calculated for all trajectories of the hydrodynamical models considered. 

Several sensitivity tests of the nucleosynthesis results with respect to the assumptions made for the reaction network and evolution of the heating rate, temperature and ejecta expansion model were performed in \cref{ch:sensitivity}. In summary, the tests did not significantly impact the heating rates, the shape of the final mass fraction distributions or the prediction of the heaviest r-process elements, validating the main assumptions that our results rely on. 

The main conclusions from our study based on the dynamical NS-NS merger models including a proper description of neutrino interactions can be summarized as follows \citep[see also][]{kullmann2021,just2021}:
\begin{itemize}
\item Solar-like final compositions for $A \ga 90$ nuclei are found for all four hydrodynamical models when including weak nucleonic interactions, compared to $A \ga 140$ when ignoring such interactions. 
\item The impact of including nucleonic weak processes is most prominent on the composition of the material ejected in the polar regions, which leads to higher (lower) initial $Y_e $ ($X_n^0$) values.
\item Uncertainties in the initial electron fraction impact the heating rate and the composition, particularly the amount of lanthanides and actinides ejected.
\item The role of fission is reduced when weak nucleonic processes are taken into account. Fission recycling still occurs in parts of the ejecta; however, it does not dominate the shaping of the final abundance distribution. 
Spontaneous fission and the $\alpha$-decay chains of $A>210$ nuclei contribute significantly less to the heating rate compared to the `no neutrino' case since a smaller amount of heavy r-process elements are produced.
\item The high-velocity component of the ejecta (also when including neutrino reactions) is found to include a non-negligible amount of free neutrons, which impacts the heating rate. 
\item Compared to simulations ignoring neutrinos, the early kilonova light curve is generally brighter due to an enhanced heating rate and lower lanthanide mass fractions (which lead to reduced opacity); however, at late times ($t\ga 10$~d), the light curve is dimmer.  
\end{itemize}

Further, the conclusions of the nuclear uncertainty study can be summarized as follows \citep[see also][]{kullmann2022}:
\begin{itemize}
\item The global shape of the r-process abundance distribution is affected by the nuclear physics uncertainties to a minor extent for the astrophysical scenarios studied here. Our models for the total ejecta reproduce the solar system distribution well for $A>90$ nuclei and yield a significant amount of Th and U, irrespective of the adopted nuclear physics model.
\item Varying the nuclear physics inputs can have a significant local impact on the abundance of specific individual species for the total ejecta but also for the dynamical or BH-torus models separately. Deviations up to a factor of $\sim20$ can be found for specific individual abundances of elements with $A>90$, particularly around the second and third r-process peaks. For the actinides, the factor between the largest and smallest yields ranges between 5 to 7. 
\item No significant impact is found on the radioactive heating rate (before thermalization) at early times ($t<10$~d) when the nuclear physics input is varied between the input sets considered. For the cases where fission plays a significant role, i.e., for the NS-BH case, the impact can be large at late times ($t>5$~d). 
\item The mean stellar ages inferred by Th/U cosmochronometry for the six metal-poor r-process-enhanced stars vary between 11 and 19~Gyr, where the uncertainties connected to the nuclear physics inputs are found to be around $\sim2$~Gyr. 
These uncertainties remain considerably smaller than those stemming from the observationally derived abundances ($\sim3-7$~Gyr). 
\item Some (but not all) of the stars are predicted by our method to be older than the age of the Universe, currently estimated to be 13.8~Gyr \citep{Planck2020}. This incompatibility could be due to additional sources of uncertainty not taken into account herein, such as uncertainties stemming from other aspects of the r-process modelling than the nuclear physics input or the underestimation of the uncertainties discussed here. It is also possible that these particular stars were enriched by another r-process site, i.e., not by NS mergers.
\end{itemize}

The nuclear physics uncertainties affecting the abundance distributions are relatively small compared to variations related to the different ejecta components (for the cases considered here). This strengthens the reliability of nucleosynthesis predictions based on hydrodynamical models, which further supports the notion that NS mergers can be significant sources of r-process elements in the Universe.
However, several potential improvements related to astrophysical and nucleosynthesis modelling exist. Let us discuss the latter first. 
As for most codes, a variety of possible improvements exist for the reaction network code \texttt{rpro}. First, the user interface should be improved if the code becomes available for others in the future. Only minor changes are required if a ``black box'' interface is desired; however, a greater amount of development (and documentation) is needed if a user should want more advanced options and insight into the inner workings of the code. 
Other numerical aspects of the code, such as the ODE solver used to evolve the abundances in time, could be considered for replacement. Such changes would require benchmark tests of the new implementation to ensure improved computing times and reliable results. 
A feature that could significantly enhance the user experience\footnote{Perhaps the author's greatest regret is that this was not prioritized early in the PhD project.} would be the possibility to restart \texttt{rpro} calculations if the program stops for whatever reason. This would vastly reduce the time spent restarting programs when calculations exceed the maximum CPU time on the supercomputers, in case of power outlets or other sudden cluster failures and ease the debugging when encountering problems in a calculation. 

The study of nuclear uncertainties was based on r-process calculations for a subset of all trajectories to reduce the computational demand. As a result, the amount of mass sampled from the complete trajectory set varied from 15 to 50 per cent (except for the NS-BH model, which used 1 per cent, see \cref{sec_sel_subs}). The subset of trajectories selected by our method successfully reproduced properties of the entire trajectory set, except for model M3A8m3a5-v2. Therefore, a possible advancement in a future study would be to improve the algorithm used to select the subsets or to be less restrictive in the subset sizes, leading to longer computing times. 

Tremendous advances have been made in experimental as well as theoretical nuclear physics in the last decades. Nevertheless, the push towards obtaining new experimental data and improved nuclear physics models predicting various properties of nuclei will continue to be invaluable for r-process modelling.
The r-process requires an enormous amount of nuclear physics inputs, particularly for neutron-rich nuclei for which no (or very little) experimental information exists. Moreover, even if the newest and most advanced experimental facilities will reach further into the neutron-rich region, many of these nuclei are outside their reach, also in the foreseeable future. Thus, r-process calculations will continue to rely on theoretical models in the years to come. 
Regarding theoretical nuclear models applied to astrophysics, a future aim should be to develop fully consistent models for all nuclear physics properties (e.g., masses, $\beta$-decay, fission) on the basis of models as microscopic as possible, which are currently unavailable. 

Let us turn our attention to the hydrodynamical simulations, which predict the thermodynamical conditions of the r-process.
The hydrodynamical models combined into the total ejecta were not completely consistent in terms of the estimated remnant BH and torus mass, and the simulations of the NS-NS dynamical ejecta ended before the remnant collapsed into a BH. These limitations can be overcome by extended simulation times for a larger number of models, covering an array of model properties. However, new long-term hydrodynamical simulations that cover both the dynamical and BH-torus remnant phases are being developed, thus making the limitations mentioned above irrelevant. 
Therefore, a future prospect would be to base the nucleosynthesis calculations of a nuclear physics uncertainty study on such a long-term simulation. 
In addition, we do not include the possible ejecta component from a longer-lived HMNS. This ejecta component is also expected to contribute significantly to the Galactic r-process enrichment. However, more work is needed to determine the composition of the HMNS ejecta, which will need to be included in future studies.

The kilonova results resented herein with respect to the dynamical ejecta cannot be used to compare directly with observations such as the kilonova observed after the 2017 GW detection of two merging NSs or similar future events. 
The modelling of the kilonova resulting from the total ejecta was outside the scope of this work. 
However, the approach taken here \citep[i.e.][]{just2021} is unlike many other kilonova studies since it is based on hydrodynamical simulations and detailed nucleosynthesis calculations (and not the bulk ejecta properties). Such a modelling approach will be valuable to include in the next generation of kilonova models to advance our ability to explain future light curve observations.

Our results for the total NS-NS or NS-BH merger ejecta reproduce the observational solar r-process distribution relatively well. 
Such nucleosynthesis yields are used as input in Galactic chemical evolution models, which model the enrichment history and evolution of the Galaxy. However, whether NS mergers are the dominant r-process site remains an open question. Thus, Galactic chemical evolution models helps to clarify the role of NS mergers as a contributor to the enrichment of r-process elements in the Galaxy, and for this, r-process yields are an indispensable input. Thus, by further improving our predictions of the r-process yields and their uncertainties, r-process modelling help advance our understanding of the evolution and enrichment of our Galaxy and, therefore, also the origin of the elements observed here on earth.



    \backmatter         

	\bibliographystyle{yahapj}
	\bibliography{bibliography}
	

\chapter*{List of publications}
\addcontentsline{toc}{chapter}{List of publications}

\begin{enumerate}
\item Goriely, S. \& \textbf{Kullmann, I.} 2023, \textit{R-process Nucleosynthesis in Neutron-star Merger Ejecta and Nuclear Dependences}, ed. I. Tanihata, H. Toki, \& T. Kajino, \textit{Handbook of Nuclear Physics}, First edition, chapter 5, section 15 (Springer), accepted for online publication November 2023 \\
\url{https://doi.org/10.1007/978-981-15-8818-1_91-1}

\item Larsen, A. C., Tveten, G. M., Renstrøm, T., Utsunomiya, H., Algin, E., Ari-izumi, T., Ay, K. O., Bello Garrote, F. L., Crespo Campo, L., Furmyr, F., Goriely, S., Görgen, A., Guttormsen, M., Ingeberg, V. W., Kheswa, B. V., \textbf{Kullmann, I. K. B.}, et al. 2023 \textit{New experimental constraint on the $^{185}$W(n, $\gamma$) $^{186}$W cross section}, Physical Review C,  \\
\url{https://arxiv.org/abs/2301.13301}

\item Goriely, S., Choplin, A., Ryssens, W., \textbf{Kullmann, I.} 2022, \textit{Progress in Nuclear Astrophysics: a multi-disciplinary field with still many open questions}, INPC conference proceeding, submitted

\item Ryssens, W., Scamps, G., Grams, G., \textbf{Kullmann, I.}, Bender, M., and Goriely, S., \textit{The mass of odd-odd nuclei in microscopic mass models}, Journal of Physics: conference series, submitted \\
\url{https://arxiv.org/abs/2211.03667}

\item \textbf{Kullmann, I.}, Goriely, S., Just, O., Bauswein, A., \& Janka, H. T. 2022, \textit{Impact of systematic nuclear uncertainties on composition and decay heat of dynamical and disk ejecta in compact binary mergers }, MNRAS submitted \\
\url{https://arxiv.org/abs/2207.07421}

\item Guttormsen, M. Ay, K. O., Ozgur, M., Algin, E., Larsen, A. C., Bello Garrote, F. L., Berg, H. C., Crespo Campo, L., Dahl-Jacobsen, T., Furmyr, F. W., Gjestvang, D., G\"orgen, A., Hagen, T. W., Ingeberg, V. W., Kheswa, B. V., \textbf{Kullmann, I. K. B.}, Klintefjord, M., Markova, M., Midtb\o{}, J. E., Modamio, V., Paulsen, W., Pedersen, L. G., Renstr\o{}m, T., Sahin, E., Siem, S., Tveten, G. M. and Wiedeking, M. 2022, \textit{Evolution of the $\ensuremath{\gamma}$-ray strength function in neodymium isotopes}, Phys. Rev. C, 106, 034314 \\
\url{https://doi.org/10.1103/PhysRevC.106.034314}

\item \textbf{Kullmann, I.}, Goriely, S., Just, O., Ardevol-Pulpillo, R.,  Bauswein, A., Janka, H-T. 2021, \textit{Dynamical ejecta of neutron star mergers with nucleonic weak processes I: Nucleosynthesis}, Monthly Notices of the Royal Astronomical Society, 510, 2804 \\
\url{https://doi.org/10.1093/mnras/stab3393}

\item Just, O., \textbf{Kullmann, I.}, Goriely, S., Bauswein, A., Janka, H-T., Collins, C. E. 2021, \textit{Dynamical ejecta of neutron star mergers with nucleonic weak processes II: Kilonova emission}, Monthly Notices of the Royal Astronomical Society, 510, 2820 \\
\url{https://doi.org/10.1093/mnras/stab3327}

\item \textbf{Kullmann, I. K. B.}, Larsen, A. C., Renstrøm, T., Beckmann, K. S., Bello Garrote, F. L., Crespo Campo, L., Görgen, A., Guttormsen, M., Midtbø, J. E., Sahin, E., Siem, S., Tveten, G. M., Zeiser, F. 2019, \textit{First Experimental Constraint on the $^{191}$Os (n,{$\gamma$}) Reaction Rate Relevant to s-process Nucleosynthesis}, Physical Review C, 99, 065806 \\
\url{https://doi.org/10.1103/PhysRevC.99.065806}
\end{enumerate}


\end{document}